\documentclass[12pt,a4paper]{report}
\usepackage{graphics}
\usepackage{amsmath}
\usepackage{amssymb}
\title{\bf{Higgs Phenomenology in the} \\ 
\bf{Two Higgs Doublet Model of type II} \\
\vspace{1.6in} \sl{A dissertation submitted
to} \\ \sl{the University of San Francisco-Quito} \\ \sl{in partial
fulfillment of} \\ \sl{the requirements for the degree of}  \\
\sl{Doctor of Philosophy}}
\author{By \\Carlos A. Mar\'{\i}n}
\date{\small{20 July 2004}\vspace{0.4in} \\ \normalsize{Supervisor: Dr. 
Bruce Hoeneisen}}
\begin{document}
\maketitle
\pagenumbering{roman}
\tableofcontents
\chapter{Introduction}
\typeout{Introduction}
\pagenumbering{arabic}
The Standard Model of quarks and leptons is based on some basic 
principles: special relativity, locality, quantum mechanics, local 
symmetries and renormalizability \cite{FW}. Therefore the predictions of 
the Standard Model \textquotedblleft{are precise and unambiguous, and 
generally cannot be 
modified \textquoteleft{a little bit}' except in very limited specific 
ways. This feature 
makes the experimental success especially meaningful, since it becomes 
hard to imagine that the theory could be approximately right without in 
some sense being exactly right.}"\cite{FW} The Standard Model predicts
the existence of a massive spin zero boson called Higgs particle. The 
Higgs mechanism is responsible for the masses of the weak interaction  
gauge bosons $W^{\pm}$ and  $Z^0$, and is also sufficient to give masses 
to the leptons and quarks. The 
discovey of the Standard Model Higgs is then  one of the principal goals 
of experimental and theoretical particle physicists. We could say that
the Higgs mechanism is a cornerstone of the Standard Model.

Among the extensions of the Standard Model that respect its principles and 
symmetries, that are compatible with present data within a region of 
parameter space, and are of interest at the large particle colliders, is 
the addition of a second doublet of Higgs fields. Higgs doublets can be 
added to the Standard Model without upsetting the $Z/W$ mass ratio; higher 
dimensional representations upset this ratio \cite{Br-Hoen}.

In the Two Higgs Doublet Model there are two choices for the Higgs-quark 
interactions. In Model I, the quarks and leptons do not couple to the 
first Higgs doublet ($\Phi_1$), but couple to the second Higgs doublet 
($\Phi_2$). In Model II, $\Phi_1$ couples only to down-type quarks and 
leptons and $\Phi_2$ couples only to up-type quarks and neutrinos. If we 
consider the neutrinos as massless particles, there are no couplings 
between neutrinos and neutral Higgs bosons. The Model II choice for the 
Higgs-fermion couplings is the required structure for the Minimal 
Supersymmetric Model.

After the electroweak symmetry-breaking mechanism, three of the eight 
degrees of freedom are absorbed by the $W^{\pm}$ and $Z^0$ gauge bosons, 
leading to the existence of five elementary Higgs particles. The physical 
spectrum of the Two Higgs Doublet Model (Model II) contains five Higgs 
bosons: one pseudoscalar $A^{0}$ (CP-odd scalar), two neutral scalars 
$H^{0}$ and $h^{0}$ (CP-even scalars), and two charged scalars $H^{+}$
and $H^{-}$. In the most general model, the masses of the 
Higgs bosons, the mixing angle $\alpha$
$(- \pi/2 < \alpha \le 0)$ between the two neutral scalar Higgs fields, 
and the ratio of the vacuum expectation values of the two neutral 
components of the Higgs doublets, $\tan\beta$ ($ 0 \le \beta < 
\frac{\pi}{2}$), are all independent parameters of the theory 
\cite{VBRP}. However, in the Minimal Supersymmetric Model the conditions 
on the potential imposed by supersymmetry reduces the number of parameteres 
to two, which may be chosen to be $\tan\beta$ and $m_{H}$ \cite{VBRP}.

\begin{eqnarray*}
\Phi_1 = \left( \begin{array}{c}
\Phi_{1}^{o*} \\-\Phi_{1}^{-}\end{array} \right),
\qquad
\Phi_2 =   
\left(\begin{array}{c} \Phi_{2}^{+} \\ 
\Phi_{2}^{o}\end{array}\right),
\qquad
\tan\beta \equiv
\frac{\langle \Phi_{2}^{o} \rangle}{\langle \Phi_{1}^{o*} \rangle}.
\end{eqnarray*}

In this thesis we study  the Two Higgs Doublet Model 
of type II, set limits to the parameter $\tan\beta$ as a function of 
the mass of the charged Higgs $m_{H}$, and find interesting discovery 
channels in hadron colliders or muon colliders.

All the analysis in this thesis are based on the 
\textquotedblleft{tree-level Higgs
potential}" \cite{VBRP}.

The plan in my thesis is the following:

In the second Chapter\cite{CM-BH1}, using the experimental data on meson 
decay rates, 
mixing and CP violation in the $K^0$ and $B^0$ systems, we set limits to 
the parameter $\tan\beta$ as a 
function of the mass of the charged Higgs $m_{H}$. Recent measurements of 
$\sin(2\beta_{CKM})$ by the B-factories Belle \cite{BfB} and BaBar 
\cite{BBar} permit us to set more stringent limits on $\tan\beta$. 
$\beta_{CKM}$ is an angle of the \textquotedblleft{unitarity triangle}". 
\cite{untrian}

In the third Chapter\cite{CM-BH2}, we present graphically the 
corresponding limits on 
$m_{H^0}$, $m_{h^0}$ and $m_{A^0}$ as a function of the mass of the 
charged Higgs, without considering the influence of 
radiative corrections. Then we calculate production cross sections, decay 
rates and branching fractions of the Higgs particles. Next, we obtain the 
running coupling constants and discuss Grand Unification. Finally, in the 
Conclusions, we list interesting discovery channels.

In Chapter four \cite{CM-3}, we analyze the possibility of the 
construction of a $\mu^{-} \mu^{+}$ collider to detect charged 
or neutral Higgs bosons. The reason for this is that in a muon collider, 
the signal could be cleaner than in a hadron collider. Some of the
production cross sections that we study are: 
$\mu^- \mu^+ \rightarrow h^0 Z^0, H^0
Z^0, H^- H^+, A^0Z^0$ and $H^{\mp} W^{\pm}$. Then, we compare  the 
channel 
$\mu^- \mu^+ \rightarrow H^{\mp} W^{\pm}$
(at $\sqrt{s} = 500\textrm{GeV/c}$ and for large values of $\tan\beta$)
with the production processes  $p\bar{p}
\rightarrow H^{\mp} W^{\pm} X$ (at the Tevatron) and $p p
\rightarrow H^{\mp} W^{\pm} X$ (at the LHC), taking into account the
$t \bar{t}$ background, to check the feasibility of 
detecting $H^{\pm}$ using a muon collider. The influence of radiative 
corrections in the masses of the Higgs bosons is considered in all the 
calculations. Finally, in the Conclusions, we also check the process 
$\mu^- \mu^+ \rightarrow A^0 h^0$. 

\noindent$\bf{NOTE:}$ The results presented in this thesis (width 
decays,
production cross sections, etc.) are calculated in detail in reference
\cite{MarinNotas}.

\chapter{Limits on the Two Higgs Doublet Model
 from meson decay, mixing and CP violation}
\typeout{Limits on the Two Higgs Doublet Model
 from meson decay, mixing and CP violation}
\thispagestyle{empty}
\begin{abstract}
\noindent
We calculate the rate of $\pi^{+}$, $K^{+}$, $D^{+}$ and $B^{+}
\rightarrow
\mu^{+} \nu_{\mu}$ decays, the branching ratio corresponding to $H^{+}
\rightarrow \tau^{+} \nu_{\tau}$, and the box diagrams of $B^{o}
\leftrightarrow
\bar{B}^{o}$, $K^{o} \leftrightarrow \bar{K}^{o}$ and
$D^{o} \leftrightarrow \bar{D}^{o}$ mixing in the Two Higgs Doublet
Model (Model II). Using the experimental data on meson decay rates,
mixing, and CP violation in the $K^o$ and $B^o$ systems
we set competitive upper and lower limits to the parameter
$\tan \beta$ as a function of the mass of the charged Higgs $m_H$.
\end{abstract}
\pagenumbering{arabic}
\section{Introduction}
The Standard Model of quarks and leptons is here to stay.
This theory is based on principles: special relativity, locality,
quantum mechanics, local symmetries and renormalizability\cite{Wilczek}.
Therefore the predictions of the Standard Model
\textquotedblleft{are precise and unambiguous, and generally cannot
be modified \textquoteleft{a little bit}' except in very limited specific 
ways. This
feature makes the experimental success especially
meaningful, since it becomes hard to imagine that the
theory could be approximately right without
in some sense being exactly right.}"\cite{Wilczek}
Among the extensions of the Standard Model that respect its
principles and symmetries, that are compatible with present data
within a region of parameter space, and are of interest at the
large particle colliders, is the addition of a second doublet of
Higgs fields. Higgs doublets can be added to the Standard
Model without upsetting the $Z/W$ mass ratio; higher dimensional
representations upset this ratio.
A second Higgs doublet could make the three running coupling
constants of the Standard Model meet at the Grand Unified Theory
(GUT) scale.
A second Higgs doublet is necessary in
Supersymmetric extensions of the
Standard Model\cite{Peccei}. In this article we explore the limits that present
data place on the parameters of the Two Higgs Doublet Model
(Model II).\cite{VB} In particular we consider meson decay, mixing
and CP violation.

All of our analysis is based on the
\textquotedblleft{tree-level Higgs potential}"\cite{VB}. The physical 
spectrum
of the Two Higgs Doublet Model (Model II)
contains five Higgs bosons: one pseudoscalar $A^{o}$ (CP-odd
scalar), two neutral scalars $H^{o}$ and $h^{o}$ (CP-even scalars), and
two charged scalars $H^{+}$ and $H^{-}$. In the most general model, the 
masses of the Higgs
bosons, the mixing angle $\alpha$ between the two neutral scalar Higgs
fields, and the ratio of the vacuum expectation values of the two
neutral components of the Higgs doublets, $\tan \beta > 0$, are all 
independent parameters of the theory. However, in the Minimal 
Supersymmetric Model the conditions on the potential imposed by 
supersymmetry reduces the number of parameters to two, which may be chosen 
to be $\tan\beta$ and $m_{H}$ \cite{VB}.

\begin{eqnarray*} \Phi_1 = \left( 
\begin{array}{c} \Phi_{1}^{o*} \\-\Phi_{1}^{-}\end{array} \right),
\qquad
\Phi_2 =
\left(\begin{array}{c} \Phi_{2}^{+} \\
\Phi_{2}^{o}\end{array}\right),
\qquad
\tan\beta \equiv
\frac{\langle \Phi_{2}^{o} \rangle}{\langle \Phi_{1}^{o*} \rangle}.
\end{eqnarray*}
Using the experimental data on meson decay rates, mixing and CP violation
we set limits to the parameter $\tan\beta$ as a function of the mass of the
charged Higgs $m_H$. Recent measurements of $\sin(2\beta_{CKM})$ by the 
B-factories
Belle\cite{Belle} and BaBar\cite{BaBar}
permit us to set more stringent limits on $\tan\beta$.
$\beta_{CKM}$ is an angle of the
\textquotedblleft{unitarity triangle}".\cite{5}

\section{Feynman rules of the charged Higgs
in the Two Higgs Doublet Model.}

The effective Lagrangian corresponding to the $H^{\pm} f \bar{f}'$ vertex 
is:
\begin{eqnarray}
\mathcal{L}_{H^{\pm}ff'} & = &
\frac{g}{2 \sqrt{2} m_{W}} \{ H^{+} V_{f f'} \bar{u}_{f} \left(
A + B \gamma^{5} \right) v_{\bar{f}'}
\nonumber \\ & &
+ H^- V_{f f'}^* \bar{u}_{f'} (A - B \gamma^5 ) v_{\bar{f}} \}
\label{LA}
\end{eqnarray}
where
\begin{equation}
A \equiv \left( m_{f'} \tan \beta + m_{f} cot \beta \right)
\label{AA}
\end{equation}
\noindent and
\begin{equation}
B \equiv \left( m_{f'} \tan \beta - m_{f} cot \beta \right),
\label{BB}
\end{equation}  
\noindent
$f =$ fermion (quark or lepton) and $ \bar{f'} =$ antifermion (antiquark 
or
antilepton). $V_{f f'}$ is  an element of the CKM matrix.

The charged-Higgs propagator is: $i / \left( \mathrm{K}^{2} - m_{H}^{2} + 
i
\varepsilon \right)$.

\section{Theory}
Consider the $\left( B^{o} , \bar{B}^{o} \right)$ system. $B^{o}
\leftrightarrow \bar{B}^{o}$ mixing occurs because of the box diagrams
illustrated in Figure \ref{Feynman_diagram1}.
The difference in mass of the two eigenstates
that diagonalize the hamiltonian can be written in the form
\begin{eqnarray}
\lefteqn{
\Delta m_{B} = \frac{\beta_{B} G_{F}^{2} m_{W}^{2}
f_{B}^{2} m_{B}}
{6 \pi^{2}}
}
\nonumber \\ & &
\times\left\vert \sum_{i,j} \xi_{i} \xi_{j} \left[ S^{WW} - 2
\cot^{2}\beta \cdot S^{HW} + \frac{1}{4} \cot^{4} \beta \cdot S^{HH} \right]
\right\vert.
\label{mixing}
\end{eqnarray}
The functions
\begin{eqnarray*}
S^{WW} \left( x_{W}^{i} , x_{W}^{j} \right),
\textrm{ }
S^{HW}\left(x_{W}^{i}, x_{W}^{j}, x_{H}^{i}, x_{H}^{j},
x_{H}^{W} \right)
\textrm{ and }
S^{HH} \left(x_{H}^{i}, x_{H}^{j}, x_{H}^{W} \right)
\end{eqnarray*}
are obtained from
the box diagrams
and are written in Appendix A. The Feynman
rules for $H^{\pm}$ are listed in Section 2.2. We have derived\cite{2}
$S^{WW}$ in agreement with the literature\cite{3}. The derivation of
$S^{HW}$ and $S^{HH}$ is given in Appendices B and C \cite{4}. The 
variables of these functions are
\begin{eqnarray*}
x_{W}^{i} \equiv
\frac{m_{i}^{2}}{m_{W}^{2}},
\qquad
x_{H}^{i} \equiv
\frac{m_{i}^{2}}{m_{H}^{2}},
\qquad
\textrm{and}
\qquad
x_{H}^{W} \equiv
\frac{m_{W}^{2}}{m_{H}^{2}}
\end{eqnarray*}
where $i = u, c, t$.
$\xi_{i} \equiv V_{ib} V_{id}^{*}$.
The notation for the remaining symbols in (\ref{mixing}) is standard\cite{5}.
To obtain the Standard Model\cite{3}, omit $S^{HW}$ and
$S^{HH}$. $\beta_{B}$ is a factor of order 1. Estimates of  $\beta_{B}$
using \textquotedblleft{vacuum intermediate state insertion}"\cite{3},
\textquotedblleft{PCAC and vacuum
saturation}"\cite{3}, \textquotedblleft{bag model}"\cite{3},
\textquotedblleft{QCD corrections}"\cite{6,7},
and the \textquotedblleft{free particles in a box}"\cite{2}
models span the range $\approx
0.4$ to $\approx 1$. $f_{B}$ is the decay constant that appears in the
decay rate for $B^{+} \rightarrow \mu^{+} \nu_{\mu}$\cite{5} which at
tree level in the Two Higgs Doublet Model (Model II) is:
\begin{figure}
\begin{center}
\scalebox{1}
{\includegraphics{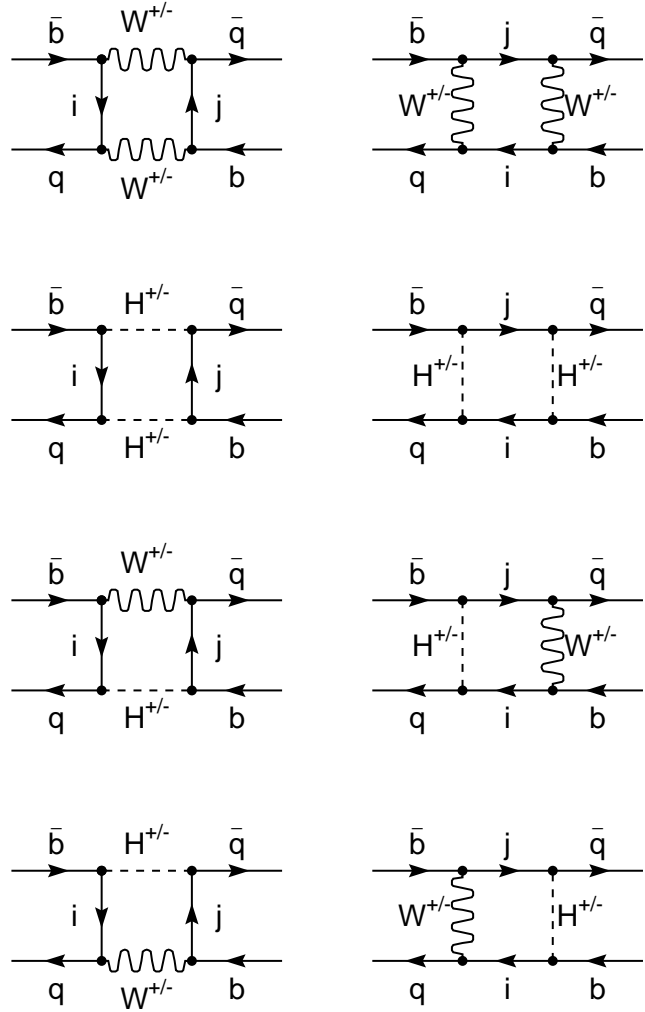}}
\caption{Feynman diagrams corresponding to $B^{o} \leftrightarrow
\bar{B}^{o}$ mixing in the Two Higgs Doublet Model. $q = d$ or $s$ and
$i, j = u, c, t$. The diagrams on the right side interfer with a
\textquotedblleft{-}" sign.}
\label{Feynman_diagram1}
\end{center}
\end{figure}
\begin{equation}
\Gamma_{B^{+}} = \frac {\mathopen{\vert} V_{ub}
\mathclose{\vert}^{2}}{8 \pi} G_{F}^{2} m_{\mu}^{2} m_{B^{+}}
\left(1-\frac{m_{\mu}^{2}}{m_{B^{+}}^{2}}\right)^{2} \left[ f_{B} - g_{B}
\frac{m_{B^{+}}^{2}}{m_{H}^{2}} \tan^{2}\beta \right]^{2}
\label{decay}
\end{equation}
In the derivation of (\ref{decay}) we have substituted
\begin{eqnarray*}
\bar{v}\left(\bar{\textrm{b}}\right) \gamma^{\mu} \left(1 - \gamma^{5}
\right) u \left( \textrm{u} \right) \rightarrow p^{\mu} f_{B}, \\
\bar{v} \left(\bar{\textrm{b}} \right) \left(1 - \gamma^{5}
\right) u \left( \textrm{u} \right) \rightarrow - \frac{m_{B^{+}}^{2}}{m_{b}} g_{B}
\end{eqnarray*}
which defines the decay constants $f_{B}$ and
$g_{B}$. $\bar{v} \left( \bar{\textrm{b}} \right)$ and $u\left(\textrm{u}\right)$ are spinors,
see Section 2. We expect $f_{B} \approx g_{B}$: for a scalar meson with
the quark and antiquark at rest $f_{B} = \frac{m_{B^{+}}}{m_{b}} g_{B}$.
The decays $B^{+}\rightarrow \mu^{+} \nu_{\mu}$ and $D^{+}\rightarrow
\mu^{+} \nu_{\mu}$ are not yet accessible to experiment so that $f_{B}$
and
$f_{D}$ are unknown. $f_{B}$ is estimated using sum rules\cite{8}, or the
$B^{*} - B$ mass difference\cite{9}, or a phenomenological model\cite{10},
or the MIT bag model\cite{11}. These estimates span the range
$ \approx 0.06 \mathrm{GeV}$ to $\approx 0.2 \mathrm{GeV}$ with the 
convention used in
reference \cite{5} and in Equation (\ref{decay}).

In the \textquotedblleft{free particles in a box}"\cite{2} model
$\beta_B = 1$ (after correcting \cite{2} by a color factor $4/3$)
and the volume of the box, \textit{i.e.} the meson, is
$V = 8 / \left( \beta_B m_B f_B^2 \right)$.

For the $\left( B_{s}^{o},
\bar{B}_{s}^{o} \right)$ system: $\xi_{i} \equiv V_{ib}
V_{is}^{*}$ where $ i = u, c, t$; in (\ref{mixing}) replace subscript $B$ by
$B_{s}$. For the $\left( K^{o}, \bar{K}^{o} \right)$
system:
$\xi_{i} \equiv V_{is} V_{id}^{*}$ where $ i = u, c, t$; in (\ref{mixing}) replace
subscript $B$ by $K$. The CP violation parameter $\varepsilon$\cite{3,5}
in the $\left( K^{o}, \bar{K}^{o} \right)$ system in the Two Higgs Doublet
Model is given by:
\begin{equation}
\varepsilon = e^{i \frac{\pi}{4}} \cdot \frac{Im
\left(\sum_{i,j} \xi_{i} \xi_{j} \left[ S^{WW} - 2
\cot^{2}\beta \cdot S^{HW} + \frac{1}{4} \cot^{4} \beta \cdot
S^{HH}\right]\right)}{2 \sqrt{2} \cdot \left\vert \sum_{i,j} \xi_{i}
\xi_{j} \left[ S^{WW} - 2 \cot^{2}\beta \cdot S^{HW} + \frac{1}{4} \cot^{4}
\beta \cdot S^{HH} \right] \right\vert}
\label{epsilon}
\end{equation}
For the $\left( D^{o} , \bar{D}^{o} \right)$ system: $\xi_{i} \equiv V_{ci}
V_{ui}^{*}$ where $i = d, s, b$; in (\ref{mixing}) replace subscript $B$ by $D$
and replace $\cot \beta$ by $\tan \beta$
(leave $\tan \beta$ as is in (\ref{decay})).

The branching ratio for $H^{+} \rightarrow \tau^{+} \nu_{\tau}$
for $m_H < m_t$ is given by
\begin{equation}
B \left( H^{+} \rightarrow \tau^{+} \nu_{\tau} \right)
\approx \\
\frac { m_{\tau}^{2} \tan^{2}\beta}
{\left| V_{cs} \right|^2 a +
\left| V_{cb} \right|^2 b
+ m_\tau^2 \tan^2 \beta }
\label{BR}
\end{equation}
with $a \equiv 3 \left[ m_s^2 \tan^2 \beta + m_c^2 \cot^2 \beta \right]$ and
$b \equiv 3 \left[ m_b^2 \tan^2 \beta + m_c^2 \cot^2 \beta \right]$.
From the measured limit\cite{ALEPH} on $m_H$ as a function
of the branching ratio and (\ref{BR}) we obtain a lower
bound of $m_H$ for each $\tan\beta$.

Let us finally mention that the time-dependent CP-violating
asymmetry $A \equiv (\Gamma - \bar{\Gamma})/(\Gamma + \bar{\Gamma})$,
where $\Gamma$ ($\bar{\Gamma}$) is the rate of the decay
$B^o \rightarrow J/\psi + K_s$ ($\bar{B^o} \rightarrow J/\psi + K_s$),
measured by CDF, Belle and BaBar is given by
$\sin(2\beta_{CKM}) \cdot \sin(\Delta M t)$ in both the
Standard Model and in the Two Higgs Doublet Model (Model II).
This is because the dominating terms of
$\xi_i \xi_j S^{HW}$ and $\xi_i \xi_j S^{HH}$
have $i = j = t$.

\section{Limits}
All experimental data are taken from \cite{5}.
In order to obtain limits
we assume conservatively $0.4 < \beta_x < 1.8$, and $f_x = g_x$
with $x =B$, $B_s$, $D$, $K$, $\pi$.
These assumptions are not critical since the upper (lower)
limits on $\tan\beta$
depend on terms $\propto \tan^4 \beta$ ($\propto \cot^4 \beta$)
in (\ref{mixing}) or (\ref{decay}).
We take the magnitude of the elements of the CKM matrix
from \cite{5} and leave the phase $\angle V_{ub}$ as a free parameter.
The following calculations are made for each $(m_H, \tan\beta)$.
The measured value of the parameter $\varepsilon$ determines the
phase $\angle V_{ub}$
of the CKM matrix, and hence $\beta_{CKM}$. This phase is required to
be within the experimental bounds:
$0.325 < \tan(\beta_{CKM}) < 0.862$ at $95\%$ confidence level \cite{5}.
The measured decay rates $\Gamma_K$ and $\Gamma_\pi$
determine $f_K$ and $f_\pi$ using (\ref{decay}). The experimental
upper bounds on $\Gamma_B$ and $\Gamma_D$ determine
upper bounds on $f_B$ and $f_D$ using (\ref{decay}).
The measured $\Delta m_B$ and $\Delta m_K$ determine
$\beta_B f_B^2$ and $\beta_K f_K^2$ using (\ref{mixing}).
The experimental upper bound on $\Delta m_D$ determines
an upper bound on $\beta_D f_D^2$. The experimental lower
bound on $\Delta m_{Bs}$ determines a lower bound on
$\beta_{Bs} f_{Bs}^2$.
From the preceding information we obtain $\beta_K$ and a lower
bound on $\beta_B$. Then the requirements
$0.4 < \beta_K < 1.8$, $\beta_B < 1.8$ and
$0.325 < \tan(\beta_{CKM}) < 0.862$ place limits on
$\tan\beta$ for each $m_H$ as listed in Table \ref{limits_table}.
The confidence level of these limits is $95\%$.
It turns out that the lower limit on $\tan\beta$ is determined by
the experimental lower limit of $\tan(\beta_{CKM})$, and the upper
limit on $\tan\beta$ is determined by $\beta_B < 1.8$.
\begin{table}
\begin{center}
\begin{tabular}{|l|l|}
\hline
$m_H = 100$GeV &  $1.74$ $ < \tan \beta < 67$ \\
$m_H = 200$GeV &  $1.36$ $ < \tan \beta < 134$ \\
$m_H = 300$GeV &  $1.13$ $ < \tan \beta < 202$ \\
$m_H = 1000$GeV & $0.58$ $ < \tan \beta < 672$ \\
\hline
\end{tabular}
\end{center}
\caption{Limits on $\tan \beta$ for several $m_H$ from
measurements of meson decay, mixing and CP violation.
These limits correspond to $95\%$ confidence.}
\label{limits_table}
\end{table}

\begin{figure}
\begin{center}
\scalebox{0.8}
{\includegraphics{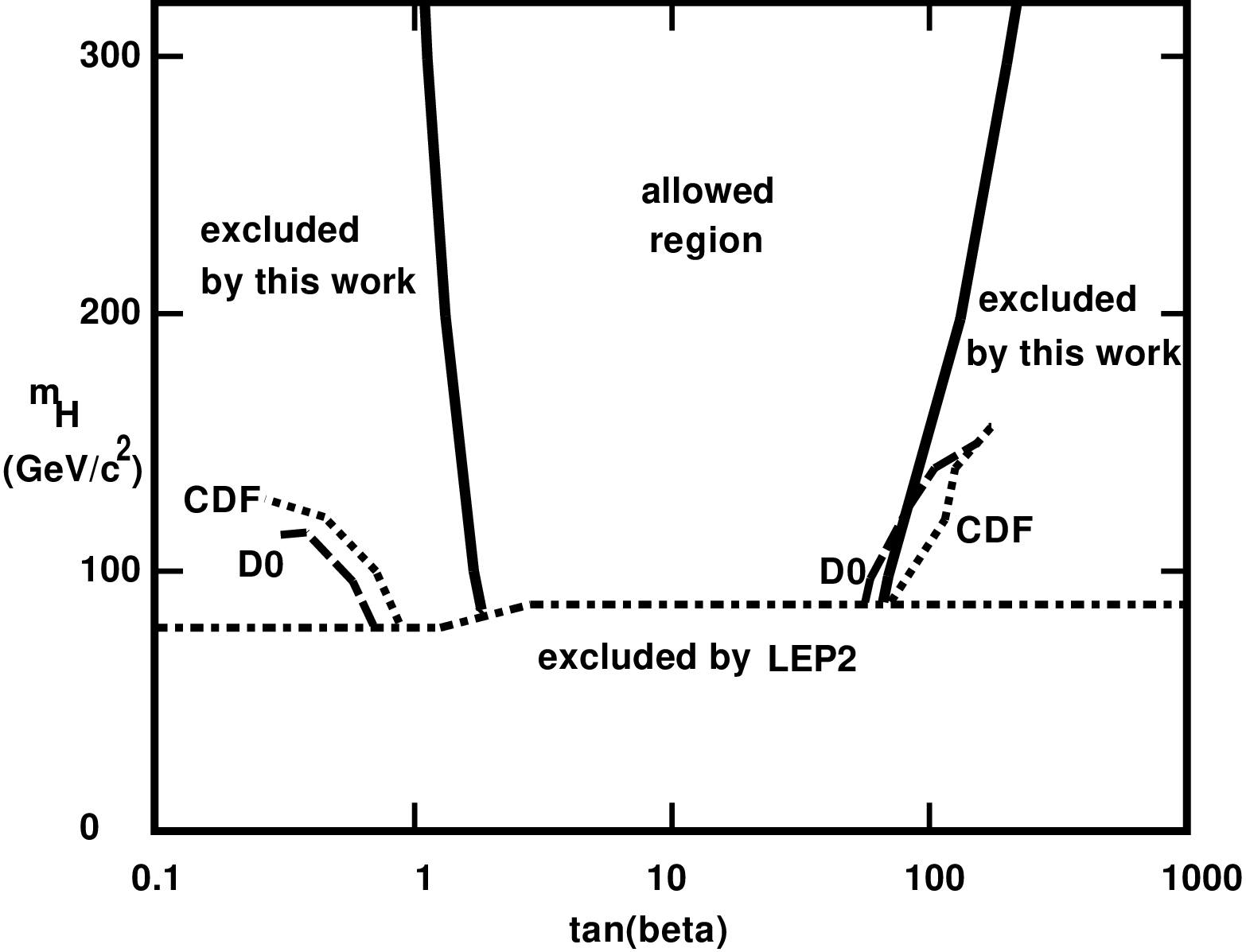}}
\caption{Lower and upper limits on $\tan\beta$ as a
function of the mass of the charged Higgs $m_H$
from meson decay, mixing and CP violation
(continuous curve) compared to
limits obtained by CDF\cite{CDF}, D0\cite{D0} and LEP2\cite{LEP2}, all
at $95\%$ confidence.}
\label{limits_fig}
\end{center}
\end{figure}

\section{Conclusions}
Using measured meson decay rates, mixing and CP violation we
have obtained lower and upper bounds of $\tan\beta$
for each $m_H$. These limits are compared with the
results of direct searches in Figures \ref{limits_fig}.
Note that the measurements of $\sin(2\beta_{CKM})$ by the
Belle and BaBar collaborations have raised the lower bound on $\tan\beta$
by a factor $\approx 5$ with respect to our previous
calculation \cite{revista_colombiana}. It is important to mention that
an indirect limit by the CLEO collaboration \cite{cleo} obtained from the 
measurements of the $b \rightarrow s \gamma$ transition, limits the Two 
Higgs Doublet Model of type II to have a charged Higgs mass in excess
of about $264 \textrm{GeV/c}^{2}$ (it is a slow function of $\tan\beta$).

\chapter{Mass constraints, production cross sections, and decay rates
in the Two Higgs Doublet Model of type II}
\typeout{Mass constraints, production cross sections, and decay rates
in the Two Higgs Doublet Model of type II}
\thispagestyle{empty}
\begin{abstract}
\noindent
We calculate masses, production cross sections, and decay rates
in the Two Higgs Doublet Model of type II.
We also discuss running coupling constants and Grand Unification.
The most interesting production channels are
$gg \rightarrow h^0, H^0, A^0$ on mass shell, and
$q \bar{q}, g g \rightarrow h^0 Z$ and
$q \bar{q'} \rightarrow h^0 W^\pm$ 
in the continuum
(tho there may be peaks at $m_{A^0}$).
The most interesting decays are
$h^0, H^0, A^0 \rightarrow b \bar{b}$-jets and
$\tau^+ \tau^-$, and, if above threshold,
$H^0 \rightarrow 
Z Z$, $W^+ W^-$ and $h^0 h^0$.
The following final states should be compared with the 
Standard Model cross section:
$b \bar{b} Z$, $b \bar{b} W^\pm$, 
$\tau^+ \tau^- Z$, $\tau^+ \tau^- W^\pm$,
$b \bar{b}$, $\tau^+ \tau^-$, $Z Z$,
$W^+ W^-$, 3 and 4 $b$-jets, $2 \tau^+ + 2 \tau^-$,
$b \bar{b} \tau^+ \tau^-$, $Z W^+ W^-$, $3 Z$,
$Z Z W^\pm$ and $3 W^\pm$.
Mass peaks should be searched in the following
channels:
$Z b \bar{b}$, $Z Z$, $Z Z Z$, $b \bar{b}$, $4 b$-jets 
and, just in case, $Z \gamma$.
\end{abstract}

\pagenumbering{arabic}
\section{Introduction}
Among the extensions of the Standard Model that respect its
principles and symmetries, and are compatible with present data
within a region of parameter space, and are of interest at the
large particle colliders, is the addition of a second doublet of
Higgs fields.
In this article we consider the Two Higgs Doublet
Model of type II\cite{1}.
The Higgs sector of the Minimal Supersymmetric Standard Model
(MSSM) is of this type (tho the model of type II does not require 
Supersymmetry). 
The physical spectrum of the model
contains five Higgs bosons: one pseudoscalar $A^{o}$ (CP-odd
scalar), two neutral scalars $H^{o}$ and $h^{o}$ (CP-even scalars), and
two charged scalars $H^{+}$ and $H^{-}$. The masses of the charged Higgs
bosons $m_H$, and the ratio of the vacuum expectation values of the two
neutral components of the Higgs doublets, $\tan \beta > 0$, are free
parameters of the theory.

In \cite{TwoHiggs} we obtained limits in the $(m_H, \tan\beta)$ plane
using measured decay rates, mixing and CP violation of mesons.
In this article we present 
graphically the corresponding limits on $m_{H^0}$, $m_{h^0}$ 
and $m_{A^0}$. Then we calculate production cross sections,
decay rates and branching fractions of the
Higgs particles. Next, we obtain the running coupling constants
and discuss Grand Unification.
Finally, in the Conclusions, we list interesting
discovery channels.

\section{Masses}
The masses of the neutral Higgs particles as a function of the
masses of the charged Higgs $m_H$, $\tan\beta$ 
and the masses of $Z$ and $W$, calculated at tree level, are:
\begin{equation}
m_{A^0}^2 = m_H^2 - m_W^2,
\label{A}
\end{equation}

\begin{eqnarray}
\lefteqn{
2 m_{H^0}^2 =  m_H^2 - m_W^2 + m_Z^2
}
\nonumber \\ & & 
+ \left[ \left( m_H^2 - m_W^2 + m_Z^2 \right)^2
- 4 m_Z^2 \left( m_H^2 - m_W^2 \right)
\left( \frac{\tan^2\beta - 1}{\tan^2\beta + 1} 
\right)^2 \right]^{\frac{1}{2}}
\label{H0}
\end{eqnarray}

\begin{eqnarray}
\lefteqn{
2 m_{h^0}^2  =  m_H^2 - m_W^2 + m_Z^2
}
\nonumber \\ & & 
- \left[ \left( m_H^2 - m_W^2 + m_Z^2 \right)^2
- 4 m_Z^2 \left( m_H^2 - m_W^2 \right)
\left( \frac{\tan^2\beta - 1}{\tan^2\beta + 1} 
\right)^2 \right]^{\frac{1}{2}}
\label{h0}
\end{eqnarray}
We have re derived these equations in agreement with
the literature.\cite{1}

\begin{figure}
\begin{center}
\scalebox{0.8}
{\includegraphics{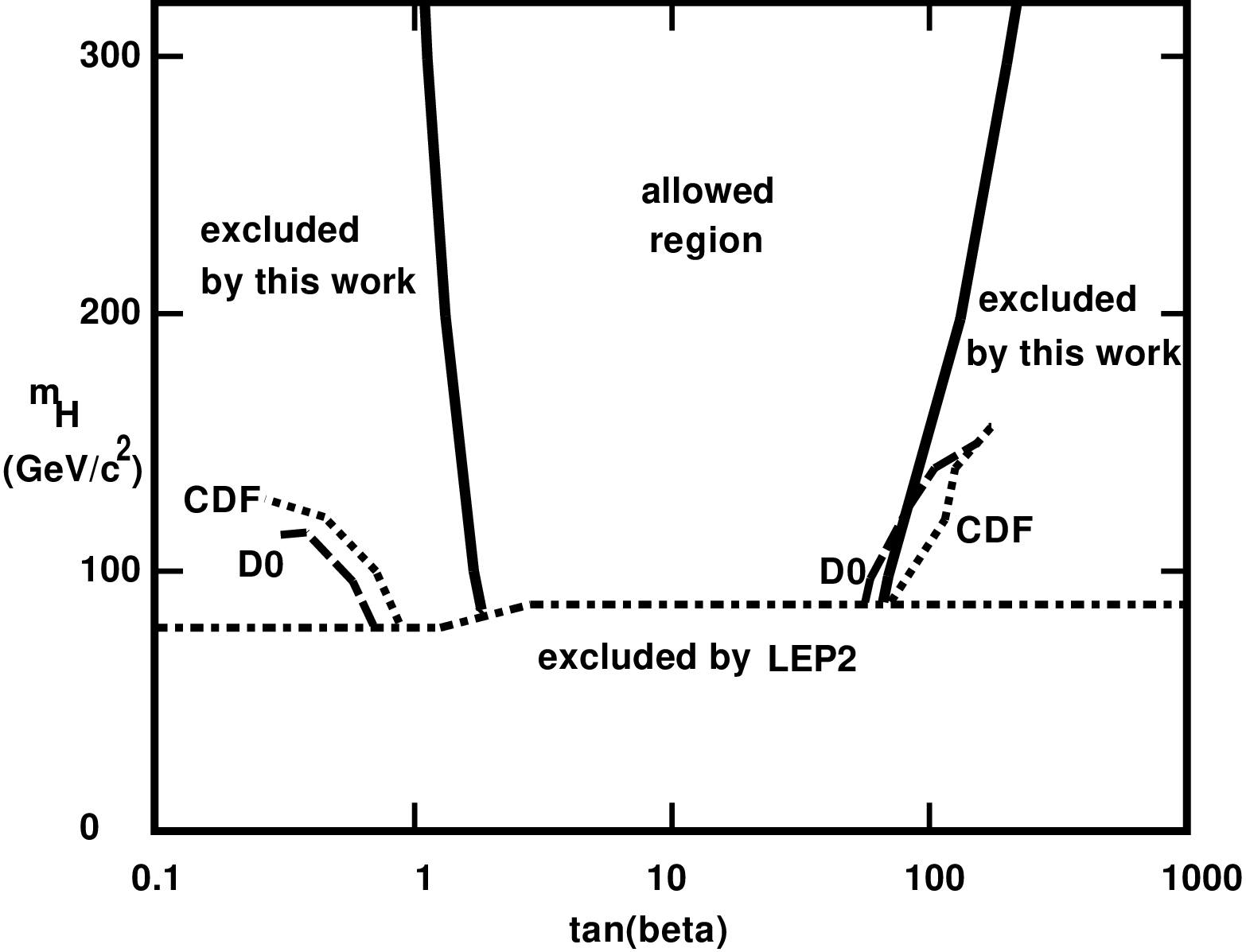}}
\caption{Lower and upper limits on $\tan\beta$ as a
function of the mass of the charged Higgs $m_H$
from meson decay, mixing and CP violation
(continuous curve) compared
to limits obtained by CDF\cite{CDFF}, D0\cite{D0F} and LEP2\cite{LEP2C}, 
all
at $95\%$ confidence. Taken from \cite{TwoHiggs}.}
\label{llimits_fig}
\end{center}
\end{figure}

\begin{figure}
\begin{center}
\scalebox{0.5}
{\includegraphics{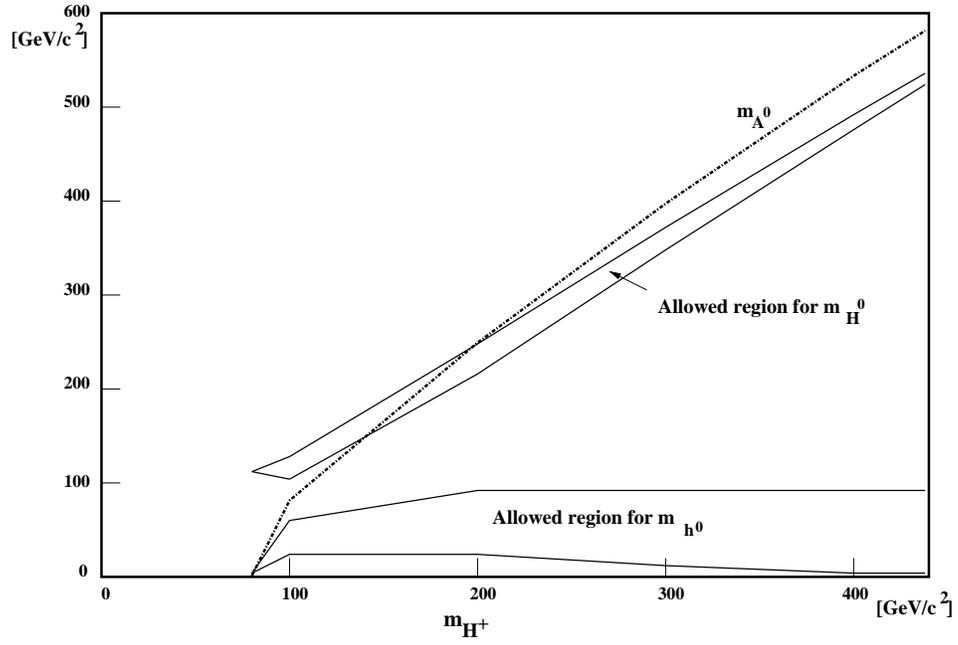}}
\caption{Allowed regions of the masses of the neutral
Higgs $h^0$, $H^0$ and $A^0$ as a function of the mass
$m_H$ of the charged Higgs $H^\pm$.
From Figure \ref{llimits_fig} and the tree level Equations \ref{A},
\ref{H0} and \ref{h0}. Radiative corrections raise the
allowed region of $h^0$.\cite{PDG}}
\label{fig3m}
\end{center}
\end{figure}

In Chapter two \cite{TwoHiggs} we obtained the limits
in the $(\tan\beta, m_H)$ plane
shown in Figure \ref{llimits_fig}. From that figure and Equations
\ref{A}, \ref{H0} and \ref{h0}
we obtain the limits on the masses of the neutral Higgs particles
shown in Figure \ref{fig3m}.

Radiative corrections can be very large. In the MSSM the largest
contributions arise from the incomplete cancellation between top and stop
loops. The corresponding plot similar to 
Figure \ref{fig3m} with radiative corrections
can be found in \cite{PDG}.


\section{Feynman rules}
The Lagrangian for the $V H H$ interaction is:\cite{1}
\begin{eqnarray}
\mathcal{L}_{VHH} & = &
\frac{-ig}{2} W_\mu^+ 
\cdot H^- \overleftrightarrow{\partial}^\mu
\left[ H^0 \sin(\alpha - \beta) + h^0 \cos(\alpha - \beta) + i A^0 \right] 
+ \mathrm{h.c.}
\nonumber \\ & &
- \frac{ig}{2 \cos\theta_W} Z_\mu
\{ i A^0 \overleftrightarrow{\partial}^\mu
\left[ H^0 \sin(\alpha - \beta) + h^0 \cos(\alpha - \beta) \right]
\nonumber \\ & &
- \left( 2 \sin^2\theta_W - 1 \right) 
\cdot H^- \overleftrightarrow{\partial}^\mu H^+
\}
\label{VHH}
\end{eqnarray}
where
\begin{equation}
A \overleftrightarrow{\partial}^\mu B =
A(\partial^\mu B) - (\partial^\mu A) B.
\label{AB}
\end{equation}
The Lagrangian for the $VVH$ interaction is:
\begin{eqnarray}
\mathcal{L}_{VVH} & = &
\left( g m_W W_\mu^+ W^{-\mu} +
\frac{g m_Z}{2 \cos\theta_W} Z_\mu Z^\mu \right)
\nonumber \\ & & 
\times
\left[ H^0 \cos(\beta - \alpha) + h^0 \sin(\beta - \alpha) \right].
\label{VVH}
\end{eqnarray}
There are no vertices $Z H^0 H^0$, 
$Z h^0 h^0$, $Z A^0 A^0$,
$Z W^+ H^-$ or $Z H^0 h^0$. 
The interactions of neutral Higgs bosons with up and down quarks
are given by:
\begin{eqnarray}
\mathcal{L}_{AHhff'} & = &
\frac{-g m_f}{2 m_W \sin\beta}
\left[ \bar{u}_f v_{\bar{f}} ( H^0 \sin\alpha + h^0 \cos\alpha) 
- i \cos\beta \cdot \bar{u}_f \gamma^5 v_{\bar{f}} A^0 \right]
\nonumber \\ & &
- \frac{g m_{f'}}{2 m_W \cos\beta}
\nonumber \\ & &
\times
\left[ \bar{u}_{f'} v_{\bar{f'}} (H^0 \cos\alpha - h^0 \sin\alpha )
- i \sin\beta \cdot \bar{u}_{f'} \gamma^5 v_{\bar{f'}} A^0 \right]
\label{hff}
\end{eqnarray}
where $f = u, c, t, \nu_e, \nu_\mu, \nu_\tau$ and
$f' = d, s, b, e^-, \mu^-, \tau^-$.
The Lagrangian corresponding to the $H^{\pm} f \bar{f}'$ vertex is:
\begin{eqnarray}
\mathcal{L}_{H^{\pm}ff'} & = &
\frac{g}{2 \sqrt{2} m_{W}} \{ H^{+} V_{f f'} \bar{u}_{f} \left(
A + B \gamma^{5} \right) v_{\bar{f}'} 
\nonumber \\ & &
+ H^- V_{f f'}^* \bar{u}_{f'} (A - B \gamma^5 ) v_{\bar{f}} \}
\label{L}
\end{eqnarray}
where
$A \equiv \left( m_{f'} \tan \beta + m_{f} \cot \beta \right)$
and $B \equiv \left( m_{f'} \tan \beta - m_{f} \cot \beta \right)$.
$V_{f f'}$ is  an element of the CKM matrix.
The Lagrangian corresponding to three Higgs bosons is:
\begin{eqnarray}
\mathcal{L}_{3h} & = &
-g H^0 \{ H^+ H^- \left[ m_W \cos(\beta - \alpha)
- \frac{m_Z}{2 \cos\theta_W} \cos(2 \beta) \cos(\beta + \alpha) \right]
\nonumber \\ & &
+ \frac{m_Z}{4 \cos\theta_W} H^0 H^0 \cos(2 \alpha) \cos(\beta + \alpha)
\nonumber \\ & &
+ \frac{m_Z}{4 \cos\theta_W} h^0 h^0 
\left[ 2 \sin(2 \alpha) \sin(\beta + \alpha) - \cos(\beta 
+ \alpha) \cos(2\alpha) \right]
\nonumber \\ & &
- \frac{m_Z}{4 \cos\theta_W} A^0 A^0 \cos(2 \beta) \cos(\beta + \alpha)
\}
\nonumber \\ & &
-g h^0 \{
H^+ H^- \left[ m_W \sin(\beta - \alpha) 
+ \frac{m_Z}{2 \cos\theta_W} \cos(2 \beta) \sin(\beta + \alpha) \right]
\nonumber \\ & &
+ \frac{m_Z}{4 \cos\theta_W} h^0 h^0 \cos(2 \alpha) \sin(\beta + \alpha)
\nonumber \\ & &
- \frac{m_Z}{4 \cos\theta_W} H^0 H^0 \left[ 2 \sin(2\alpha) \cos(\beta+\alpha)
+ \sin(\beta+\alpha) \cos(2 \alpha) \right]
\nonumber \\ & &
+ \frac{m_Z}{4 \cos\theta_W} A^0 A^0 \cos(2 \beta) \sin(\beta + \alpha)
\}.
\label{3h}
\end{eqnarray}
Vertexes with four partons including two Higgs bosons are
\begin{eqnarray}
\mathcal{L}_{4} & = & e^2 A_\mu A^\mu H^+ H^- +
\frac{e g \cos \left( 2 \theta_W \right)}{\cos \theta_W} 
A_\mu Z^\mu H^+ H^- \nonumber \\
& & - \frac{e g}{2} \sin \left( \beta - \alpha \right) A_\mu W^{\pm \mu} 
H^0 H^\mp + \frac{e g}{2} \cos \left( \beta - \alpha \right)
A_\mu W^{\pm \mu} h^0 H^\mp \nonumber \\
& & \pm \frac{i g e}{2} A_\mu W^{\pm \mu} A^0 H^\mp,
\label{L4}
\end{eqnarray}
The $H^+H^- \gamma$ vertex is
\begin{equation}
\mathcal{L}_{H^+H^- \gamma} = - i g \sin \theta_W A_\mu H^- 
\overleftrightarrow{\partial^\mu} H^+
\label{HHgamma}
\end{equation}
The Higgs propagators are: $i / \left( k^{2} - m^{2} + i
\varepsilon \right)$.

\begin{figure}
\begin{center}
{\includegraphics{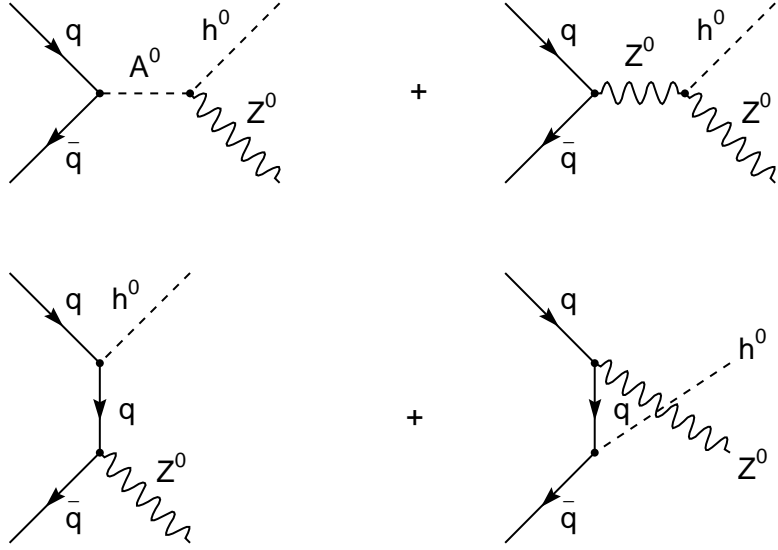}}
\caption{Feynman diagrams corresponding to the
production of $h^0$ in the
\newline channel
$q \bar{q} \rightarrow h^0 Z^0$.}
\label{nhiggs2}
\end{center}
\end{figure}

Feynman diagrams corresponding to the production of $Z h^0$ are shown
in Figures \ref{nhiggs2}, \ref{nhiggs3} and \ref{nhiggs4}.
Note that the invariant mass of $Z h^0$ can have a resonance at
$m_{A^0}$ which is an interesting experimental signature.
Feynman diagrams corresponding to the production of $W^\pm h^0$ or
$W^\pm H^0$ are shown in Figure \ref{wh_fig}.

\begin{figure}
\begin{center}
{\includegraphics{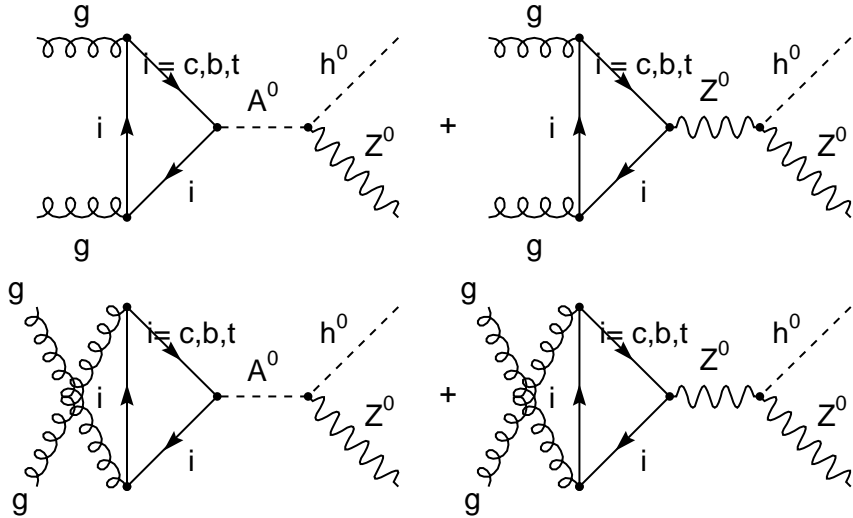}}
\caption{Feynman diagrams corresponding to the 
production of $h^0$ in the 
\newline channel
$g g \rightarrow h^0 Z^0$. Continued in
Figure \ref{nhiggs4}.}
\label{nhiggs3}
\end{center}
\end{figure}
\begin{figure}
\begin{center}
{\includegraphics{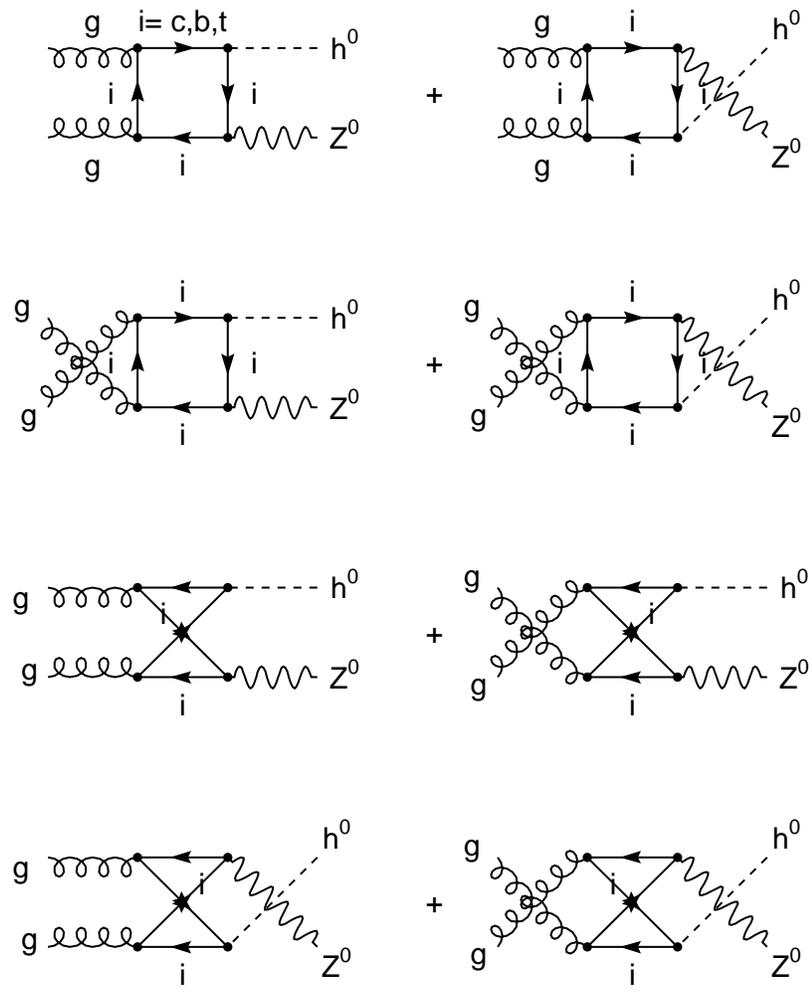}}
\caption{Continued from Figure \ref{nhiggs3}.}
\label{nhiggs4}
\end{center}
\end{figure}
\begin{figure}
\begin{center}
{\includegraphics{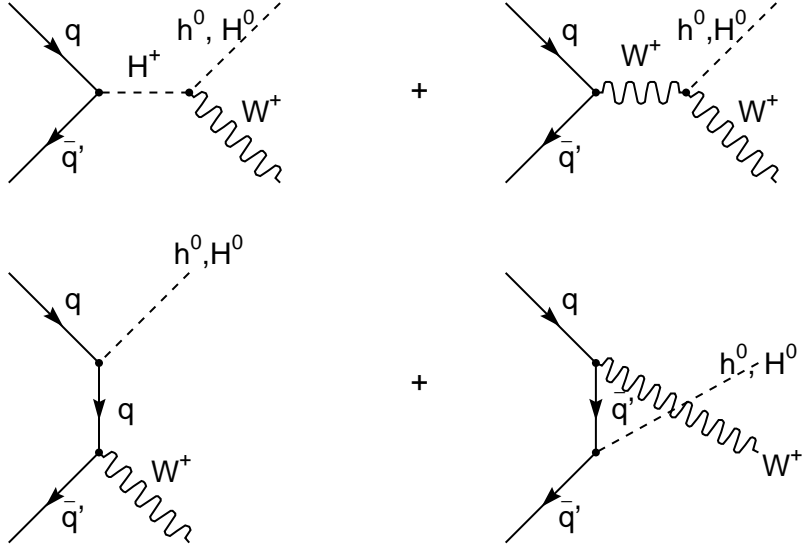}}
\caption{Feynman diagrams corresponding to the 
production of $h^0 W^\pm$ and $H^0 W^\pm$.}
\label{wh_fig}
\end{center}
\end{figure}

\section{Decay rates of $h^0$}
\begin{figure}
\begin{center}
{\includegraphics{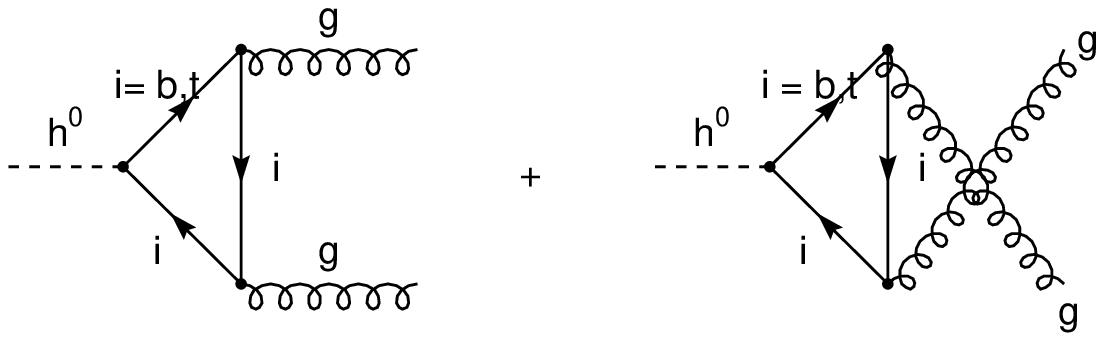}}
\caption{Feynman diagrams corresponding to the decay
$h^0 \rightarrow g g$.}
\label{nhiggs}
\end{center}
\end{figure}

Calculating the Feynman diagrams of Figure \ref{nhiggs} we obtain the
decay rate corresponding to $h^0 \rightarrow g g$:
\begin{eqnarray}
\Gamma(h^0 \rightarrow g g) & = &
\frac{\sqrt{2} G_F \alpha_s^2 m_{h^0}^3}{64 \pi^3}
\frac{\cos^2{\alpha}}{\sin^2\beta}
| \tan\beta \tan\alpha \tau_b \left[ \left( \tau_b - 1 \right)
f(\tau_b) + 2 \right]
\nonumber \\ & &
- \tau_t \left[ \left( \tau_t - 1 \right) 
f(\tau_t) + 2 \right] | ^2
\label{2g}
\end{eqnarray}
where
\begin{equation}
\tau_i = \frac{4 m_i^2}{m_{h^0}^2},
\label{tau}
\end{equation}
(note that $\tau_i$ will change from Section to Section),
\begin{equation}
f(\tau_i) = \left\{ \begin{array}{ll}
-2 \left[ \arcsin \left( \tau_i^{-1/2} \right) \right]^2 &
\mbox{if $\tau_i > 1$} \\
\frac{1}{2} \left[ \ln \left( \frac{1 + \left( 1- \tau_i \right)^{1/2}}
{1 - \left( 1 - \tau_i \right)^{1/2}} \right) - i \pi \right]^2 &
\mbox{if $\tau_i \le 1$,}
\end{array} \right. 
\label{fftau}
\end{equation}

\begin{equation}
\tan\alpha = - \left\{ \frac{1 + F}{1 - F} \right\}^\frac{1}{2},
\label{tan_alpha}
\end{equation}
\begin{equation}
F = \frac{1 - \tan^2\beta}{\left( 1 + \tan^2\beta \right) G}
\left[ 1 - \frac{m_Z^2}{m_H^2} - \frac{m_W^2}{m_H^2} \right],
\label{F}
\end{equation}
\begin{equation}
G = \left[ \left( 1 + \frac{m_Z^2}{m_H^2} - \frac{m_W^2}{m_H^2} \right)^2
- 4 \frac{m_Z^2}{m_H^2} 
\left( 1 - \frac{m_W^2}{m_H^2} \right)
\left( \frac{\tan^2\beta - 1}{\tan^2\beta + 1} \right)^2
\right]^\frac{1}{2}.
\label{G}
\end{equation}

\begin{figure}
\begin{center}
{\includegraphics{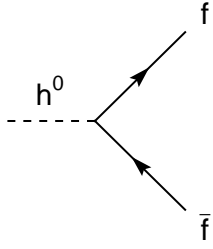}}
\caption{Feynman diagram corresponding to the decays
$h^0 \rightarrow b \bar{b}, c \bar{c}, \tau \bar{\tau}$.}
\label{nhiggs5}
\end{center}
\end{figure}

Calculating the Feynman diagrams of Figure \ref{nhiggs5} we obtain:
\begin{equation}
\Gamma\left( h^0 \rightarrow c \bar{c} \right) =
\frac{3 G_F m_c^2 m_{h^0} \cos^2\alpha}{\sqrt{2} \cdot 
4 \pi \sin^2\beta}
\left( 1 - \frac{4 m_c^2}{m_{h^0}^2} \right)^\frac{3}{2},
\label{ccbar}
\end{equation}
\begin{equation}
\Gamma\left( h^0 \rightarrow b \bar{b} \right) =
\frac{3 G_F m_b^2 m_{h^0} \sin^2\alpha}{\sqrt{2} \cdot 
4 \pi \cos^2\beta}
\left( 1 - \frac{4 m_b^2}{m_{h^0}^2} \right)^\frac{3}{2},
\label{bbbar}
\end{equation}
and
\begin{equation}
\Gamma\left( h^0 \rightarrow \tau^- \tau^+ \right) =
\frac{G_F m_\tau^2 m_{h^0} \sin^2\alpha}{\sqrt{2} \cdot 
4 \pi \cos^2\beta}
\left( 1 - \frac{4 m_\tau^2}{m_{h^0}^2} \right)^\frac{3}{2}.
\label{tautaubar}
\end{equation}


\section{Branching fractions of $h^0$}
From the preceding decay rates we obtain the following
branching fractions for the case
$m_b, m_c, m_\tau \ll m_{h^0} < 120$GeV/c$^2$:
\begin{equation}
B \left( h^0 \rightarrow b \bar{b} \right) =
\frac{3 m_b^2 \sin^2\alpha}{3 m_b^2 \sin^2\alpha + 
3 m_c^2 \cos^2\alpha \cot^2\beta 
+ m_\tau^2 \sin^2\alpha + J}
\label{Bbbar}
\end{equation}
and
\begin{equation}
B \left( h^0 \rightarrow \tau^+ \tau^- \right) =
\frac{m_\tau^2 \sin^2\alpha}{3 m_b^2 \sin^2\alpha + 
3 m_c^2 \cos^2\alpha \cot^2\beta 
+ m_\tau^2 \sin^2\alpha + J}
\label{Btau}
\end{equation}
where
\begin{eqnarray}
J & = & \frac{\alpha_s^2 m_{h^0}^2}{8\pi^2} \frac{\cos^2\alpha}{\tan^2\beta}
\vert 
\tan\beta \tan\alpha \left[ \tau_b \left( -
\frac{1}{2} \left\{ \ln \left( \frac{\tau_b}{4} \right)
+ i\pi \right\}^2 + 2 \right) \right] 
\nonumber \\ & &
- \tau_t \left\{ \left( \tau_t - 1 \right) 
f \left( \tau_t \right) + 2 \right\} \vert^2.
\label{J}
\end{eqnarray}
For $90 < m_H < 1000$GeV/c$^2$, $B \left( h^0 \rightarrow b \bar{b} \right)$
varies from 0.856 to 0.944. Neglecting $B\left( h^0 \rightarrow g g \right)$
and the contribution of $c \bar{c}$ we obtain
\begin{equation}
B \left( h^0 \rightarrow b \bar{b} \right) =
\frac{3m_b^2}{3m_b^2 + m_\tau^2} = 0.944
\label{Bbbar_short}
\end{equation}
and
\begin{equation}
B \left( h^0 \rightarrow \tau^+ \tau^- \right) =
\frac{m_\tau^2}{3 m_b^2 + m_\tau^2} = 0.056.
\label{Btau_short}
\end{equation}

\section{Decay rates of $H^{\pm}$}
\begin{figure}
\begin{center}
{\includegraphics{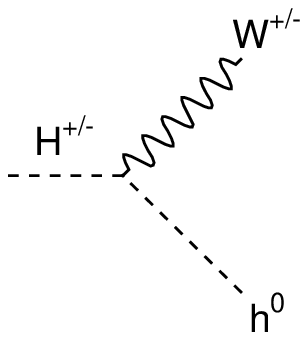}}
\caption{Feynman diagram corresponding to the decay
$H^\pm \rightarrow W^\pm h^0$.}
\label{H_Wh_fig}
\end{center}
\end{figure}

\begin{figure}
\begin{center}
{\includegraphics{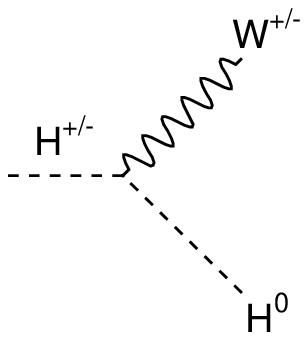}}
\caption{Feynman diagram corresponding to the decay
$H^\pm \rightarrow W^\pm H^0$.}
\label{H__WH_fig}
\end{center}
\end{figure}

\begin{figure}
\begin{center}
{\includegraphics{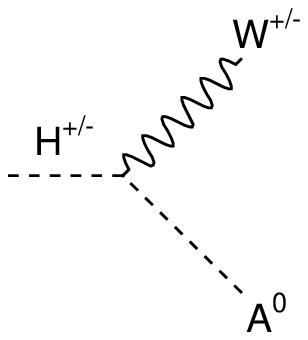}}
\caption{Feynman diagram corresponding to the decay
$H^\pm \rightarrow W^\pm A^0$.}
\label{Hpm_Wpm_A_fig}
\end{center}
\end{figure}

The tree level Feynman diagram of Figure \ref{H_Wh_fig} gives 
the following decay rate:
\begin{eqnarray}
\Gamma \left( H^{\pm} \rightarrow W^{\pm} h^0 \right) & = &
\frac{\sqrt{2} G_F \cos^2{\alpha}}
{16 \pi m_H^3 \left( 1 + \tan^2{\beta} \right)}
\left[ 1 + \tan{\beta} \tan{\alpha} \right]^2 
\nonumber \\ & &
\times\Lambda^{3/2} \left(m_H^2, m_W^2, m_{h^0}^2 \right)
\label{H_Wh}
\end{eqnarray}
where
\begin{equation}
\Lambda(a, b, c) = a^2 + b^2 + c^2 - 2ab - 2bc - 2ca.
\label{Lambda}
\end{equation}
Similarly from the Feynman diagrams of Figures \ref{H__WH_fig} 
and \ref{Hpm_Wpm_A_fig} we obtain
\begin{eqnarray}
\Gamma \left( H^{\pm} \rightarrow W^{\pm} H^0 \right) & = &
\frac{\sqrt{2} G_F \left( \tan\beta - \tan\alpha \right)^2 }
{16 \pi m_H^3 \left( 1 + \tan^2{\beta} \right)
\left( 1 + \tan^2{\alpha} \right)}
\nonumber \\ & &
\times\Lambda^{3/2} \left(m_H^2, m_W^2, m_{H^0}^2 \right),
\label{H__WH}
\end{eqnarray}
\begin{equation}
\Gamma \left( H^\pm \rightarrow W^\pm A^0 \right) =
\frac{\sqrt{2} G_F}{16 \pi m_H^3}
\Lambda^{3/2} \left( m_{A^0}^2, m_W^2, m_H^2 \right).
\label{H_WA}
\end{equation}


\section{Decays of $H^0$.}
\begin{figure}
\begin{center}
{\includegraphics{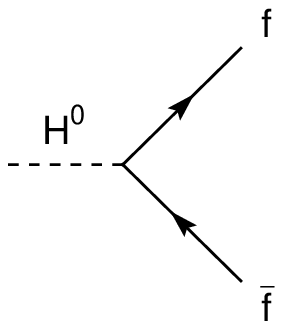}}
\caption{Feynman diagram corresponding to the decay
$H^0 \rightarrow f \bar{f}$.}
\label{H0_ff_fig}
\end{center}
\end{figure}

The tree level Feynman diagrams of Figure \ref{H0_ff_fig}
give the following decay rates:
\begin{equation}
\Gamma(H^0 \rightarrow f \bar{f}) =
\frac{\sqrt{2} G_F m_f^2 m_{H^0}}{8 \pi}
\left( 1 - \frac{4 m_f^2}{m_{H^0}^2} \right)^{3/2} N_f B_f^2
\label{H0_ff}
\end{equation}
where $N_f = 3$ for quarks, $N_f = 1$ for leptons,
$B_f^2 = \sin^2\alpha ( 1 + \cot^2\beta )$ for $f = u, c, t$, and
$B_f^2 = \cos^2\alpha ( 1 + \tan^2\beta )$ for $f = d, s, b, 
e^-, \mu^-, \tau^-$. 

\begin{figure}
\begin{center}
{\includegraphics{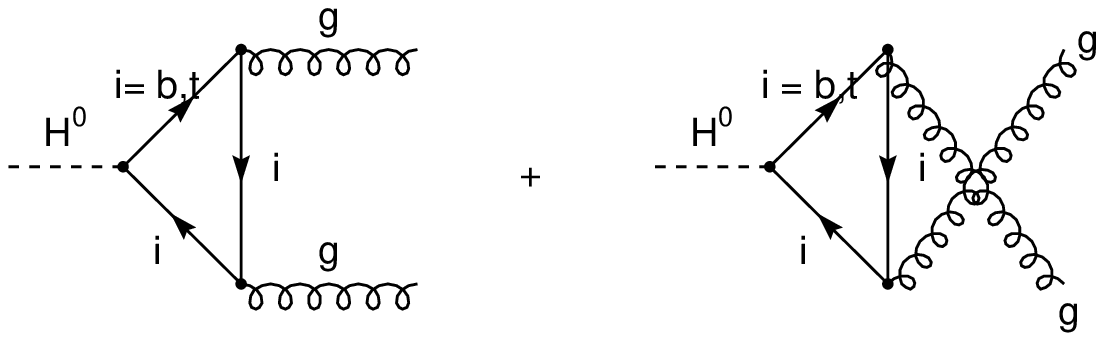}}
\caption{Feynman diagrams corresponding to the decay
$H^0 \rightarrow g g$.}
\label{H0_gg_fig}
\end{center}
\end{figure}

\begin{figure}
\begin{center}
{\includegraphics{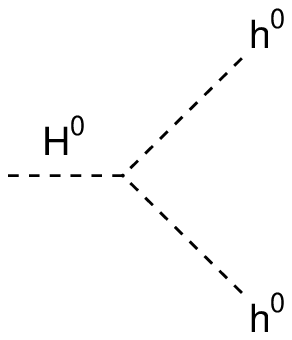}}
\caption{Feynman diagram corresponding to the decay
$H^0 \rightarrow h^0 h^0$.}
\label{H_hh_fig}
\end{center}
\end{figure}

\begin{figure}
\begin{center}
{\includegraphics{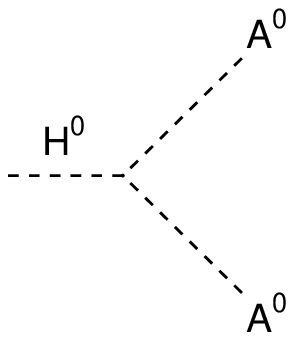}}
\caption{Feynman diagram corresponding to the decay
$H^0 \rightarrow A^0 A^0$.}
\label{H_AA_fig}
\end{center}
\end{figure}

\begin{figure}
\begin{center}
{\includegraphics{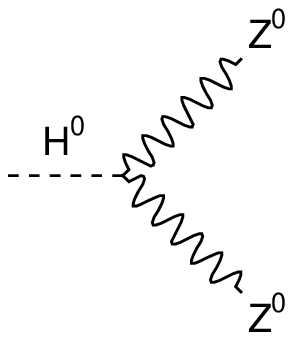}}
\caption{Feynman diagram corresponding to the decay
$H^0 \rightarrow Z Z$.}
\label{H0_ZZ_fig}
\end{center}
\end{figure}

From the Feynman diagrams of Figure \ref{H0_gg_fig} we obtain
\begin{eqnarray}
\Gamma(H^0 \rightarrow g g) & = &
\frac{\sqrt{2} G_F \alpha_s^2 m_{H^0}^3}{64 \pi^3}
\sin^2\alpha ( 1+ \cot^2\beta ) 
\nonumber \\ & &
\times| \frac{\tan\beta}{\tan\alpha} \tau_b
\left[ ( \tau_b - 1 ) f(\tau_b) + 2 \right] 
\nonumber \\ & &
+ { \tau_t \left[ ( \tau_t - 1 ) f(\tau_t) + 2 \right]} |^2 
\label{H_gg}
\end{eqnarray}
where
\begin{equation}
\tau_i = \frac{4 m_i^2}{m_{H^0}^2}.
\label{tau_H0}
\end{equation}
From the Feynman diagram of Figure \ref{H_hh_fig} we obtain
\begin{eqnarray}
\lefteqn{
\Gamma \left( H^0 \rightarrow h^0 h^0 \right)  = 
\frac{\sqrt{2} G_F m_Z^4}{32 \pi m_{H^0}}
}
\nonumber \\ & &
\times\left( 1 - \frac{4 m_{h^0}^2}{m_{H^0}^2} \right)^{1/2}
\frac{\left( 1 - \tan^2\alpha \right)^2}
{\left( 1 + \tan^2\alpha \right)^3 \left( 1 + \tan^2\beta \right)}
\nonumber \\ & &
\times\left[ \frac{4 \tan\alpha}{1 - \tan^2\alpha}
\left( \tan\alpha + \tan\beta \right)
- \left( 1 - \tan\alpha \tan\beta \right) \right]^2.
\label{H_hh}
\end{eqnarray}
From the Feynman diagrams of Figures \ref{H_AA_fig} 
and \ref{H0_ZZ_fig} we obtain
\begin{eqnarray}
\lefteqn{
\Gamma \left( H^0 \rightarrow A^0 A^0 \right)  =
\frac{\sqrt{2} G_F m_Z^4}{32 \pi m_{H^0}}
\left( 1 - \frac{4 m_{A^0}^2}{m_{H^0}^2} \right)^{1/2}
}
\nonumber \\ & &
\times\frac{\left( \tan^2\beta - 1 \right)^2}{ \left( 1 + \tan^2\alpha 
\right)
\left( 1 + \tan^2\beta \right)^3}
\left[ 1 - \tan\alpha \tan\beta \right]^2,
\label{H_AA}
\end{eqnarray}
\begin{eqnarray}
\Gamma \left( H^0 \rightarrow Z Z \right) & = &
\frac{\sqrt{2} G_F \left( 1 + \tan\beta \tan\alpha \right)^2 m_{H^0}^3}
{32 \pi \left( 1 + \tan^2\beta \right) \left( 1 + \tan^2\alpha \right)}
\nonumber \\ & &
\times \left( 12 x^2 - 4 x + 1 \right) \left( 1 - 4 x \right)^{1/2}
\label{H_ZZ}
\end{eqnarray}
where
\begin{equation}
x = \frac{m_Z^2}{m_{H^0}^2}.
\label{x}
\end{equation}

\begin{figure}
\begin{center}
{\includegraphics{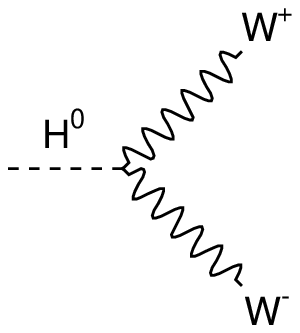}}
\caption{Feynman diagram corresponding to the decay
$H^0 \rightarrow W^+ W^-$.}
\label{H0_WW_fig}
\end{center}
\end{figure}

Similarly, from the diagram of Figure \ref{H0_WW_fig} we obtain
\begin{eqnarray}
\Gamma \left( H^0 \rightarrow W^+ W^- \right) & = &
\frac{\sqrt{2} G_F \left( 1 + \tan\beta \tan\alpha \right)^2 m_{H^0}^3}
{16 \pi \left( 1 + \tan^2\beta \right) \left( 1 + \tan^2\alpha \right)}
\nonumber \\ & &
\times \left( 12 y^2 - 4y + 1 \right) \left( 1 - 4 y \right)^{1/2}
\label{H_WW}
\end{eqnarray}
where
\begin{equation}
y = \frac{m_W^2}{m_{H^0}^2}.
\label{y}
\end{equation}

\begin{figure}
\begin{center}
{\includegraphics{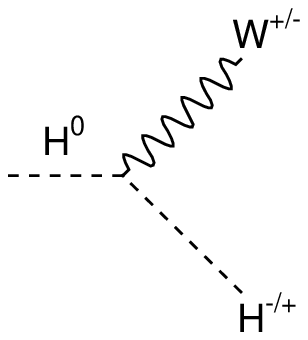}}
\caption{Feynman diagram corresponding to the decay
$H^0 \rightarrow W^\pm H^\mp$.}
\label{H_WH_fig}
\end{center}
\end{figure}

From the Feynman diagram of Figure \ref{H_WH_fig} we obtain
\begin{eqnarray}
\Gamma \left( H^0 \rightarrow W^\pm H^\mp \right) & = &
\frac{\sqrt{2} G_F \left( \tan\alpha - \tan\beta \right)^2}
{16 \pi m_{H^0}^3 \left( 1 + \tan^2\alpha \right) \left( 1 + \tan^2\beta \right)}
\nonumber \\ & &
\times \Lambda^{3/2} \left( m_{H^0}^2, m_W^2, m_H^2 \right).
\label{H_WH}
\end{eqnarray}

\begin{figure}
\begin{center}
{\includegraphics{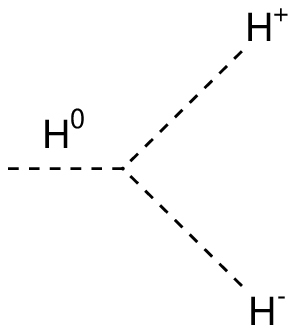}}
\caption{Feynman diagram corresponding to the decay
$H^0 \rightarrow H^+ H^-$.}
\label{H_HH_fig}
\end{center}
\end{figure}

From the Feynman diagram of Figure \ref{H_HH_fig} we obtain
\begin{eqnarray}
\lefteqn{
\Gamma \left( H^0 \rightarrow H^+ H^- \right)  = 
\frac{\sqrt{2} G_F m_W^4}{4 \pi m_{H^0} \left( 1 + \tan^2\beta \right)
\left( 1 + \tan^2\alpha \right)}
}
\nonumber \\ & &
\times \left[ \left( 1 + \tan\beta \tan\alpha \right) 
- \frac{ \left(1 - \tan\beta \tan\alpha \right)}{2 \cos^2\theta_W}
\frac{1 - \tan^2\beta}{1 + \tan^2\beta} \right]^2
\nonumber \\ & &
\times \left( 1 - \frac{4 m_H^2}{m_{H^0}^2} \right)^{1/2}.
\label{H_HH}
\end{eqnarray}

\begin{figure}
\begin{center}
{\includegraphics{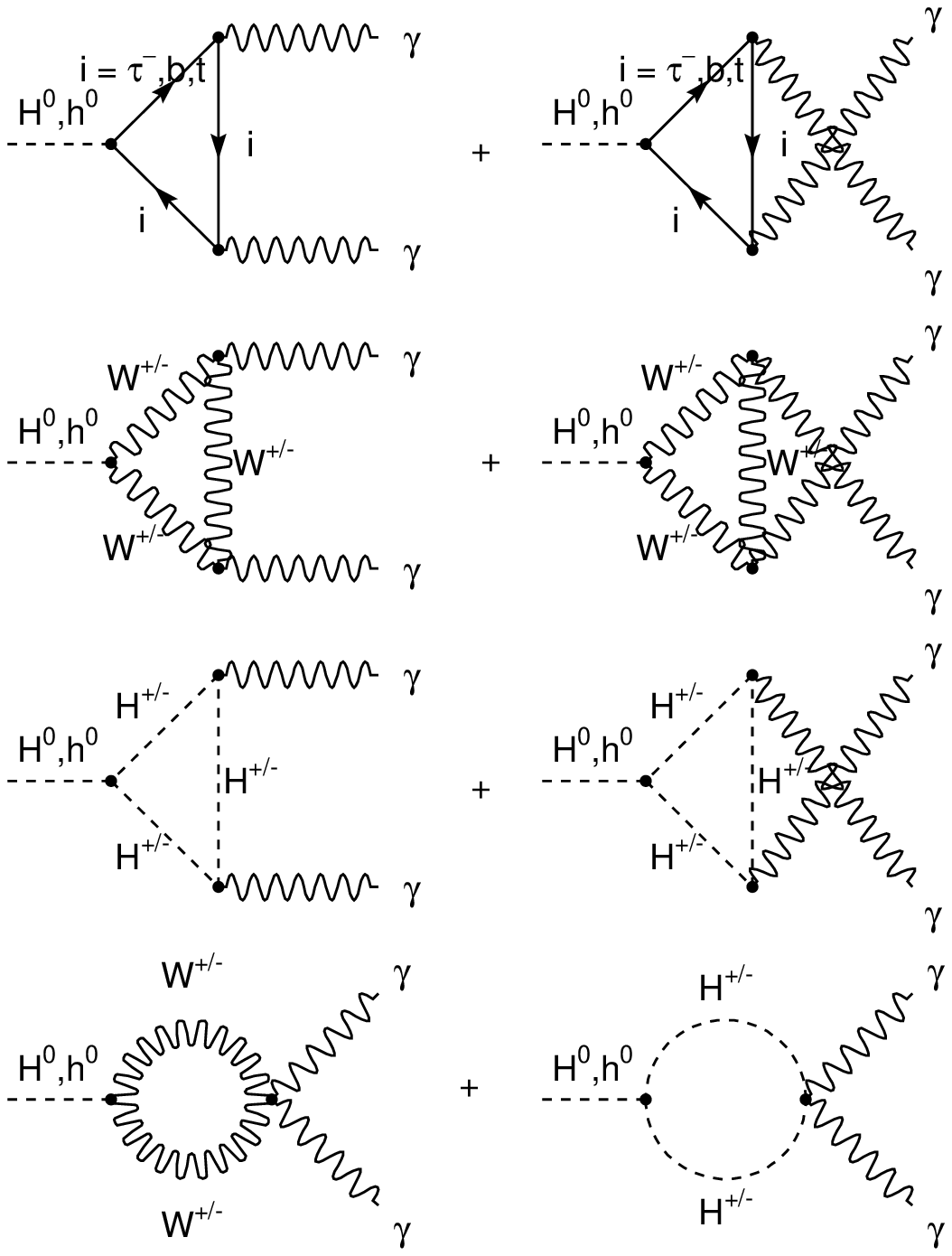}}
\caption{Feynman diagrams corresponding
to $h^0, H^0 \rightarrow \gamma \gamma$.}
\label{hhgamma_fig}
\end{center}
\end{figure}

The Feynman diagrams corresponding to 
$h^0, H^0 \rightarrow \gamma \gamma$ are shown in 
Figure \ref{hhgamma_fig}.
\begin{equation}
\Gamma ( H^0 \rightarrow h^0 Z) = 0.
\label{HhZ}
\end{equation}


\section{Decay rates of $A^0$}
\begin{figure}
\begin{center}
{\includegraphics{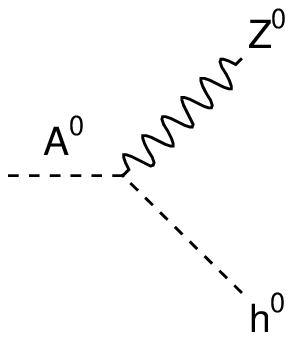}}
\caption{Feynman diagram corresponding to the decay
$A^0 \rightarrow Z h^0$.}
\label{A_Zh_fig}
\end{center}
\end{figure}

\begin{figure}
\begin{center}
{\includegraphics{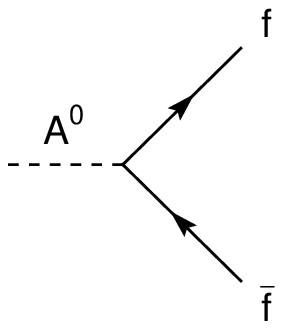}}
\caption{Feynman diagram corresponding to the decay
$A^0 \rightarrow f \bar{f}$.}
\label{A_ff_fig}
\end{center}
\end{figure}

From the tree level Feynman diagram of Figure \ref{A_Zh_fig} 
we obtain
\begin{eqnarray}
\lefteqn{
\Gamma \left( A^0 \rightarrow Z h^0 \right) =
\frac{\sqrt{2} G_F \cos^2{\alpha}}
{16 \pi m_{A^0}^3 \left( 1 + \tan^2{\beta} \right)}
}
\nonumber \\ & &
\times\left[ 1 + \tan{\beta} \tan{\alpha} \right]^2 
\Lambda^{3/2} \left( m_{A^0}^2, m_{h^0}^2, m_Z^2 \right)
\label{A_Zh}
\end{eqnarray}
From the tree level diagram of Figure \ref{A_ff_fig} we obtain
\begin{equation}
\Gamma \left( A^0 \rightarrow f \bar{f} \right) =
\frac{\sqrt{2} G_F}{8 \pi} m_{A^0} m_f^2 A_f^2
\left(1 - \frac{4 m_f^2}{m_{A^0}^2} \right)^{1/2} N_f
\label{A_ff}
\end{equation}
where
$N_f = 3$ for quarks, $N_f = 1$ for leptons,
$A_f = \cot{\beta}$ for $f = u, c, t$, 

\noindent and
$A_f = \tan{\beta}$ for $f = d, s, b, e^-, \mu^-, \tau^-$.

From the Feynman diagrams shown in Figure \ref{A_Zgamma_fig} we
obtain (see Appendix D):
\begin{eqnarray}
\lefteqn{
\Gamma \left( A^0 \rightarrow Z \gamma \right) =
\frac{\sqrt{2} G_F \alpha_{em}^2 m_{A^0}^3}{512 \pi^3 \sin^2{\theta_W}
\cos^2{\theta_W}}
\left( 1 - \frac{m_Z^2}{m_{A^0}^2} \right)^3
}
\nonumber \\ & &
\times| \tan{\beta}
\{ \left( \frac{1}{2} - \frac{2}{3} \sin^2{\theta_W} \right)
I \left( \tau_b, \Lambda_b \right)
\nonumber \\ & &
+ \left( \frac{1}{2} - 2 \sin^2{\theta_W} \right)
I \left( \tau_\tau, \Lambda_\tau \right) \}
\nonumber \\ & &
+ 2 \cot{\beta} \left( \frac{1}{2} - \frac{4}{3} \sin^2{\theta_W} \right)
I \left( \tau_t, \Lambda_t \right) |^2
\label{A_Zgamma}
\end{eqnarray}
where
\begin{equation}
\tau_i = \frac{4 m_i^2}{m_{A^0}^2}, \qquad
\Lambda_i = \frac{4 m_i^2}{m_Z^2},
\label{Lambda_i}
\end{equation}
and
\begin{equation}
I \left( \tau_i, \Lambda_i \right) =
\frac{\tau_i \Lambda_i}{\Lambda_i - \tau_i}
\left\{ f(\tau_i) - f(\Lambda_i) \right\}.
\label{I}
\end{equation}

From the Feynman diagrams of Figure \ref{A_gg} we obtain the decay rate:
\begin{equation}
\Gamma ( A^0 \rightarrow g g ) =
\frac{\sqrt{2} G_F \alpha_s^2 m_{A^0}^3}{128 \pi^3}
|\tan\beta \tau_b f(\tau_b) +
\cot\beta \tau_t f(\tau_t) |^2.
\label{A_to_2g}
\end{equation}

\begin{equation}
\Gamma \left( A^0 \rightarrow h^0 h^0 \right) = 
\Gamma \left( A^0 \rightarrow H^0 H^0 \right) = 0.
\label{A_hh}
\end{equation}

From the Feynman diagram \ref{A_WH_fig} we obtain
\begin{equation}
\Gamma \left( A^0 \rightarrow W^\pm H^\mp \right) =
\frac{\sqrt{2} G_F}{16 \pi m_{A^0}^3}
\Lambda^{3/2} \left(m_{A^0}^2, m_W^2, m_H^2 \right). 
\label{A_WH}
\end{equation}

From the Feynman diagram \ref{A_ZH_fig} we obtain
\begin{eqnarray}
\Gamma \left( A^0 \rightarrow Z H^0 \right) & = &
\frac{\sqrt{2} G_F \left( \tan\beta - \tan\alpha \right)^2}
{16 \pi m_{A^0}^3 \left( 1 + \tan^2\alpha \right)
\left( 1 + \tan^2\beta \right)}
\nonumber \\ & &
\times \Lambda^{3/2} \left( m_{A^0}^2, m_{H^0}^2, m_Z^2 \right).
\label{A_ZH}
\end{eqnarray}

From the Feynman diagrams of \ref{Agammagamma_fig}
we obtain:
\begin{eqnarray}
\Gamma \left( A^0 \rightarrow \gamma \gamma \right) & = &
\frac{\sqrt{2} G_F \alpha^2_{em} m^3_{A^0}}{256 \pi^3} 
\left| \tan\beta \left[ \frac{1}{3} \tau_b f(\tau_b) + 
\tau_\tau f(\tau_\tau) \right] \right.
\nonumber \\ & &
\left. + \cot\beta \left[ \frac{4}{3} \tau_t f(\tau_t) \right] 
\right|^2
\label{Agammagamma}
\end{eqnarray}

\begin{figure}
\begin{center}
{\includegraphics{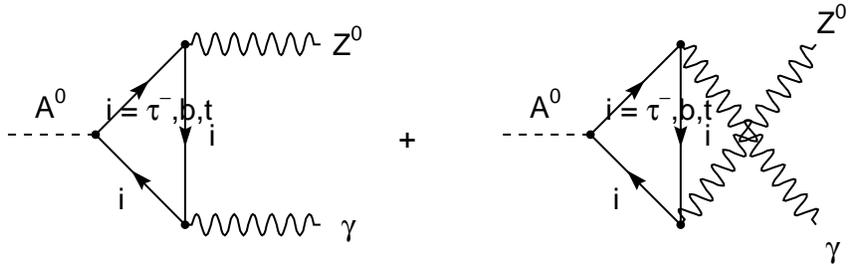}}
\caption{Feynman diagrams corresponding to the 
decay $A^0 \rightarrow Z \gamma$.}
\label{A_Zgamma_fig}
\end{center}
\end{figure}
\begin{figure}
\begin{center}
{\includegraphics{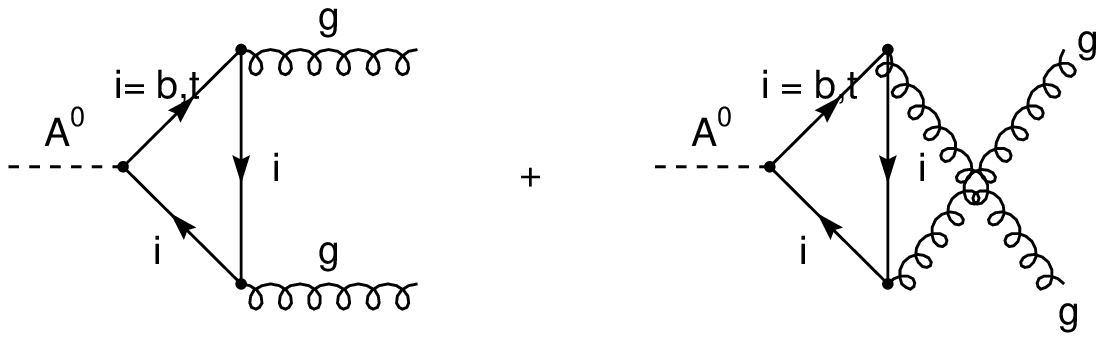}}
\caption{Feynman diagrams of $A^0 \rightarrow g g$.}
\label{A_gg}
\end{center}
\end{figure}
\begin{figure}
\begin{center}
{\includegraphics{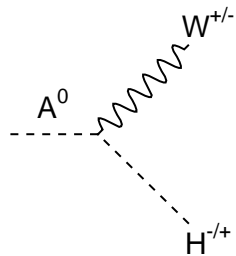}}
\caption{Feynman diagram of $A^0 \rightarrow W^\pm H^\mp$.}
\label{A_WH_fig}
\end{center}
\end{figure}
\begin{figure}
\begin{center}
{\includegraphics{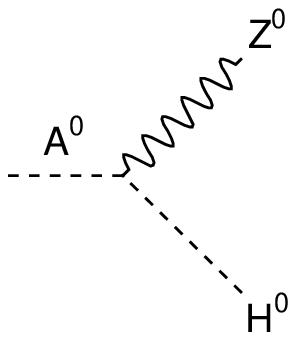}}
\caption{Feynman diagram of $A^0 \rightarrow Z H^0$.}
\label{A_ZH_fig}
\end{center}
\end{figure}
\begin{figure}
\begin{center}
{\includegraphics{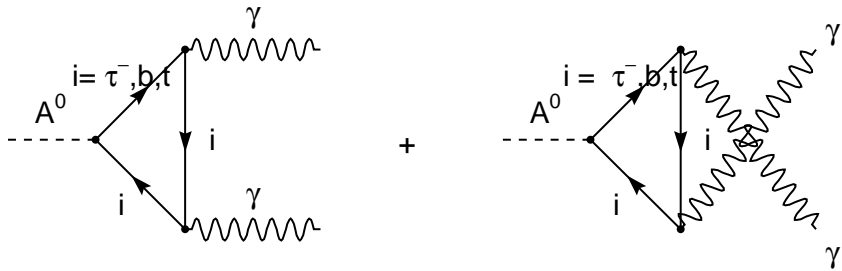}}
\caption{Feynman diagrams for $A^0 \rightarrow \gamma \gamma$.}
\label{Agammagamma_fig}
\end{center}
\end{figure}

\section{Decay $Z \rightarrow h^0 \gamma$}

\begin{figure}
\begin{center}
{\includegraphics{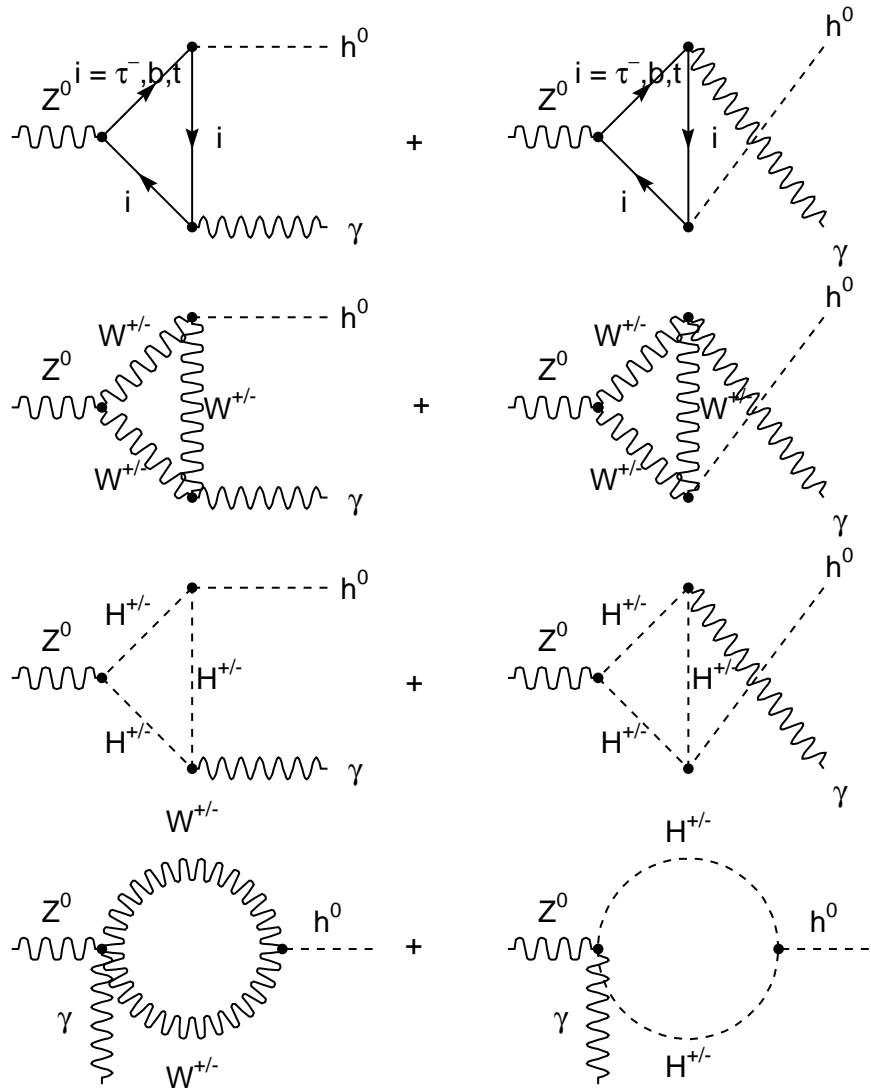}}
\caption{Feynman diagrams for $Z \rightarrow h^0 \gamma$.}
\label{Z_hgamma_fig}
\end{center}
\end{figure}

From the Feynman diagrams of Figure \ref{Z_hgamma_fig} we obtain:
\begin{eqnarray}
\Gamma \left( Z \rightarrow h^0 \gamma \right) & = &
\frac{\sqrt{2} G_F \alpha_{em}^2 m_Z^3}{64 \pi^3 \sin^2{\theta_W} \cos^2{\theta_W}}
\left( 1 - \frac{m_{h^0}^2}{m_Z^2} \right)^3 
\cos^2\alpha \cos^2\beta
\nonumber \\ & & 
\times\mathopen{\vert} \frac{1}{\cos\beta \sin\beta}
[ \tan{\alpha} \tan{\beta} \left( \frac{1}{2}
- \frac{2}{3} \sin^2{\theta_W} \right) F(\tau_b, \Lambda_b)
\nonumber \\ & &
+ \tan{\alpha} \tan{\beta}
\left( \frac{1}{2} - 2 \sin^2{\theta_W} \right) F(\tau_\tau, \Lambda_\tau) 
\nonumber \\ & &
- 2 \left( \frac{1}{2} - \frac{4}{3} \sin^2{\theta_W} \right) 
F(\tau_t, \Lambda_t) ]
\nonumber \\ & &
+ \left[ \tan\beta - \tan\alpha +
\frac{1 - \tan^2\beta}{1 + \tan^2\beta} \frac{\tan\beta + \tan\alpha}
{2 \cos^2\theta_W} \right]
\nonumber \\ & &
\times
\frac{m_W^2}{2 m_H^2} (1 - 2 \sin^2\theta_W) I(\tau_H, \Lambda_H)
\nonumber \\ & &
- \frac{1}{2} \left( \tan \beta - \tan \alpha \right) \cos^2 \theta_W
[ 4 \left( 3 - \tan^2 \theta_W \right) K( \tau_W, \Lambda_W)
\nonumber \\ & &
+ \left\{ \left( 1 + \frac{2}{\tau_W} \right) \tan^2 \theta_W
- \left( 5 + \frac{2}{\tau_W} \right) \right\} 
\nonumber \\ & &
\times I(\tau_W, \Lambda_W) ]
\mathclose{\vert}^2
\label{Z_hgamma}
\end{eqnarray}
where
\begin{equation}
\tau_i = \frac{4 m_i^2}{m_{h^0}^2}, \qquad
\Lambda_i = \frac{4 m_i^2}{m_Z^2}, \qquad
\tau_H = \frac{4 m_H^2}{m_{h^0}^2}, \qquad
\Lambda_H = \frac{4 m_H^2}{m_Z^2},
\label{Lambda_H}
\end{equation}
\begin{eqnarray} 
F(\tau_i, \Lambda_i) & = &
- \frac{1}{2} \frac{\tau_i \Lambda_i}{\tau_i - \Lambda_i}
- \frac{\tau_i^2 \Lambda_i}{\left( \tau_i - \Lambda_i \right)^2}
\left\{ g(\tau_i) - g(\Lambda_i) \right\}
\nonumber \\ & &
+ \frac{1}{4} \frac{\tau_i \Lambda_i}{\tau_i - \Lambda_i}
\left[ 1 + \frac{\tau_i \Lambda_i}{\tau_i - \Lambda_i} \right]
\left\{ f(\tau_i) - f(\Lambda_i) \right\},
\label{F_tau_Lambda}
\end{eqnarray}
\begin{equation}
g(x) = \left\{ \begin{array}{ll}
\left( x - 1 \right)^{1/2} \arcsin{ \left( x^{-1/2} \right) } &
\mbox{if $x \ge 1$} \\
\frac{1}{2} \left( 1 - x \right)^{1/2}
\left[ \ln{ \left\{ \frac{1 + (1 - x)^{1/2}}{1 - (1 - x)^{1/2}} \right\} }
- i \pi \right] &
\mbox{if $x < 1$,}
\end{array} \right.
\label{f1}
\end{equation}
\begin{eqnarray}
I(\tau_H, \Lambda_H) & = &
- \frac{1}{2} \frac{\tau_H \Lambda_H}{(\tau_H - \Lambda_H )}
- \frac{\tau_H^2 \Lambda_H}{(\tau_H - \Lambda_H)^2}
\left[ g(\tau_H) - g(\Lambda_H) \right]
\nonumber \\ & &
+ \frac{1}{4} \frac{\tau_H^2 \Lambda_H^2}{(\tau_H - \Lambda_H)^2}
\left[ f(\tau_H) - f(\Lambda_H) \right],
\label{I2}
\end{eqnarray}
\begin{equation}
\tau_W = \frac{4 m_W^2}{m_{h^0}^2}, \qquad
\Lambda_W = \frac{4 m_W^2}{m_Z^2},
\label{tau_W}
\end{equation}
\begin{equation}
K( \tau_W, \Lambda_W ) = - \frac{\tau_W \Lambda_W}{4 ( \tau_W - \Lambda_W )}
[ f(\tau_W) - f( \Lambda_W ) ].
\label{K}
\end{equation}
The decay width of Equation \ref{Z_hgamma} turns out to be negligible compared to 
the full width of $Z$ so we can not use it to constrain the mass of $h^0$.

\section{Vertex with four particles}
\begin{figure}
\begin{center}
{\includegraphics{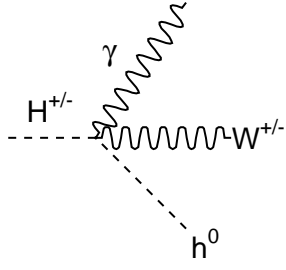}}
\caption{Diagram for $H^\pm \rightarrow W^\pm \gamma h^0$.}
\label{hwgammah}
\end{center}
\end{figure}

The decay rate corresponding to the Feynman diagram \ref{hwgammah} is:
\begin{eqnarray}
\lefteqn{
\Gamma \left( H^\pm \rightarrow W^\pm \gamma h^0 \right)
 = \frac{3 G_F^2 \sin^2 \theta_W \left( 1 + \tan \beta \tan \alpha \right)^2
m_W^5}{32 \left( 1 + \tan^2 \beta \right)
\left( 1 + \tan^2 \alpha \right) \pi^3 \left( x_H^W \right)^{1/2}} 
} 
\nonumber \\
& & \times \{ \frac{1}{2} \Lambda^{\frac{1}{2}} \left( 1, x_H^W, x_H^{h^0} \right)
\left( 1 + x_H^W + x_H^{h^0} \right) 
\nonumber \\
& & + \left( 2 x_H^W x_H^{h^0} - x_H^W - x_H^{h^0} \right)
\times \ln \left| \frac{\Lambda^{\frac{1}{2}} \left( 1, x_H^W, x_H^{h^0} \right)
+ 1 - x_H^W - x_H^{h^0}}
{2 \left( x_H^W x_H^{h^0} \right)^{1/2}} \right| 
\nonumber \\
& & - \left| x_H^{h^0} - x_H^W \right| \times
\nonumber \\
& & \ln \left| \frac{ 
\left| x_H^{h^0} - x_H^W \right|
\Lambda^{\frac{1}{2}} \left( 1, x_H^W, x_H^{h^0} \right) 
- \left( x_H^W + x_H^{h^0} \right) +
\left( x_H^W - x_H^{h^0} \right)^2}
{2 \left( x_H^W x_H^{h^0} \right)^{1/2}} \right| \}
\label{H_Wgh}
\end{eqnarray}
where
\begin{equation}
x_H^W = \frac{m_W^2}{m_{H}^2}, \qquad
x_H^{h^0} = \frac{m_{h^0}^2}{m_{H}^2}.
\label{xWHxhH}
\end{equation}

\begin{figure}
\begin{center}
{\includegraphics{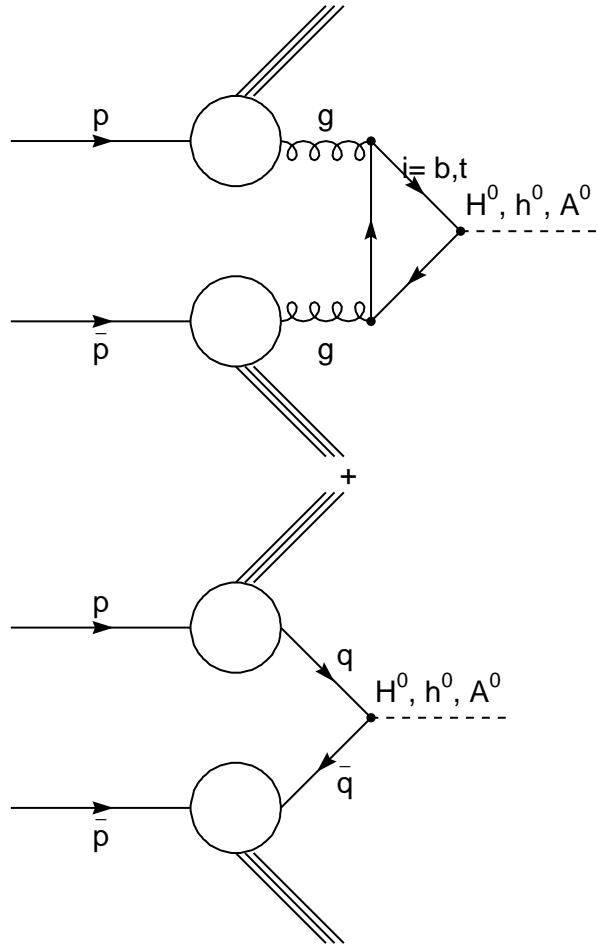}}
\caption{Feynman diagrams for 
$p \bar{p} \rightarrow A X$ with
$A \equiv h^0, H^0, A^0$.}
\label{ppbar_A_fig}
\end{center}
\end{figure}

\section{Production of $h^0$, $H^0$ and $A^0$}
From the Feynman diagrams in Figure \ref{ppbar_A_fig} we obtain 
\begin{eqnarray}
\lefteqn{
\sigma \left( p \bar{p} \rightarrow A X \right)  = 
\frac{\pi^2 \Gamma \left( A \rightarrow 2g \right) \gamma_A}
{8 m_A^3}
}
\nonumber \\ & &
\times\int_{\gamma_A}^1 \frac{dx_a}{x_a} g \left( x_a, m^2 \right) 
g \left( \frac{\gamma_A}{x_a}, m^2 \right)
+ \frac{4 \pi^2 \gamma_A}{3 m_A^3}
\nonumber \\ & &
\times\left[ \sum_{q=u,d,s,c,b} {\Gamma \left( A \rightarrow q \bar{q} 
\right)
\int_{\gamma_A}^1 {\frac{dx_a}{x_a} f_q \left( x_a, m^2 \right) 
f_q \left( \frac{\gamma_A}{x_a}, m^2 \right)} } \right]
\label{sigma_pp_AX}
\end{eqnarray}
where $A \equiv h^0, H^0, A^0$ and $\gamma_A \equiv m_A^2/s$.
Here $f_q$ is the unpolarized parton distribution function for quark
or anti-quark
$q$ and $g$ is the parton distribution function for gluons.
$m^2$ is the factorization scale.
$\Gamma(h^0 \rightarrow g g)$ is given by (\ref{2g}),
$\Gamma(H^0 \rightarrow g g)$ by (\ref{H_gg}),
$\Gamma ( A^0 \rightarrow g g )$ by (\ref{A_to_2g}),
$\Gamma\left( h^0 \rightarrow c \bar{c} \right)$ by (\ref{ccbar}),
$\Gamma\left( h^0 \rightarrow b \bar{b} \right)$ by (\ref{bbbar}),
$\Gamma(H^0 \rightarrow q \bar{q})$ by (\ref{H0_ff}),
and finally, $\Gamma \left( A^0 \rightarrow q \bar{q} \right)$ is given
by (\ref{A_ff}).

\section{Production of $h^0 Z^0 X$}
A production channel with interesting experimental signature
is 

\noindent $p \bar{p} \rightarrow h^0 Z^0 X$. 
The differential cross section obtained from the 
Feynman diagrams in Figure 
\ref{nhiggs2} is
\begin{eqnarray}
\frac{d^2 \sigma}{dy d \left( p_T \right)^2} & = &
\sum_{f} { \int_{x_{amin}}^1 {
dx_a f_f \left( x_a, m_a^2 \right) f_f \left( x_b, m_b^2 \right)
\frac{x_b \hat{s}}{m_{h^0}^2 - \hat{u}} 
}}
\nonumber \\
& & \times \frac{d \sigma}{d \hat{t}} \left( f \bar{f} \rightarrow h^0 Z^0 \right)
\label{pp_hZX}
\end{eqnarray}
where $f$ is $q$ or $g$,
\begin{equation}
x_{amin} = \frac{\sqrt{s} m_T e^y + m_{h^0}^2 - m_Z^2}
{s - \sqrt{s} m_T e^{-y}},
\label{xamin}
\end{equation}
\begin{equation}
m_T = \left( m_Z^2 + p_T^2 \right)^{\frac{1}{2}},
\label{mT}
\end{equation}
\begin{equation}
x_b = \frac{x_a \sqrt{s} m_T e^{-y} + m_{h^0}^2 - m_Z^2}
{x_a s - \sqrt{s} m_T e^y},
\label{xb}
\end{equation}
\begin{equation}
\hat{s} = x_a x_b s,
\label{s_hat}
\end{equation}
\begin{equation}
p_T^2 = \frac{\Lambda \left( \hat{s}, m_{h^0}^2, m_Z^2 \right) \sin^2 \theta}
{4 \hat{s}},
\label{pT2}
\end{equation}
\begin{equation}
\hat{u} = \frac{1}{2} \left( m_{h^0}^2 + m_Z^2 -\hat{s} 
- \cos \theta \Lambda^{1/2} ( \hat{s}, m_{h^0}^2, m_Z^2 ) 
\right)
\label{uhat}
\end{equation}
and
\begin{equation}
\hat{u} \hat{t} = m_{h^0}^2 m_Z^2 + \hat{s} p_T^2.
\label{ut}
\end{equation}
$y$ is the rapidity, $\theta$ is the angle of
dispersion, and $p_T$ is the transverse momentum
of $Z^0$.
For the light quarks $u$, $d$ and $s$ we obtain
\begin{eqnarray}
\frac{d \sigma}{d \hat{t}} & = & \frac{1}{48 \pi \hat{s}}
G_F^2 m_Z^4 \frac{\sin^2 \left( \beta - \alpha \right)}
{\left( \hat{s} - m_Z^2 \right)^2 + m_Z^2 \Gamma_Z^2}
\left[ \left( g_V^f \right)^2 + \left( g_A^f \right)^2 \right]
\nonumber \\
& & \times
\left[ 8 m_Z^2 + \frac{\Lambda \left( \hat{s}, m_{h^0}^2, m_Z^2 \right)}
{\hat{s}} \sin^2 \theta \right]
\label{ds_dthat}
\end{eqnarray}
where $g_A^f \equiv t_{3L} \left( f \right)$ and
$g_V^f \equiv t_{3L} \left( f \right) - 2 q_f \sin^2 \theta_W$.
Coefficients in Equation (\ref{ds_dthat}) are given in Table \ref{c}.
The Standard Model cross section is obtained by
omitting the factor $\sin^2 \left( \beta - \alpha \right)$ in
Equation (\ref{ds_dthat}). The contributions to the cross section
from the heavy quarks $c$ and $b$ are negligible. 
$\Gamma_Z$ is the total decay width of the $Z^0$.
\begin{table}
\begin{center}
\begin{tabular}{|c|c|c|c|c|}
\hline
$f$ & $t_{3L} \left( f \right)$ & $q_f$ & $g_A^f$ & $g_V^f$ \\
\hline
$e^-$, $\mu^-$, $\tau^-$ & $- \frac{1}{2}$ & $-1$ & 
$- \frac{1}{2}$ & $- \frac{1}{2} + 2 \sin^2 \theta_W$ \\
\hline
$u$, $c$, $t$ & $\frac{1}{2}$ & $\frac{2}{3}$ & $\frac{1}{2}$ & 
$\frac{1}{2} - \frac{4}{3} \sin^2 \theta_W$ \\
\hline
$d$, $s$, $b$ & $-\frac{1}{2}$ & $-\frac{1}{3}$ & $-\frac{1}{2}$ &
$-\frac{1}{2} + \frac{2}{3} \sin^2 \theta_W$ \\
\hline
\end{tabular}
\end{center}
\caption{Coefficients in Equation (\ref{ds_dthat}).}
\label{c}
\end{table}

\begin{figure}
\begin{center}
{\includegraphics{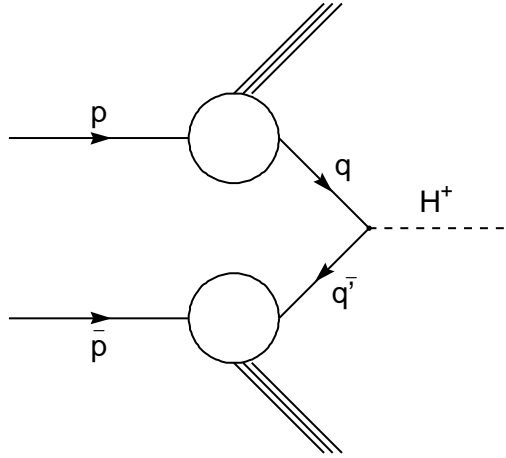}}
\caption{Feynman diagram for 
$p \bar{p} \rightarrow H^+ X$.}
\label{ppbar_Hp_fig}
\end{center}
\end{figure}

\begin{figure}
\begin{center}
{\includegraphics{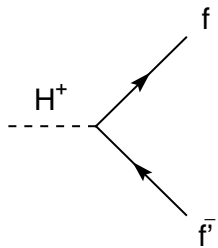}}
\caption{Feynman diagram for 
$H^+ \rightarrow f \bar{f}'$.}
\label{Hffbar_fig}
\end{center}
\end{figure}

\section{Production of $H^+$}
From the diagrams in Figures \ref{ppbar_Hp_fig} and \ref{Hffbar_fig} we obtain
\begin{eqnarray}
\lefteqn{
\sigma ( p \bar{p} \rightarrow H^+ X ) = \frac{4 \pi^2 \gamma_H}{3 m_H^3}
\sum_{q, q'}{ \Gamma ( H^+ \rightarrow q \bar{q}' ) }
}
\nonumber \\
& & 
\int_{\gamma_H}^1 {
\frac{dx_a}{x_a} \left[
f_q^p(x_a, m^2) \cdot f_{\bar{q}'}^{\bar{p}} ( \frac{\gamma_H}{x_a}, m^2)
+ f_{\bar{q}'}^p (x_a, m^2) \cdot f_q^{\bar{p}} ( \frac{\gamma_H}{x_a}, m^2 )
\right] }
\label{ppbar_HpX}
\end{eqnarray}
where
\begin{equation}
\gamma_H = \frac{m_H^2}{s}
\label{gamma_H}
\end{equation}
and
\begin{eqnarray}
\lefteqn{
\Gamma (H^+ \rightarrow f \bar{f}' ) = 
\Gamma (H^- \rightarrow f' \bar{f}) }
\nonumber \\
& & = \frac{ \sqrt{2} G_F \left| V_{f f'} \right|^2 N_c}
{16 \pi m_H} \cdot
\Lambda^{1/2} \left( \frac{m_f^2}{m_H^2}, \frac{m_{f'}^2}{m_H^2}, 1 \right)
\nonumber \\
& & \times \left\{ A^2 \left[ m_H^2 - \left( m_f + m_{f'} \right)^2 \right]
+ B^2 \left[ m_H^2 - \left( m_f - m_{f'} \right)^2 \right] \right\}
\label{gamma_H_ff}
\end{eqnarray}
with
$N_c = 3$ for quarks, $N_c = 1$ for leptons,
\begin{equation}
A = m_{f'} \tan \beta + m_f \cot \beta
\label{AAA}
\end{equation}
\begin{equation}
B = m_{f'} \tan \beta - m_f \cot \beta
\label{B}
\end{equation}
$f = u, c, t, \nu_e, \nu_\mu, \nu_\tau$,
and $f' = d, s, b, e^-, \mu^-, \tau^-$.\cite{hernandez}

\section{Production of $h^0 W^+ X$}
Let us now consider the channel $p \bar{p} \rightarrow h^0 W^+ X$.
We obtain:
\begin{eqnarray}
\lefteqn{
\frac{d^2 \sigma}{dy d \left( p_T \right)^2}  = 
\sum_{q, q'} { \int_{x_{amin}}^1 {
dx_a [ f_{q}^p \left( x_a, m_a^2 \right) f_{\bar{q'}}^{\bar{p}} 
\left( x_b, m_b^2 \right) }}
}
\nonumber \\ & &
+ f_{\bar{q'}}^p \left( x_a, m_a^2 \right) 
f_{q}^{\bar{p}}
\left( x_b, m_b^2 \right) ]
\frac{x_b \hat{s}}{m_{h^0}^2 - \hat{u}} 
\frac{d \sigma}{d \hat{t}} \left( q  \bar{q'} \rightarrow h^0 W^+ \right)
\label{pp_hWX}
\end{eqnarray}
where
\begin{equation}
f_{q'}^p = f_{\bar{q'}}^p, \qquad
f_{\bar{q}}^{\bar{p}} = f_q^{\bar{p}}, \qquad
f_{\bar{q}}^p = f_q^p, \qquad
f_{q'}^{\bar{p}} = f_{\bar{q'}}^{\bar{p}},
\label{ff}
\end{equation}
\begin{equation}
x_{amin} = \frac{\sqrt{s} m_T e^y + m_{h^0}^2 - m_W^2}
{s - \sqrt{s} m_T e^{-y}},
\label{xamin2}
\end{equation}
\begin{equation}
m_T = \left( m_W^2 + p_T^2 \right)^{\frac{1}{2}},
\label{mT2}
\end{equation}
\begin{equation}
x_b = \frac{x_a \sqrt{s} m_T e^{-y} + m_{h^0}^2 - m_W^2}
{x_a s - \sqrt{s} m_T e^y},
\label{xb2}
\end{equation}
\begin{equation}
\hat{s} = x_a x_b s,
\label{s_hat2}
\end{equation}
\begin{equation}
p_T^2 = \frac{\Lambda \left( \hat{s}, m_{h^0}^2, m_W^2 \right) \sin^2 \theta}
{4 \hat{s}},
\label{pT22}
\end{equation}
\begin{equation}
\hat{u} = \frac{1}{2} \left[ m_{h^0}^2 + m_W^2 -\hat{s} 
- \cos \theta \Lambda^{1/2} ( \hat{s}, m_{h^0}^2, m_W^2 ) 
\right],
\label{uhat2}
\end{equation}
\begin{equation}
\hat{t} = \frac{1}{2} \left[ m_{h^0}^2 + m_W^2 -\hat{s} 
+ \cos \theta \Lambda^{1/2} ( \hat{s}, m_{h^0}^2, m_W^2 ) 
\right],
\label{that2}
\end{equation}
\begin{equation}
\cos \theta = \left( 1 -
\frac{4 \hat{s} p_T^2}{\Lambda (\hat{s}, m_{h^0}^2, m_W^2 )} \right)^{1/2}
\label{costheta2}
\end{equation}
and
\begin{equation}
\hat{u} \hat{t} = m_{h^0}^2 m_W^2 + \hat{s} p_T^2.
\label{ut2}
\end{equation}
$y$ is the rapidity of $W^+$ and $p_T$ is the transverse momentum of $W^+$.
From the Feynman diagrams of Figure \ref{wh_fig} we obtain
for $f \bar{f'} \rightarrow h^0 W^+$:
\begin{eqnarray}
\lefteqn{
\frac{d \sigma}{d \hat{t}}  =  \frac{1}{16 \pi \hat{s}^2}
\left| V_{f f'} \right|^2 G_F^2
\{ 
\left| C_{H^+} \right|^2 \hat{s}
\Lambda 
}
\nonumber \\ & & 
\times \left[ m_{f'}^2 \tan^2 \beta + m_f^2 \cot^2 \beta \right]
+ m_W^4 \left| C_W \right|^2 
\left[ 8 \hat{s} m_W^2 + \Lambda \sin^2 \theta \right]
\nonumber \\ & & 
- 2 C_{H^+} \Re{ (C_W) }
[ m_{f'}^2 \tan \beta ( \hat{s} \Lambda 
+ 2 m_W^2 \hat{u} \left( \hat{s} - m_{h^0}^2 \right)
\nonumber \\ & &
+ 2 m_W^4 \left( 2 m_{h^0}^2 - \hat{t} \right) )
- m_f^2 \cot \beta ( \hat{s} \Lambda + 2 m_W^2 \hat{t}
\left( \hat{s} - m_{h^0}^2 \right) 
\nonumber \\ & &
+ 2 m_W^4 \left( 2 m_{H^0}^2 - \hat{u} \right) 
) ]
\nonumber \\ & &
+ \frac{1}{2} m_W^2 \Lambda \sin^2 \theta
\left[ \frac{m_f^2 C_f^2}{\hat{t}^2} + \frac{m_{f'}^2 C_{f'}^2}{\hat{u}}
\right] +
\hat{s} \left[ m_f^2 C_f^2 + m_{f'}^2 C_{f'}^2 \right]
\nonumber \\ & &
+ 2 C_{H^+} \hat{s} 
\left[ m_{h^0}^2 m_W^2 - \frac{1}{4} \Lambda \sin^2 \theta \right]
\left[ \frac{m_f^2 \cot \beta C_f}{\hat{t}} - 
\frac{m_{f'}^2 \tan \beta C_{f'}}{\hat{u}} \right]
\nonumber \\ & &
+ 2 C_{H^+} \hat{s}
\left[ m_{f'}^2 \tan \beta C_{f'} \hat{u} - m_f^2 \cot \beta C_f \hat{t} \right]
\nonumber \\ & &
-2 \Re{(C_W)} \left[ \frac{1}{2} \Lambda \sin^2 \theta
\left( m_W^2 + \frac{\hat{s}}{2} \right)
- \hat{s} m_{h^0}^2 m_W^2 + 4 \hat{s} m_W^4 + 4 m_W^6 \right]
\nonumber \\ & &
\times \left[ \frac{m_f^2 C_f}{\hat{t}} + \frac{m_{f'}^2 C_{f'}}{\hat{u}} \right]
- 2 \Re{(C_W)} m_f^2 C_f \left[ -2 m_W^4 + \hat{t} 
\left( \hat{s} - 2 m_W^2 \right) \right]
\nonumber \\ & &
- 2 \Re{(C_W)} m_{f'}^2 C_{f'} \left[ -2 m_W^4 + \hat{u} 
\left( \hat{s} - 2 m_W^2 \right) \right]
\}
\label{pp_h0W}
\end{eqnarray}
where $\Lambda$ stands for $\Lambda ( \hat{s}, m_{h^0}^2, m_W^2 )$,
\begin{equation}
C_{H^+} = \frac{\cos \left( \beta - \alpha \right)}{\hat{s} - m_H^2},
\label{CHp}
\end{equation}
\begin{equation}
C_W = \frac{\sin \left( \beta - \alpha \right)
\left( \hat{s} - m_W^2 - i m_W \Gamma_W \right)}
{\left( \hat{s} - m_W^2 \right)^2 + m_W^2 \Gamma_W^2},
\label{CW}
\end{equation}
\begin{equation}
C_f = - \frac{\cos \alpha}{\sin \beta}, \qquad
C_{f'} = \frac{\sin \alpha}{\cos \beta}.
\label{Cf}
\end{equation}
For $p \bar{p} \rightarrow h^0 W^- X$ interchange
$\hat{u} \leftrightarrow \hat{t}$.

For the Standard Model we obtain the differential
cross section (\ref{pp_h0W}) with $h^0$ replaced
by the Standard Model Higgs, $C_{H^+} = 0$, 
$C_f = C_{f'} = -1$, and $\sin(\beta - \alpha) = 1$ in (\ref{CW}).

\section{Numerical examples}
Two sensitive channels for the search of the Standard Model Higgs
are $p \bar{p} \rightarrow h^0 Z X$ and
$p \bar{p} \rightarrow h^0 W^\pm X$.
The cross section for $p \bar{p} \rightarrow h^0 Z X$
off resonance
in the Doublet model differs from the Standard Model
by a factor $\sin^2 (\beta - \alpha)$
(see Equation (\ref{ds_dthat}))
and it will
be hard to obtain both $m_{h^0}$ and
$\tan(\beta)$. We are therefore interested
in the production of $h^0 Z$ on resonance.
In particular
$p \bar{p} \rightarrow A^0$ followed by
$A^0 \rightarrow h^0 Z \rightarrow b \bar{b} l^- l^+$
where $l = \mu, e$. A peak should be observed in
the $h^0 Z$ invariant mass.
From Equation (\ref{sigma_pp_AX})
we obtain the cross sections listed in Tables
\ref{numbers_A0_200} and \ref{numbers_A0_250}.

\begin{table}
\begin{center}
\begin{tabular}{|c|c|c|c|}
\hline
partons & $\tan(\beta) = 100$ & $\tan(\beta) = 10$ & $\tan(\beta) = 2$ \\
\hline
$g g$       & 0.20E+1 & 0.13E-1 & 0.35E-1 \\
$b \bar{b}$ & 0.31E+2 & 0.31E+0 & 0.12E-1 \\
$c \bar{c}$ & 0.73E-7 & 0.73E-5 & 0.18E-3 \\
$s \bar{s}$ & 0.20E+0 & 0.20E-2 & 0.81E-4 \\
$d \bar{d}$ & 0.10E-1 & 0.10E-3 & 0.41E-5 \\
$u \bar{u}$ & 0.18E-9 & 0.18E-7 & 0.46E-6 \\
\hline
\end{tabular}
\end{center}
\caption{Production cross section [pb] for
$p \bar{p} \rightarrow A^0$ from the indicated
partons. $m_{A^0} = 200$GeV/c$^2$, $\sqrt{s} = 1960$GeV/c$^2$.}
\label{numbers_A0_200}
\end{table}

\begin{table}
\begin{center}
\begin{tabular}{|c|c|c|c|}
\hline
partons & $\tan(\beta) = 100$ & $\tan(\beta) = 10$ & $\tan(\beta) = 2$ \\
\hline
$g g$       & 0.42E+0 & 0.22E-2 & 0.19E-1 \\
$b \bar{b}$ & 0.82E+1 & 0.82E-1 & 0.33E-2 \\
$c \bar{c}$ & 0.19E-7 & 0.19E-5 & 0.49E-4 \\
$s \bar{s}$ & 0.57E-1 & 0.57E-3 & 0.23E-4 \\
$d \bar{d}$ & 0.44E-2 & 0.44E-4 & 0.18E-5 \\
$u \bar{u}$ & 0.89E-10 & 0.89E-8 & 0.22E-6 \\
\hline
\end{tabular}
\end{center}
\caption{Production cross section [pb] for
$p \bar{p} \rightarrow A^0$ from the indicated
partons. $m_{A^0} = 250$GeV/c$^2$, $\sqrt{s} = 1960$GeV/c$^2$.}
\label{numbers_A0_250}
\end{table}

Let us now consider the decays of $A^0$. As an example
we take $m_{h^0} = 120$GeV/c$^2$, $m_{H^0} = 250$GeV/c$^2$,
$m_H = 200$GeV/c$^2$ and $m_{A^0} = 250$GeV/c$^2$.
The corresponding branching fractions are listed in
Table \ref{BR_A0}. From Tables \ref{numbers_A0_250} and
\ref{BR_A0} we obtain a production cross section times
branching fraction for the process
$p \bar{p} \rightarrow A^0 \rightarrow h^0 Z$
of 0.018pb for $\tan(\beta) = 2$, and
0.0045pb for $\tan(\beta) = 10$.

\begin{table}
\begin{center}
\begin{tabular}{|c|c|c|c|}
\hline
partons & $\tan(\beta) = 100$ & $\tan(\beta) = 10$ & $\tan(\beta) = 2$ \\
\hline
$A \rightarrow gg$            & 3.0E-4 & 1.5E-4 & 2.0E-3 \\
$A \rightarrow b \bar{b}$     & 1.0E+0 & 9.4E-1 & 5.9E-2 \\
$A \rightarrow c \bar{c}$     & 8.0E-10 & 7.5E-6 & 2.9E-4 \\
$A \rightarrow s \bar{s}$     & 8.0E-4 & 7.5E-4 & 4.7E-5 \\
$A \rightarrow Z h^0$         & 6.1E-6 & 5.5E-2 & 9.4E-1 \\
$A \rightarrow Z \gamma$      & 1.9E-8 & 6.0E-9 & 1.2E-6 \\
$A \rightarrow \gamma \gamma$ & 1.2E-7 & 8.1E-8 & 7.5E-6 \\
\hline
\end{tabular}
\end{center}
\caption{Branching fractions for $A^0$ assuming
$m_H = 200$GeV/c$^2$, $m_{H^0} = 250$GeV/c$^2$, $m_{A^0} = 250$GeV/c$^2$
and $m_{h^0} = 120$GeV/c$^2$.}
\label{BR_A0}
\end{table}

From Equations (\ref{ppbar_HpX}) and (\ref{gamma_H_ff}) we obtain the production
cross sections for $p \bar{p} \rightarrow H^+ X$ shown in
Table \ref{sigmaH}.
\begin{table}
\begin{center}
\begin{tabular}{|c|c|c|c|}
\hline
partons & $\tan(\beta) = 100$ & $\tan(\beta) = 10$ & $\tan(\beta) = 2$ \\
\hline
$u \bar{d}$ & 0.82E-2 & 0.82E-4 & 0.34E-5 \\
$u \bar{s}$ & 0.66E-1 & 0.66E-3 & 0.26E-4 \\
$u \bar{b}$ & 0.89E-2 & 0.89E-4 & 0.36E-5 \\
$c \bar{s}$ & 0.31E-1 & 0.32E-3 & 0.91E-4 \\
$c \bar{b}$ & 0.24E-1 & 0.24E-3 & 0.96E-5 \\
\hline
\end{tabular}
\end{center}
\caption{Production cross section [pb] for
$p \bar{p} \rightarrow H^+ X$ from the indicated
partons. $m_{H} = 250$GeV/c$^2$, $\sqrt{s} = 1960$GeV/c$^2$.}
\label{sigmaH}
\end{table}

\begin{figure}
\begin{center}
{\includegraphics{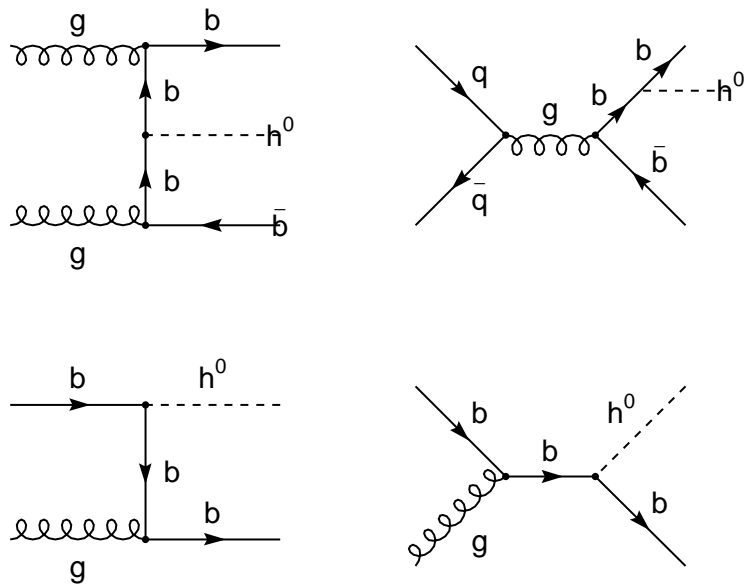}}
\caption{Some Feynman diagrams for the production of
three or more $b$-jets.}
\label{gluonb_fig}
\end{center}
\end{figure}

Other channels of experimental interest are the production of
3 or more $b$-jets as in Figure \ref{gluonb_fig}. Some numerical calculations using the 
CompHEP program\cite{comphep} are presented in Table \ref{comphep_tab}.

\begin{table}
\begin{center}
\begin{tabular}{|c|c|c|c|}
\hline
process & $\tan(\beta) = 100$ & $\tan(\beta) = 50$ & $\tan(\beta) = 2$ \\
\hline
$b +  g \rightarrow b + h^0$ & 0.021 & 0.021 & 0.011 \\
$u + \bar{u} \rightarrow b + \bar{b} + h^0$ & 0.002 & 0.001 & 0.0004 \\
$d + \bar{d} \rightarrow b + \bar{b} + h^0$ & 0.0005 & 0.0005 & 0.0001 \\
$g + g \rightarrow b + \bar{b} + h^0$ & 0.015 & 0.015 & 0.008 \\
\hline
\end{tabular}
\end{center}
\caption{Production cross section [pb] for
$p \bar{p} \rightarrow b h^0 X$ from the indicated
processes. $m_{h^0} = 120$GeV/c$^2$, 
$m_{A^0} = 250$GeV/c$^2$,
$\sqrt{s} = 1960$GeV/c$^2$.}
\label{comphep_tab}
\end{table}

\section{Running coupling constants and Grand 
\newline
Unification}

The coupling constants of the Two Higgs Doublet Model of type II
are $g_s(\mu)$ for SU(3), $g(\mu)$ for SU(2), and $g'(\mu)$ for
U(1). 
These coupling constants
depend on the energy scale $\mu$ as follows:
\begin{equation}
\frac{1}{g_s^2(\mu)} = \frac{1}{g_s^2(m_x)} + 
\frac{1}{8 \pi^2} \left( -11 + \frac{4}{3} n_F \right) 
\ln{ \left( \frac{m_x}{\mu} \right) },
\label{gs}
\end{equation}
\begin{equation}
\frac{1}{g^2(\mu)} = \frac{1}{g^2(m_x)} + 
\frac{1}{8 \pi^2} \left( - \frac{22}{3} + \frac{4}{3} n_F + \frac{1}{6} n_S\right) 
\ln{ \left( \frac{m_x}{\mu} \right) },
\label{g}
\end{equation}
\begin{equation}
\frac{1}{g'^2(\mu)} = \frac{1}{g'^2(m_x)} + 
\frac{1}{8 \pi^2} \left( \frac{20}{9} n_F + \frac{1}{6} n_S \right) 
\ln{ \left( \frac{m_x}{\mu} \right) },
\label{g_prime}
\end{equation}
where $n_F$ is the number of 
families of quarks and leptons, and $n_S$ is the number of
higgs doublets. For the Two Higgs Doublet Model of type II considered
in this article, $n_F = 3$ and $n_S = 2$. In terms of the
elementary electric charge and the Weinberg angle,
$g(m_Z) = e(m_Z)/\sin \theta_W(m_Z)$, 
$g'(m_Z) = e(m_Z)/\cos \theta_W(m_Z)$. The fine structure 
constant is $\alpha(m_Z) = e^2(m_Z)/(4 \pi)$.

Let us now assume that a Grand Unified Theory (GUT) breaks its symmetry
to SU(3)$\times$SU(2)$\times$U(1) at the energy scale $m_x$.
At this scale we take
\begin{equation}
g_s^2(m_x) = g^2(m_x) = \frac{5}{3} g'^2(m_x),
\label{gmx}
\end{equation}
and obtain
\begin{equation}
\sin^2 \theta_W = \frac{11 + \frac{1}{2} n_S + \frac{5e^2}{3g_s^2} 
\left( 22 - \frac{1}{5} n_S \right)}
{ 66 + n_S },
\label{sin2tW}
\end{equation}
\begin{equation}
\ln{ \left( \frac{m_x}{m_Z} \right) } =
\frac{24 \pi^2}{e^2}
\frac{1 - \frac{8e^2}{3g_s^2} }{66 + n_S},
\label{mx}
\end{equation}
with all running couplings evaluated at $m_Z$.

The corresponding equations of the Minimum Supersymmetry Model\cite{MSSM}
are
\begin{equation}
\sin^2 \theta_W = \frac{18 + 3n_S + \frac{e^2}{g_s^2} 
\left( 60 - 2n_S \right) }
{108 + 6n_S},
\label{sin2tW_MSSM}
\end{equation}
\begin{equation}
\ln{ \left( \frac{m_x}{m_Z} \right) } =
\frac{8 \pi^2}{e^2}
\left[ \frac{1 - \frac{8e^2}{3g_s^2} }{18+n_S} \right].
\label{mx_MSSM}
\end{equation}

Some numerical results are presented in Table \ref{sint_mx}.
From the Table we conclude that the Two Higgs Doublet Model 
of type II is in disagreement with the measured value
of $\sin^2 \theta_W(m_Z)$, and with the non-observation of
proton decay ($m_x$ is too low). Raising the number of doublets to $\approx 7$
would bring $\sin^2 \theta_W(m_Z)$ into agreement with 
observations, but $m_x$ is still too low. The
MSSM with $n_S = 2$ (which includes the Two Higgs Doublet Model
of type II) is in agreement with both the observed
$\sin^2 \theta_W(m_Z)$, and with the non-observation of
proton decay.

\begin{table}
\begin{center}
\begin{tabular}{|c|c|c|c|c|}
\hline
 & \multicolumn{2}{|c|}{Doublet Model}
 & \multicolumn{2}{|c|}{MSSM} \\
\hline
$n_S$ & $\sin^2 \theta_W(m_Z)$ & $m_x$ & $\sin^2 \theta_W(m_Z)$ & $m_x$ \\
\hline
0 & 0.2037 & $1.0\cdot10^{15}$ & 0.2037 & $8.0\cdot10^{17}$ \\ 
2 & 0.2118 & $4.2\cdot10^{14}$ & 0.2311 & $2.0\cdot10^{16}$ \\ 
4 & 0.2194 & $1.8\cdot10^{14}$ & 0.2536 & $1.0\cdot10^{15}$ \\ 
6 & 0.2266 & $8.3\cdot10^{13}$ & 0.2722 & $8.3\cdot10^{13}$ \\ 
8 & 0.2334 & $3.9\cdot10^{13}$ & 0.2880 & $1.0\cdot10^{13}$ \\ 
\hline
\end{tabular}
\end{center}
\caption{Predicted $\sin^2 \theta_W(m_Z)$ and $m_x$ for the
Two Higgs Doublet Model of type II, and the Minimum Supersymmetric
Model as a function of the number of doublets $n_S$.}
\label{sint_mx}
\end{table}

\section{Conclusions}
One of the major efforts at the Fermilab
Tevatron in Run II, and at the future LHC, is the search for the
Standard Model Higgs $h_{SM}$. The four channels with largest
production cross section are\cite{PDG}
$gg \rightarrow h_{SM}$, $q \bar{q'} \rightarrow h_{SM} W$,
$q \bar{q} \rightarrow h_{SM} Z$ and $qq \rightarrow h_{SM} qq$.
The decay modes of $h_{SM}$ with largest branching fraction\cite{PDG}
are $b \bar{b}$ for $m_{h_{SM}} \lesssim 137$GeV/c$^2$ and $W^+ W^-$
for $m_{h_{SM}} \gtrsim 137$GeV/c$^2$.

The search for the Standard Model Higgs will also constrain 
or discover particles of the
Two Higgs Doublet Model of type II. 

The most interesting production channels are
$gg \rightarrow h^0, H^0, A^0$ on mass shell, and
$q \bar{q}, g g \rightarrow h^0 Z$ and
$q \bar{q'} \rightarrow h^0 W^\pm$ 
in the continuum
(tho there may be peaks at $m_{A^0}$).
The most interesting decays are
$h^0, H^0, A^0 \rightarrow b \bar{b}$-jets and
$\tau^+ \tau^-$, and, if above threshold,
$H^0 \rightarrow 
Z Z$, $W^+ W^-$ and $h^0 h^0$.
The following final states should be compared with the 
Standard Model cross section:
$b \bar{b} Z$, $b \bar{b} W^\pm$, 
$\tau^+ \tau^- Z$, $\tau^+ \tau^- W^\pm$,
$b \bar{b}$, $\tau^+ \tau^-$, $Z Z$,
$W^+ W^-$, 3 and 4 $b$-jets, $2 \tau^+ + 2 \tau^-$,
$b \bar{b} \tau^+ \tau^-$, $Z W^+ W^-$, $3 Z$,
$Z Z W^\pm$ and $3 W^\pm$.
Mass peaks should be searched in the following
channels:
$Z b \bar{b}$, $Z Z$, $Z Z Z$, $b \bar{b}$, $4 b$-jets 
and, just in case, $Z \gamma$.

We have discussed the masses of the Higgs
particles in the Two Higgs Doublet Model of type II, and have
calculated several relevant production cross sections and decay rates.
We have also discussed running coupling constants and
Grand Unification. If the Two Higgs Doublet Model of type II
is part of a Grand Unified Theory, then it does not
agree with the observed $\sin^2 \theta_W$ nor with the
non-observation of proton decay. The
MSSM with $n_S = 2$ (which includes the Two Higgs Doublet Model
of type II) is in agreement with both the observed
$\sin^2 \theta_W(m_Z)$, and with the non-observation of
proton decay.

\chapter{Higgs production at a muon collider in the Two Higgs Doublet 
Model of type II}
\typeout{Higgs production at a muon collider in the Two Higgs Doublet
Model of type II}
\thispagestyle{empty}
\begin{abstract}
\noindent
We calculate Higgs production cross sections at a muon collider
in the Two Higgs Doublet Model of type II. The most interesting
productions channels are $\mu^- \mu^+ \rightarrow h^0 Z^0, H^0
Z^0, H^- H^+, A^0Z^0$ and $H^{\mp} W^{\pm}$. The last channel is 
compared with the production processes $p \bar{p} \rightarrow
H^{\mp} W^{\pm} X$ and $p p \rightarrow H^{\mp} W^{\pm} X$ at the 
Tevatron and LHC energies, respectively, for large values of 
$\tan\beta$.
\end{abstract}
\pagenumbering{arabic}
\section{Introduction}
In this article we calculate neutral and charged Higgs production
cross sections at a muon collider in the Two Higgs Doublet Model
of type II. The Higgs sector of the Minimal Supersymmetric
Standard Model (MSSM) is of 
this type (tho the model of type II does not require Supersymmetry).
Higgs doublets can be added to the Standard Model without upsetting the
$Z/W$ mass ratio. Higher dimensional representations upset this ratio
\cite{B-H}. Adding a second complex doublet to the Standard Model results 
in five Higgs bosons: one pseudoscalar $A^0$ (CP-odd scalar), two neutral
scalars $H^0$ and $h^0$ (CP-even scalars), and two charged scalars $H^+$ 
and $H^-$. In the Standard Model we only have a single neutral Higgs.

In recent years, some papers have appeared,
suggesting the possibility of the 
construction of a $\mu^- \mu^+$ collider to detect charged or neutral
Higgs bosons [\cite{mu1}, \cite{mu2}]. The main reason is that in
a muon collider, the signal could be cleaner than in a hadron
collider. In this paper, we analyze this possibility studying some 
production cross sections like:
 $\mu^- \mu^+ \rightarrow h^0 Z^0, H^0   
Z^0, H^- H^+, 
\newline
\noindent A^0Z^0$ and $H^{\mp} W^{\pm}$ (Sections 4.2-4.6). 

In Sections 4.5,4.6,4.8,4.9 we will focus our interest in the production 
of charged Higgs bosons. There are three ways of producing $H^{\pm}$.
One is via $p \bar{p}$ or $pp$ interactions in a hadron collider. In 
hadron colliders, the signals are overwhelmed by  backgrounds due 
basically to $t\bar{t}$ production \cite{S.Moretti}. The other ways 
to produce charged Higgs are  $e^- e^+$ or $\mu^- \mu^+$ colliders 
, in which backgrounds are considerably less. In some 
processes like $\mu^- \mu^+ \rightarrow H^- H^+$ and $e^- e^+ \rightarrow
H^- H^+$, there is no difference between the cross sections obtained in an 
$e^- e^+$ collider or a $\mu^- \mu^+$ collider. However, in reactions like
$\mu^- \mu^+ \rightarrow H^{\mp} W^{\pm}$ and $e^- e^+ \rightarrow H^{\mp} 
W^{\pm}$, the total cross section is proportional to the square of the 
mass of the fermion and then $e^- e^+$ interactions give us very small 
cross sections. This motivated us to compare in Section
4.9  the channel $\mu^- \mu^+ \rightarrow H^{\mp} W^{\pm}$
(at $\sqrt{s} = 500\textrm{GeV/c}$ and for large values of $\tan\beta$) 
with the production processes  $p 
\bar{p} 
\rightarrow H^{\mp} W^{\pm} X$ (at the Tevatron) and $p p 
\rightarrow H^{\mp} W^{\pm} X$ (at the LHC), to check the feasibility
of detecting $H^{\pm}$ using a muon collider.

The influence of radiative corrections in the masses of the Higgs 
bosons is  considered in all the calculations. 

\section{Higgs bosons masses and radiative corrections}
The masses of the neutral Higgs particles, calculated at tree level,
are \cite{V.Barger}:

\begin{equation}
m_{A^0}^2 = m_H^2 - m_W^2
\label{mAo}
\end{equation}

\begin{equation}
m_{H^0}^2 = \frac{1}{2} \left[ m_Z^2 + m_{A^0}^2
+ \left[ \left(m_Z^2 - m_{A^0}^2 \right)^2 +
4 m_{A^0}^2 m_Z^2 \sin^2 2\beta \right]^{1/2}
\right]
\label{mHo}
\end{equation}

\begin{equation}
m_{h^0}^2 = \frac{1}{2} \left[ m_Z^2 + m_{A^0}^2
- \left[ \left(m_Z^2 - m_{A^0}^2 \right)^2 +
4 m_{A^0}^2 m_Z^2 \sin^2 2\beta \right]^{1/2}
\right]
\label{mhowrc}
\end{equation}

\noindent with $ 0 \le \beta < \frac{\pi}{2}$

From these relations, the Higgs bosons masses satisfy the
bounds:

\begin{equation}
m_{A^0} < m_H
\label{mA<mH}
\end{equation}

\begin{equation}
m_H > m_W
\label{mH>mW}
\end{equation}

\begin{equation}
m_{h^0} \le m_Z
\label{mh<mZ}
\end{equation}

\begin{equation}
m_Z \le m_{H^0} \le \sec\theta_W m_H
\label{mHbounds}
\end{equation}

The bound given by (\ref{mh<mZ}) practically has been
excluded by the present limits on $m_{h^0}$ obtained by LEP
and CDF \cite{LEP_CDF}.

The mixing angle $\alpha$ $( - \pi/2 < \alpha \le 0)$
between the two neutral scalar Higgs
fields $H^0$, $h^0$ is given by

\begin{equation}
\tan\alpha = -\left[ \frac{1 + F}{1 - F} \right]^{1/2}
\label{ttan_alpha}
\end{equation}
\begin{equation}
F = \frac{\left( 1 - \tan^2\beta \right)}
{ \left( 1 + \tan^2 \beta \right) G}
\left[ 1 - \frac{m_Z^2}{m_H^2} - \frac{m_W^2}{m_H^2} \right]
\label{FF}
\end{equation}

\begin{equation}
G = \left[ \left( 1 - \frac{m_W^2}{m_H^2}
+ \frac{m_Z^2}{m_H^2} \right)^2 -
4 \left(\frac{m_Z^2}{m_H^2} \right)
\left( 1 - \frac{m_W^2}{m_H^2} \right)
\left( \frac{1 - \tan^2\beta}{1 + \tan^2\beta}
\right)^2 \right]^{1/2}
\label{G*}  
\end{equation}

In terms of $m_H$ and $G$ Equations (\ref{mHo}) and
(\ref{mhowrc}) are:

\begin{equation}
m_{H^0}^2 = \frac{1}{2} m_H^2 \left[ 1 - \frac{m_W^2}{m_H^2} +
\frac{m_Z^2}{m_H^2} + G \right]
\label{mHoG}
\end{equation}

\begin{equation}
m_{h^0}^2 = \frac{1}{2} m_H^2 \left[ 1 - \frac{m_W^2}{m_H^2} +
\frac{m_Z^2}{m_H^2} - G \right]
\label{mhoG}
\end{equation}

Taking into account radiative corrections, (\ref{mHo})
and (\ref{mhowrc}) can be written as [see \cite{Weinberg},
\cite{Zhou}]:

\begin{eqnarray}
\lefteqn{
m_{H^0}^2 = \frac{1}{2} \{ m_A^2 + m_Z^2 + \Delta_t
+ \Delta_b
}
\nonumber \\ & &
+ \{ \left( \left( m_A^2 - m_Z^2 \right) \cos 2\beta
+ \Delta_t - \Delta_b \right)^2
\nonumber \\ & &
+
\left( m_A^2 + m_Z^2 \right)^2 \sin^2 2\beta \}^{1/2}
\}
\label{radiative_mHo}
\end{eqnarray}

\begin{eqnarray}
\lefteqn{
m_{h^0}^2 = \frac{1}{2} \{ m_A^2 + m_Z^2 + \Delta_t
+ \Delta_b
}
\nonumber \\ & &
- \{ \left( \left( m_A^2 - m_Z^2 \right) \cos 2\beta
+ \Delta_t - \Delta_b \right)^2
\nonumber \\ & &
+
\left( m_A^2 + m_Z^2 \right)^2 \sin^2 2\beta \}^{1/2}
\}
\label{radiative_mho}
\end{eqnarray}

\noindent where:
\begin{equation} 
\Delta_b = \frac{3 \sqrt{2} m_b^4 G_F \left( 1 +
\tan^2\beta \right)}{2 \pi^2 }
ln\left( \frac{M_{sb}^2}{m_b^2} \right)  
\label{Delta_b}
\end{equation}

\noindent and
\begin{equation}
\Delta_t = \frac{3 \sqrt{2} m_t^4 G_F \left( 1 +
\tan^2\beta \right)}{2 \pi^2 \tan^2\beta}
ln\left( \frac{M_{st}^2}{m_t^2} \right)
\label{Delta_t}
\end{equation}

\noindent $M_{sb}$ and $M_{st}$ are the masses of the sbottom and stop
(the scalar superpartners of the bottom and top quarks).

Equation (\ref{mAo}) is practically unaffected by radiative 
corrections. According to (\ref{radiative_mho})
 $m_{h^0}$ increases
as the value of $m_A$ increases. Then, for very large values of
$m_A$ we can set an upper bound for $m_{h^0}$:

\begin{equation}
m_{h^0}^{2} \le m_{h^0}^{2} (m_{A^0} \rightarrow \infty) =
m_Z^2 \left( \frac{1 - \tan^2\beta}{1 + \tan^2\beta} \right)^2 
+ \frac{ \Delta_t \tan^2\beta}{\left( 1 + \tan^2\beta \right)}
+ \frac{\Delta_b}{\left( 1 + \tan^2\beta \right)}
\label{mhupperlimit}
\end{equation} 

Taking $m_b = 4.3$ $\textrm{GeV/c}^2$, $m_t = 174.3 \textrm{GeV/c}^2$,
$M_{st} \sim M_{sb} \sim 1 \textrm{TeV}$ \cite{Weinberg} and $m_Z = 
91.1876 \textrm{GeV/c}^2$
we obtain:

\begin{equation}
\Delta_b = 1.123 \times 10^{-6} 
\left( 1 + \tan^2\beta \right) m_Z^2
\label{Db}
\end{equation}

\begin{equation}
\Delta_t = 0.9723 m_Z^2 \frac{\left(1 + \tan^2\beta \right)}
{\tan^2\beta}
\label{Dt}
\end{equation}

The contribution of the b-quark loop is negligible.
Using Equations (\ref{Db}) and (\ref{Dt}), 
(\ref{mhupperlimit}) can be expressed as:

\begin{equation}
m_{h^0} \le m_Z \left[ \left( \frac{ 1 - \tan^2\beta}
{1 + \tan^2\beta} \right)^2 + 0.9723 \right]^{1/2}
\label{mholimit}
\end{equation}

For large values of $\tan\beta$ ($\tan\beta \rightarrow
\infty$) we obtain the limit

\begin{equation}
m_{h^0} \le 1.4044 m_Z = 128.062 \textrm{GeV/c}^2
\label{mhofinallimit}
\end{equation}

The upper bound on $m_{h^0}$ is raised by radiative corrections from
$m_Z$ to 128.062 $\textrm{GeV/c}^2$ for stop masses of order 1 TeV.

Considering radiative corrections, we can write, for the masses of
the neutral Higgs scalars:

\begin{equation}
m_{H^0}^2 = \frac{1}{2} m_H^2 \left[ 1 - \frac{m_W^2}{m_H^2} +
\frac{m_Z^2}{m_H^2} +\frac{\Delta_t}{m_H^2} + G_{rc} \right]
\label{mHoGrc}
\end{equation}

\begin{equation}
m_{h^0}^2 = \frac{1}{2} m_H^2 \left[ 1 - \frac{m_W^2}{m_H^2} +
\frac{m_Z^2}{m_H^2} + \frac{\Delta_t}{m_H^2} - G_{rc} \right]
\label{mhoGrc}
\end{equation}

\begin{eqnarray}
\lefteqn{
G_{rc} = [ \left( 1 - \frac{m_W^2}{m_H^2}
+ \frac{m_Z^2}{m_H^2} \right)^2 -
4 \left(\frac{m_Z^2}{m_H^2} \right)
\left( 1 - \frac{m_W^2}{m_H^2} \right)   
\left( \frac{1 - \tan^2\beta}{1 + \tan^2\beta}
\right)^2
}
\nonumber \\ & &
+ 2 \left(\frac{\Delta_t}{m_H^2} \right) 
 \left( 1 - \frac{m_W^2}{m_H^2}
- \frac{m_Z^2}{m_H^2} \right)
\left( \frac{1 - \tan^2\beta}{1 + \tan^2\beta}
\right)
 + \left( \frac{\Delta_t}{m_H^2}
 \right)^2 ]^{1/2}
\label{Grc}
\end{eqnarray}

With radiative corrections, the value of the $\alpha$ parameter 
is:

\begin{equation}
\tan\alpha = -\left[ \frac{1 + F_{rc}}{1 - F_{rc}} \right]^{1/2}
\label{tan_alpharc}
\end{equation}

\begin{equation}
F_{rc} = \frac{\left[\left(\frac{ 1 - \tan^2\beta }
{ 1 + \tan^2 \beta }\right)
\left( 1 - \frac{m_Z^2}{m_H^2} - \frac{m_W^2}{m_H^2} \right)
+ \frac{\Delta_t}{m_H^2} \right]}{G_{rc}}
\label{Frc}
\end{equation}

Additionally we have:

\begin{equation}
\sin2\alpha = - \frac{2 \tan\beta}{\left(1 + \tan^2\beta
\right)} \frac{\left( 1 - \frac{m_W^2}{m_H^2} + \frac{m_Z^2}
{m_H^2} \right)}{G_{rc}}
\label{sin2alpha}
\end{equation}

\section{Production of $h^0$ , $H^0$}
\begin{figure}
\begin{center}
{\includegraphics{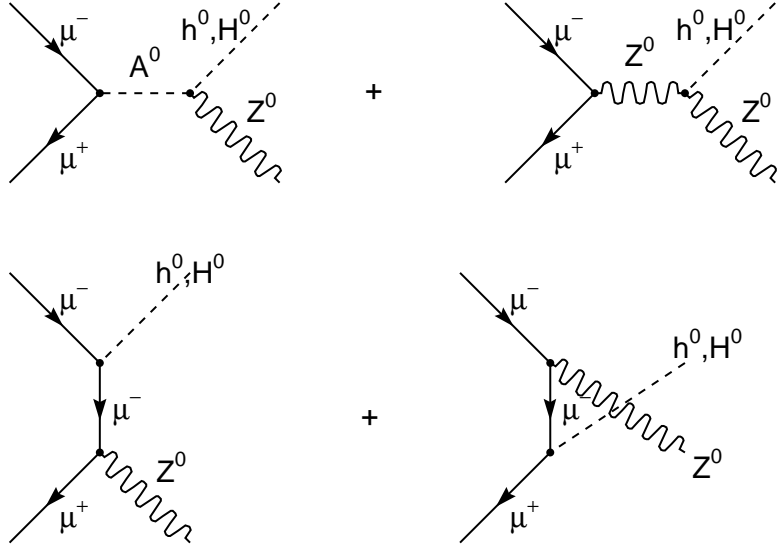}}
\caption{Feynman diagrams corresponding to the production
of $h^0$ or $H^0$ in the channel $\mu^- \mu^+ \rightarrow h^0 Z^0$.}
\label{mumu_hZ_fig}
\end{center}
\end{figure}

From the Feynman diagrams in Figure \ref{mumu_hZ_fig} and the
corresponding Feynman rules given in reference \cite{M_H}, we obtain
the differential cross section for the reaction  $\mu^- \mu^+ 
\rightarrow h^0 Z^0$ in the center of mass system

\begin{eqnarray}
\lefteqn{
\frac{d \sigma}{d \Omega} (\mu^- \mu^+ \rightarrow h^0 Z^0) =
\frac{1}{ 64 \pi^2 s^2} G_F^2 m_Z^4 \left| C_Z \right|^2 
\Lambda^{1/2} ( s, m_{h^0}^2, m_Z^2 )
}
\nonumber \\
& &
\left[  \left( g_A^\mu \right)^2 + \left( g_V^\mu \right)^2 
\right] \left[ 8 s m_Z^2 + \Lambda ( s, m_{h^0}^2, m_Z^2 ) 
sin^2 \theta \right]
\label{muon_hZ}
\end{eqnarray}
where 
\begin{eqnarray}
\lefteqn{
g_A^\mu = - \frac{1}{2}
}
\nonumber \\ & 
g_V^\mu = - \frac{1}{2} + 2 \sin^2 \theta_W
\label{gA,gV}
\end{eqnarray}

\begin{equation}
\Lambda (a, b, c) = a^2 + b^2 + c^2 - 2ab - 2ac - 2bc
\label{Lambdar}
\end{equation}
\begin{equation}
 C_Z  = \frac{\sin \left( \beta - \alpha \right)}
{\left( s - m_Z^2 + i m_Z \Gamma_Z \right)}
\label{CZ}
\end{equation}
$\Gamma_Z$ is the total decay width of the $Z^0$ and $\theta$
is the scattering angle in the center of mass system. 

The total cross section corresponding to $\mu^- \mu^+ \rightarrow h^0 
Z^0$ is obtained integrating  Equation (\ref{muon_hZ}):

\begin{eqnarray}
\lefteqn{
\sigma (\mu^- \mu^+ \rightarrow h^0 Z^0) = \frac{G_F^2 m_Z^4 
\left( \tan\beta - \tan\alpha \right)^2}
{48 \pi s^2 \left( 1 + \tan^2\alpha \right) \left( 1 + 
\tan^2 \beta \right)}
}
\nonumber \\ & &
\times
\frac{\left( 1 - 4 \sin^2 \theta_W + 8 \sin^4 \theta_W \right)}
{\left[ \left( s - m_Z^2 \right)^2 + m_Z^2 \Gamma_Z^2 \right]}
\left[ 12 s m_Z^2 + \Lambda ( s, m_{h^0}^2, m_Z^2 ) \right]
\nonumber \\ & &
\times\Lambda^{1/2} ( s, m_{h^0}^2, m_Z^2 ) \times 
\left( 3.8938 \times 10^{11} \right) \textrm{fb}
\label{sigma_muhz}
\end{eqnarray}

\begin{figure}
\begin{center}
\input{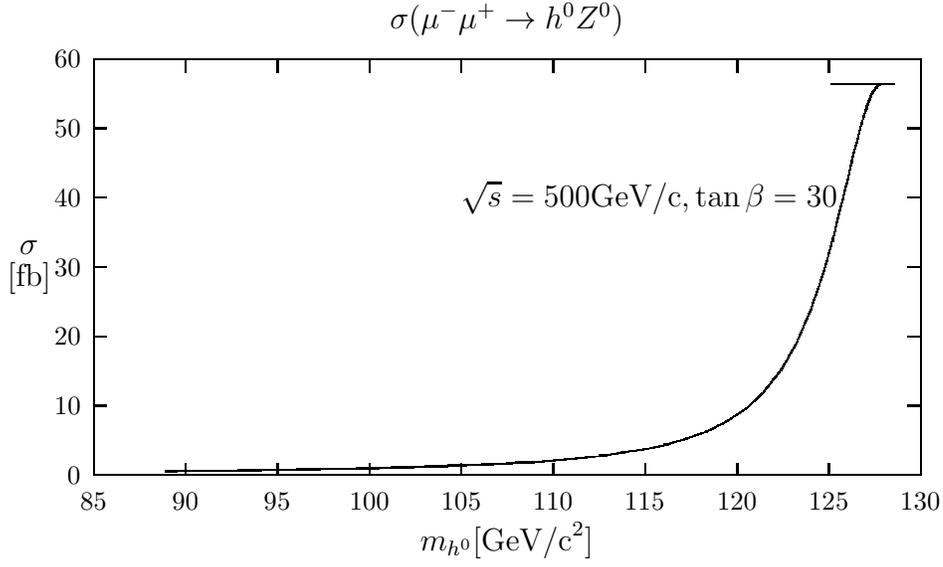}
\caption{Total cross section for the process
$\mu^- \mu^+ \rightarrow h^0 Z^0$ as a function
of $m_{h^0}$. We have taken
$\sqrt{s} = 500 \textrm{GeV/c}$ and $\tan\beta= 30$.}
\label{mumuhz_graph1}
\end{center}
\end{figure}

In Figures \ref{mumuhz_graph1} and \ref{mumuhz_graph2}
, the total cross section for 
$\mu^- \mu^+ \rightarrow h^0 Z^0$, is plotted as a function
$m_{h^0}$ for several values of  
$\sqrt{s}$ and $\tan\beta$. These total cross sections were
plotted considering the radiative corrections of the masses
given by Equations
(\ref{mhoGrc}), (\ref{Grc}), (\ref{tan_alpharc}) and (\ref{Frc}). 
According to these graphs, the total cross section becomes important
in the mass interval $118 \leq m_{h^0} \leq 128 [\textrm{GeV/c}^2]$.

\begin{figure}
\begin{center}
\setlength{\unitlength}{0.240900pt}
\ifx\plotpoint\undefined\newsavebox{\plotpoint}\fi
\sbox{\plotpoint}{\rule[-0.200pt]{0.400pt}{0.400pt}}%
\begin{picture}(1500,900)(0,0)
\font\gnuplot=cmr10 at 10pt
\gnuplot
\sbox{\plotpoint}{\rule[-0.200pt]{0.400pt}{0.400pt}}%
\put(161.0,123.0){\rule[-0.200pt]{4.818pt}{0.400pt}}
\put(141,123){\makebox(0,0)[r]{0}}
\put(1419.0,123.0){\rule[-0.200pt]{4.818pt}{0.400pt}}
\put(161.0,188.0){\rule[-0.200pt]{4.818pt}{0.400pt}}
\put(141,188){\makebox(0,0)[r]{10}}
\put(1419.0,188.0){\rule[-0.200pt]{4.818pt}{0.400pt}}
\put(161.0,254.0){\rule[-0.200pt]{4.818pt}{0.400pt}}
\put(141,254){\makebox(0,0)[r]{20}}
\put(1419.0,254.0){\rule[-0.200pt]{4.818pt}{0.400pt}}
\put(161.0,319.0){\rule[-0.200pt]{4.818pt}{0.400pt}}
\put(141,319){\makebox(0,0)[r]{30}}
\put(1419.0,319.0){\rule[-0.200pt]{4.818pt}{0.400pt}}
\put(161.0,385.0){\rule[-0.200pt]{4.818pt}{0.400pt}}
\put(141,385){\makebox(0,0)[r]{40}}
\put(1419.0,385.0){\rule[-0.200pt]{4.818pt}{0.400pt}}
\put(161.0,450.0){\rule[-0.200pt]{4.818pt}{0.400pt}}
\put(141,450){\makebox(0,0)[r]{50}}
\put(1419.0,450.0){\rule[-0.200pt]{4.818pt}{0.400pt}}
\put(161.0,515.0){\rule[-0.200pt]{4.818pt}{0.400pt}}
\put(141,515){\makebox(0,0)[r]{60}}
\put(1419.0,515.0){\rule[-0.200pt]{4.818pt}{0.400pt}}
\put(161.0,581.0){\rule[-0.200pt]{4.818pt}{0.400pt}}
\put(141,581){\makebox(0,0)[r]{70}}
\put(1419.0,581.0){\rule[-0.200pt]{4.818pt}{0.400pt}}
\put(161.0,646.0){\rule[-0.200pt]{4.818pt}{0.400pt}}
\put(141,646){\makebox(0,0)[r]{80}}
\put(1419.0,646.0){\rule[-0.200pt]{4.818pt}{0.400pt}}
\put(161.0,712.0){\rule[-0.200pt]{4.818pt}{0.400pt}}
\put(141,712){\makebox(0,0)[r]{90}}
\put(1419.0,712.0){\rule[-0.200pt]{4.818pt}{0.400pt}}
\put(161.0,777.0){\rule[-0.200pt]{4.818pt}{0.400pt}}
\put(141,777){\makebox(0,0)[r]{100}}
\put(1419.0,777.0){\rule[-0.200pt]{4.818pt}{0.400pt}}
\put(161.0,123.0){\rule[-0.200pt]{0.400pt}{4.818pt}}
\put(161,82){\makebox(0,0){85}}
\put(161.0,757.0){\rule[-0.200pt]{0.400pt}{4.818pt}}
\put(303.0,123.0){\rule[-0.200pt]{0.400pt}{4.818pt}}
\put(303,82){\makebox(0,0){90}}
\put(303.0,757.0){\rule[-0.200pt]{0.400pt}{4.818pt}}
\put(445.0,123.0){\rule[-0.200pt]{0.400pt}{4.818pt}}
\put(445,82){\makebox(0,0){95}}
\put(445.0,757.0){\rule[-0.200pt]{0.400pt}{4.818pt}}
\put(587.0,123.0){\rule[-0.200pt]{0.400pt}{4.818pt}}
\put(587,82){\makebox(0,0){100}}
\put(587.0,757.0){\rule[-0.200pt]{0.400pt}{4.818pt}}
\put(729.0,123.0){\rule[-0.200pt]{0.400pt}{4.818pt}}
\put(729,82){\makebox(0,0){105}}
\put(729.0,757.0){\rule[-0.200pt]{0.400pt}{4.818pt}}
\put(871.0,123.0){\rule[-0.200pt]{0.400pt}{4.818pt}}
\put(871,82){\makebox(0,0){110}}
\put(871.0,757.0){\rule[-0.200pt]{0.400pt}{4.818pt}}
\put(1013.0,123.0){\rule[-0.200pt]{0.400pt}{4.818pt}}
\put(1013,82){\makebox(0,0){115}}
\put(1013.0,757.0){\rule[-0.200pt]{0.400pt}{4.818pt}}
\put(1155.0,123.0){\rule[-0.200pt]{0.400pt}{4.818pt}}
\put(1155,82){\makebox(0,0){120}}
\put(1155.0,757.0){\rule[-0.200pt]{0.400pt}{4.818pt}}
\put(1297.0,123.0){\rule[-0.200pt]{0.400pt}{4.818pt}}
\put(1297,82){\makebox(0,0){125}}
\put(1297.0,757.0){\rule[-0.200pt]{0.400pt}{4.818pt}}
\put(1439.0,123.0){\rule[-0.200pt]{0.400pt}{4.818pt}}
\put(1439,82){\makebox(0,0){130}}
\put(1439.0,757.0){\rule[-0.200pt]{0.400pt}{4.818pt}}
\put(161.0,123.0){\rule[-0.200pt]{307.870pt}{0.400pt}}
\put(1439.0,123.0){\rule[-0.200pt]{0.400pt}{157.549pt}}
\put(161.0,777.0){\rule[-0.200pt]{307.870pt}{0.400pt}}
\put(40,450){\makebox(0,0){\shortstack{$\sigma$ \\ $[\textrm{fb}]$}}}
\put(800,21){\makebox(0,0){$m_{h^0} [\textrm{GeV/c}^2]$}}
\put(800,839){\makebox(0,0){$\sigma(\mu^- \mu^+ \rightarrow h^0 Z^0)$}}
\put(729,385){\makebox(0,0)[l]{$\sqrt{s} = 400 \textrm{GeV/c}, 
\tan\beta=50$}}
\put(161.0,123.0){\rule[-0.200pt]{0.400pt}{157.549pt}}
\put(1279,737){\makebox(0,0)[r]{ }}
\put(1299.0,737.0){\rule[-0.200pt]{24.090pt}{0.400pt}}
\put(274,125){\usebox{\plotpoint}}
\put(274,124.67){\rule{9.154pt}{0.400pt}}
\multiput(274.00,124.17)(19.000,1.000){2}{\rule{4.577pt}{0.400pt}}
\put(499,125.67){\rule{8.913pt}{0.400pt}}
\multiput(499.00,125.17)(18.500,1.000){2}{\rule{4.457pt}{0.400pt}}
\put(312.0,126.0){\rule[-0.200pt]{45.048pt}{0.400pt}}
\put(608,126.67){\rule{8.672pt}{0.400pt}}
\multiput(608.00,126.17)(18.000,1.000){2}{\rule{4.336pt}{0.400pt}}
\put(536.0,127.0){\rule[-0.200pt]{17.345pt}{0.400pt}}
\put(716,127.67){\rule{8.432pt}{0.400pt}}
\multiput(716.00,127.17)(17.500,1.000){2}{\rule{4.216pt}{0.400pt}}
\put(751,128.67){\rule{8.672pt}{0.400pt}}
\multiput(751.00,128.17)(18.000,1.000){2}{\rule{4.336pt}{0.400pt}}
\put(644.0,128.0){\rule[-0.200pt]{17.345pt}{0.400pt}}
\put(822,129.67){\rule{8.432pt}{0.400pt}}
\multiput(822.00,129.17)(17.500,1.000){2}{\rule{4.216pt}{0.400pt}}
\put(857,130.67){\rule{8.191pt}{0.400pt}}
\multiput(857.00,130.17)(17.000,1.000){2}{\rule{4.095pt}{0.400pt}}
\put(891,132.17){\rule{7.100pt}{0.400pt}}
\multiput(891.00,131.17)(20.264,2.000){2}{\rule{3.550pt}{0.400pt}}
\put(926,133.67){\rule{8.191pt}{0.400pt}}
\multiput(926.00,133.17)(17.000,1.000){2}{\rule{4.095pt}{0.400pt}}
\put(960,135.17){\rule{6.900pt}{0.400pt}}
\multiput(960.00,134.17)(19.679,2.000){2}{\rule{3.450pt}{0.400pt}}
\multiput(994.00,137.61)(7.383,0.447){3}{\rule{4.633pt}{0.108pt}}
\multiput(994.00,136.17)(24.383,3.000){2}{\rule{2.317pt}{0.400pt}}
\multiput(1028.00,140.61)(7.383,0.447){3}{\rule{4.633pt}{0.108pt}}
\multiput(1028.00,139.17)(24.383,3.000){2}{\rule{2.317pt}{0.400pt}}
\multiput(1062.00,143.60)(4.722,0.468){5}{\rule{3.400pt}{0.113pt}}
\multiput(1062.00,142.17)(25.943,4.000){2}{\rule{1.700pt}{0.400pt}}
\multiput(1095.00,147.59)(2.932,0.482){9}{\rule{2.300pt}{0.116pt}}
\multiput(1095.00,146.17)(28.226,6.000){2}{\rule{1.150pt}{0.400pt}}
\multiput(1128.00,153.59)(2.145,0.488){13}{\rule{1.750pt}{0.117pt}}
\multiput(1128.00,152.17)(29.368,8.000){2}{\rule{0.875pt}{0.400pt}}
\multiput(1161.00,161.58)(1.210,0.493){23}{\rule{1.054pt}{0.119pt}}
\multiput(1161.00,160.17)(28.813,13.000){2}{\rule{0.527pt}{0.400pt}}
\multiput(1192.00,174.58)(0.866,0.495){33}{\rule{0.789pt}{0.119pt}}
\multiput(1192.00,173.17)(29.363,18.000){2}{\rule{0.394pt}{0.400pt}}
\multiput(1223.00,192.58)(0.535,0.497){53}{\rule{0.529pt}{0.120pt}}
\multiput(1223.00,191.17)(28.903,28.000){2}{\rule{0.264pt}{0.400pt}}
\multiput(1253.58,220.00)(0.497,0.806){53}{\rule{0.120pt}{0.743pt}}
\multiput(1252.17,220.00)(28.000,43.458){2}{\rule{0.400pt}{0.371pt}}
\multiput(1281.58,265.00)(0.496,1.408){45}{\rule{0.120pt}{1.217pt}}
\multiput(1280.17,265.00)(24.000,64.475){2}{\rule{0.400pt}{0.608pt}}
\multiput(1305.58,332.00)(0.496,2.204){37}{\rule{0.119pt}{1.840pt}}
\multiput(1304.17,332.00)(20.000,83.181){2}{\rule{0.400pt}{0.920pt}}
\multiput(1325.58,419.00)(0.494,3.097){27}{\rule{0.119pt}{2.527pt}}
\multiput(1324.17,419.00)(15.000,85.756){2}{\rule{0.400pt}{1.263pt}}
\multiput(1340.58,510.00)(0.492,3.420){19}{\rule{0.118pt}{2.755pt}}
\multiput(1339.17,510.00)(11.000,67.283){2}{\rule{0.400pt}{1.377pt}}
\multiput(1351.59,583.00)(0.485,3.696){11}{\rule{0.117pt}{2.900pt}}
\multiput(1350.17,583.00)(7.000,42.981){2}{\rule{0.400pt}{1.450pt}}
\multiput(1358.59,632.00)(0.477,3.493){7}{\rule{0.115pt}{2.660pt}}
\multiput(1357.17,632.00)(5.000,26.479){2}{\rule{0.400pt}{1.330pt}}
\multiput(1363.61,664.00)(0.447,4.258){3}{\rule{0.108pt}{2.767pt}}
\multiput(1362.17,664.00)(3.000,14.258){2}{\rule{0.400pt}{1.383pt}}
\put(1366.17,684){\rule{0.400pt}{2.700pt}}
\multiput(1365.17,684.00)(2.000,7.396){2}{\rule{0.400pt}{1.350pt}}
\put(1368.17,697){\rule{0.400pt}{1.700pt}}
\multiput(1367.17,697.00)(2.000,4.472){2}{\rule{0.400pt}{0.850pt}}
\put(1370.17,705){\rule{0.400pt}{1.300pt}}
\multiput(1369.17,705.00)(2.000,3.302){2}{\rule{0.400pt}{0.650pt}}
\put(1371.67,711){\rule{0.400pt}{1.204pt}}
\multiput(1371.17,711.00)(1.000,2.500){2}{\rule{0.400pt}{0.602pt}}
\put(1372.67,716){\rule{0.400pt}{0.723pt}}
\multiput(1372.17,716.00)(1.000,1.500){2}{\rule{0.400pt}{0.361pt}}
\put(1373.67,719){\rule{0.400pt}{0.482pt}}
\multiput(1373.17,719.00)(1.000,1.000){2}{\rule{0.400pt}{0.241pt}}
\put(787.0,130.0){\rule[-0.200pt]{8.431pt}{0.400pt}}
\put(1374.67,723){\rule{0.400pt}{0.482pt}}
\multiput(1374.17,723.00)(1.000,1.000){2}{\rule{0.400pt}{0.241pt}}
\put(1375.0,721.0){\rule[-0.200pt]{0.400pt}{0.482pt}}
\put(1376,725.67){\rule{0.241pt}{0.400pt}}
\multiput(1376.00,725.17)(0.500,1.000){2}{\rule{0.120pt}{0.400pt}}
\put(1376.0,725.0){\usebox{\plotpoint}}
\put(1377,727){\usebox{\plotpoint}}
\put(1377,727.67){\rule{0.241pt}{0.400pt}}
\multiput(1377.00,727.17)(0.500,1.000){2}{\rule{0.120pt}{0.400pt}}
\put(1377.0,727.0){\usebox{\plotpoint}}
\put(1378,729){\usebox{\plotpoint}}
\put(1378,729){\usebox{\plotpoint}}
\put(1378.0,729.0){\usebox{\plotpoint}}
\put(1378.0,730.0){\usebox{\plotpoint}}
\put(1379.0,730.0){\usebox{\plotpoint}}
\put(1379.0,731.0){\usebox{\plotpoint}}
\put(1380.0,731.0){\usebox{\plotpoint}}
\put(1380.0,732.0){\rule[-0.200pt]{0.482pt}{0.400pt}}
\put(1382.0,732.0){\usebox{\plotpoint}}
\end{picture}
\caption{Total cross section for the process
$\mu^- \mu^+ \rightarrow h^0 Z^0$ as a function
of $m_{h^0}$. We have taken
$\sqrt{s} = 400 \textrm{GeV/c}$ and $\tan\beta= 50$.}
\label{mumuhz_graph2}
\end{center}
\end{figure}

The Standard Model cross section is:

\begin{eqnarray}
\sigma (\mu^- \mu^+ \rightarrow h_{SM}^0 Z^0)_{SM} = \frac{G_F^2 m_Z^4
}
{48 \pi s^2}
\frac{\left( 1 - 4 \sin^2 \theta_W + 8 \sin^4 \theta_W \right)}
{\left[ \left( s - m_Z^2 \right)^2 + m_Z^2 \Gamma_Z^2 \right]}
\nonumber \\
\times \left[ 12 s m_Z^2 + 
\Lambda ( s, m_{h_{SM}^0}^2, m_Z^2 ) \right]
\nonumber \\ 
\times\Lambda^{1/2} ( s, m_{h_{SM}^0}^2, m_Z^2 ) \times
\left( 3.8938 \times 10^{11} \right) \textrm{fb}
\label{sigmaSM_muhz}
\end{eqnarray}

\noindent where $h_{SM}$ is the Standard Model Higgs boson.
 
\begin{figure}
\begin{center}
\input{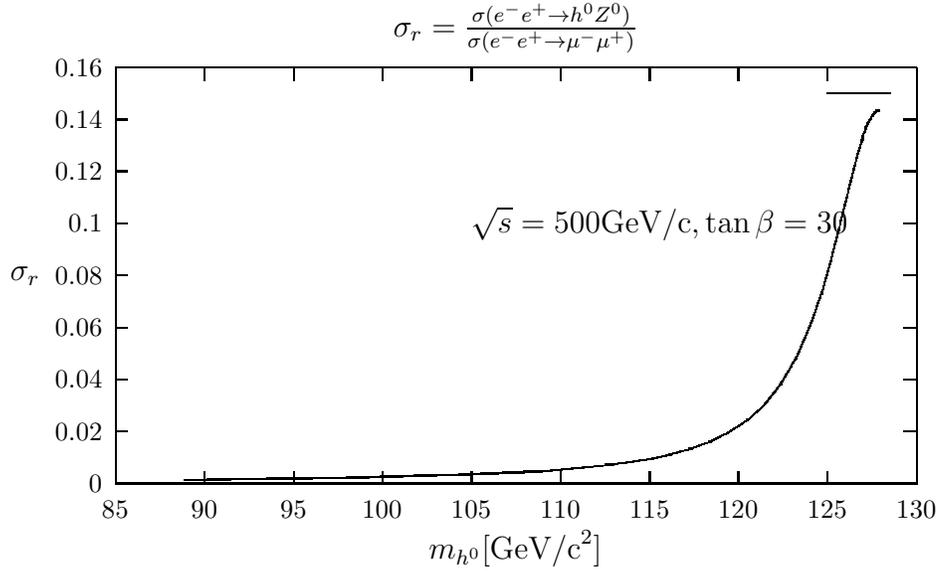}
\caption{Total cross section
$\sigma(e^- e^+ \rightarrow h^0 Z^0)$ compared with the
cross section $\sigma(e^- e^+ \rightarrow
\mu^- \mu^+)$ as a function
of $m_{h^0}$. We have taken
$\sqrt{s} = 500 \textrm{GeV/c}$ and $\tan\beta= 30$.}
\label{annihi1}
\end{center}
\end{figure}

The production  cross section corresponding to
$e^- e^+ \rightarrow h^0 Z^0$
is given by an expression identical to (\ref{sigma_muhz}).
In terms of the cross section
\newline
$\sigma \left( e^- e^+ \rightarrow \mu^- \mu^+ \right)$ we can write:
\begin{eqnarray}
\lefteqn{
\frac{\sigma \left( e^- e^+ \rightarrow h^0 Z^0 \right) }
{\sigma \left( e^- e^+ \rightarrow \mu^- \mu^+ \right)} =
\frac{1}{128 s} \frac{\left( \tan\beta - \tan\alpha \right)
^2}{\left( 1 + \tan^2\alpha \right) \left( 1 + 
\tan^2\beta \right) } \Lambda^{1/2} ( s, m_{h^0}^2, m_Z^{2} ) 
}
\nonumber \\ & &
\times
\frac{ \left( 1 - 4 \sin^2\theta_W
+ 8 \sin^4\theta_W \right)}{\sin^4\theta_W 
\left( 1 - \sin^2\theta_W \right)^2}
 \frac{\left[ 12 s m_Z^2 + \Lambda ( s, m_{h^0}^2, m_Z^2 )
\right] }
{\left[ \left( s - m_Z^2 \right)^2 + m_Z^2 \Gamma_Z^2
\right]}
\label{annihilation1}
\end{eqnarray} 
Equation (\ref{annihilation1}) is plotted in Figure 
\ref{annihi1}, as a function of
 $m_{h^0}$ for $\sqrt{s} = 500 \textrm{GeV/c}$ and $\tan\beta= 30$.

The total cross section corresponding to 
$\mu^- \mu^+ \rightarrow H^0 Z^0$ is obtained from Equation
(\ref{sigma_muhz}) replacing $\left(\tan\beta - \tan\alpha
\right)^2$ by $\left( 1 + \tan\beta \tan\alpha \right)^2$
in the numerator and $m_{h^0}$ by $m_{H^0}$. This production 
cross section is plotted in Figures \ref{mumu_HZ1},
\ref{mumu_HZrc} as a 
function of $m_{H^0}$ for $\sqrt{s} = 500 \textrm{GeV/c}$ and 
$\tan\beta = 30$, without and with mass radiative corrections, 
respectively.
In Figure \ref{muhzee_figure} we show the ratio
between the production cross section
$\sigma \left(e^- e^+ \rightarrow H^0 Z^0 \right)$
and the cross section 
$\sigma \left(e^- e^+ \rightarrow \mu^- \mu^+ \right)$ in terms
of $m_{H^0}$. The radiatively corrected masses total cross section is 
shown 
in Figure \ref{muhzeerc_figure}. Figures \ref{mumu_HZrc} and
\ref{muhzeerc_figure} show the importance of the radiative corrections
of the masses
in the processes $\mu^- \mu^+ \rightarrow H^0 Z^0$ and $e^- e^+ 
\rightarrow H^0 Z^0$. 

\begin{figure}
\begin{center}
\setlength{\unitlength}{0.240900pt}
\ifx\plotpoint\undefined\newsavebox{\plotpoint}\fi
\sbox{\plotpoint}{\rule[-0.200pt]{0.400pt}{0.400pt}}%
\begin{picture}(1500,900)(0,0)
\font\gnuplot=cmr10 at 10pt
\gnuplot
\sbox{\plotpoint}{\rule[-0.200pt]{0.400pt}{0.400pt}}%
\put(141.0,123.0){\rule[-0.200pt]{4.818pt}{0.400pt}}
\put(121,123){\makebox(0,0)[r]{0}}
\put(1419.0,123.0){\rule[-0.200pt]{4.818pt}{0.400pt}}
\put(141.0,205.0){\rule[-0.200pt]{4.818pt}{0.400pt}}
\put(121,205){\makebox(0,0)[r]{5}}
\put(1419.0,205.0){\rule[-0.200pt]{4.818pt}{0.400pt}}
\put(141.0,287.0){\rule[-0.200pt]{4.818pt}{0.400pt}}
\put(121,287){\makebox(0,0)[r]{10}}
\put(1419.0,287.0){\rule[-0.200pt]{4.818pt}{0.400pt}}
\put(141.0,368.0){\rule[-0.200pt]{4.818pt}{0.400pt}}
\put(121,368){\makebox(0,0)[r]{15}}
\put(1419.0,368.0){\rule[-0.200pt]{4.818pt}{0.400pt}}
\put(141.0,450.0){\rule[-0.200pt]{4.818pt}{0.400pt}}
\put(121,450){\makebox(0,0)[r]{20}}
\put(1419.0,450.0){\rule[-0.200pt]{4.818pt}{0.400pt}}
\put(141.0,532.0){\rule[-0.200pt]{4.818pt}{0.400pt}}
\put(121,532){\makebox(0,0)[r]{25}}
\put(1419.0,532.0){\rule[-0.200pt]{4.818pt}{0.400pt}}
\put(141.0,614.0){\rule[-0.200pt]{4.818pt}{0.400pt}}
\put(121,614){\makebox(0,0)[r]{30}}
\put(1419.0,614.0){\rule[-0.200pt]{4.818pt}{0.400pt}}
\put(141.0,695.0){\rule[-0.200pt]{4.818pt}{0.400pt}}
\put(121,695){\makebox(0,0)[r]{35}}
\put(1419.0,695.0){\rule[-0.200pt]{4.818pt}{0.400pt}}
\put(141.0,777.0){\rule[-0.200pt]{4.818pt}{0.400pt}}
\put(121,777){\makebox(0,0)[r]{40}}
\put(1419.0,777.0){\rule[-0.200pt]{4.818pt}{0.400pt}}
\put(141.0,123.0){\rule[-0.200pt]{0.400pt}{4.818pt}}
\put(141,82){\makebox(0,0){50}}
\put(141.0,757.0){\rule[-0.200pt]{0.400pt}{4.818pt}}
\put(401.0,123.0){\rule[-0.200pt]{0.400pt}{4.818pt}}
\put(401,82){\makebox(0,0){100}}
\put(401.0,757.0){\rule[-0.200pt]{0.400pt}{4.818pt}}
\put(660.0,123.0){\rule[-0.200pt]{0.400pt}{4.818pt}}
\put(660,82){\makebox(0,0){150}}
\put(660.0,757.0){\rule[-0.200pt]{0.400pt}{4.818pt}}
\put(920.0,123.0){\rule[-0.200pt]{0.400pt}{4.818pt}}
\put(920,82){\makebox(0,0){200}}
\put(920.0,757.0){\rule[-0.200pt]{0.400pt}{4.818pt}}
\put(1179.0,123.0){\rule[-0.200pt]{0.400pt}{4.818pt}}
\put(1179,82){\makebox(0,0){250}}
\put(1179.0,757.0){\rule[-0.200pt]{0.400pt}{4.818pt}}
\put(1439.0,123.0){\rule[-0.200pt]{0.400pt}{4.818pt}}
\put(1439,82){\makebox(0,0){300}}
\put(1439.0,757.0){\rule[-0.200pt]{0.400pt}{4.818pt}}
\put(141.0,123.0){\rule[-0.200pt]{312.688pt}{0.400pt}}
\put(1439.0,123.0){\rule[-0.200pt]{0.400pt}{157.549pt}}
\put(141.0,777.0){\rule[-0.200pt]{312.688pt}{0.400pt}}
\put(40,450){\makebox(0,0){\shortstack{$\sigma$ \\ $[\textrm{fb}]$}}}
\put(790,21){\makebox(0,0){$m_{H^0} [\textrm{GeV/c}^2]$}}
\put(790,839){\makebox(0,0){$\sigma(\mu^- \mu^+ \rightarrow H^0 Z^0)$}}
\put(660,450){\makebox(0,0)[l]{$\sqrt{s} = 500 \textrm{GeV/c}, 
\tan\beta=30$}}
\put(141.0,123.0){\rule[-0.200pt]{0.400pt}{157.549pt}}
\put(1279,737){\makebox(0,0)[r]{ }}
\put(1299.0,737.0){\rule[-0.200pt]{24.090pt}{0.400pt}}
\put(366,760){\usebox{\plotpoint}}
\put(366.17,655){\rule{0.400pt}{21.100pt}}
\multiput(365.17,716.21)(2.000,-61.206){2}{\rule{0.400pt}{10.550pt}}
\multiput(368.60,608.92)(0.468,-15.981){5}{\rule{0.113pt}{11.100pt}}
\multiput(367.17,631.96)(4.000,-86.961){2}{\rule{0.400pt}{5.550pt}}
\multiput(372.60,503.07)(0.468,-14.518){5}{\rule{0.113pt}{10.100pt}}
\multiput(371.17,524.04)(4.000,-79.037){2}{\rule{0.400pt}{5.050pt}}
\multiput(376.59,417.69)(0.477,-8.948){7}{\rule{0.115pt}{6.580pt}}
\multiput(375.17,431.34)(5.000,-67.343){2}{\rule{0.400pt}{3.290pt}}
\multiput(381.59,343.33)(0.477,-6.721){7}{\rule{0.115pt}{4.980pt}}
\multiput(380.17,353.66)(5.000,-50.664){2}{\rule{0.400pt}{2.490pt}}
\multiput(386.59,290.41)(0.482,-3.926){9}{\rule{0.116pt}{3.033pt}}
\multiput(385.17,296.70)(6.000,-37.704){2}{\rule{0.400pt}{1.517pt}}
\multiput(392.59,249.73)(0.482,-2.841){9}{\rule{0.116pt}{2.233pt}}
\multiput(391.17,254.36)(6.000,-27.365){2}{\rule{0.400pt}{1.117pt}}
\multiput(398.59,220.50)(0.482,-1.937){9}{\rule{0.116pt}{1.567pt}}
\multiput(397.17,223.75)(6.000,-18.748){2}{\rule{0.400pt}{0.783pt}}
\multiput(404.59,199.88)(0.482,-1.485){9}{\rule{0.116pt}{1.233pt}}
\multiput(403.17,202.44)(6.000,-14.440){2}{\rule{0.400pt}{0.617pt}}
\multiput(410.59,184.26)(0.482,-1.033){9}{\rule{0.116pt}{0.900pt}}
\multiput(409.17,186.13)(6.000,-10.132){2}{\rule{0.400pt}{0.450pt}}
\multiput(416.59,172.82)(0.482,-0.852){9}{\rule{0.116pt}{0.767pt}}
\multiput(415.17,174.41)(6.000,-8.409){2}{\rule{0.400pt}{0.383pt}}
\multiput(422.59,163.65)(0.482,-0.581){9}{\rule{0.116pt}{0.567pt}}
\multiput(421.17,164.82)(6.000,-5.824){2}{\rule{0.400pt}{0.283pt}}
\multiput(428.00,157.93)(0.581,-0.482){9}{\rule{0.567pt}{0.116pt}}
\multiput(428.00,158.17)(5.824,-6.000){2}{\rule{0.283pt}{0.400pt}}
\multiput(435.00,151.94)(0.774,-0.468){5}{\rule{0.700pt}{0.113pt}}
\multiput(435.00,152.17)(4.547,-4.000){2}{\rule{0.350pt}{0.400pt}}
\multiput(441.00,147.94)(0.774,-0.468){5}{\rule{0.700pt}{0.113pt}}
\multiput(441.00,148.17)(4.547,-4.000){2}{\rule{0.350pt}{0.400pt}}
\multiput(447.00,143.95)(1.355,-0.447){3}{\rule{1.033pt}{0.108pt}}
\multiput(447.00,144.17)(4.855,-3.000){2}{\rule{0.517pt}{0.400pt}}
\put(454,140.17){\rule{1.300pt}{0.400pt}}
\multiput(454.00,141.17)(3.302,-2.000){2}{\rule{0.650pt}{0.400pt}}
\put(460,138.17){\rule{1.300pt}{0.400pt}}
\multiput(460.00,139.17)(3.302,-2.000){2}{\rule{0.650pt}{0.400pt}}
\put(466,136.17){\rule{1.300pt}{0.400pt}}
\multiput(466.00,137.17)(3.302,-2.000){2}{\rule{0.650pt}{0.400pt}}
\put(472,134.67){\rule{1.686pt}{0.400pt}}
\multiput(472.00,135.17)(3.500,-1.000){2}{\rule{0.843pt}{0.400pt}}
\put(479,133.67){\rule{1.445pt}{0.400pt}}
\multiput(479.00,134.17)(3.000,-1.000){2}{\rule{0.723pt}{0.400pt}}
\put(485,132.17){\rule{1.300pt}{0.400pt}}
\multiput(485.00,133.17)(3.302,-2.000){2}{\rule{0.650pt}{0.400pt}}
\put(497,130.67){\rule{1.445pt}{0.400pt}}
\multiput(497.00,131.17)(3.000,-1.000){2}{\rule{0.723pt}{0.400pt}}
\put(503,129.67){\rule{1.686pt}{0.400pt}}
\multiput(503.00,130.17)(3.500,-1.000){2}{\rule{0.843pt}{0.400pt}}
\put(510,128.67){\rule{1.445pt}{0.400pt}}
\multiput(510.00,129.17)(3.000,-1.000){2}{\rule{0.723pt}{0.400pt}}
\put(491.0,132.0){\rule[-0.200pt]{1.445pt}{0.400pt}}
\put(522,127.67){\rule{1.445pt}{0.400pt}}
\multiput(522.00,128.17)(3.000,-1.000){2}{\rule{0.723pt}{0.400pt}}
\put(516.0,129.0){\rule[-0.200pt]{1.445pt}{0.400pt}}
\put(540,126.67){\rule{1.445pt}{0.400pt}}
\multiput(540.00,127.17)(3.000,-1.000){2}{\rule{0.723pt}{0.400pt}}
\put(528.0,128.0){\rule[-0.200pt]{2.891pt}{0.400pt}}
\put(559,125.67){\rule{1.445pt}{0.400pt}}
\multiput(559.00,126.17)(3.000,-1.000){2}{\rule{0.723pt}{0.400pt}}
\put(546.0,127.0){\rule[-0.200pt]{3.132pt}{0.400pt}}
\put(589,124.67){\rule{1.445pt}{0.400pt}}
\multiput(589.00,125.17)(3.000,-1.000){2}{\rule{0.723pt}{0.400pt}}
\put(565.0,126.0){\rule[-0.200pt]{5.782pt}{0.400pt}}
\put(642,123.67){\rule{1.445pt}{0.400pt}}
\multiput(642.00,124.17)(3.000,-1.000){2}{\rule{0.723pt}{0.400pt}}
\put(595.0,125.0){\rule[-0.200pt]{11.322pt}{0.400pt}}
\put(782,122.67){\rule{1.204pt}{0.400pt}}
\multiput(782.00,123.17)(2.500,-1.000){2}{\rule{0.602pt}{0.400pt}}
\put(648.0,124.0){\rule[-0.200pt]{32.281pt}{0.400pt}}
\put(787.0,123.0){\rule[-0.200pt]{153.694pt}{0.400pt}}
\end{picture}
\caption{Total cross section for the process
$\mu^- \mu^+ \rightarrow H^0 Z^0$ as a function
of $m_{H^0}$. The radiative corrections of the masses were not 
taken into account.
We have taken
$\sqrt{s} = 500 \textrm{GeV/c}$ and $\tan\beta= 30$.}
\label{mumu_HZ1}
\end{center}
\end{figure}
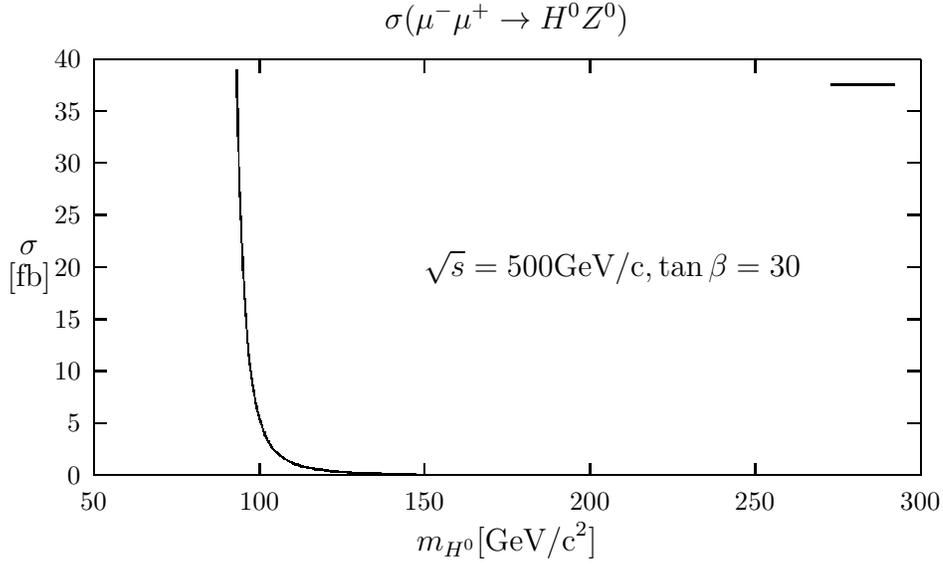

\begin{figure}
\begin{center}
\setlength{\unitlength}{0.240900pt}
\ifx\plotpoint\undefined\newsavebox{\plotpoint}\fi
\sbox{\plotpoint}{\rule[-0.200pt]{0.400pt}{0.400pt}}%
\begin{picture}(1500,900)(0,0)
\font\gnuplot=cmr10 at 10pt
\gnuplot
\sbox{\plotpoint}{\rule[-0.200pt]{0.400pt}{0.400pt}}%
\put(141.0,123.0){\rule[-0.200pt]{4.818pt}{0.400pt}}
\put(121,123){\makebox(0,0)[r]{0}}
\put(1419.0,123.0){\rule[-0.200pt]{4.818pt}{0.400pt}}
\put(141.0,232.0){\rule[-0.200pt]{4.818pt}{0.400pt}}
\put(121,232){\makebox(0,0)[r]{10}}
\put(1419.0,232.0){\rule[-0.200pt]{4.818pt}{0.400pt}}
\put(141.0,341.0){\rule[-0.200pt]{4.818pt}{0.400pt}}
\put(121,341){\makebox(0,0)[r]{20}}
\put(1419.0,341.0){\rule[-0.200pt]{4.818pt}{0.400pt}}
\put(141.0,450.0){\rule[-0.200pt]{4.818pt}{0.400pt}}
\put(121,450){\makebox(0,0)[r]{30}}
\put(1419.0,450.0){\rule[-0.200pt]{4.818pt}{0.400pt}}
\put(141.0,559.0){\rule[-0.200pt]{4.818pt}{0.400pt}}
\put(121,559){\makebox(0,0)[r]{40}}
\put(1419.0,559.0){\rule[-0.200pt]{4.818pt}{0.400pt}}
\put(141.0,668.0){\rule[-0.200pt]{4.818pt}{0.400pt}}
\put(121,668){\makebox(0,0)[r]{50}}
\put(1419.0,668.0){\rule[-0.200pt]{4.818pt}{0.400pt}}
\put(141.0,777.0){\rule[-0.200pt]{4.818pt}{0.400pt}}
\put(121,777){\makebox(0,0)[r]{60}}
\put(1419.0,777.0){\rule[-0.200pt]{4.818pt}{0.400pt}}
\put(141.0,123.0){\rule[-0.200pt]{0.400pt}{4.818pt}}
\put(141,82){\makebox(0,0){120}}
\put(141.0,757.0){\rule[-0.200pt]{0.400pt}{4.818pt}}
\put(326.0,123.0){\rule[-0.200pt]{0.400pt}{4.818pt}}
\put(326,82){\makebox(0,0){140}}
\put(326.0,757.0){\rule[-0.200pt]{0.400pt}{4.818pt}}
\put(512.0,123.0){\rule[-0.200pt]{0.400pt}{4.818pt}}
\put(512,82){\makebox(0,0){160}}
\put(512.0,757.0){\rule[-0.200pt]{0.400pt}{4.818pt}}
\put(697.0,123.0){\rule[-0.200pt]{0.400pt}{4.818pt}}
\put(697,82){\makebox(0,0){180}}
\put(697.0,757.0){\rule[-0.200pt]{0.400pt}{4.818pt}}
\put(883.0,123.0){\rule[-0.200pt]{0.400pt}{4.818pt}}
\put(883,82){\makebox(0,0){200}}
\put(883.0,757.0){\rule[-0.200pt]{0.400pt}{4.818pt}}
\put(1068.0,123.0){\rule[-0.200pt]{0.400pt}{4.818pt}}
\put(1068,82){\makebox(0,0){220}}
\put(1068.0,757.0){\rule[-0.200pt]{0.400pt}{4.818pt}}
\put(1254.0,123.0){\rule[-0.200pt]{0.400pt}{4.818pt}}
\put(1254,82){\makebox(0,0){240}}
\put(1254.0,757.0){\rule[-0.200pt]{0.400pt}{4.818pt}}
\put(1439.0,123.0){\rule[-0.200pt]{0.400pt}{4.818pt}}
\put(1439,82){\makebox(0,0){260}}
\put(1439.0,757.0){\rule[-0.200pt]{0.400pt}{4.818pt}}
\put(141.0,123.0){\rule[-0.200pt]{312.688pt}{0.400pt}}
\put(1439.0,123.0){\rule[-0.200pt]{0.400pt}{157.549pt}}
\put(141.0,777.0){\rule[-0.200pt]{312.688pt}{0.400pt}}
\put(40,450){\makebox(0,0){\shortstack{$\sigma$ \\ $[\textrm{fb}]$}}}
\put(790,21){\makebox(0,0){$m_{H^0} [\textrm{GeV/c}^2]$}}
\put(790,839){\makebox(0,0){$\sigma(\mu^- \mu^+ \rightarrow H^0 Z^0)$}}
\put(419,341){\makebox(0,0)[l]{$\sqrt{s} = 500 \textrm{GeV/c}, \tan\beta = 
30$}}
\put(141.0,123.0){\rule[-0.200pt]{0.400pt}{157.549pt}}
\put(1279,737){\makebox(0,0)[r]{ }}
\put(1299.0,737.0){\rule[-0.200pt]{24.090pt}{0.400pt}}
\put(217,731){\usebox{\plotpoint}}
\put(217,731){\usebox{\plotpoint}}
\put(217.0,730.0){\usebox{\plotpoint}}
\put(217.0,730.0){\usebox{\plotpoint}}
\put(218,723.67){\rule{0.241pt}{0.400pt}}
\multiput(218.00,724.17)(0.500,-1.000){2}{\rule{0.120pt}{0.400pt}}
\put(218.0,725.0){\rule[-0.200pt]{0.400pt}{1.204pt}}
\put(218.67,718){\rule{0.400pt}{0.482pt}}
\multiput(218.17,719.00)(1.000,-1.000){2}{\rule{0.400pt}{0.241pt}}
\put(219.0,720.0){\rule[-0.200pt]{0.400pt}{0.964pt}}
\put(219.67,710){\rule{0.400pt}{0.723pt}}
\multiput(219.17,711.50)(1.000,-1.500){2}{\rule{0.400pt}{0.361pt}}
\put(220.0,713.0){\rule[-0.200pt]{0.400pt}{1.204pt}}
\put(220.67,702){\rule{0.400pt}{0.964pt}}
\multiput(220.17,704.00)(1.000,-2.000){2}{\rule{0.400pt}{0.482pt}}
\put(221.0,706.0){\rule[-0.200pt]{0.400pt}{0.964pt}}
\put(221.67,689){\rule{0.400pt}{1.686pt}}
\multiput(221.17,692.50)(1.000,-3.500){2}{\rule{0.400pt}{0.843pt}}
\put(222.67,680){\rule{0.400pt}{2.168pt}}
\multiput(222.17,684.50)(1.000,-4.500){2}{\rule{0.400pt}{1.084pt}}
\put(223.67,668){\rule{0.400pt}{2.891pt}}
\multiput(223.17,674.00)(1.000,-6.000){2}{\rule{0.400pt}{1.445pt}}
\put(224.67,653){\rule{0.400pt}{3.614pt}}
\multiput(224.17,660.50)(1.000,-7.500){2}{\rule{0.400pt}{1.807pt}}
\put(226.17,633){\rule{0.400pt}{4.100pt}}
\multiput(225.17,644.49)(2.000,-11.490){2}{\rule{0.400pt}{2.050pt}}
\put(228.17,607){\rule{0.400pt}{5.300pt}}
\multiput(227.17,622.00)(2.000,-15.000){2}{\rule{0.400pt}{2.650pt}}
\put(230.17,573){\rule{0.400pt}{6.900pt}}
\multiput(229.17,592.68)(2.000,-19.679){2}{\rule{0.400pt}{3.450pt}}
\multiput(232.61,548.79)(0.447,-9.393){3}{\rule{0.108pt}{5.833pt}}
\multiput(231.17,560.89)(3.000,-30.893){2}{\rule{0.400pt}{2.917pt}}
\multiput(235.60,508.83)(0.468,-7.207){5}{\rule{0.113pt}{5.100pt}}
\multiput(234.17,519.41)(4.000,-39.415){2}{\rule{0.400pt}{2.550pt}}
\multiput(239.59,460.99)(0.477,-6.165){7}{\rule{0.115pt}{4.580pt}}
\multiput(238.17,470.49)(5.000,-46.494){2}{\rule{0.400pt}{2.290pt}}
\multiput(244.59,408.36)(0.482,-4.921){9}{\rule{0.116pt}{3.767pt}}
\multiput(243.17,416.18)(6.000,-47.182){2}{\rule{0.400pt}{1.883pt}}
\multiput(250.59,356.72)(0.485,-3.772){11}{\rule{0.117pt}{2.957pt}}
\multiput(249.17,362.86)(7.000,-43.862){2}{\rule{0.400pt}{1.479pt}}
\multiput(257.59,308.39)(0.485,-3.239){11}{\rule{0.117pt}{2.557pt}}
\multiput(256.17,313.69)(7.000,-37.693){2}{\rule{0.400pt}{1.279pt}}
\multiput(264.59,269.50)(0.489,-1.893){15}{\rule{0.118pt}{1.567pt}}
\multiput(263.17,272.75)(9.000,-29.748){2}{\rule{0.400pt}{0.783pt}}
\multiput(273.59,237.79)(0.489,-1.485){15}{\rule{0.118pt}{1.256pt}}
\multiput(272.17,240.39)(9.000,-23.394){2}{\rule{0.400pt}{0.628pt}}
\multiput(282.59,213.08)(0.489,-1.077){15}{\rule{0.118pt}{0.944pt}}
\multiput(281.17,215.04)(9.000,-17.040){2}{\rule{0.400pt}{0.472pt}}
\multiput(291.59,194.82)(0.489,-0.844){15}{\rule{0.118pt}{0.767pt}}
\multiput(290.17,196.41)(9.000,-13.409){2}{\rule{0.400pt}{0.383pt}}
\multiput(300.58,180.76)(0.491,-0.547){17}{\rule{0.118pt}{0.540pt}}
\multiput(299.17,181.88)(10.000,-9.879){2}{\rule{0.400pt}{0.270pt}}
\multiput(310.00,170.93)(0.626,-0.488){13}{\rule{0.600pt}{0.117pt}}
\multiput(310.00,171.17)(8.755,-8.000){2}{\rule{0.300pt}{0.400pt}}
\multiput(320.00,162.93)(0.721,-0.485){11}{\rule{0.671pt}{0.117pt}}
\multiput(320.00,163.17)(8.606,-7.000){2}{\rule{0.336pt}{0.400pt}}
\multiput(330.00,155.93)(1.044,-0.477){7}{\rule{0.900pt}{0.115pt}}
\multiput(330.00,156.17)(8.132,-5.000){2}{\rule{0.450pt}{0.400pt}}
\multiput(340.00,150.93)(1.044,-0.477){7}{\rule{0.900pt}{0.115pt}}
\multiput(340.00,151.17)(8.132,-5.000){2}{\rule{0.450pt}{0.400pt}}
\multiput(350.00,145.95)(2.025,-0.447){3}{\rule{1.433pt}{0.108pt}}
\multiput(350.00,146.17)(7.025,-3.000){2}{\rule{0.717pt}{0.400pt}}
\multiput(360.00,142.95)(2.025,-0.447){3}{\rule{1.433pt}{0.108pt}}
\multiput(360.00,143.17)(7.025,-3.000){2}{\rule{0.717pt}{0.400pt}}
\put(370,139.17){\rule{2.300pt}{0.400pt}}
\multiput(370.00,140.17)(6.226,-2.000){2}{\rule{1.150pt}{0.400pt}}
\put(381,137.17){\rule{2.100pt}{0.400pt}}
\multiput(381.00,138.17)(5.641,-2.000){2}{\rule{1.050pt}{0.400pt}}
\put(391,135.17){\rule{2.100pt}{0.400pt}}
\multiput(391.00,136.17)(5.641,-2.000){2}{\rule{1.050pt}{0.400pt}}
\put(401,133.67){\rule{2.650pt}{0.400pt}}
\multiput(401.00,134.17)(5.500,-1.000){2}{\rule{1.325pt}{0.400pt}}
\put(412,132.67){\rule{2.409pt}{0.400pt}}
\multiput(412.00,133.17)(5.000,-1.000){2}{\rule{1.204pt}{0.400pt}}
\put(422,131.67){\rule{2.409pt}{0.400pt}}
\multiput(422.00,132.17)(5.000,-1.000){2}{\rule{1.204pt}{0.400pt}}
\put(432,130.67){\rule{2.409pt}{0.400pt}}
\multiput(432.00,131.17)(5.000,-1.000){2}{\rule{1.204pt}{0.400pt}}
\put(442,129.67){\rule{2.650pt}{0.400pt}}
\multiput(442.00,130.17)(5.500,-1.000){2}{\rule{1.325pt}{0.400pt}}
\put(222.0,696.0){\rule[-0.200pt]{0.400pt}{1.445pt}}
\put(463,128.67){\rule{2.409pt}{0.400pt}}
\multiput(463.00,129.17)(5.000,-1.000){2}{\rule{1.204pt}{0.400pt}}
\put(473,127.67){\rule{2.650pt}{0.400pt}}
\multiput(473.00,128.17)(5.500,-1.000){2}{\rule{1.325pt}{0.400pt}}
\put(453.0,130.0){\rule[-0.200pt]{2.409pt}{0.400pt}}
\put(504,126.67){\rule{2.650pt}{0.400pt}}
\multiput(504.00,127.17)(5.500,-1.000){2}{\rule{1.325pt}{0.400pt}}
\put(484.0,128.0){\rule[-0.200pt]{4.818pt}{0.400pt}}
\put(535,125.67){\rule{2.409pt}{0.400pt}}
\multiput(535.00,126.17)(5.000,-1.000){2}{\rule{1.204pt}{0.400pt}}
\put(515.0,127.0){\rule[-0.200pt]{4.818pt}{0.400pt}}
\put(586,124.67){\rule{2.409pt}{0.400pt}}
\multiput(586.00,125.17)(5.000,-1.000){2}{\rule{1.204pt}{0.400pt}}
\put(545.0,126.0){\rule[-0.200pt]{9.877pt}{0.400pt}}
\put(668,123.67){\rule{2.409pt}{0.400pt}}
\multiput(668.00,124.17)(5.000,-1.000){2}{\rule{1.204pt}{0.400pt}}
\put(596.0,125.0){\rule[-0.200pt]{17.345pt}{0.400pt}}
\put(909,122.67){\rule{2.409pt}{0.400pt}}
\multiput(909.00,123.17)(5.000,-1.000){2}{\rule{1.204pt}{0.400pt}}
\put(678.0,124.0){\rule[-0.200pt]{55.648pt}{0.400pt}}
\put(919.0,123.0){\rule[-0.200pt]{118.041pt}{0.400pt}}
\end{picture}
\caption{Radiatively corrected masses total cross section for the process
$\mu^- \mu^+ \rightarrow H^0 Z^0$ as a function
of $m_{H^0}$. We have taken
$\sqrt{s} = 500 \textrm{GeV/c}$ and $\tan\beta= 30$.}
\label{mumu_HZrc}
\end{center} 
\end{figure}
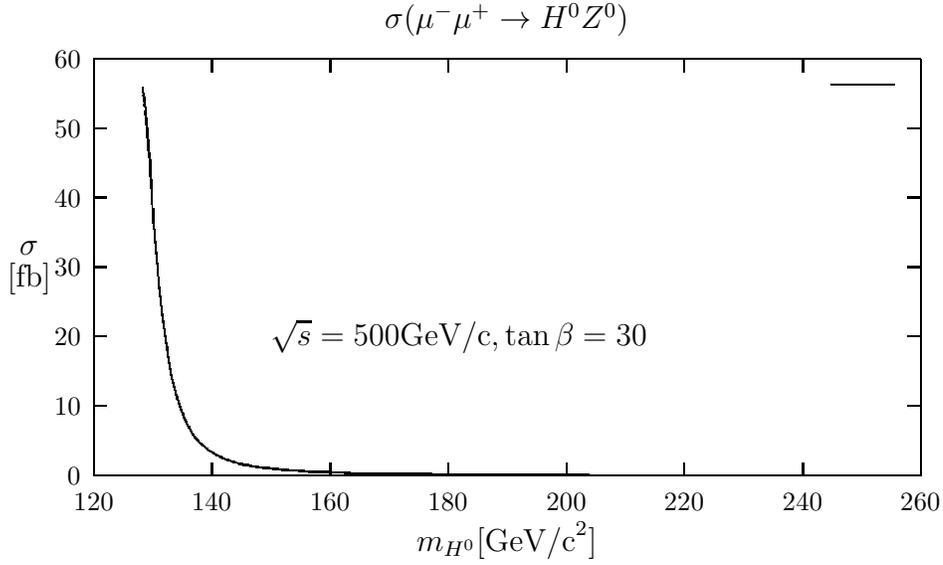

\begin{figure}
\begin{center}
\setlength{\unitlength}{0.240900pt}
\ifx\plotpoint\undefined\newsavebox{\plotpoint}\fi
\sbox{\plotpoint}{\rule[-0.200pt]{0.400pt}{0.400pt}}%
\begin{picture}(1500,900)(0,0)
\font\gnuplot=cmr10 at 10pt
\gnuplot
\sbox{\plotpoint}{\rule[-0.200pt]{0.400pt}{0.400pt}}%
\put(181.0,123.0){\rule[-0.200pt]{4.818pt}{0.400pt}}
\put(161,123){\makebox(0,0)[r]{0}}
\put(1419.0,123.0){\rule[-0.200pt]{4.818pt}{0.400pt}}
\put(181.0,188.0){\rule[-0.200pt]{4.818pt}{0.400pt}}
\put(161,188){\makebox(0,0)[r]{0.01}}
\put(1419.0,188.0){\rule[-0.200pt]{4.818pt}{0.400pt}}
\put(181.0,254.0){\rule[-0.200pt]{4.818pt}{0.400pt}}
\put(161,254){\makebox(0,0)[r]{0.02}}
\put(1419.0,254.0){\rule[-0.200pt]{4.818pt}{0.400pt}}
\put(181.0,319.0){\rule[-0.200pt]{4.818pt}{0.400pt}}
\put(161,319){\makebox(0,0)[r]{0.03}}
\put(1419.0,319.0){\rule[-0.200pt]{4.818pt}{0.400pt}}
\put(181.0,385.0){\rule[-0.200pt]{4.818pt}{0.400pt}}
\put(161,385){\makebox(0,0)[r]{0.04}}
\put(1419.0,385.0){\rule[-0.200pt]{4.818pt}{0.400pt}}
\put(181.0,450.0){\rule[-0.200pt]{4.818pt}{0.400pt}}
\put(161,450){\makebox(0,0)[r]{0.05}}
\put(1419.0,450.0){\rule[-0.200pt]{4.818pt}{0.400pt}}
\put(181.0,515.0){\rule[-0.200pt]{4.818pt}{0.400pt}}
\put(161,515){\makebox(0,0)[r]{0.06}}
\put(1419.0,515.0){\rule[-0.200pt]{4.818pt}{0.400pt}}
\put(181.0,581.0){\rule[-0.200pt]{4.818pt}{0.400pt}}
\put(161,581){\makebox(0,0)[r]{0.07}}
\put(1419.0,581.0){\rule[-0.200pt]{4.818pt}{0.400pt}}
\put(181.0,646.0){\rule[-0.200pt]{4.818pt}{0.400pt}}
\put(161,646){\makebox(0,0)[r]{0.08}}
\put(1419.0,646.0){\rule[-0.200pt]{4.818pt}{0.400pt}}
\put(181.0,712.0){\rule[-0.200pt]{4.818pt}{0.400pt}}
\put(161,712){\makebox(0,0)[r]{0.09}}
\put(1419.0,712.0){\rule[-0.200pt]{4.818pt}{0.400pt}}
\put(181.0,777.0){\rule[-0.200pt]{4.818pt}{0.400pt}}
\put(161,777){\makebox(0,0)[r]{0.1}}
\put(1419.0,777.0){\rule[-0.200pt]{4.818pt}{0.400pt}}
\put(181.0,123.0){\rule[-0.200pt]{0.400pt}{4.818pt}}
\put(181,82){\makebox(0,0){100}}
\put(181.0,757.0){\rule[-0.200pt]{0.400pt}{4.818pt}}
\put(321.0,123.0){\rule[-0.200pt]{0.400pt}{4.818pt}}
\put(321,82){\makebox(0,0){120}}
\put(321.0,757.0){\rule[-0.200pt]{0.400pt}{4.818pt}}
\put(461.0,123.0){\rule[-0.200pt]{0.400pt}{4.818pt}}
\put(461,82){\makebox(0,0){140}}
\put(461.0,757.0){\rule[-0.200pt]{0.400pt}{4.818pt}}
\put(600.0,123.0){\rule[-0.200pt]{0.400pt}{4.818pt}}
\put(600,82){\makebox(0,0){160}}
\put(600.0,757.0){\rule[-0.200pt]{0.400pt}{4.818pt}}
\put(740.0,123.0){\rule[-0.200pt]{0.400pt}{4.818pt}}
\put(740,82){\makebox(0,0){180}}
\put(740.0,757.0){\rule[-0.200pt]{0.400pt}{4.818pt}}
\put(880.0,123.0){\rule[-0.200pt]{0.400pt}{4.818pt}}
\put(880,82){\makebox(0,0){200}}
\put(880.0,757.0){\rule[-0.200pt]{0.400pt}{4.818pt}}
\put(1020.0,123.0){\rule[-0.200pt]{0.400pt}{4.818pt}}
\put(1020,82){\makebox(0,0){220}}
\put(1020.0,757.0){\rule[-0.200pt]{0.400pt}{4.818pt}}
\put(1159.0,123.0){\rule[-0.200pt]{0.400pt}{4.818pt}}
\put(1159,82){\makebox(0,0){240}}
\put(1159.0,757.0){\rule[-0.200pt]{0.400pt}{4.818pt}}
\put(1299.0,123.0){\rule[-0.200pt]{0.400pt}{4.818pt}}
\put(1299,82){\makebox(0,0){260}}
\put(1299.0,757.0){\rule[-0.200pt]{0.400pt}{4.818pt}}
\put(1439.0,123.0){\rule[-0.200pt]{0.400pt}{4.818pt}}
\put(1439,82){\makebox(0,0){280}}
\put(1439.0,757.0){\rule[-0.200pt]{0.400pt}{4.818pt}}
\put(181.0,123.0){\rule[-0.200pt]{303.052pt}{0.400pt}}
\put(1439.0,123.0){\rule[-0.200pt]{0.400pt}{157.549pt}}
\put(181.0,777.0){\rule[-0.200pt]{303.052pt}{0.400pt}}
\put(40,450){\makebox(0,0){$\sigma_{ro}$}}
\put(810,21){\makebox(0,0){$m_{H^0} [\textrm{GeV/c}^2]$}}
\put(810,839){\makebox(0,0){$\sigma_{ro}=\frac{\sigma(e^-e^+ \rightarrow H^0 Z^0)}{\sigma(e^- e^+ \rightarrow \mu^- \mu^+)}$}}
\put(600,450){\makebox(0,0)[l]{$\sqrt{s} = 500 \textrm{GeV/c}, \tan\beta = 
30$}}
\put(181.0,123.0){\rule[-0.200pt]{0.400pt}{157.549pt}}
\put(1279,737){\makebox(0,0)[r]{ }}
\put(1299.0,737.0){\rule[-0.200pt]{24.090pt}{0.400pt}}
\put(321,772){\usebox{\plotpoint}}
\multiput(321.59,746.20)(0.485,-8.120){11}{\rule{0.117pt}{6.214pt}}
\multiput(320.17,759.10)(7.000,-94.102){2}{\rule{0.400pt}{3.107pt}}
\multiput(328.59,638.02)(0.485,-8.502){11}{\rule{0.117pt}{6.500pt}}
\multiput(327.17,651.51)(7.000,-98.509){2}{\rule{0.400pt}{3.250pt}}
\multiput(335.59,528.39)(0.485,-7.739){11}{\rule{0.117pt}{5.929pt}}
\multiput(334.17,540.69)(7.000,-89.695){2}{\rule{0.400pt}{2.964pt}}
\multiput(342.59,430.90)(0.485,-6.290){11}{\rule{0.117pt}{4.843pt}}
\multiput(341.17,440.95)(7.000,-72.948){2}{\rule{0.400pt}{2.421pt}}
\multiput(349.59,352.88)(0.485,-4.688){11}{\rule{0.117pt}{3.643pt}}
\multiput(348.17,360.44)(7.000,-54.439){2}{\rule{0.400pt}{1.821pt}}
\multiput(356.59,295.15)(0.485,-3.315){11}{\rule{0.117pt}{2.614pt}}
\multiput(355.17,300.57)(7.000,-38.574){2}{\rule{0.400pt}{1.307pt}}
\multiput(363.59,253.76)(0.485,-2.476){11}{\rule{0.117pt}{1.986pt}}
\multiput(362.17,257.88)(7.000,-28.879){2}{\rule{0.400pt}{0.993pt}}
\multiput(370.59,223.13)(0.485,-1.713){11}{\rule{0.117pt}{1.414pt}}
\multiput(369.17,226.06)(7.000,-20.065){2}{\rule{0.400pt}{0.707pt}}
\multiput(377.59,201.55)(0.485,-1.255){11}{\rule{0.117pt}{1.071pt}}
\multiput(376.17,203.78)(7.000,-14.776){2}{\rule{0.400pt}{0.536pt}}
\multiput(384.59,185.74)(0.485,-0.874){11}{\rule{0.117pt}{0.786pt}}
\multiput(383.17,187.37)(7.000,-10.369){2}{\rule{0.400pt}{0.393pt}}
\multiput(391.59,174.21)(0.485,-0.721){11}{\rule{0.117pt}{0.671pt}}
\multiput(390.17,175.61)(7.000,-8.606){2}{\rule{0.400pt}{0.336pt}}
\multiput(398.00,165.93)(0.492,-0.485){11}{\rule{0.500pt}{0.117pt}}
\multiput(398.00,166.17)(5.962,-7.000){2}{\rule{0.250pt}{0.400pt}}
\multiput(405.00,158.93)(0.581,-0.482){9}{\rule{0.567pt}{0.116pt}}
\multiput(405.00,159.17)(5.824,-6.000){2}{\rule{0.283pt}{0.400pt}}
\multiput(412.00,152.93)(0.710,-0.477){7}{\rule{0.660pt}{0.115pt}}
\multiput(412.00,153.17)(5.630,-5.000){2}{\rule{0.330pt}{0.400pt}}
\multiput(419.00,147.95)(1.355,-0.447){3}{\rule{1.033pt}{0.108pt}}
\multiput(419.00,148.17)(4.855,-3.000){2}{\rule{0.517pt}{0.400pt}}
\multiput(426.00,144.95)(1.355,-0.447){3}{\rule{1.033pt}{0.108pt}}
\multiput(426.00,145.17)(4.855,-3.000){2}{\rule{0.517pt}{0.400pt}}
\multiput(433.00,141.95)(1.355,-0.447){3}{\rule{1.033pt}{0.108pt}}
\multiput(433.00,142.17)(4.855,-3.000){2}{\rule{0.517pt}{0.400pt}}
\put(440,138.17){\rule{1.500pt}{0.400pt}}
\multiput(440.00,139.17)(3.887,-2.000){2}{\rule{0.750pt}{0.400pt}}
\put(447,136.17){\rule{1.500pt}{0.400pt}}
\multiput(447.00,137.17)(3.887,-2.000){2}{\rule{0.750pt}{0.400pt}}
\put(454,134.67){\rule{1.686pt}{0.400pt}}
\multiput(454.00,135.17)(3.500,-1.000){2}{\rule{0.843pt}{0.400pt}}
\put(461,133.67){\rule{1.686pt}{0.400pt}}
\multiput(461.00,134.17)(3.500,-1.000){2}{\rule{0.843pt}{0.400pt}}
\put(468,132.67){\rule{1.686pt}{0.400pt}}
\multiput(468.00,133.17)(3.500,-1.000){2}{\rule{0.843pt}{0.400pt}}
\put(475,131.67){\rule{1.686pt}{0.400pt}}
\multiput(475.00,132.17)(3.500,-1.000){2}{\rule{0.843pt}{0.400pt}}
\put(482,130.67){\rule{1.686pt}{0.400pt}}
\multiput(482.00,131.17)(3.500,-1.000){2}{\rule{0.843pt}{0.400pt}}
\put(489,129.67){\rule{1.686pt}{0.400pt}}
\multiput(489.00,130.17)(3.500,-1.000){2}{\rule{0.843pt}{0.400pt}}
\put(502,128.67){\rule{1.686pt}{0.400pt}}
\multiput(502.00,129.17)(3.500,-1.000){2}{\rule{0.843pt}{0.400pt}}
\put(509,127.67){\rule{1.686pt}{0.400pt}}
\multiput(509.00,128.17)(3.500,-1.000){2}{\rule{0.843pt}{0.400pt}}
\put(496.0,130.0){\rule[-0.200pt]{1.445pt}{0.400pt}}
\put(530,126.67){\rule{1.686pt}{0.400pt}}
\multiput(530.00,127.17)(3.500,-1.000){2}{\rule{0.843pt}{0.400pt}}
\put(516.0,128.0){\rule[-0.200pt]{3.373pt}{0.400pt}}
\put(551,125.67){\rule{1.686pt}{0.400pt}}
\multiput(551.00,126.17)(3.500,-1.000){2}{\rule{0.843pt}{0.400pt}}
\put(537.0,127.0){\rule[-0.200pt]{3.373pt}{0.400pt}}
\put(586,124.67){\rule{1.686pt}{0.400pt}}
\multiput(586.00,125.17)(3.500,-1.000){2}{\rule{0.843pt}{0.400pt}}
\put(558.0,126.0){\rule[-0.200pt]{6.745pt}{0.400pt}}
\put(649,123.67){\rule{1.686pt}{0.400pt}}
\multiput(649.00,124.17)(3.500,-1.000){2}{\rule{0.843pt}{0.400pt}}
\put(593.0,125.0){\rule[-0.200pt]{13.490pt}{0.400pt}}
\put(817,122.67){\rule{1.686pt}{0.400pt}}
\multiput(817.00,123.17)(3.500,-1.000){2}{\rule{0.843pt}{0.400pt}}
\put(656.0,124.0){\rule[-0.200pt]{38.785pt}{0.400pt}}
\put(824.0,123.0){\rule[-0.200pt]{129.604pt}{0.400pt}}
\end{picture}
\caption{Total cross section for the process
$e^- e^+ \rightarrow H^0 Z^0$ compared with the
cross section $\sigma(e^- e^+
\rightarrow \mu^- \mu^+)$ as a function
of $m_{H^0}$. We have taken
$\sqrt{s} = \textrm{500 GeV/c}$ and $\tan\beta= 30$. The radiative 
corrections of the masses
were not taken into account.}
\label{muhzee_figure}  
\end{center}
\end{figure}
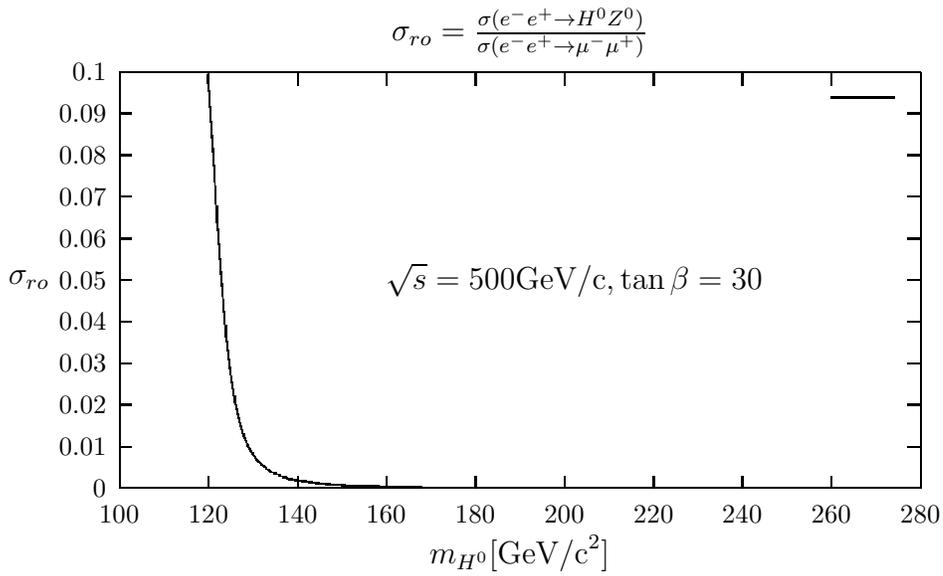   

\begin{figure} 
\begin{center}
\input{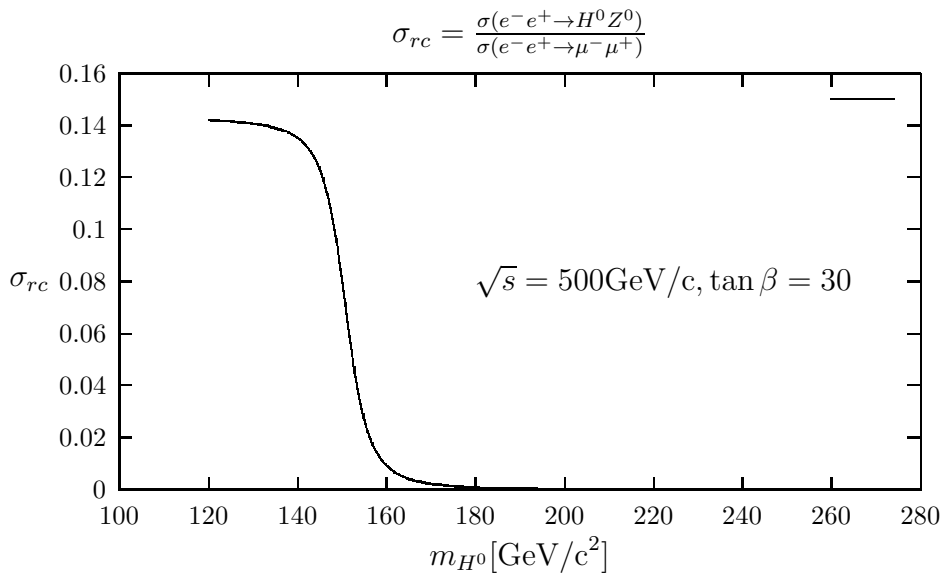}
\caption{Radiatively corrected masses total cross section for the process
$e^- e^+ \rightarrow H^0 Z^0$ compared with the
cross section $\sigma(e^- e^+
\rightarrow \mu^- \mu^+)$ as a function
of $m_{H^0}$. We have taken
$\sqrt{s} = 500 \textrm{GeV/c}$ and $\tan\beta= 30$.}
\label{muhzeerc_figure}
\end{center}   
\end{figure}

\section{Production of $A^0$}
From the Feynman diagrams of Figure \ref{mumu_AZ_fig} and the Feynman 
rules given in \cite{M_H}, we obtain the
differential cross section for the production process
$\mu^- \mu^+ \rightarrow A^0 Z^0$ in the center of mass system:
\begin{figure}  
\begin{center}   
{\includegraphics{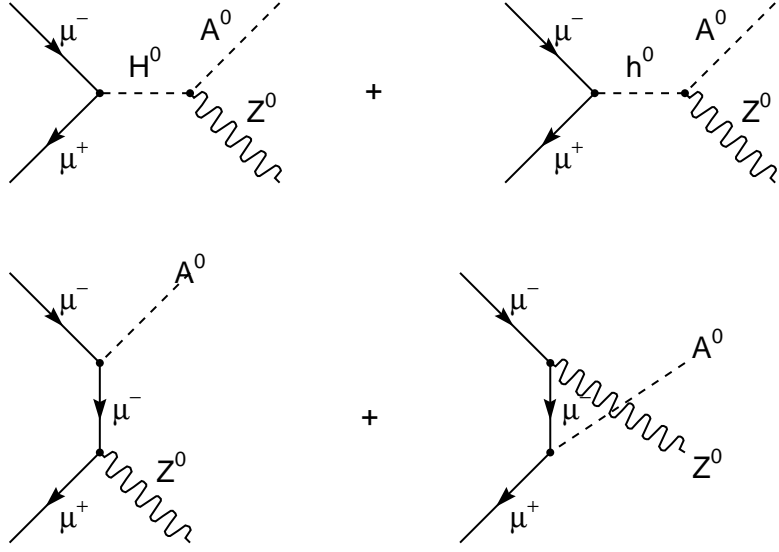}}
\caption{Feynman diagrams corresponding to the production
of $A^0$ in the channel $\mu^- \mu^+ \rightarrow A^0 Z^0$.}
\label{mumu_AZ_fig}
\end{center}
\end{figure}  

\begin{eqnarray}
\lefteqn{
\frac{d \sigma}{d \Omega} (\mu^- \mu^+ \rightarrow A^0 Z^0) =
\frac{1}{64 \pi^2 s} \Lambda^{1/2}\left( s, m_{A^0}^2, 
m_Z^2 \right) G_F^2 m_{\mu}^2 
}
\nonumber \\ & &
\times\{  C_{Hb}^2 
\Lambda \left( s, m_{A^0}^2, m_Z^2 \right)
+ \left[  \left( g_A^\mu \right)^2 + \left( g_V^\mu \right)^2
\right] 
\nonumber \\ & &
[ \tan^2\beta \left( 1 +
 \frac{\Lambda \left( s, m_{A^0}^2, m_Z^2 \right)m_Z^2
\sin^2\theta}{2s t^2}\right)
\nonumber \\ & &
 + \tan^2\beta 
\left( 1 +  \frac{\Lambda \left( s, m_{A^0}^2, m_Z^2 \right)m_Z^2
\sin^2\theta}{2s u^2}\right) ]
\nonumber \\ & &
+ 2 g_A^\mu \tan\beta C_{Hb}
\left[\frac{m_{A^0}^2 m_Z^2}{t} - 
\frac{\Lambda \left( s, m_{A^0}^2, m_Z^2 \right) \sin^2\theta}
{4t} - t \right]
\nonumber \\ & &
+ 2 g_A^\mu \tan\beta C_{Hb}
\left[\frac{m_{A^0}^2 m_Z^2}{u} -      
\frac{\Lambda \left( s, m_{A^0}^2, m_Z^2 \right) \sin^2\theta}
{4u} - u \right] 
\nonumber \\ & &
+ 2\tan^2\beta
 \left[  \left( g_V^\mu \right)^2 - \left( g_A^\mu \right)^2
\right]
[\frac{m_{A^0}^2 m_Z^2}{ut} -
\frac{\Lambda \left( s, m_{A^0}^2, m_Z^2 \right) \sin^2\theta}
{4ut}
\nonumber \\ & &
 + \frac{ \Lambda
 \left( s, m_{A^0}^2, m_Z^2 \right) m_Z^2 \sin^2\theta }
{2sut} ] \}
\label{diffmumu_AZ}
\end{eqnarray}
\noindent
where $g_A^\mu$ and $g_V^\mu$ are given by 
(\ref{gA,gV}); $s$,$t$,$u$ are the
Mandelstam invariant variables and

\begin{equation}
C_{Hb} = \frac{ \left( \frac{1}{2} \sin 2 \alpha +
\tan\beta \sin^2\alpha \right) }{\left (s - m_{h^0}^2
\right) }
- \frac{ \left( \frac{1}{2} \sin2\alpha - \tan\beta
\cos^2\alpha \right) }{ \left( s - m_{H^0}^2 \right)}
\label{CHb}
\end{equation}

To obtain the total cross section, we integrate
Equation (\ref{diffmumu_AZ}) over the solid angle $\Omega$.
\begin{eqnarray}
\lefteqn{
\sigma  (\mu^- \mu^+ \rightarrow A^0 Z^0) = 
\frac{G_F^2 m_{\mu}^2}{16 \pi s^2}
\{\Lambda^{1/2} \left( s, m_{A^0}^2, m_Z^2 \right)
[ s C_{Hb}^2 \Lambda \left( s, m_{A^0}^2, m_Z^2 \right)
}
\nonumber \\ & &
+ 4 \tan^2\beta \sin^2\theta_W \left(1 -2 \sin^2\theta_W
\right) \left( s - 2 m_Z^2 \right) + 2 s \tan\beta  C_{Hb}
\nonumber \\ & &
\times \left( m_{A^0}^2 + m_Z^2 - s \right)
+ \left( 1 - 4 \sin^2\theta_W + 8 \sin^4\theta_W \right)
\nonumber \\ & &
\times\tan^2\beta \left( s - 4 m_Z^2 \right) ]
+ 4 m_Z^2 \tan\beta f\left( s, m_{A^0}^2, m_Z^2 \right)
\nonumber \\ & &
\times[ - s C_{Hb} m_{A^0}^2 + \frac{1}{2} \tan\beta \left(1 - 
4 \sin^2\theta_W + 8 \sin^4\theta_W \right)
\left( m_{A^0}^2 + m_Z^2 - s \right)
\nonumber \\ & &
- 4 \frac{\sin^2\theta_W \left( 1 - 2 \sin^2\theta_W\right)
\tan\beta m_{A^0}^2 \left( s - m_Z^2 \right)}
{\left( m_{A^0}^2 + m_Z^2 - s \right)}]\}
\nonumber \\ & &
\times
\left( 3.8938 \times 10^{11} \right) \textrm{fb}
\label{sigmamumu_AZ}
\end{eqnarray}

\noindent where

\begin{equation}
f \left( s, m_{A^0}^2, m_Z^2 \right) \equiv
\ln\left| \frac{m_{A^0}^2 + m_Z^2 - s +
\Lambda^{1/2} \left( s, m_{A^0}^2, m_Z^2 \right)}
{ m_{A^0}^2 + m_Z^2 - s -
\Lambda^{1/2} \left( s, m_{A^0}^2, m_Z^2 \right)}
\right|
\label{f}
\end{equation}

Note that if $m_{A^0} = \sqrt{s} - m_Z$, then we have,
$\Lambda \left( s, m_{A^0}^2, m_Z^2 \right) = 0$ and
$f \left( s, m_{A^0}^2, m_Z^2 \right)$ = 0. Therefore
$\sigma  (\mu^- \mu^+ \rightarrow A^0 Z^0) =0$.

\begin{figure}
\begin{center}
\input{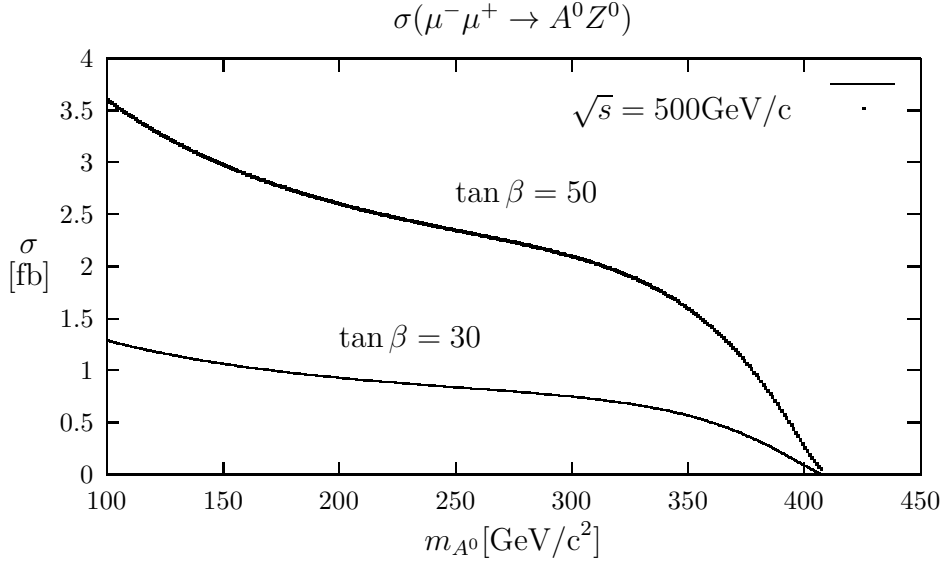}
\caption{Total cross section for the process
$\mu^- \mu^+ \rightarrow A^0 Z^0$ as a function
of $m_{A^0}$. We have taken
$\sqrt{s} = 500 \textrm{GeV/c}$ and $\tan\beta= 30, 50$. The total 
cross section is not affected by radiative corrections of the masses.}
\label{muaz_figure}
\end{center}
\end{figure}

Figure \ref{muaz_figure} shows the total 
cross section $\sigma  (\mu^- \mu^+ \rightarrow A^0 Z^0)$
as a function of $m_{A^0}$ for $\sqrt{s} = 500 \textrm{GeV/c}$ and 
$\tan\beta= 30, 50$. The total cross section is not
affected by radiative corrections of the masses. From Figure 
\ref{muaz_figure} we can 
see that cross sections are important for large values of $\tan\beta$.

The total cross section corresponding to 
$e^- e^+ \rightarrow A^0 Z^0$ can be obtained from Equation
(\ref{sigmamumu_AZ}) replacing $m_{\mu}$ by $m_{e}$:
\begin{equation}
\frac{\sigma(e^- e^+ \rightarrow A^0 Z^0)}
{\sigma(\mu^- \mu^+ \rightarrow A^0 Z^0)} = 
\left(\frac{m_{e}}{m_{\mu}}\right)^2 = 2.34 \cdot 10^{-5}.
\label{sigmaeeAZ}
\end{equation}
 
\section{Production of $H^{\pm}$}

\begin{figure}
\begin{center}
{\includegraphics{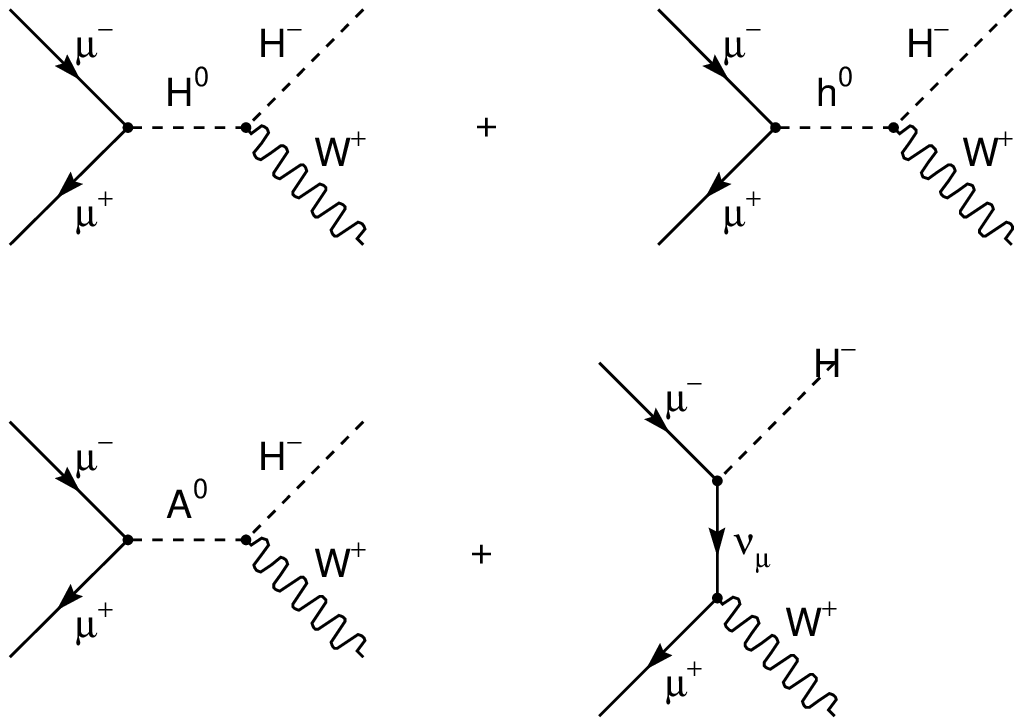}}
\caption{Feynman diagrams corresponding to the production
of $H^-$ in the channel $\mu^- \mu^+ \rightarrow H^-
W^+$.}
\label{mumu_HW_fig}
\end{center}
\end{figure}

From the Feynman diagrams of Figure \ref{mumu_HW_fig}
we obtain the differential cross section
in the center of mass system for the process
$\mu^- \mu^+ \rightarrow H^- W^+$:

\begin{eqnarray}
\lefteqn{
\frac{d \sigma}{d \Omega} (\mu^- \mu^+ \rightarrow H^- W^+) =
\frac{1}{64 \pi^2 s} \Lambda^{1/2}\left( s, m_H^2,
m_W^2 \right) G_F^2 m_{\mu}^2
}
\nonumber \\ & &
\times\{ \left[ C_{Hb}^2 + C_{Ab}^2 \right] \Lambda\left( s, m_H^2, 
m_W^2 \right) 
+ 2 \left( \frac{\tan\beta}{t} \right)^2
\nonumber \\ & &
\times\left[ \frac{\Lambda\left( s, m_H^2, m_W^2 \right)
\sin^2\theta m_W^2}{2s} + t^2 \right]
- 2  \left( \frac{\tan\beta}{t} \right) \left(C_{Ab} + C_{Hb} 
\right)
\nonumber \\ & &
\times\left[ -t^2 - \frac{1}{4} 
\Lambda\left( s, m_H^2, m_W^2 \right) \sin^2\theta
+ m_W^2 m_H^2 \right] \}
\label{diffmumu_HW}
\end{eqnarray}

\noindent where $C_{Hb}$ is given by Equation (\ref{CHb}) and 
\begin{equation}
C_{Ab} = \frac{\tan\beta}{\left( s - m_{A^0}^2 \right)}
\label{CAb}
\end{equation}

The differential cross section corresponding to
$\mu^- \mu^+ \rightarrow H^+
W^-$ is obtained from  (\ref{diffmumu_HW}) by replacing $t$ by
$u$.

The integration of (\ref{diffmumu_HW})
over the solid angle $\Omega$ give us the total cross
section:

\begin{eqnarray}
\lefteqn{
\sigma (\mu^- \mu^+ \rightarrow H^- W^+) =
\frac{G_F^2 m_{\mu}^2}{16 \pi s^2} \{ s 
\Lambda^{3/2}\left( s, m_H^2,m_W^2 \right)
\left[ C_{Hb}^2 + C_{Ab}^2 \right]
}
\nonumber \\ & &
+ 2 \tan\beta \Lambda^{1/2}\left( s, m_H^2,m_W^2 \right)
[ \tan\beta \left( s - 4 m_W^2 \right) 
\nonumber \\ & &
+ \left( C_{Ab} +
C_{Hb} \right) s \left( m_H^2 + m_W^2 - s \right)]
+ 4 m_W^2 \tan\beta f \left( s, m_H^2, m_W^2 \right)
\nonumber \\ & &
\left[ \tan\beta \left( m_H^2 + m_W^2 - s \right)
- \left( C_{Ab} + C_{Hb} \right) s m_H^2 \right] \}
\nonumber \\ & &
\times
\left( 3.8938 \times 10^{11} \right) \textrm{fb}
\label{sigmamumu_HW}
\end{eqnarray}

\noindent where 

\begin{equation}
f \left( s, m_H^2, m_W^2 \right) =
\ln\left| \frac{m_H^2 + m_W^2 - s +
\Lambda^{1/2} \left( s, m_H^2, m_W^2 \right)}
{ m_H^2 + m_W^2 - s -
\Lambda^{1/2} \left( s, m_H^2, m_W^2 \right)}
\right|
\label{fW}
\end{equation}

For the process $ \mu^- \mu^+ \rightarrow H^+ W^-$ we obtain:

\begin{equation}
\sigma (\mu^- \mu^+ \rightarrow H^+ W^-) =
\sigma (\mu^- \mu^+ \rightarrow H^- W^+)
\label{H+/-W-/+}
\end{equation}

\noindent and then

\begin{equation}
\sigma (\mu^- \mu^+ \rightarrow H^{\pm} W^{\mp}) = 2
\sigma (\mu^- \mu^+ \rightarrow H^- W^+)
\label{sigmaH+/-W-/+}
\end{equation}

Observe that $\sigma (\mu^- \mu^+ \rightarrow H^{\pm} W^{\mp})
= 0$ if $m_H = \sqrt{s} - m_W$.

\begin{figure}
\begin{center}
\input{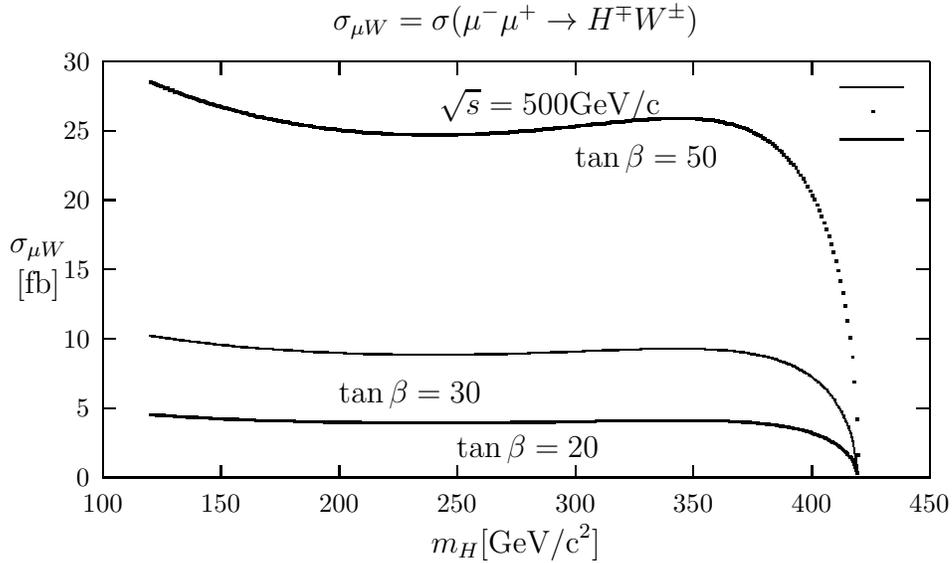}
\caption{Total cross section for the process
$\mu^- \mu^+ \rightarrow H^{\mp} W^{\pm}$ as a function
of $m_H$. We have taken
$\sqrt{s} = 500 \textrm{GeV/c}$ and $\tan\beta= 20, 30, 50$. The radiative 
corrections of the masses
are negligible.}
\label{muhw_figure}
\end{center}
\end{figure}

The total cross section corresponding to $\mu^- \mu^+ 
\rightarrow H^{\mp} W^{\pm}$ is given in Figure \ref{muhw_figure} for 
$\sqrt{s} = 500 \textrm{GeV/c}$ and $\tan\beta = 20, 30 , 50$. 
This total cross
section is not affected by radiative corrections of the masses. From 
Figure 
\ref{muhw_figure} we see that 
$\sigma(\mu^- \mu^+ \rightarrow H^{\mp} W^{\pm}) \gtrsim 5 \textrm{fb}$ 
for 
$\tan\beta \geq 20$ in the mass interval
$100 \leq m_{H} \leq 400 [\textrm{GeV/c}^2]$.

For the process
$e^- e^+ \rightarrow H^{\mp} W^{\pm}$, the total cross section 
is obtained from Equations (\ref{sigmamumu_HW}), (\ref{sigmaH+/-W-/+})  
replacing $m_{\mu}$ by $m_{e}$. This cross section is smaller than the
one ploted in Figure \ref{muhw_figure} by a factor  
$ m_{e}^2/m_{\mu}^2 = 2.34\cdot10^{-5}$. 

Equations (\ref{sigmamumu_AZ}) and (\ref{sigmamumu_HW}) are in agreement
with the cross sections calculated in \cite{mu2}.

\section{Production of charged Higgs boson pairs}

\begin{figure}
\begin{center}
{\includegraphics{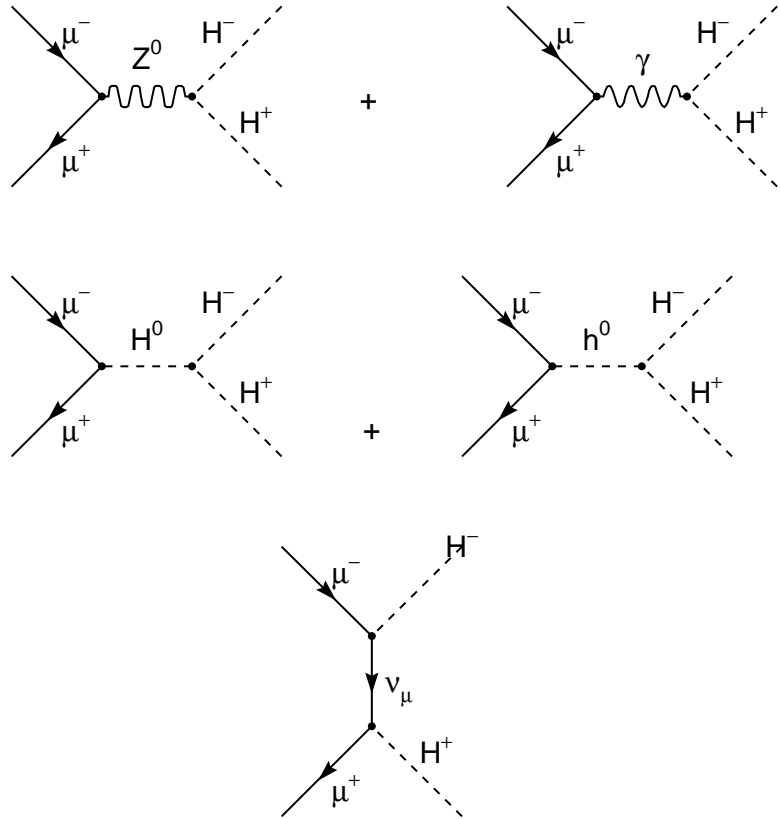}}
\caption{Feynman diagrams corresponding to the production
of charged Higgs boson pairs
in the channel $\mu^- \mu^+ \rightarrow H^-
H^+$.}
\label{mumu_HH_fig}
\end{center}
\end{figure}

From the Feynman diagrams of Figure \ref{mumu_HH_fig}, the differential 
cross section in the center of mass system corresponding to
$\mu^- \mu^+ \rightarrow H^- H^+$ is

\begin{eqnarray}
\lefteqn{
\frac{d \sigma}{d \Omega} (\mu^- \mu^+ \rightarrow H^- H^+) =
\frac{G_F^2 m_W^4}{8 \pi^2 s} \left( 1 - 4 \frac{m_H^2}{s}
\right)^{1/2} \{ s \left(s - 4m_H^2 \right) \sin^2\theta
}
\nonumber \\ & &
\left[ \frac{1}{8} \left|C_1 \right|^2\left[ 
\left( g_A^\mu \right)^2 + \left( g_V^\mu \right)^2
\right] + 2 \left( \frac{\sin^2\theta_W}{s} \right)^2
- \left( \frac{\sin^2\theta_W}{s} \right) \Re{(C_1)}
g_V^{\mu} \right] 
\nonumber \\ & &
+ 2 m_{\mu}^2 [ \left(s - 4 m_H^2 
\right) 
[ \frac{\left| C_1 \right|^2}{4}
\left( \cos^2\theta \left( g_V^{\mu} \right)^2
-\sin^2\theta \left( g_A^{\mu} \right)^2 \right)
\nonumber \\ & &
+ 4 \left( \frac{\sin^2\theta_W}{s} \right)^2
\cos^2\theta
- 2  \left( \frac{\sin^2\theta_W}{s} \right)
\Re{(C_1)} g_V^{\mu} \cos^2\theta ]
\nonumber \\ & &
+ \frac{1}{4} \left( C^{Hh} \right)^2 s 
+ \left( s \left( s - 4 m_H^2 \right) \right)^{1/2}
\nonumber \\ & &
\times\cos\theta C^{Hh} \left( \frac{1}{2} g_V^{\mu}
\Re{(C_1)} - 2 \left( \frac{\sin^2\theta_W}{s} \right)
\right) ] \}
\label{diffmumu_HH}
\end{eqnarray}

\noindent where $g_A^{\mu}$ and $g_V^{\mu}$ are given by Equation
(\ref{gA,gV}),

\begin{equation}
C_1 \equiv \frac{\cos(2\theta_W)}{\cos^2\theta_W}
\frac{1}{\left( s - m_Z^2 + i m_Z \Gamma_Z \right)},
\label{C1}
\end{equation}

\begin{equation}
C^{Hh} = \frac{a_1}{\left(s - m_{H^0}^2 \right)}
- \frac{a_2}{\left( s - m_{h^0}^2 \right)},
\label{CHh}
\end{equation}

\begin{eqnarray}
\lefteqn{
a_1 = [ \cos^2\alpha + \frac{\tan\beta \sin2\alpha}{2}
- \frac{m_Z}{2m_W \cos\theta_W}
}
\nonumber \\ & &
\frac{\left( 1 - \tan^2\beta \right)}
{\left( 1 + \tan^2\beta \right)}
\left( \cos^2\alpha - \frac{\tan\beta \sin2\alpha}{2}
\right) ],
\label{a1}
\end{eqnarray}

\begin{eqnarray}
\lefteqn{
a_2 = [ \frac{\tan\beta \sin2\alpha}{2}
-\sin^2\alpha + \frac{m_Z}{2m_W \cos\theta_W}
}
\nonumber \\ & &
\frac{\left( 1 - \tan^2\beta \right)}
{\left( 1 + \tan^2\beta \right)}
\left( \sin^2\alpha + \frac{\tan\beta \sin2\alpha}{2}
\right) ]
\label{a2}
\end{eqnarray}

The integration of (\ref{diffmumu_HH}) give us the total 
cross section for the process $\mu^- \mu^+ \rightarrow H^- H^+$:

\begin{eqnarray}
\lefteqn{
\sigma\left(\mu^+ \mu^- \rightarrow H^+ H^- \right) =
\sigma\left(\mu^- \mu^+ \rightarrow H^- H^+ \right) =
\frac{2m_W^4 G_F^2 \sin^4\theta_W}{3\pi s}
}
\nonumber \\ & &
\left( 1 - \frac{4 m_H^2}{s} \right)^{3/2}
\{ [ 1 + \frac{\left(1 -2 \sin^2\theta_W \right)^2
\left(1 + \left(4\sin^2\theta_W -1 \right)^2
\right)  }
{64 \sin^4\theta_W \cos^4\theta_W}
\nonumber \\ & &
\times\frac{1}{\left[ \left( 1 - \frac{m_Z^2}{s}\right)^2
+ \left(\frac{M_Z\Gamma_Z}{s}\right)^2 \right]}
- \frac{\left(1 -2 \sin^2\theta_W\right)
\left(4\sin^2\theta_W -1\right)}
{4 \sin^2\theta_W\cos^2\theta_W}
\nonumber \\ & &
\times\frac{\left(1- \frac{m_Z^2}{s} \right)}
{\left[ \left( 1 - \frac{m_Z^2}{s}\right)^2  
+ \left(\frac{M_Z\Gamma_Z}{s}\right)^2 \right]} ]
+ \frac{m_{\mu}^2}{s} [ \frac{\left(\left(4
\sin^2\theta_W -1\right)^2-2\right)}
{32\sin^4\theta_W\cos^4\theta_W}
\nonumber \\ & &
\times\frac{\left(1 - 2\sin^2\theta_W
\right)^2}{\left[\left( 1 - \frac{m_Z^2}{s}\right)^2
+ \left(\frac{M_Z\Gamma_Z}{s}\right)^2 \right]}
+2 - \frac{\left( 1 - 2 \sin^2\theta_W\right)
\left(4 \sin^2\theta_W -1\right)}
{2\sin^2\theta_W\cos^2\theta_W}
\nonumber \\ & &
\times\frac{\left(1 - \frac{m_Z^2}{s}\right)}
{\left[\left( 1 - \frac{m_Z^2}{s}\right)^2
+ \left(\frac{M_Z\Gamma_Z}{s}\right)^2 \right]}
+ \frac{3}{4}\frac{s^2 (C^{Hh})^2}
{\left(1-4\frac{m_H^2}{s}\right)} ] \}
\nonumber \\ & &
\times
\left( 3.8938 \times 10^{11} \right) \textrm{fb}
\label{sigmamumu_HH}
\end{eqnarray}

Neglecting the mass of the muon we can write:
\begin{eqnarray}
\lefteqn{
\sigma\left(\mu^+ \mu^- \rightarrow H^+ H^- \right) =
\sigma\left(\mu^- \mu^+ \rightarrow H^- H^+ \right) =
\frac{2m_W^4 G_F^2 \sin^4\theta_W}{3\pi s}
}
\nonumber \\ & &
\left( 1 - \frac{4 m_H^2}{s} \right)^{3/2}
\{  1 + \frac{\left(1 -2 \sin^2\theta_W \right)^2
\left(1 + \left(4\sin^2\theta_W -1 \right)^2 
\right)  }
{64 \sin^4\theta_W \cos^4\theta_W}
\nonumber \\ & &
\times\frac{1}{\left[ \left( 1 - \frac{m_Z^2}{s}\right)^2
+ \left(\frac{M_Z\Gamma_Z}{s}\right)^2 \right]}
- \frac{\left(1 -2 \sin^2\theta_W\right)
\left(4\sin^2\theta_W -1\right)}
{4 \sin^2\theta_W\cos^2\theta_W}
\nonumber \\ & &
\times\frac{\left(1- \frac{m_Z^2}{s} \right)}
{\left[ \left( 1 - \frac{m_Z^2}{s}\right)^2
+ \left(\frac{M_Z\Gamma_Z}{s}\right)^2 \right]} \}
\times
\left( 3.8938 \times 10^{11} \right) \textrm{fb}
\label{sigmareducedmumu_HH}
\end{eqnarray}

\begin{figure}
\begin{center}
\input{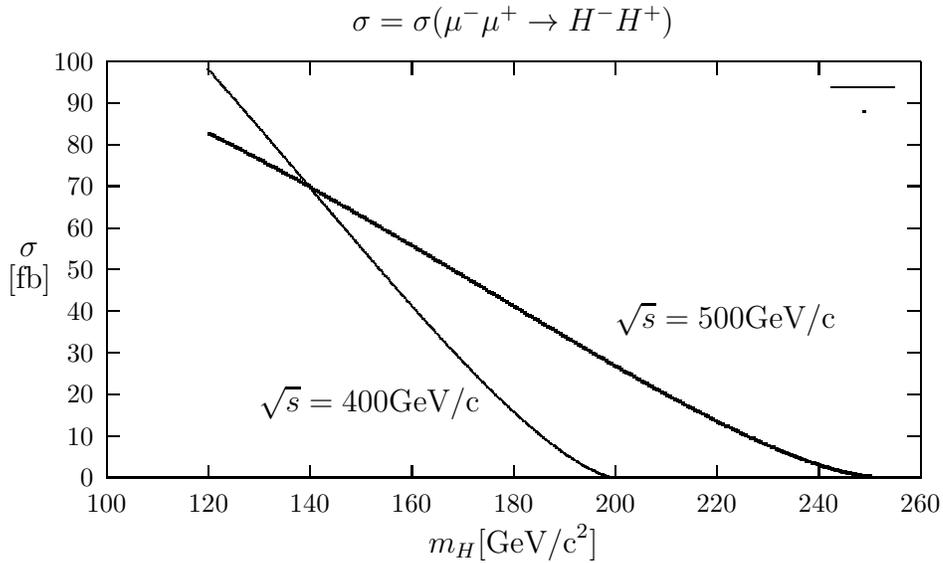}
\caption{Total cross section for the process
$\mu^- \mu^+ \rightarrow H^- H^+$ as a function
of $m_H$. We have taken
$\sqrt{s} = 400, 500 \textrm{GeV/c}$. The total cross section is
practically independent of $\tan\beta$. The radiative 
corrections of the masses
are negligible.}
\label{sigmamumu->HH_fig}
\end{center}
\end{figure}

In the last approximation there is no difference with the total cross
section corresponding to the process $e^- e^+ \rightarrow H^- H^+$.
In Figure \ref{sigmamumu->HH_fig} we have plotted 
the total cross section given by Equation
(\ref{sigmamumu_HH}) as a function of the mass of the
charged Higgs for $\sqrt{s}= 400, 500 \textrm{GeV/c}$. The total
cross section is practically independent of $\tan\beta$.
The radiative corrections of the masses are also negligible. 

\begin{figure}
\begin{center}
\input{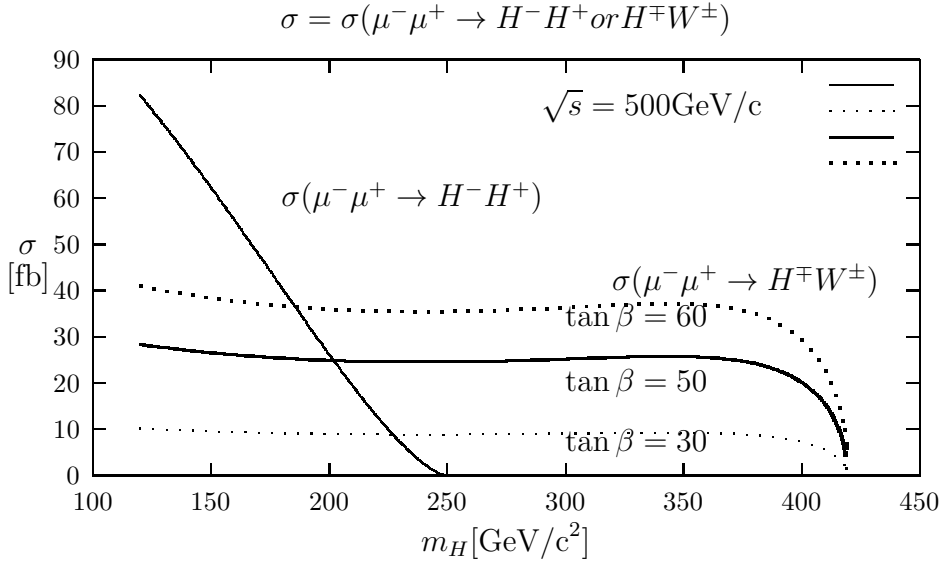}
\caption{Total cross section for the process
$\mu^- \mu^+ \rightarrow H^- H^+$ as a function
of $m_H$ compared with $\mu^- \mu^+ \rightarrow
H^{\mp} W^{\pm}$. We have taken
$\sqrt{s} = 500 \textrm{GeV/c}$ and $ \tan\beta= 30, 50, 60$.}
\label{sigmamumuhwhh_fig}
\end{center}
\end{figure}

In Figure \ref{sigmamumuhwhh_fig} we have plotted the total
cross section corresponding to the process $\mu^- \mu^+ \rightarrow
H^- H^+$ as a function of $m_H$ compared with $\mu^- \mu^+ \rightarrow
H^{\mp} W^{\pm}$. We have taken $\sqrt{s} = 500 \textrm{GeV/c}$.

\section{$\mu^- \mu^+ \rightarrow t \bar{t}$
annihilation}

The main background in the processes 
$\mu^- \mu^+ \rightarrow H^{\pm} W^{\mp}$
, assuming $H^+ \rightarrow t \bar{b}$
or $H^- \rightarrow \bar{t} b$ decays, comes from 
$t \bar{t}$ production.

\begin{figure}
\begin{center}
{\includegraphics{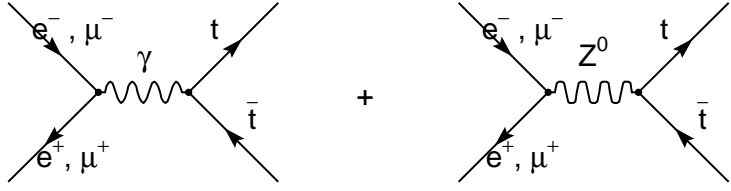}}
\caption{Feynman diagrams corresponding to 
$  e^- e^+
\rightarrow t \bar{t}$ and $ \mu^- \mu^+
\rightarrow t \bar{t}$ annihilation.}
\label{mumuttbar}
\end{center}
\end{figure}

To lowest order in $e^2$ the Feynman diagrams corresponding
to the process $\mu^- \mu^+ \rightarrow t \bar{t}$  
are given in Figure \ref{mumuttbar}. The corresponding total cross
section is (see reference \cite{CM}):

\begin{eqnarray}
\lefteqn{
\frac{\sigma \left( \mu^- \mu^+ \rightarrow t \bar{t} \right)}
{\sigma_0} = \frac{3}{4} \{ m_t^2 s \{
\left[\frac{4}{3s} + \frac{\left( \frac{8}{3} \sin^2\theta_W
-1 \right)}{2\cos^2\theta_W \left( s - m_Z^2 \right)} \right]^2
}
\nonumber \\ & &
+ \left[ \frac{4}{3s} - \frac{\left(\frac{14}{3} \sin^2\theta_W
- \frac{16}{3} \sin^4\theta_W -1 \right)}{\sin^2\left
(2\theta_W\right) \left(s - m_Z^2 \right)} \right]^2 \}
\nonumber \\ & &
+ 2\left[ \frac{2}{3} + \frac{\left( \frac{8}{3}
\sin^2\theta_W -1 \right)s}{4 \cos^2\theta_W
\left( s - m_Z^2 \right)}\right]^2
+ \left[ \frac{s^2\left( 1 - \frac{4 m_t^2}{s} \right)}
{8 \cos^4\theta_W\left( s - m_Z^2\right)^2}\right]
\nonumber \\ & &
+ 2\left[ \frac{2}{3} + \frac{\left(2\sin^2\theta_W -1 \right)
\left(\frac{4}{3}\sin^2\theta_W - \frac{1}{2}\right)s}
{\sin^2\left(2\theta_W\right) \left(s - m_Z^2\right)}\right]^2
\nonumber \\ & &
+ \frac{\left(1 - \frac{4 m_t^2}{s} \right)
\left(2 \sin^2\theta_W -1 \right)^2 s^2}
{2\sin^4\left(2\theta_W\right) \left(s - m_Z^2\right)^2}
\} \left( 1 - \frac{4 m_t^2}{s} \right)^{1/2}
\label{mumuttbarsigma}
\end{eqnarray}

\noindent where

\begin{equation}
\sigma_0 = \sigma\left( e^- e^+ \rightarrow \mu^- \mu^+ 
\right) = \frac{4 \pi \alpha_{em}^2}{3s}
\end{equation}

In (\ref{mumuttbarsigma}) we have neglected $\Gamma_Z$ that is very small
for large values of $\sqrt{s}$. 

Taking $\sin^2\theta_W = 0.231$, $m_Z = 91.1876 \textrm{GeV/c}^2$,
$m_t= 174.3 \textrm{GeV/c}^2$ and
$\sqrt{s} = 500 \textrm{GeV/c}$ we get $\sigma\left(\mu^- \mu^+
\rightarrow t \bar{t}\right) = 495.1 \textrm{fb}$.

The total cross section corresponding to $e^- e^+ \rightarrow t \bar{t}$ 
is given by the same Equation (\ref{mumuttbarsigma}).

\section{$H^{\mp}W^{\pm}$ production at a Hadron Collider}
\subsection{$q \bar{q} \rightarrow H^- W^+ $ interaction}
From the Feynman diagrams of Figures \ref{qqHW_Feynman}
and \ref{qqHW_Feynman2} we obtain:

\begin{eqnarray}
\lefteqn{
\frac{d\sigma_I}{d\hat{t}}\left( q \bar{q}
\rightarrow H^- W^+ \right) = \frac{G_F^2}{48 \pi
\hat{s}} \{ m_q^2 \Lambda \left( \hat{s}, m_H^2, m_W^2 \right)
\left[ (\hat{C_{Hb}})^2 + (\hat{C_{Ab}})^2 \right]
}
\nonumber \\ & &
+2 \sum_{i,j = u,c,t} V_{iq}V_{jq}^*
[ m_q^2 c_{t_{1i}}c_{t_{1j}} \left( \hat{t}^2 +
\frac{\Lambda \left( \hat{s}, m_H^2, m_W^2 \right)
\sin^2\theta m_W^2}{2 \hat{s}} \right)
\nonumber \\ & &
+ m_i^2 m_j^2 c_{t_{2i}}c_{t_{2j}} 
\left( 2 m_W^2 + \frac{\Lambda 
\left( \hat{s}, m_H^2, m_W^2 \right) \sin^2\theta}
{4 \hat{s}} \right) ] + m_q^2 [ - 2 m_H^2m_W^2 
+ 2 \hat{t}^2
\nonumber \\ & &
+ \frac{1}{2} \Lambda \left( \hat{s}, m_H^2, m_W^2 \right)
\sin^2\theta ] \left( \hat{C_{Hb}} + \hat{C_{Ab}} \right)
\sum_{i= u, c, t} \Re{(V_{iq})} c_{t_{1i}} \}
\label{diffqqbar_H-W+}
\end{eqnarray}

\noindent for $ q = d, s, b$. 
In Equation (\ref{diffqqbar_H-W+}),
{ $\hat{C_{Hb}}$ and $\hat{C_{Ab}}$ are given by Equations
(\ref{CHb}) and (\ref{CAb}) replacing $s$ by $\hat{s}$.
$V_{iq}$ are elements of the CKM matrix.

\begin{equation}
c_{t_{1i}} = \frac{\tan\beta}{\left(\hat{t} - m_i^2\right)},
\label{ct1i}
\end{equation}

\noindent and

\begin{equation}
c_{t_{2i}} = \frac{\cot\beta}{\left( \hat{t} 
- m_i^2 \right)}.
\label{ct2i}
\end{equation}

On the other hand,

\begin{eqnarray}
\lefteqn{
\frac{d\sigma_{II}}{d\hat{t}}\left( q \bar{q}
\rightarrow H^- W^+ \right) = \frac{G_F^2}{48 \pi
\hat{s}} \{ m_q^2 \Lambda \left( \hat{s}, m_H^2, m_W^2 \right)
\left( \hat{C_{Ht}}^2 + \hat{C_{At}}^2 \right)
}
\nonumber \\ & &
+2 \sum_{i,j = d,s,b} V_{qi}^*V_{qj}
[ m_q^2 c_{u_{2i}}c_{u_{2j}} \left( \hat{u}^2 +
\frac{\Lambda \left( \hat{s}, m_H^2, m_W^2 \right) 
\sin^2\theta m_W^2}{2 \hat{s}} \right)
\nonumber \\ & &
+ m_i^2 m_j^2 c_{u_{1i}}c_{u_{1j}}
\left( 2 m_W^2 + \frac{\Lambda
\left( \hat{s}, m_H^2, m_W^2 \right) \sin^2\theta}
{4 \hat{s}} \right) ] + m_q^2 [ - 2 m_H^2m_W^2
+ 2 \hat{u}^2
\nonumber \\ & &
+ \frac{1}{2} \Lambda \left( \hat{s}, m_H^2, m_W^2 \right)
\sin^2\theta ] \left( \hat{C_{At}} - \hat{C_{Ht}} \right)
\sum_{i= d, s, b} \Re{(V_{qi})} c_{u_{2i}} \}
\label{diffqqbarIII_H-W+}
\end{eqnarray}

\noindent for $q = u,c$.

\begin{eqnarray}
\lefteqn{
\hat{C_{Ht}} = [ \frac{\left( \frac{1}{2} \sin2\alpha
-\sin^2\alpha \left(\tan\beta \right)^{-1} \right)}
{\left( \hat{s} - m_{H^0}^2 \right)}
}
\nonumber \\ & &
- \frac{ \left(\frac{1}{2} \sin2\alpha + \cos^2\alpha
\left( \tan\beta \right)^{-1}\right)}
{\left( \hat{s} - m_{h^0}^2 \right)} ],
\label{CHt}
\end{eqnarray}

\begin{equation}
\hat{C_{At}} = \frac{\cot\beta}{\left( \hat{s} - m_{A^0}^2
\right) },
\label{CAt}
\end{equation}

\begin{equation}
c_{u_{1i}} = \frac{\tan\beta}{\left(\hat{u} - m_i^2 \right)}
\label{Cu1i}
\end{equation}

\noindent and

\begin{equation}
c_{u_{2i}} = \frac{\cot\beta}{\left(\hat{u} - m_i^2 \right)}.
\label{Cu2i}  
\end{equation}

\begin{figure}
\begin{center}
{\includegraphics{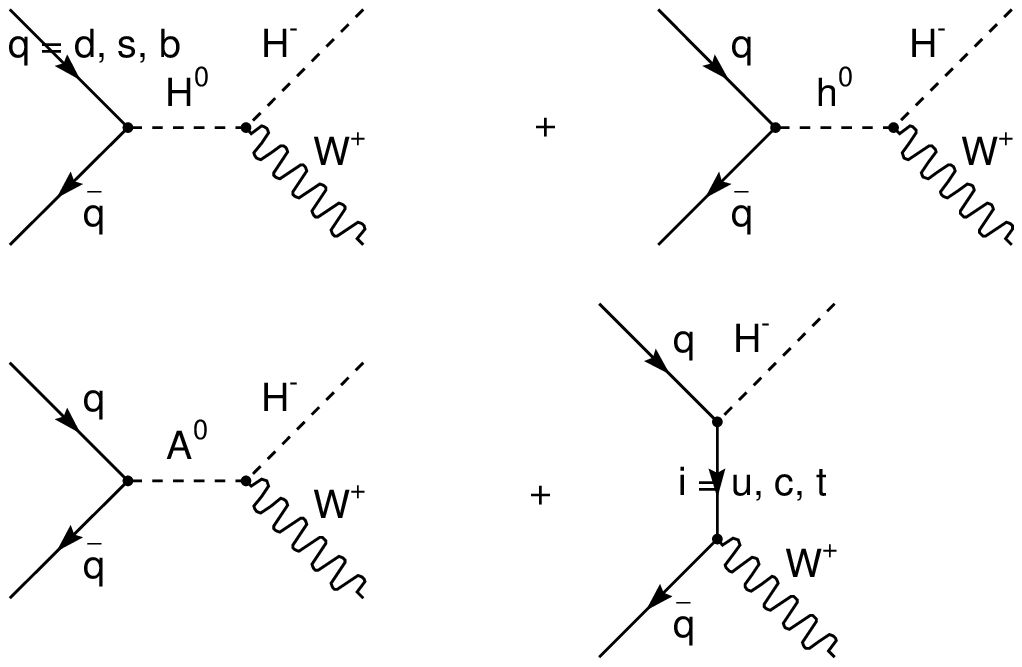}}
\caption{Feynman diagrams corresponding to the process
$\left( q \bar{q}
\rightarrow H^- W^+ \right)$ for $q = d, s, b$.}
\label{qqHW_Feynman}
\end{center}
\end{figure}

\begin{figure}
\begin{center}
{\includegraphics{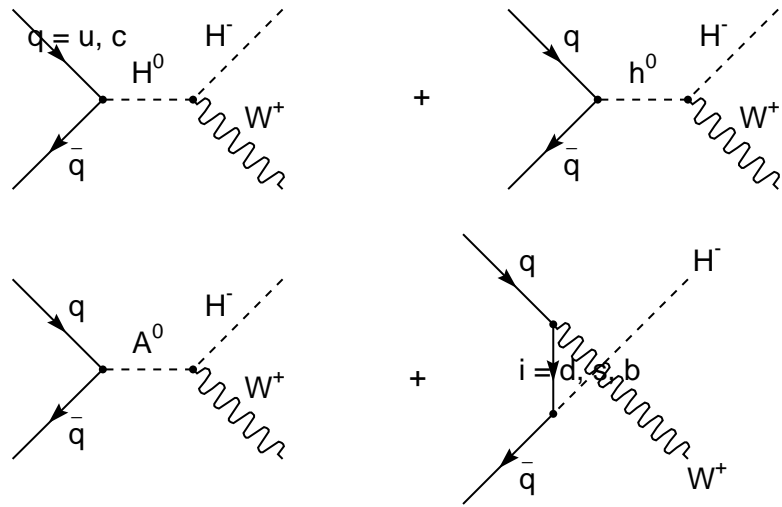}}
\caption{Feynman diagrams corresponding to the process
$\left( q \bar{q}
\rightarrow H^- W^+ \right)$ for $q = u, c$.}
\label{qqHW_Feynman2}
\end{center}
\end{figure}

The differential cross section corresponding to the process
$q \bar{q} \rightarrow H^+ W^-$ for $q = d, s, b$ is obtained
from Equation (\ref{diffqqbar_H-W+}) with the replacement $\hat{t}
\rightarrow \hat{u}$. For $ q = u, c$ we change $\hat{u}$ by
$\hat{t}$ in (\ref{diffqqbarIII_H-W+}).

\subsection{ $g g \rightarrow H^- W^+$ interaction}
The differential cross section corresponding to the sum of the
triangle diagrams in Figure \ref{gghw_Feynman} is given by:

\begin{eqnarray}
\lefteqn{
\frac{d \sigma_{\triangle}}{d\hat{t}} \left( g g \rightarrow
H^- W^+ \right) = \frac{\alpha_s^2 G_F^2}
{4096 \pi^3} 
}
\nonumber \\ & &
\times \Lambda \left( \hat{s}, m_H^2, m_W^2 \right)
\{ \left| \sum_{i = b,t} \left[\hat{C_{Hi}}\left(2\tau_i +
\tau_i \left( \tau_i - 1 \right) f(\tau_i)\right) \right]
\right|^2 
\nonumber \\ & &
+ \frac{1}{2} \left|\sum_{i = b,t} \hat{C_{Ai}} \tau_i 
f(\tau_i) \right|^2 \}
\label{ggH-W+_triangle}
\end{eqnarray} 

\noindent where 
\begin{equation}
\tau_i = \frac{4 m_i^2}{\hat{s}}
\end{equation}

\noindent and

\begin{equation}
f(\tau_i) = \left\{ \begin{array}{ll}
-2 \left[ \arcsin \left( \tau_i^{-1/2} \right) \right]^2 &
\mbox{if $\tau_i > 1$} \\
\frac{1}{2} \left[ \ln \left( \frac{1 + \left( 1- \tau_i \right)^{1/2}}
{1 - \left( 1 - \tau_i \right)^{1/2}} \right) - i \pi \right]^2 &
\mbox{if $\tau_i \le 1$}
\end{array} \right.
\label{ftau}
\end{equation}

\begin{figure}
\begin{center}
{\includegraphics{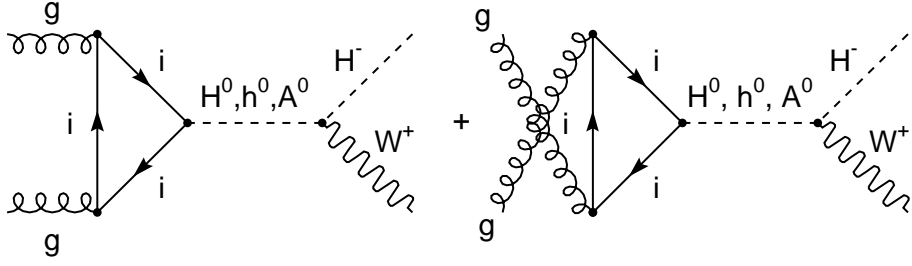}}
\caption{Triangle diagrams corresponding to the process
$ g g \rightarrow H^- W^+ $. $i = b, t$.}
\label{gghw_Feynman}
\end{center}
\end{figure}

Due to charge-conjugation invariance 

\begin{equation}
\frac{d \sigma_{\triangle}}{d\hat{t}} \left( g g \rightarrow
H^- W^+ \right) = \frac{d \sigma_{\triangle}}{d\hat{t}} \left( g g 
\rightarrow
H^+ W^- \right). 
\end{equation}

Equations (\ref{diffqqbar_H-W+}), (\ref{diffqqbarIII_H-W+})
and (\ref{ggH-W+_triangle}) are in agreement with the differential cross
sections calculated in reference \cite{Barrientos_K}. 
In this reference, the differential cross section 
corresponding to the sum of the box diagrams of Figures
\ref{gghwboxI} and \ref{gghwboxII}, also has been calculated 
with the aid of the computer packages FEYNARTS, FEYNCALC and FF.
According to the analisis presented in \cite{Barrientos_K},
the dominant subprocesses of $W^{\pm}H^{\mp}$ associated production
are $b\bar{b} \rightarrow W^{\pm}H^{\mp}$ at the tree level and 
$g g \rightarrow W^{\pm} H^{\mp}$ at one loop.

\begin{figure}
\begin{center}
{\includegraphics{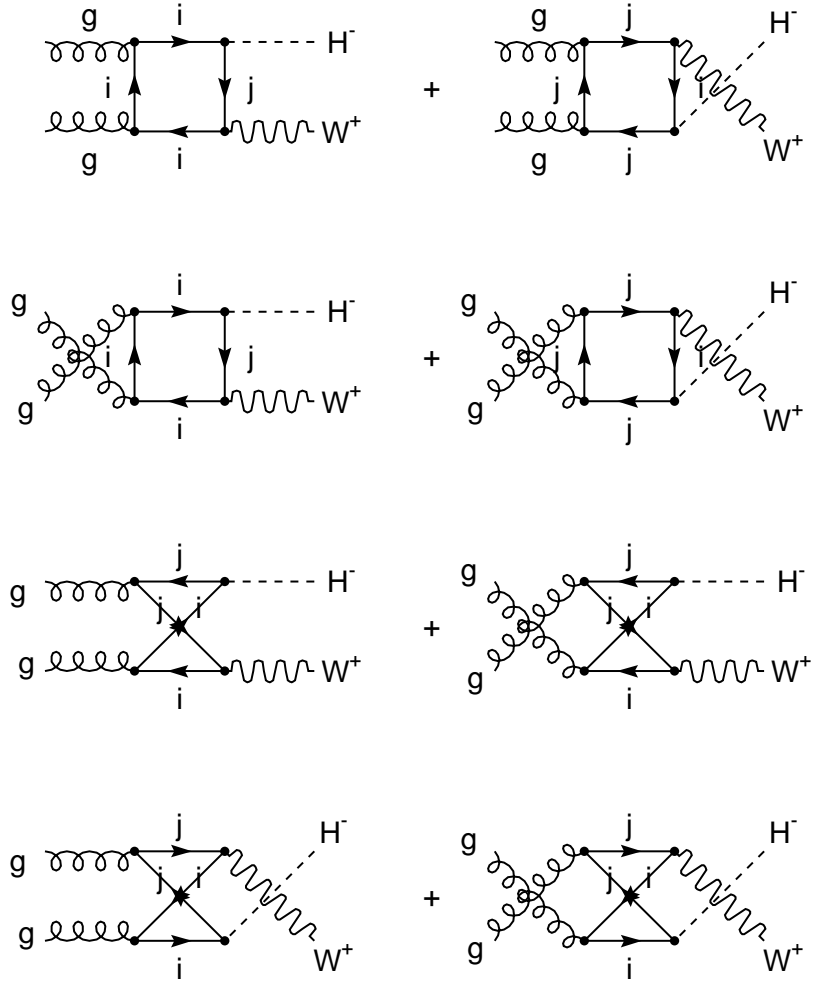}}
\caption{Box diagrams corresponding to the process
$ g g \rightarrow H^- W^+ $. $i = d,s,b$ ; $j=u,c,t$.
Continued in Figure \ref{gghwboxII}.}
\label{gghwboxI}
\end{center}  
\end{figure}

\begin{figure}
\begin{center}
{\includegraphics{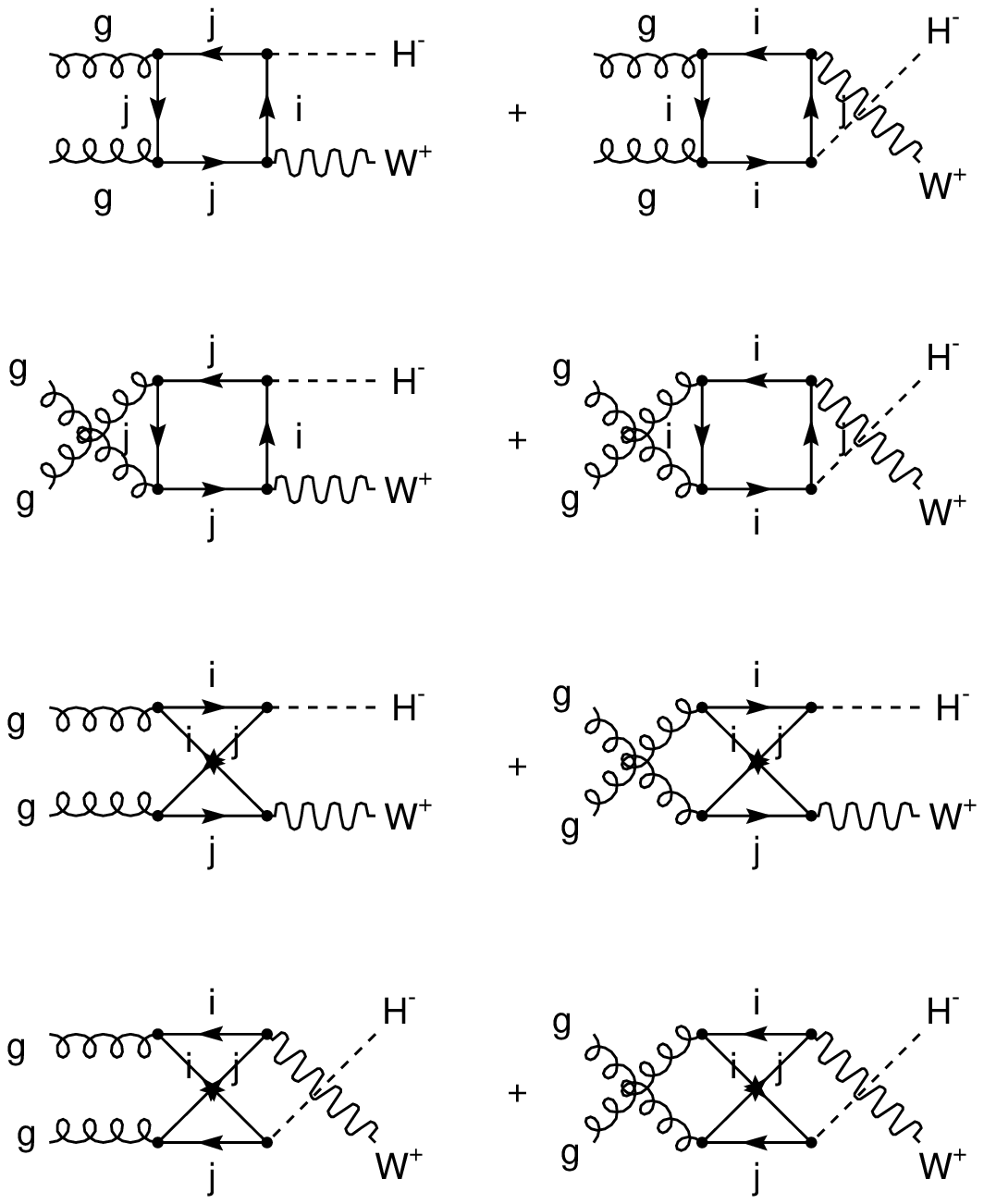}}
\caption{
Continued from Figure \ref{gghwboxI}.}
\label{gghwboxII}
\end{center}
\end{figure}
\subsection{Differential cross section $p \bar{p}
\rightarrow H^{\mp} W^{\pm} X$}

The differential cross section corresponding to the channel
$p \bar{p} \rightarrow H^{\mp} W^{\pm} X$ is:

\begin{eqnarray}
\frac{d^2 \sigma}{dy d \left( p_T \right)^2}
\left(p \bar{p}
\rightarrow H^{\mp} W^{\pm} X\right) & = & 
\sum_{f} { \int_{x_{amin}}^1 {
dx_a f_f \left( x_a, m_a^2 \right) f_f \left( x_b, m_b^2 \right)
}}
\nonumber \\ & &
\times\frac{x_b \hat{s}}{\left(m_H^2 - \hat{u}\right)}
\frac{d \sigma}{d \hat{t}} \left( f \bar{f} \rightarrow 
H^{\mp} W^{\pm} 
\right)
\label{pp_HWX}  
\end{eqnarray}
where $f$ is $q$ or $g$,
\begin{equation}
x_{amin} = \frac{\sqrt{s} m_T e^y + m_H^2 - m_W^2}
{s - \sqrt{s} m_T e^{-y}},
\label{xaminHW}
\end{equation}
\begin{equation}
m_T = \left( m_W^2 + p_T^2 \right)^{\frac{1}{2}},
\label{mTW}
\end{equation}
\begin{equation}
x_b = \frac{x_a \sqrt{s} m_T e^{-y} + m_H^2 - m_W^2}
{x_a s - \sqrt{s} m_T e^y},
\label{xbHW}
\end{equation}
\begin{equation}
\hat{s} = x_a x_b s,
\label{s_hatHW}
\end{equation}
\begin{equation}
p_T^2 = \frac{\Lambda \left( \hat{s}, m_H^2, m_W^2 \right) \sin^2 
\theta}
{4 \hat{s}},
\label{pT2HW}
\end{equation}
\begin{equation}
\hat{u} = \frac{1}{2} \left( m_H^2 + m_W^2 -\hat{s}
- \cos \theta \Lambda^{1/2} ( \hat{s}, m_H^2, m_W^2 )
\right),
\label{uhatHW}  
\end{equation}
\begin{equation}
\hat{t} = \frac{1}{2} \left( m_H^2 + m_W^2 -\hat{s}
+ \cos \theta \Lambda^{1/2} ( \hat{s}, m_H^2, m_W^2 )
\right),
\label{thatHW}  
\end{equation}
\begin{equation}
\hat{u} \hat{t} = m_H^2 m_W^2 + \hat{s} p_T^2,
\label{utHW}
\end{equation}  
and
\begin{equation}
\cos\theta = \left( 1 - \frac{4 \hat{s} p_T^2}
{ \Lambda^{1/2} ( \hat{s}, m_H^2, m_W^2 )}
\right)^{1/2}
\label{costhetaHW}
\end{equation}

\noindent $y$ is the rapidity of $W^{\pm}$, $\theta$ is the angle of
dispersion in the center of mass system,
$p_T$ is the transverse 
momentum of $W^{\pm}$, $f_f$ are the unpolarized parton 
distribution functions for quarks (antiquarks) or gluons.
Finally, $m_a^2$ or $m_b^2$ represent the factorization scale.

A similar expression is valid for the reaction $p p \rightarrow
H^{\mp} W^{\pm} + X$.

In Figure \ref{LHC_figure} (taken from reference \cite{Barrientos_K})
the total cross section $\sigma$ of $pp \rightarrow W^{\pm} H^{\mp} + X$
via $b\bar{b}$ annihilation and $gg$ fusion is plotted as a function of
$m_H$ at LHC energies ($\sqrt{s} = 14 \textrm{TeV/c}$) for $\tan\beta 
=30$. Other 
contributions are negligible.

In Figure \ref{Tevatron_figure} (taken from reference 
\cite{Barrientos_K})
the total cross section $\sigma$ of $p\bar{p} \rightarrow W^{\pm} H^{\mp} 
+ X$
via $b\bar{b}$ annihilation and $gg$ fusion is plotted as a function of
$m_H$ at the Tevatron energy ($\sqrt{s} = 2 \textrm{TeV/c}$) for 
$\tan\beta 
=30$. The
contributions of the other partons are negligible.
\begin{figure}
\begin{center}
\setlength{\unitlength}{0.240900pt}
\ifx\plotpoint\undefined\newsavebox{\plotpoint}\fi
\sbox{\plotpoint}{\rule[-0.200pt]{0.400pt}{0.400pt}}%
\begin{picture}(1500,900)(0,0)
\font\gnuplot=cmr10 at 10pt
\gnuplot
\sbox{\plotpoint}{\rule[-0.200pt]{0.400pt}{0.400pt}}%
\put(181.0,123.0){\rule[-0.200pt]{4.818pt}{0.400pt}}
\put(161,123){\makebox(0,0)[r]{0.1}}
\put(1419.0,123.0){\rule[-0.200pt]{4.818pt}{0.400pt}}
\put(181.0,172.0){\rule[-0.200pt]{2.409pt}{0.400pt}}
\put(1429.0,172.0){\rule[-0.200pt]{2.409pt}{0.400pt}}
\put(181.0,201.0){\rule[-0.200pt]{2.409pt}{0.400pt}}
\put(1429.0,201.0){\rule[-0.200pt]{2.409pt}{0.400pt}}
\put(181.0,221.0){\rule[-0.200pt]{2.409pt}{0.400pt}}
\put(1429.0,221.0){\rule[-0.200pt]{2.409pt}{0.400pt}}
\put(181.0,237.0){\rule[-0.200pt]{2.409pt}{0.400pt}}
\put(1429.0,237.0){\rule[-0.200pt]{2.409pt}{0.400pt}}
\put(181.0,250.0){\rule[-0.200pt]{2.409pt}{0.400pt}}
\put(1429.0,250.0){\rule[-0.200pt]{2.409pt}{0.400pt}}
\put(181.0,261.0){\rule[-0.200pt]{2.409pt}{0.400pt}}
\put(1429.0,261.0){\rule[-0.200pt]{2.409pt}{0.400pt}}
\put(181.0,271.0){\rule[-0.200pt]{2.409pt}{0.400pt}}
\put(1429.0,271.0){\rule[-0.200pt]{2.409pt}{0.400pt}}
\put(181.0,279.0){\rule[-0.200pt]{2.409pt}{0.400pt}}
\put(1429.0,279.0){\rule[-0.200pt]{2.409pt}{0.400pt}}
\put(181.0,287.0){\rule[-0.200pt]{4.818pt}{0.400pt}}
\put(161,287){\makebox(0,0)[r]{1}}
\put(1419.0,287.0){\rule[-0.200pt]{4.818pt}{0.400pt}}
\put(181.0,336.0){\rule[-0.200pt]{2.409pt}{0.400pt}}
\put(1429.0,336.0){\rule[-0.200pt]{2.409pt}{0.400pt}}
\put(181.0,365.0){\rule[-0.200pt]{2.409pt}{0.400pt}}
\put(1429.0,365.0){\rule[-0.200pt]{2.409pt}{0.400pt}}
\put(181.0,385.0){\rule[-0.200pt]{2.409pt}{0.400pt}}
\put(1429.0,385.0){\rule[-0.200pt]{2.409pt}{0.400pt}}
\put(181.0,401.0){\rule[-0.200pt]{2.409pt}{0.400pt}}
\put(1429.0,401.0){\rule[-0.200pt]{2.409pt}{0.400pt}}
\put(181.0,414.0){\rule[-0.200pt]{2.409pt}{0.400pt}}
\put(1429.0,414.0){\rule[-0.200pt]{2.409pt}{0.400pt}}
\put(181.0,425.0){\rule[-0.200pt]{2.409pt}{0.400pt}}
\put(1429.0,425.0){\rule[-0.200pt]{2.409pt}{0.400pt}}
\put(181.0,434.0){\rule[-0.200pt]{2.409pt}{0.400pt}}
\put(1429.0,434.0){\rule[-0.200pt]{2.409pt}{0.400pt}}
\put(181.0,443.0){\rule[-0.200pt]{2.409pt}{0.400pt}}
\put(1429.0,443.0){\rule[-0.200pt]{2.409pt}{0.400pt}}
\put(181.0,450.0){\rule[-0.200pt]{4.818pt}{0.400pt}}
\put(161,450){\makebox(0,0)[r]{10}}
\put(1419.0,450.0){\rule[-0.200pt]{4.818pt}{0.400pt}}
\put(181.0,499.0){\rule[-0.200pt]{2.409pt}{0.400pt}}
\put(1429.0,499.0){\rule[-0.200pt]{2.409pt}{0.400pt}}
\put(181.0,528.0){\rule[-0.200pt]{2.409pt}{0.400pt}}
\put(1429.0,528.0){\rule[-0.200pt]{2.409pt}{0.400pt}}
\put(181.0,548.0){\rule[-0.200pt]{2.409pt}{0.400pt}}
\put(1429.0,548.0){\rule[-0.200pt]{2.409pt}{0.400pt}}
\put(181.0,564.0){\rule[-0.200pt]{2.409pt}{0.400pt}}
\put(1429.0,564.0){\rule[-0.200pt]{2.409pt}{0.400pt}}
\put(181.0,577.0){\rule[-0.200pt]{2.409pt}{0.400pt}}
\put(1429.0,577.0){\rule[-0.200pt]{2.409pt}{0.400pt}}
\put(181.0,588.0){\rule[-0.200pt]{2.409pt}{0.400pt}}
\put(1429.0,588.0){\rule[-0.200pt]{2.409pt}{0.400pt}}
\put(181.0,598.0){\rule[-0.200pt]{2.409pt}{0.400pt}}
\put(1429.0,598.0){\rule[-0.200pt]{2.409pt}{0.400pt}}
\put(181.0,606.0){\rule[-0.200pt]{2.409pt}{0.400pt}}
\put(1429.0,606.0){\rule[-0.200pt]{2.409pt}{0.400pt}}
\put(181.0,614.0){\rule[-0.200pt]{4.818pt}{0.400pt}}
\put(161,614){\makebox(0,0)[r]{100}}
\put(1419.0,614.0){\rule[-0.200pt]{4.818pt}{0.400pt}}
\put(181.0,663.0){\rule[-0.200pt]{2.409pt}{0.400pt}}
\put(1429.0,663.0){\rule[-0.200pt]{2.409pt}{0.400pt}}
\put(181.0,692.0){\rule[-0.200pt]{2.409pt}{0.400pt}}
\put(1429.0,692.0){\rule[-0.200pt]{2.409pt}{0.400pt}}
\put(181.0,712.0){\rule[-0.200pt]{2.409pt}{0.400pt}}
\put(1429.0,712.0){\rule[-0.200pt]{2.409pt}{0.400pt}}
\put(181.0,728.0){\rule[-0.200pt]{2.409pt}{0.400pt}}
\put(1429.0,728.0){\rule[-0.200pt]{2.409pt}{0.400pt}}
\put(181.0,741.0){\rule[-0.200pt]{2.409pt}{0.400pt}}
\put(1429.0,741.0){\rule[-0.200pt]{2.409pt}{0.400pt}}
\put(181.0,752.0){\rule[-0.200pt]{2.409pt}{0.400pt}}
\put(1429.0,752.0){\rule[-0.200pt]{2.409pt}{0.400pt}}
\put(181.0,761.0){\rule[-0.200pt]{2.409pt}{0.400pt}}
\put(1429.0,761.0){\rule[-0.200pt]{2.409pt}{0.400pt}}
\put(181.0,770.0){\rule[-0.200pt]{2.409pt}{0.400pt}}
\put(1429.0,770.0){\rule[-0.200pt]{2.409pt}{0.400pt}}
\put(181.0,777.0){\rule[-0.200pt]{4.818pt}{0.400pt}}
\put(161,777){\makebox(0,0)[r]{1000}}
\put(1419.0,777.0){\rule[-0.200pt]{4.818pt}{0.400pt}}
\put(181.0,123.0){\rule[-0.200pt]{0.400pt}{4.818pt}}
\put(181,82){\makebox(0,0){100}}
\put(181.0,757.0){\rule[-0.200pt]{0.400pt}{4.818pt}}
\put(307.0,123.0){\rule[-0.200pt]{0.400pt}{4.818pt}}
\put(307,82){\makebox(0,0){150}}
\put(307.0,757.0){\rule[-0.200pt]{0.400pt}{4.818pt}}
\put(433.0,123.0){\rule[-0.200pt]{0.400pt}{4.818pt}}
\put(433,82){\makebox(0,0){200}}
\put(433.0,757.0){\rule[-0.200pt]{0.400pt}{4.818pt}}
\put(558.0,123.0){\rule[-0.200pt]{0.400pt}{4.818pt}}
\put(558,82){\makebox(0,0){250}}
\put(558.0,757.0){\rule[-0.200pt]{0.400pt}{4.818pt}}
\put(684.0,123.0){\rule[-0.200pt]{0.400pt}{4.818pt}}
\put(684,82){\makebox(0,0){300}}
\put(684.0,757.0){\rule[-0.200pt]{0.400pt}{4.818pt}}
\put(810.0,123.0){\rule[-0.200pt]{0.400pt}{4.818pt}}
\put(810,82){\makebox(0,0){350}}
\put(810.0,757.0){\rule[-0.200pt]{0.400pt}{4.818pt}}
\put(936.0,123.0){\rule[-0.200pt]{0.400pt}{4.818pt}}
\put(936,82){\makebox(0,0){400}}
\put(936.0,757.0){\rule[-0.200pt]{0.400pt}{4.818pt}}
\put(1062.0,123.0){\rule[-0.200pt]{0.400pt}{4.818pt}}
\put(1062,82){\makebox(0,0){450}}
\put(1062.0,757.0){\rule[-0.200pt]{0.400pt}{4.818pt}}
\put(1187.0,123.0){\rule[-0.200pt]{0.400pt}{4.818pt}}
\put(1187,82){\makebox(0,0){500}}
\put(1187.0,757.0){\rule[-0.200pt]{0.400pt}{4.818pt}}
\put(1313.0,123.0){\rule[-0.200pt]{0.400pt}{4.818pt}}
\put(1313,82){\makebox(0,0){550}}
\put(1313.0,757.0){\rule[-0.200pt]{0.400pt}{4.818pt}}
\put(1439.0,123.0){\rule[-0.200pt]{0.400pt}{4.818pt}}
\put(1439,82){\makebox(0,0){600}}
\put(1439.0,757.0){\rule[-0.200pt]{0.400pt}{4.818pt}}
\put(181.0,123.0){\rule[-0.200pt]{303.052pt}{0.400pt}}
\put(1439.0,123.0){\rule[-0.200pt]{0.400pt}{157.549pt}}
\put(181.0,777.0){\rule[-0.200pt]{303.052pt}{0.400pt}}
\put(40,450){\makebox(0,0){\shortstack{$\sigma$ \\ $[\textrm{fb}]$}}}
\put(810,21){\makebox(0,0){$m_H [\textrm{GeV/c}^2]$}}
\put(810,839){\makebox(0,0){$\sigma = \sigma(pp \rightarrow W^{\pm} H^{\mp} + X)$}}
\put(936,741){\makebox(0,0)[l]{$\sqrt{s} = 14 \textrm{TeV/c}$}}
\put(936,663){\makebox(0,0)[l]{$\tan\beta=30$}}
\put(684,577){\makebox(0,0)[l]{$b\bar{b}$}}
\put(558,365){\makebox(0,0)[l]{$gg$}}
\put(181.0,123.0){\rule[-0.200pt]{0.400pt}{157.549pt}}
\put(1279,737){\makebox(0,0)[r]{ }}
\put(181,770){\raisebox{-.8pt}{\makebox(0,0){$\Diamond$}}}
\put(231,752){\raisebox{-.8pt}{\makebox(0,0){$\Diamond$}}}
\put(282,738){\raisebox{-.8pt}{\makebox(0,0){$\Diamond$}}}
\put(332,725){\raisebox{-.8pt}{\makebox(0,0){$\Diamond$}}}
\put(382,712){\raisebox{-.8pt}{\makebox(0,0){$\Diamond$}}}
\put(433,696){\raisebox{-.8pt}{\makebox(0,0){$\Diamond$}}}
\put(483,679){\raisebox{-.8pt}{\makebox(0,0){$\Diamond$}}}
\put(533,663){\raisebox{-.8pt}{\makebox(0,0){$\Diamond$}}}
\put(584,653){\raisebox{-.8pt}{\makebox(0,0){$\Diamond$}}}
\put(634,642){\raisebox{-.8pt}{\makebox(0,0){$\Diamond$}}}
\put(684,626){\raisebox{-.8pt}{\makebox(0,0){$\Diamond$}}}
\put(735,617){\raisebox{-.8pt}{\makebox(0,0){$\Diamond$}}}
\put(785,606){\raisebox{-.8pt}{\makebox(0,0){$\Diamond$}}}
\put(835,598){\raisebox{-.8pt}{\makebox(0,0){$\Diamond$}}}
\put(885,588){\raisebox{-.8pt}{\makebox(0,0){$\Diamond$}}}
\put(936,577){\raisebox{-.8pt}{\makebox(0,0){$\Diamond$}}}
\put(986,571){\raisebox{-.8pt}{\makebox(0,0){$\Diamond$}}}
\put(1036,564){\raisebox{-.8pt}{\makebox(0,0){$\Diamond$}}}
\put(1087,554){\raisebox{-.8pt}{\makebox(0,0){$\Diamond$}}}
\put(1137,545){\raisebox{-.8pt}{\makebox(0,0){$\Diamond$}}}
\put(1187,533){\raisebox{-.8pt}{\makebox(0,0){$\Diamond$}}}
\put(1238,523){\raisebox{-.8pt}{\makebox(0,0){$\Diamond$}}}
\put(1288,518){\raisebox{-.8pt}{\makebox(0,0){$\Diamond$}}}
\put(1338,512){\raisebox{-.8pt}{\makebox(0,0){$\Diamond$}}}
\put(1389,503){\raisebox{-.8pt}{\makebox(0,0){$\Diamond$}}}
\put(1439,496){\raisebox{-.8pt}{\makebox(0,0){$\Diamond$}}}
\put(1349,737){\raisebox{-.8pt}{\makebox(0,0){$\Diamond$}}}
\put(1279,696){\makebox(0,0)[r]{ }}
\put(181,496){\makebox(0,0){$+$}}
\put(231,474){\makebox(0,0){$+$}}
\put(282,434){\makebox(0,0){$+$}}
\put(332,401){\makebox(0,0){$+$}}
\put(382,373){\makebox(0,0){$+$}}
\put(407,385){\makebox(0,0){$+$}}
\put(433,401){\makebox(0,0){$+$}}
\put(483,381){\makebox(0,0){$+$}}
\put(533,357){\makebox(0,0){$+$}}
\put(584,339){\makebox(0,0){$+$}}
\put(634,324){\makebox(0,0){$+$}}
\put(684,299){\makebox(0,0){$+$}}
\put(735,287){\makebox(0,0){$+$}}
\put(785,271){\makebox(0,0){$+$}}
\put(835,261){\makebox(0,0){$+$}}
\put(885,237){\makebox(0,0){$+$}}
\put(936,221){\makebox(0,0){$+$}}
\put(986,212){\makebox(0,0){$+$}}
\put(1036,201){\makebox(0,0){$+$}}
\put(1087,188){\makebox(0,0){$+$}}
\put(1137,172){\makebox(0,0){$+$}}
\put(1187,152){\makebox(0,0){$+$}}
\put(1238,142){\makebox(0,0){$+$}}
\put(1288,130){\makebox(0,0){$+$}}
\put(1338,123){\makebox(0,0){$+$}}
\put(1349,696){\makebox(0,0){$+$}}
\end{picture}
\caption{Total cross section for the process
$p p \rightarrow W^{\pm} H^{\mp} + X$ as a function
of $m_H$ via $b\bar{b}$ annihilation and $gg$ fusion
at LHC energies 
($\sqrt{s} = 14 \textrm{TeV/c}$) for
$\tan\beta = 30$. Taken from \cite{Barrientos_K}.}
\label{LHC_figure}
\end{center}
\end{figure}
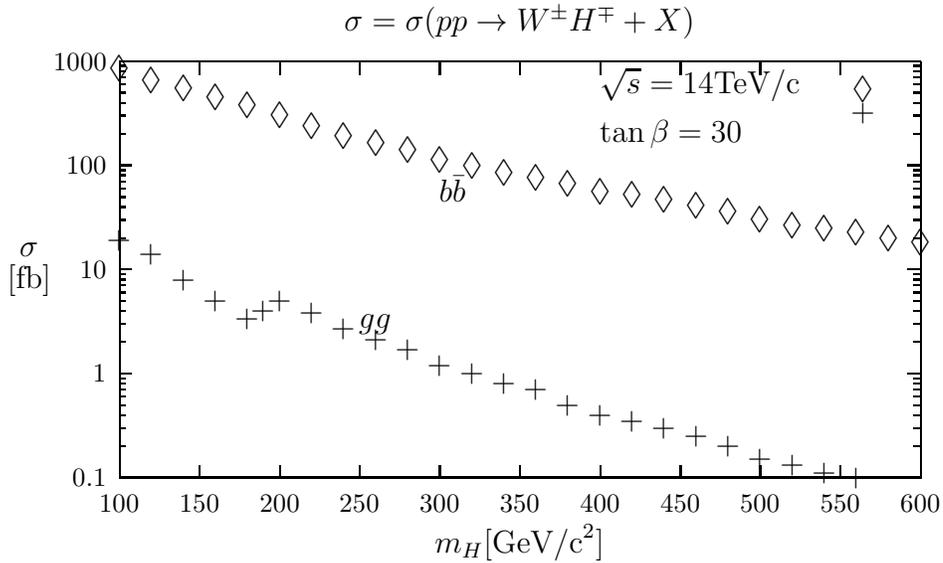
 
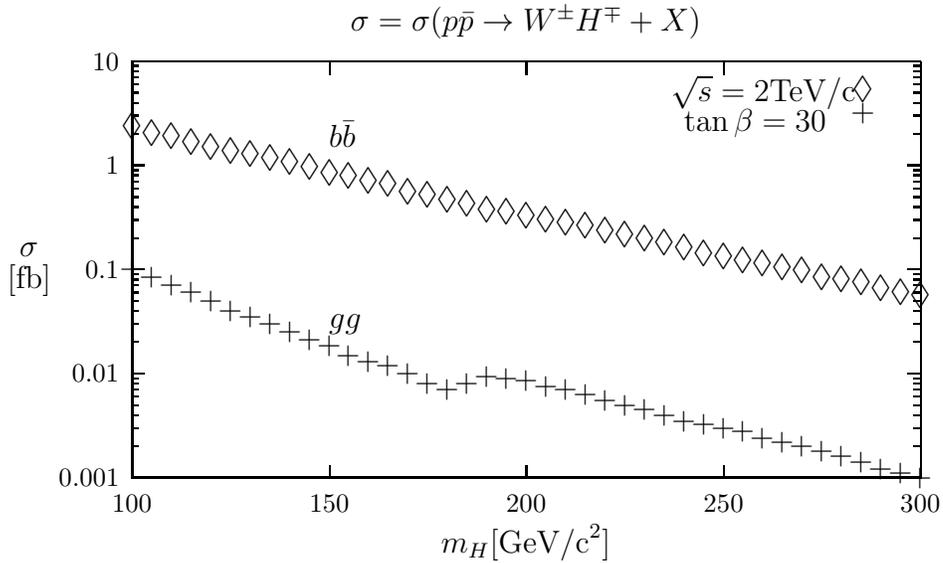
\begin{figure}
\begin{center}
\setlength{\unitlength}{0.240900pt}
\ifx\plotpoint\undefined\newsavebox{\plotpoint}\fi
\sbox{\plotpoint}{\rule[-0.200pt]{0.400pt}{0.400pt}}%
\begin{picture}(1500,900)(0,0)
\font\gnuplot=cmr10 at 10pt
\gnuplot
\sbox{\plotpoint}{\rule[-0.200pt]{0.400pt}{0.400pt}}%
\put(201.0,123.0){\rule[-0.200pt]{4.818pt}{0.400pt}}
\put(181,123){\makebox(0,0)[r]{0.001}}
\put(1419.0,123.0){\rule[-0.200pt]{4.818pt}{0.400pt}}
\put(201.0,172.0){\rule[-0.200pt]{2.409pt}{0.400pt}}
\put(1429.0,172.0){\rule[-0.200pt]{2.409pt}{0.400pt}}
\put(201.0,201.0){\rule[-0.200pt]{2.409pt}{0.400pt}}
\put(1429.0,201.0){\rule[-0.200pt]{2.409pt}{0.400pt}}
\put(201.0,221.0){\rule[-0.200pt]{2.409pt}{0.400pt}}
\put(1429.0,221.0){\rule[-0.200pt]{2.409pt}{0.400pt}}
\put(201.0,237.0){\rule[-0.200pt]{2.409pt}{0.400pt}}
\put(1429.0,237.0){\rule[-0.200pt]{2.409pt}{0.400pt}}
\put(201.0,250.0){\rule[-0.200pt]{2.409pt}{0.400pt}}
\put(1429.0,250.0){\rule[-0.200pt]{2.409pt}{0.400pt}}
\put(201.0,261.0){\rule[-0.200pt]{2.409pt}{0.400pt}}
\put(1429.0,261.0){\rule[-0.200pt]{2.409pt}{0.400pt}}
\put(201.0,271.0){\rule[-0.200pt]{2.409pt}{0.400pt}}
\put(1429.0,271.0){\rule[-0.200pt]{2.409pt}{0.400pt}}
\put(201.0,279.0){\rule[-0.200pt]{2.409pt}{0.400pt}}
\put(1429.0,279.0){\rule[-0.200pt]{2.409pt}{0.400pt}}
\put(201.0,287.0){\rule[-0.200pt]{4.818pt}{0.400pt}}
\put(181,287){\makebox(0,0)[r]{0.01}}
\put(1419.0,287.0){\rule[-0.200pt]{4.818pt}{0.400pt}}
\put(201.0,336.0){\rule[-0.200pt]{2.409pt}{0.400pt}}
\put(1429.0,336.0){\rule[-0.200pt]{2.409pt}{0.400pt}}
\put(201.0,365.0){\rule[-0.200pt]{2.409pt}{0.400pt}}
\put(1429.0,365.0){\rule[-0.200pt]{2.409pt}{0.400pt}}
\put(201.0,385.0){\rule[-0.200pt]{2.409pt}{0.400pt}}
\put(1429.0,385.0){\rule[-0.200pt]{2.409pt}{0.400pt}}
\put(201.0,401.0){\rule[-0.200pt]{2.409pt}{0.400pt}}
\put(1429.0,401.0){\rule[-0.200pt]{2.409pt}{0.400pt}}
\put(201.0,414.0){\rule[-0.200pt]{2.409pt}{0.400pt}}
\put(1429.0,414.0){\rule[-0.200pt]{2.409pt}{0.400pt}}
\put(201.0,425.0){\rule[-0.200pt]{2.409pt}{0.400pt}}
\put(1429.0,425.0){\rule[-0.200pt]{2.409pt}{0.400pt}}
\put(201.0,434.0){\rule[-0.200pt]{2.409pt}{0.400pt}}
\put(1429.0,434.0){\rule[-0.200pt]{2.409pt}{0.400pt}}
\put(201.0,443.0){\rule[-0.200pt]{2.409pt}{0.400pt}}
\put(1429.0,443.0){\rule[-0.200pt]{2.409pt}{0.400pt}}
\put(201.0,450.0){\rule[-0.200pt]{4.818pt}{0.400pt}}
\put(181,450){\makebox(0,0)[r]{0.1}}
\put(1419.0,450.0){\rule[-0.200pt]{4.818pt}{0.400pt}}
\put(201.0,499.0){\rule[-0.200pt]{2.409pt}{0.400pt}}
\put(1429.0,499.0){\rule[-0.200pt]{2.409pt}{0.400pt}}
\put(201.0,528.0){\rule[-0.200pt]{2.409pt}{0.400pt}}
\put(1429.0,528.0){\rule[-0.200pt]{2.409pt}{0.400pt}}
\put(201.0,548.0){\rule[-0.200pt]{2.409pt}{0.400pt}}
\put(1429.0,548.0){\rule[-0.200pt]{2.409pt}{0.400pt}}
\put(201.0,564.0){\rule[-0.200pt]{2.409pt}{0.400pt}}
\put(1429.0,564.0){\rule[-0.200pt]{2.409pt}{0.400pt}}
\put(201.0,577.0){\rule[-0.200pt]{2.409pt}{0.400pt}}
\put(1429.0,577.0){\rule[-0.200pt]{2.409pt}{0.400pt}}
\put(201.0,588.0){\rule[-0.200pt]{2.409pt}{0.400pt}}
\put(1429.0,588.0){\rule[-0.200pt]{2.409pt}{0.400pt}}
\put(201.0,598.0){\rule[-0.200pt]{2.409pt}{0.400pt}}
\put(1429.0,598.0){\rule[-0.200pt]{2.409pt}{0.400pt}}
\put(201.0,606.0){\rule[-0.200pt]{2.409pt}{0.400pt}}
\put(1429.0,606.0){\rule[-0.200pt]{2.409pt}{0.400pt}}
\put(201.0,614.0){\rule[-0.200pt]{4.818pt}{0.400pt}}
\put(181,614){\makebox(0,0)[r]{1}}
\put(1419.0,614.0){\rule[-0.200pt]{4.818pt}{0.400pt}}
\put(201.0,663.0){\rule[-0.200pt]{2.409pt}{0.400pt}}
\put(1429.0,663.0){\rule[-0.200pt]{2.409pt}{0.400pt}}
\put(201.0,692.0){\rule[-0.200pt]{2.409pt}{0.400pt}}
\put(1429.0,692.0){\rule[-0.200pt]{2.409pt}{0.400pt}}
\put(201.0,712.0){\rule[-0.200pt]{2.409pt}{0.400pt}}
\put(1429.0,712.0){\rule[-0.200pt]{2.409pt}{0.400pt}}
\put(201.0,728.0){\rule[-0.200pt]{2.409pt}{0.400pt}}
\put(1429.0,728.0){\rule[-0.200pt]{2.409pt}{0.400pt}}
\put(201.0,741.0){\rule[-0.200pt]{2.409pt}{0.400pt}}
\put(1429.0,741.0){\rule[-0.200pt]{2.409pt}{0.400pt}}
\put(201.0,752.0){\rule[-0.200pt]{2.409pt}{0.400pt}}
\put(1429.0,752.0){\rule[-0.200pt]{2.409pt}{0.400pt}}
\put(201.0,761.0){\rule[-0.200pt]{2.409pt}{0.400pt}}
\put(1429.0,761.0){\rule[-0.200pt]{2.409pt}{0.400pt}}
\put(201.0,770.0){\rule[-0.200pt]{2.409pt}{0.400pt}}
\put(1429.0,770.0){\rule[-0.200pt]{2.409pt}{0.400pt}}
\put(201.0,777.0){\rule[-0.200pt]{4.818pt}{0.400pt}}
\put(181,777){\makebox(0,0)[r]{10}}
\put(1419.0,777.0){\rule[-0.200pt]{4.818pt}{0.400pt}}
\put(201.0,123.0){\rule[-0.200pt]{0.400pt}{4.818pt}}
\put(201,82){\makebox(0,0){100}}
\put(201.0,757.0){\rule[-0.200pt]{0.400pt}{4.818pt}}
\put(511.0,123.0){\rule[-0.200pt]{0.400pt}{4.818pt}}
\put(511,82){\makebox(0,0){150}}
\put(511.0,757.0){\rule[-0.200pt]{0.400pt}{4.818pt}}
\put(820.0,123.0){\rule[-0.200pt]{0.400pt}{4.818pt}}
\put(820,82){\makebox(0,0){200}}
\put(820.0,757.0){\rule[-0.200pt]{0.400pt}{4.818pt}}
\put(1130.0,123.0){\rule[-0.200pt]{0.400pt}{4.818pt}}
\put(1130,82){\makebox(0,0){250}}
\put(1130.0,757.0){\rule[-0.200pt]{0.400pt}{4.818pt}}
\put(1439.0,123.0){\rule[-0.200pt]{0.400pt}{4.818pt}}
\put(1439,82){\makebox(0,0){300}}
\put(1439.0,757.0){\rule[-0.200pt]{0.400pt}{4.818pt}}
\put(201.0,123.0){\rule[-0.200pt]{298.234pt}{0.400pt}}
\put(1439.0,123.0){\rule[-0.200pt]{0.400pt}{157.549pt}}
\put(201.0,777.0){\rule[-0.200pt]{298.234pt}{0.400pt}}
\put(40,450){\makebox(0,0){\shortstack{$\sigma$ \\ $[\textrm{fb}]$}}}
\put(820,21){\makebox(0,0){$m_H [\textrm{GeV/c}^2]$}}
\put(820,839){\makebox(0,0){$\sigma = \sigma(p \bar{p} \rightarrow W^{\pm} H^{\mp} + X)$}}
\put(1048,729){\makebox(0,0)[l]{$\sqrt{s} = 2 \textrm{TeV/c}$}}
\put(1068,680){\makebox(0,0)[l]{$\tan\beta = 30$}}
\put(511,663){\makebox(0,0)[l]{$b\bar{b}$}}
\put(511,365){\makebox(0,0)[l]{$gg$}}
\put(201.0,123.0){\rule[-0.200pt]{0.400pt}{157.549pt}}
\put(1279,737){\makebox(0,0)[r]{ }}
\put(201,679){\raisebox{-.8pt}{\makebox(0,0){$\Diamond$}}}
\put(232,669){\raisebox{-.8pt}{\makebox(0,0){$\Diamond$}}}
\put(263,663){\raisebox{-.8pt}{\makebox(0,0){$\Diamond$}}}
\put(294,655){\raisebox{-.8pt}{\makebox(0,0){$\Diamond$}}}
\put(325,647){\raisebox{-.8pt}{\makebox(0,0){$\Diamond$}}}
\put(356,640){\raisebox{-.8pt}{\makebox(0,0){$\Diamond$}}}
\put(387,635){\raisebox{-.8pt}{\makebox(0,0){$\Diamond$}}}
\put(418,629){\raisebox{-.8pt}{\makebox(0,0){$\Diamond$}}}
\put(449,623){\raisebox{-.8pt}{\makebox(0,0){$\Diamond$}}}
\put(480,616){\raisebox{-.8pt}{\makebox(0,0){$\Diamond$}}}
\put(511,606){\raisebox{-.8pt}{\makebox(0,0){$\Diamond$}}}
\put(541,602){\raisebox{-.8pt}{\makebox(0,0){$\Diamond$}}}
\put(572,593){\raisebox{-.8pt}{\makebox(0,0){$\Diamond$}}}
\put(603,588){\raisebox{-.8pt}{\makebox(0,0){$\Diamond$}}}
\put(634,577){\raisebox{-.8pt}{\makebox(0,0){$\Diamond$}}}
\put(665,572){\raisebox{-.8pt}{\makebox(0,0){$\Diamond$}}}
\put(696,564){\raisebox{-.8pt}{\makebox(0,0){$\Diamond$}}}
\put(727,557){\raisebox{-.8pt}{\makebox(0,0){$\Diamond$}}}
\put(758,548){\raisebox{-.8pt}{\makebox(0,0){$\Diamond$}}}
\put(789,545){\raisebox{-.8pt}{\makebox(0,0){$\Diamond$}}}
\put(820,539){\raisebox{-.8pt}{\makebox(0,0){$\Diamond$}}}
\put(851,533){\raisebox{-.8pt}{\makebox(0,0){$\Diamond$}}}
\put(882,528){\raisebox{-.8pt}{\makebox(0,0){$\Diamond$}}}
\put(913,523){\raisebox{-.8pt}{\makebox(0,0){$\Diamond$}}}
\put(944,515){\raisebox{-.8pt}{\makebox(0,0){$\Diamond$}}}
\put(975,509){\raisebox{-.8pt}{\makebox(0,0){$\Diamond$}}}
\put(1006,503){\raisebox{-.8pt}{\makebox(0,0){$\Diamond$}}}
\put(1037,496){\raisebox{-.8pt}{\makebox(0,0){$\Diamond$}}}
\put(1068,488){\raisebox{-.8pt}{\makebox(0,0){$\Diamond$}}}
\put(1099,479){\raisebox{-.8pt}{\makebox(0,0){$\Diamond$}}}
\put(1130,474){\raisebox{-.8pt}{\makebox(0,0){$\Diamond$}}}
\put(1160,469){\raisebox{-.8pt}{\makebox(0,0){$\Diamond$}}}
\put(1191,463){\raisebox{-.8pt}{\makebox(0,0){$\Diamond$}}}
\put(1222,457){\raisebox{-.8pt}{\makebox(0,0){$\Diamond$}}}
\put(1253,453){\raisebox{-.8pt}{\makebox(0,0){$\Diamond$}}}
\put(1284,442){\raisebox{-.8pt}{\makebox(0,0){$\Diamond$}}}
\put(1315,438){\raisebox{-.8pt}{\makebox(0,0){$\Diamond$}}}
\put(1346,434){\raisebox{-.8pt}{\makebox(0,0){$\Diamond$}}}
\put(1377,425){\raisebox{-.8pt}{\makebox(0,0){$\Diamond$}}}
\put(1408,419){\raisebox{-.8pt}{\makebox(0,0){$\Diamond$}}}
\put(1439,414){\raisebox{-.8pt}{\makebox(0,0){$\Diamond$}}}
\put(1349,737){\raisebox{-.8pt}{\makebox(0,0){$\Diamond$}}}
\put(1279,696){\makebox(0,0)[r]{ }}
\put(201,450){\makebox(0,0){$+$}}
\put(232,438){\makebox(0,0){$+$}}
\put(263,425){\makebox(0,0){$+$}}
\put(294,414){\makebox(0,0){$+$}}
\put(325,401){\makebox(0,0){$+$}}
\put(356,385){\makebox(0,0){$+$}}
\put(387,375){\makebox(0,0){$+$}}
\put(418,365){\makebox(0,0){$+$}}
\put(449,352){\makebox(0,0){$+$}}
\put(480,339){\makebox(0,0){$+$}}
\put(511,330){\makebox(0,0){$+$}}
\put(541,315){\makebox(0,0){$+$}}
\put(572,305){\makebox(0,0){$+$}}
\put(603,299){\makebox(0,0){$+$}}
\put(634,287){\makebox(0,0){$+$}}
\put(665,271){\makebox(0,0){$+$}}
\put(696,261){\makebox(0,0){$+$}}
\put(727,271){\makebox(0,0){$+$}}
\put(758,281){\makebox(0,0){$+$}}
\put(789,279){\makebox(0,0){$+$}}
\put(820,275){\makebox(0,0){$+$}}
\put(851,266){\makebox(0,0){$+$}}
\put(882,261){\makebox(0,0){$+$}}
\put(913,253){\makebox(0,0){$+$}}
\put(944,244){\makebox(0,0){$+$}}
\put(975,237){\makebox(0,0){$+$}}
\put(1006,230){\makebox(0,0){$+$}}
\put(1037,221){\makebox(0,0){$+$}}
\put(1068,212){\makebox(0,0){$+$}}
\put(1099,206){\makebox(0,0){$+$}}
\put(1130,201){\makebox(0,0){$+$}}
\put(1160,196){\makebox(0,0){$+$}}
\put(1191,185){\makebox(0,0){$+$}}
\put(1222,179){\makebox(0,0){$+$}}
\put(1253,172){\makebox(0,0){$+$}}
\put(1284,165){\makebox(0,0){$+$}}
\put(1315,156){\makebox(0,0){$+$}}
\put(1346,147){\makebox(0,0){$+$}}
\put(1377,136){\makebox(0,0){$+$}}
\put(1408,130){\makebox(0,0){$+$}}
\put(1439,123){\makebox(0,0){$+$}}
\put(1349,696){\makebox(0,0){$+$}}
\end{picture}
\caption{Total cross section for the process
$p \bar{p} \rightarrow W^{\pm} H^{\mp} + X$ as a function
of $m_H$ via $b\bar{b}$ annihilation and $gg$ fusion
at the Tevatron energy
($\sqrt{s} = 2 \textrm{TeV/c}$) for
$\tan\beta = 30$. Taken from \cite{Barrientos_K}.}
\label{Tevatron_figure}
\end{center}
\end{figure}
\section{Comparison between $\mu^- \mu^+ \rightarrow H^{\mp} W^{\pm}$
and $ p \bar{p}, pp \rightarrow H^{\mp} W^{\pm}X$ for large values of 
$\tan\beta$}
Let us compare the channel  $\mu^- \mu^+ \rightarrow H^{\mp} W^{\pm}$
at $\sqrt{s} = 500 \textrm{GeV/c}$
with the processes $ p \bar{p}, pp \rightarrow H^{\mp} W^{\pm}X$ at the 
Tevatron energy ($\sqrt{s} = 2 \textrm{TeV/c}$) and LHC energies 
($\sqrt{s} = 14 
\textrm{TeV/c}$) respectively for large values of $\tan\beta$ (for example 
$\tan\beta= 30$).

At the FNAL energy  (Figure \ref{TeVmuon_figure}), we have:
$\sigma(\mu^- \mu^+ \rightarrow H^{\mp} W^{\pm})
> \sigma( p \bar{p} \rightarrow W^{\pm} H^{\mp}X)$ for $\tan\beta= 30$.

At LHC energies (Figure \ref{LHCmuon_figure}), we have:
$\sigma( p p \rightarrow W^{\pm} H^{\mp}X) > \sigma(\mu^- \mu^+ 
\rightarrow H^{\mp}W^{\pm})$ for $\tan\beta=30$.

According to Figure  \ref{muhw_figure}, $\sigma(\mu^- \mu^+
\rightarrow H^{\mp}W^{\pm}) \gtrsim 5 \textrm{fb}$ for $\tan\beta \geq 20$ 
in the 
mass interval $ 100 \leq m_{H} \leq 400 [\textrm{GeV/c}^2]$, which would 
be an observable 
number of $H^{\pm}$ for luminosities $>50 \textrm{fb}^{-1}$. In the mass 
region 
of interest shown in the figures, the dominant decay mode of $H^{\pm}$ is
$H^+ \rightarrow t\bar{b}$ or $H^- \rightarrow \bar{t} b$. So the main 
background would be from $t\bar{t}$ production. Reference 
\cite{S.Moretti} shows that such a background overwhelms the charged
Higgs boson signal in $p p \rightarrow W^{\pm} H^{\mp}X$ at the LHC.
In fact, in Section 4.7 we have shown that $\sigma(\mu^- \mu^+ \rightarrow 
t 
\bar{t}) \approx 495 \textrm{fb}$ for $\sqrt{s} = 500 \textrm{GeV/c}$. In 
the LHC
the background due to $t \bar{t}$ production is of order \cite{S.Moretti} 
800 pb (three 
orders of magnitude larger than at a muon collider with $\sqrt{s} = 500 
\textrm{GeV/c}$). At the FNAL energy ($\sqrt{s} = 2 \textrm{TeV/c}$) 
something 
similar happens
because $\sigma(p \bar{p} \rightarrow t \bar{t}) = 5.5 \textrm{pb}$ 
\cite{Abazovetal}.

In the muon collider, the signal of the charged Higgs boson is not 
overwhelmed.

Then, for large values of $\tan\beta$, the process $\mu^- \mu^+ 
\rightarrow H^{\mp} W^{\pm}$ is a very attractive channel for the 
search of $H^{\pm}$ at a $\mu^- \mu^+$ collider.
 
\begin{figure}
\begin{center}
\input{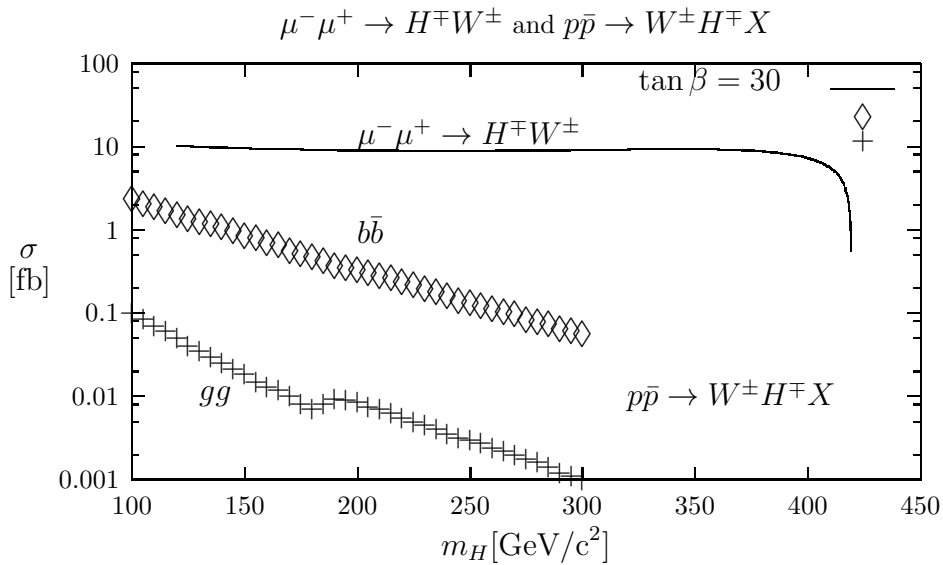}
\caption{Total cross section for the processes
$\mu^- \mu^+ \rightarrow H^{\mp} W^{\pm}$ and
$p \bar{p} \rightarrow W^{\pm} H^{\mp} + X$ (via $b\bar{b}$ annihilation 
and $gg$ fusion) as a function of $m_{H}$
at $\sqrt{s} = 500 \textrm{GeV/c}$ and
$\sqrt{s} = 2 \textrm{TeV/c}$, respectively,  for
$\tan\beta = 30$. Taken partially from \cite{Barrientos_K}.}
\label{TeVmuon_figure}
\end{center}
\end{figure}

\begin{figure}
\begin{center}
\input{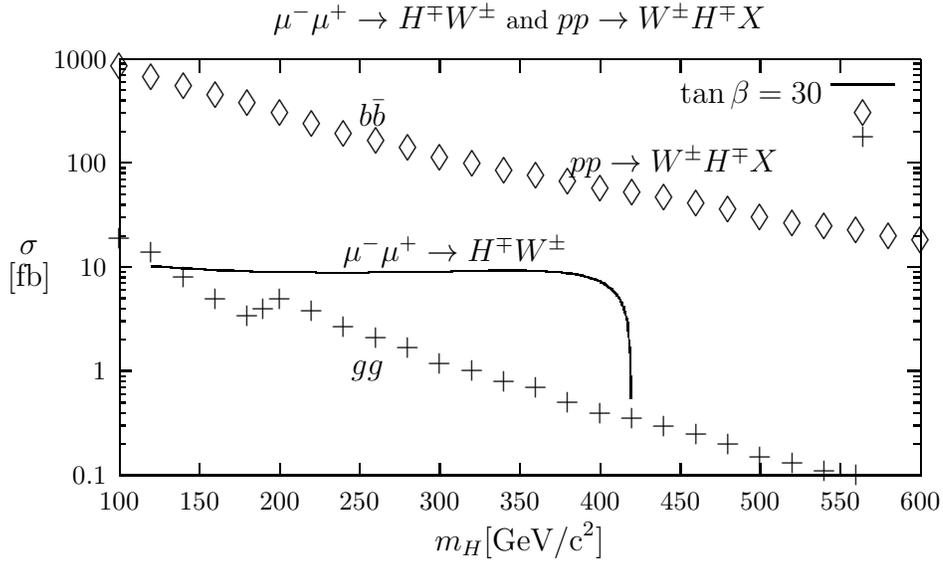}
\caption{Total cross section for the processes
$\mu^- \mu^+ \rightarrow H^{\mp} W^{\pm}$ and
$p p \rightarrow W^{\pm} H^{\mp} + X$ (via $b\bar{b}$ annihilation
and $gg$ fusion) as a function of $m_{H}$
at $\sqrt{s} = 500 \textrm{GeV/c}$ and
$\sqrt{s} = 14 \textrm{TeV/c}$, respectively,  for
$\tan\beta = 30$. Taken partially from \cite{Barrientos_K}.}
\label{LHCmuon_figure}
\end{center}
\end{figure}
\section{Conclusions}
The discovery of the Standard Model Higgs is one of the principal goals
of experimental and theoretical particle physicists. This is because the
Higgs mechanism is a cornerstone of the Standard Model. The search for the
Standard Model Higgs will also constrain or discover particles of the Two
Higgs Doublet Model of type II.

In this paper we have discussed the masses of the Higgs particles in the
Two Higgs Doublet Model of type II, and considered the influence of the 
radiative corrections on these masses. In the absence of radiative 
corrections, the Higgs boson $h^0$ obeys the bound $m_{h^0} \leq m_Z$. 
This bound practically has been excluded by the present limits on 
$m_{h^0}$ obtained by LEP and CDF \cite{LEP_CDF}. However, when the 
radiative corrections are taken into account, $m_{h^0}$ increases as the 
value of $m_{A}$ increases. As a result,  we have a new bound: $m_{h^0} 
\leq 128.062
\textrm{GeV/c}^2$ taking $M_{sb}$ (sbottom mass) and $M_{st}$ (stop mass) 
of order
$1 \textrm{TeV/c}^2$.

Considering the radiative corrections of the masses, we have calculated 
Higgs production 
cross sections at a muon collider in the Two Higgs Doublet Model of type 
II. The most interesting production channels are $\mu^- \mu^+ \rightarrow 
h^0 Z^0, H^0 Z^0$
, $H^- H^+, A^0 Z^0$ and $H^{\mp} W^{\pm}$.  In the first 
two channels the radiative corrections of the masses play an important 
role, which is 
not true for the other channels. In the reaction  $\mu^- \mu^+ \rightarrow
h^0 Z^0$, the total cross section becomes important
in the mass interval $118 \leq m_{h^0} \leq 128 [\textrm{GeV/c}^2]$.
  
The process $ \mu^- \mu^+ \rightarrow A^0 Z^0$, would provide an 
alternative way for searching the $A^0$ looking for peaks in the $b 
\bar{b}$ distribution. Another interesting channel could be $\mu^- \mu^+ 
\rightarrow A^0 h^0$. However, this is highly supressed for $m_{A} \geq 
200 \textrm{GeV/c}^2$ because the total cross section is proportional to 
the factor 
\begin{equation}
\cos^2\left( \beta - \alpha \right) = \frac{\left( 1 + \tan\beta 
\tan\alpha\right)^2}{\left(1+\tan^2\beta\right)\left(1+\tan^2\alpha\right)}
\nonumber
\end{equation}
(see the Feynman rules given in \cite{M_H}). This factor 
decreases as the mass of the $A^0$ increases.

The most attractive channel is $\mu^- \mu^+ \rightarrow H^{\mp} W^{\pm}$, 
see Figures \ref{TeVmuon_figure} and \ref{LHCmuon_figure}. In this 
reaction $\sigma(\mu^- \mu^+ \rightarrow H^{\mp} W^{\pm}) \gtrsim 5 
\textrm{fb}$ 
for
$ \tan\beta \geq 20$ in the mass interval $100 \leq m_{H} \leq 400 
[\textrm{GeV/c}^2]$, which would give an observable number of $H^{\pm}$ 
for 
luminosities $>50 \textrm{fb}^{-1}$ at $\sqrt{s} = 500 \textrm{GeV/c}$.

Because the main background in a hadron collider in the reactions 
$p\bar{p} \rightarrow W^{\pm} H^{\mp}X$ (Tevatron energy) or $ pp 
\rightarrow W^{\pm} 
H^{\mp}X$ (LHC energies) comes from $t \bar{t}$ production, the charged 
Higgs boson 
signal would be overwhelmed by such a background. In a muon collider
with $\sqrt{s} = 500 \textrm{GeV/c}$, the signal of the $H^{\pm}$ is not 
overwhelmed. This means, that for large values of $\tan\beta$, the channel
$\mu^- \mu^+ \rightarrow H^{\mp} W^{\pm}$ is a very attractive channel for 
the search of charged Higgs bosons at a $\mu^- \mu^+$ collider.

\section*{Acknowledgment}
I would like to thank Bruce Hoeneisen for the critical reading of this

\noindent manuscript.

\appendix
\pagenumbering{arabic}
\chapter{Functions $S^{WW}, S^{HW}$ and $S^{HH}$}
\typeout{Functions $S^{WW}, S^{HW}$ and $S^{HH}$}
If $i \neq j$:
\begin{eqnarray}
S^{WW}\left(x_{W}^{i}, x_{W}^{j}\right) &=& \frac{x_{W}^{i} + x_{W}^{j}
- \frac{11}{4}
x_{W}^{i}x_{W}^{j}}{\left(1-x_{W}^{i}\right)\left(1-x_{W}^{j}\right)} 
\nonumber \\ & & 
+ \frac{1}{\left(x_{W}^{i} - x_{W}^{j}\right)}
\left[ G\left(x_{W}^{i},x_{W}^{j} \right) -
G\left(x_{W}^{j},x_{W}^{i}\right)\right]
\label{SWW}
\end{eqnarray}
where
\begin{equation}
G\left(x_{W}^{i},x_{W}^{j}\right) = \frac{\left(x_{W}^{i}\right)^{2}   
\ln\left( x_{W}^{i} \right)}{\left( 1 - x_{W}^{i} \right)^{2}} \left[1 - 2
x_{W}^{j} + \frac{1}{4} x_{W}^{i} x_{W}^{j} \right].
\label{GW}
\end{equation}
If $i = j$:
\begin{eqnarray}
S^{WW}\left( x_{W}^{i}, x_{W}^{i} \right) = \frac{x_{W}^{i}}{\left( 1 -
x_{W}^{i} \right)^{2}} \left[ 3 - \frac{19}{4} x_{W}^{i} + \frac{1}{4}
\left(x_{W}^{i} \right)^{2} \right] \nonumber \\
+ \frac{2 x_{W}^{i} \ln\left(
x_{W}^{i} \right)}{\left( 1 - x_{W}^{i} \right)^{2}} \left[ 1 -
\frac{3}{4} \frac{\left( x_{W}^{i} \right)^{2}}{\left( 1 - x_{W}^{i}
\right)} \right].
\label{SWWeq} 
\end{eqnarray}
If $i \neq j$:  
\begin{equation}
S^{HH}\left(x_{H}^{i}, x_{H}^{j}, x_{H}^{W} \right) = \frac{x_{H}^{i} 
x_{H}^{j}}{x_{H}^W} \left[ \frac{J\left(x_{H}^{i}\right) -
J\left(x_{H}^{j} \right)}{x_{H}^{i} - x_{H}^{j}} \right]
\label{SHH}
\end{equation}
with
\begin{equation}
J\left( x_{H}^{i} \right) = \frac{1}{\left( 1 - x_{H}^{i}\right)} +
\frac{\left( x_{H}^{i} \right)^{2} \ln\left( x_{H}^{i} \right)}{\left(1 -
x_{H}^{i} \right)^{2}}.
\label{JH}
\end{equation}
If $ i = j$:
\begin{equation}
S^{HH}\left(x_{H}^{i}, x_{H}^{i}, x_{H}^{W} \right) =
\frac{\left(x_{H}^{i}\right)^{2}}{x_{H}^{W}} \left[\frac{1 -
\left(x_{H}^{i} \right)^{2} + 2 x_{H}^{i} \ln\left(x_{H}^{i}
\right)}{\left( 1 - x_{H}^{i} \right)^{3}} \right].
\label{SHHeq}
\end{equation}
For $ i \neq j$:
\begin{eqnarray}
\lefteqn{   
S^{HW} \left( x_{W}^{i}, x_{W}^{j}, x_{H}^{i}, x_{H}^{j}, x_{H}^{W}
\right) = \frac{x_{H}^{i} x_{H}^{j}}{\left( x_{H}^{W} - 1
\right) \left(x_{H}^{i} - 1 \right) \left( x_{H}^{j} - 1 \right)} \left[ 1
- \frac{1}{8 x_{H}^{W}} \right]
}
\nonumber \\ & &
+ \frac{x_{H}^{i} x_{H}^{j} x_{H}^{W}}{\left( x_{H}^{W} -
1\right) \left(x_{H}^{i} - x_{H}^{W} \right) \left( x_{H}^{j} - x_{H}^{W}
\right)} \left[ \frac{3}{4} \ln\left(x_{H}^{W} \right) - \frac{7}{8}
\right]     
\nonumber \\ & &
+ \frac{\left(x_{H}^{i}\right)^{2}
x_{H}^{j}}{\left(x_{H}^{i} -
x_{H}^{W} \right) \left(x_{H}^{i} - x_{H}^{j} \right) \left( x_{H}^{i} - 1
\right)} \left[ \ln\left( x_{H}^{i}\right) \left(1 - \frac{1}{4} x_{W}^{i}
\right) + \left(\frac{1}{8} x_{W}^{i} - 1\right) \right]
\nonumber \\ & &
+ \frac{\left(x_{H}^{j}\right)^{2}
x_{H}^{i}}{\left(x_{H}^{j} -
x_{H}^{W} \right) \left(x_{H}^{j} - x_{H}^{i} \right) \left( x_{H}^{j} - 1
\right)}
\nonumber \\ & &
\times\left[ \ln\left( x_{H}^{j}\right) \left(1 - \frac{1}{4}
x_{W}^{j}
\right) + \left(\frac{1}{8} x_{W}^{j} - 1\right) \right].
\label{SHW}
\end{eqnarray}  
For $ i = j$:
\begin{eqnarray}
S^{HW} \left( x_{W}^{i},x_{W}^{i},x_{H}^{i},x_{H}^{i},x_{H}^{W}
\right) = \left( x_{H}^{i} \right)^{2} \left[ \frac{ \ln \left(x_{H}^{i}
\right)}{\left( x_{H}^{W} - 1 \right) \left( x_{H}^{i} - 1 \right)^{2}}
\left( 1 - \frac{1}{4 x_{H}^{W}} \right)  \right. \nonumber \\
\left. - \frac{3}{4} \frac{x_{H}^{W} \ln \left( x_{W}^{i} \right)}{
\left(x_{H}^{W} - 1 \right) \left( x_{H}^{i} - x_{H}^{W}\right)^{2}} +
\frac{1}{  
\left( x_{H}^{i} - 1 \right) \left( x_{H}^{i} - x_{H}^{W} \right)} \left(
1 - \frac{1}{4} x_{W}^{i} \right) \right].
\label{SHWeq}   
\end{eqnarray}

\chapter{Calculation of the box diagrams corresponding to charged 
Higgs 
contributions to $B^0 - \bar{B^0}$ mixing in the \textquotedblleft{Two 
Higgs Doublet Model 
of type II}"}
\typeout{Calculation of the box diagrams corresponding to charged Higgs   
contributions to $B^0 - \bar{B^0}$ mixing in the \textquotedblleft{Two 
Higgs Doublet Model
of type II}"}

\section{Invariant amplitude $M^{HH}$}
In the unitary gauge the invariant amplitude corresponding to the box 
diagram (HH1) in Figure \ref{Feynman_dBox} is:

\begin{eqnarray}
\lefteqn{
M_{1}^{HH} = i \left(\frac{g}{2\sqrt{2}m_{W}}\right)^4\sum_{i,j}\xi_{i}
\xi_{j}\int\frac{d^{4}\mathrm{K}}{\left(2\pi\right)^4}\bar{v}(\bar{q})
}
\nonumber \\ & &
\left[m_q\tan\beta\left(1 - \gamma^5\right) + m_j \cot\beta\left(
1 + \gamma^5\right)\right]\left(\not{\mathrm{K}} + m_{j}\right)
\nonumber \\ & &
\left[m_b\tan\beta\left( 1 + \gamma^5\right) + m_j\cot\beta\left(1 - 
\gamma^5\right)\right]u(b)
\nonumber \\ & &
\cdot\bar{u}(q)\left[m_q\tan\beta\left(1 - \gamma^5\right) + 
m_i\cot\beta\left(1 + \gamma^5\right)\right]\left(\not{\mathrm{K}} + 
m_i\right)
\nonumber \\ & &
\left[m_b\tan\beta\left(1 + \gamma^5\right) + m_i \cot\beta\left(1 - 
\gamma^5\right)\right]v(\bar{b})
\nonumber \\ & &
\times\frac{1}{\left(\mathrm{K}^2 - m_i^{2}\right)\left(\mathrm{K}^2 -
m_j^{2}\right)\left(\mathrm{K}^2 - m_{H}^2\right)^2}.
\label{M1HH}
\end{eqnarray}
\noindent where $\not{\mathrm{K}} = \gamma^\mu\mathrm{K}_{\mu}$ and 
$\xi_{i} = V_{ib}V_{iq}^{*}$ ($q = d$ or $s$ and $i,j = u,c,t$). Here we 
have taken the approximation in which all external momenta are zero in the 
loop.
\begin{figure}
\begin{center}
\scalebox{0.9}
{\includegraphics{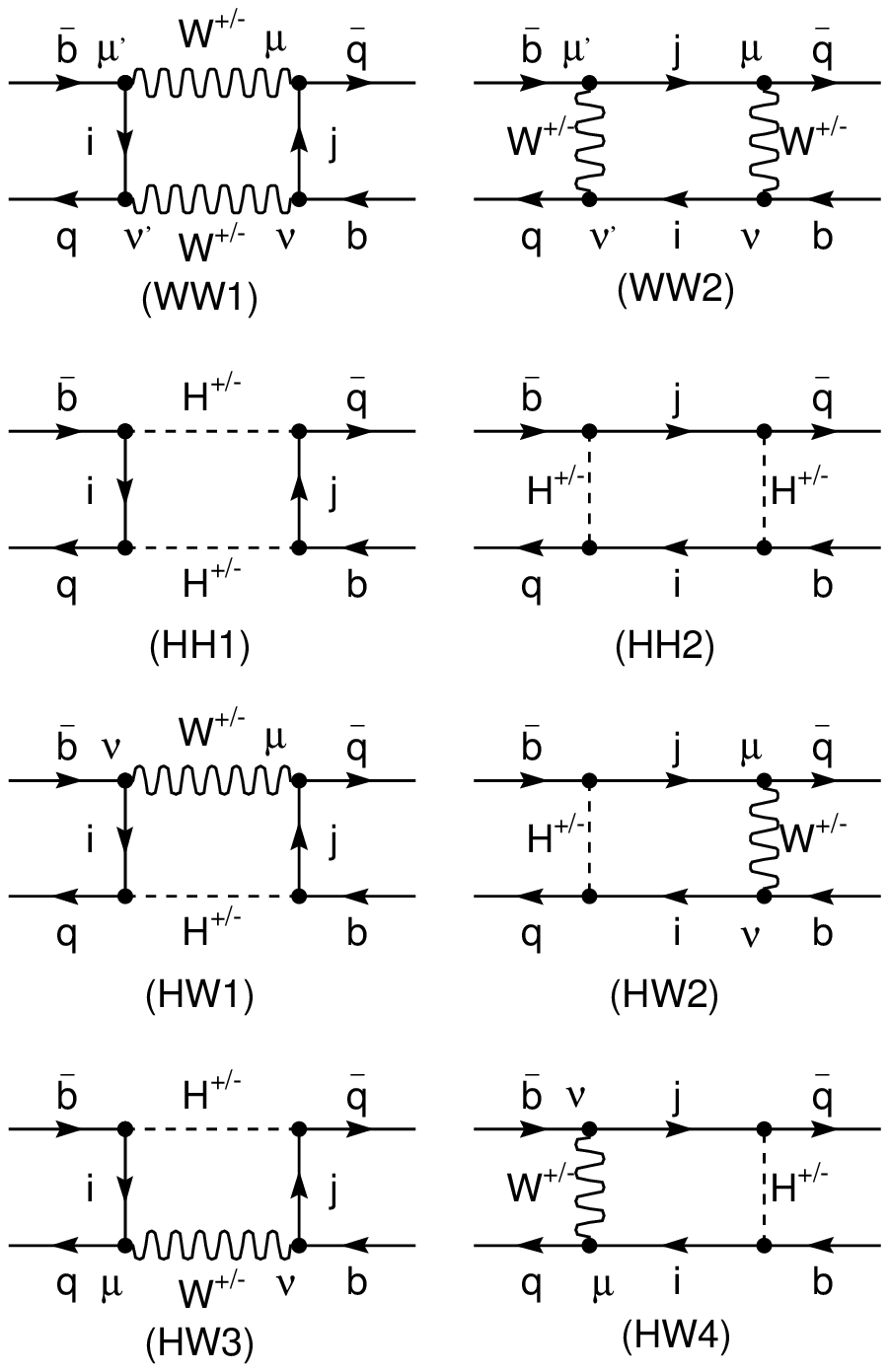}}
\caption{Feynman diagrams corresponding to $B^{o} \leftrightarrow
\bar{B}^{o}$ mixing in the Two Higgs Doublet Model. $q = d$ or $s$ and
$i, j = u, c, t$. The diagrams on the right side interfer with a
\textquotedblleft{-}" sign.}
\label{Feynman_dBox}
\end{center}
\end{figure}
\noindent Using:

$\left(1 + \gamma^5\right)\left(1 - \gamma^5\right) = 0$,
$\gamma^{\mu} \gamma^5 + \gamma^5 \gamma^{\mu} = 0$,
we obtain (in the limit $m_q
\rightarrow 0$)

\begin{eqnarray}
\lefteqn{
M_{1}^{HH} = 4i \left(\frac{g}{2\sqrt{2}m_{W}}\right)^4\sum_{i,j}
\xi_{i}\xi_{j}\{m_{i}^{2}m_{j}^{2}\cot^{4}\beta  \bar{v}(\bar{q})
\gamma^{\alpha}\left(1 - \gamma^5\right)u(b)
}
\nonumber \\ & &
\cdot\bar{u}(q)\gamma^{\beta}\left( 1 - \gamma^5\right) v(\bar{b})
I_{\alpha\beta}^{HH}(i,j) + m_{i}^{2}m_{j}^{2}m_{b}^{2} \bar{v}(\bar{q})
\left(1 + \gamma^5\right) u(b)
\nonumber \\ & &
\cdot\bar{u}(q)\left(1 + \gamma^5\right) v(\bar{b})I^{HH}(i,j) +
m_{i}^{2}m_{j}^{2}m_{b}\cot^2\beta  \bar{v}(\bar{q})
\nonumber \\ & &
\left(\gamma^{\alpha}\left(1 - \gamma^5\right) u(b)\cdot\bar{u}(q)
\left(1 + \gamma^5\right) + \left(1 + \gamma^5\right) u(b)\cdot
\bar{u}(q)\gamma^{\alpha}\left(1 - \gamma^5\right)\right)
\nonumber \\ & &
v(\bar{b})I_{\alpha}^{HH}(i,j)\}
\label{M1HHs}
\end{eqnarray}
\noindent where
\begin{eqnarray}
I_{\alpha\beta}^{HH}(i,j) \equiv \int\frac{d^{4}\mathrm{K}}{
\left(2\pi\right)^4}\frac{\mathrm{K}_{\alpha}\mathrm{K}_{\beta}}
{\left(\mathrm{K}^2 - m_{H}^{2}\right)^2\left(\mathrm{K}^{2}- m_{i}^{2}
\right)\left(\mathrm{K}^2 - m_{j}^{2}\right)}
\label{IalphabetaHH}
\end{eqnarray}
\begin{eqnarray}
I_{\alpha}^{HH}(i,j) \equiv \int\frac{d^{4}\mathrm{K}}{
\left(2\pi\right)^4}\frac{\mathrm{K}_{\alpha}}    
{\left(\mathrm{K}^2 - m_{H}^{2}\right)^2\left(\mathrm{K}^{2}- m_{i}^{2} 
\right)\left(\mathrm{K}^2 - m_{j}^{2}\right)}
\label{IalphaHH}
\end{eqnarray}
\begin{eqnarray}
I^{HH}(i,j) \equiv \int\frac{d^{4}\mathrm{K}}{
\left(2\pi\right)^4}\frac{1}                  
{\left(\mathrm{K}^2 - m_{H}^{2}\right)^2\left(\mathrm{K}^{2}- m_{i}^{2}
\right)\left(\mathrm{K}^2 - m_{j}^{2}\right)}
\label{IHH}
\end{eqnarray}
\noindent The integrals, $I_{\alpha\beta}^{HH}(i,j)$ and $I^{HH}(i,j)$, 
were 
calculated in detail in Appendix 2 of reference \cite{SMCMBH} (replacing
$m_{H}$ by $m_{W}$):

\noindent If $i \neq j$:
\begin{eqnarray}
I_{\alpha\beta}^{HH}(i,j) = \frac{-i\pi^2\eta_{\alpha\beta}}{4
\left(2\pi\right)^{4}m_{H}^2}\left[\frac{J\left(x_{H}^{i}\right)
- J\left(x_{H}^{j}\right)}{x_{H}^{i} - x_{H}^{j}}\right]
\end{eqnarray}
\noindent where
\begin{equation}
\eta_{\alpha\beta} = diag(1,-1,-1,-1),
\nonumber
\label{eta}
\end{equation}
\noindent $x_{H}^{i} = \frac{m_{i}^{2}}{m_{H}^{2}}$ and $J\left(x_{H}^{i}
\right)$ is given in Equation \ref{JH}. 

\noindent If $i = j$:
\begin{equation}
I_{\alpha\beta}^{HH}(i,i) = \frac{-i\pi^{2}\eta_{\alpha\beta}}
{4\left(2\pi\right)^{4}m_{H}^2}\left[\frac{1 - \left(x_{H}^{i}\right)^2
+ 2x_{H}^{i} ln\left(x_{H}^{i}\right)}{\left(1 - 
x_{H}^{i}\right)^3}\right].
\label{IabHHii}
\end{equation}

\noindent If $i \neq j$:

\begin{equation}
I^{HH}(i,j) = \frac{i\pi^2}{\left(2\pi\right)^{4}m_{H}^4}
\frac{1}{\left(1 - x_{H}^{i}\right)\left(1 - x_{H}^{j}\right)}
\left[F\left(x_{H}^{i},x_{H}^{j}\right)+ F\left(x_{H}^{j},x_{H}^{i}
\right) -1 \right]
\label{IHHij}
\end{equation}

\noindent where
\begin{equation}
F\left(x_{H}^{i},x_{H}^{j}\right) = - \frac{x_{H}^{i} ln\left(x_{H}^{i}
\right)\left(1 - x_{H}^{j}\right)}{\left(1 - x_{H}^{i}\right)\left(
x_{H}^{i} - x_{H}^{j}\right)}.
\label{Fxhi-xhj}
\end{equation}

\noindent If $i = j$:
\begin{equation}
I^{HH}(i,i) = \frac{-i\pi^2}{\left(2\pi\right)^{4}m_{H}^4}
\frac{1}{\left(1 - x_{H}^{i}\right)^2}\left[2 + \frac{\left(1 + x_{H}^{i}
\right)}{\left(1 - x_{H}^{i}\right)} ln\left(x_{H}^{i}\right)\right].
\label{IHHii}
\end{equation}

\noindent The value of the second integral (see Appendix C) is:

\noindent For $i \neq j$ and $ i = j$:

\begin{equation}
I_{\alpha}^{HH}(i,j) = 0.
\label{Ialphahhij}
\end{equation}

\noindent Neglecting the second term in (\ref{M1HHs}) and because 
$\frac{G_{F}}{\sqrt{2}} = \frac{g^2}{8m_{W}^{2}}$, we have

\begin{eqnarray}
\lefteqn{
M_{1}^{HH} = \frac{G_{F}^{2} m_{W}^{2}}{32\pi^2}\cot^4\beta
\sum_{i,j}\xi_{i}\xi_{j} \bar{v}(\bar{q})\gamma^{\mu}\left(1
- \gamma^5\right) u(b)
}
\nonumber \\ & &
\cdot \bar{u}(q)\gamma_{\mu}\left(1 - \gamma^5\right) v(\bar{b})
S^{HH}\left(x_{H}^{i},x_{H}^{j},x_{H}^{W}\right).
\label{M1HHfinal}
\end{eqnarray}

\noindent $S^{HH}\left(x_{H}^{i},x_{H}^{j},x_{H}^{W}\right)$ for $i \neq 
j$
and $i = j$ are given in (\ref{SHH}) and (\ref{SHHeq}), respectively. 

\noindent Note that:
\begin{equation}
\lim_{x_{H}^{i}\rightarrow 
0}S^{HH}\left(x_{H}^{i},x_{H}^{i},x_{H}^{W}\right) = 0.
\label{limitSHH}
\end{equation}

In a similar way, we can chow that the invariant amplitude corresponding 
to the diagram (HH2) in Figure \ref{Feynman_dBox} is:

\begin{eqnarray}
\lefteqn{
M_{2}^{HH} =  \frac{G_{F}^{2} m_{W}^{2}}{32\pi^2}\cot^4\beta
\sum_{i,j}\xi_{i}\xi_{j} \bar{v}(\bar{q})\gamma^{\mu}\left(1
- \gamma^5\right) v(\bar{b})
}
\nonumber \\ & &
\cdot \bar{u}(q)\gamma_{\mu}\left(1 - \gamma^5\right) u(b)
S^{HH}\left(x_{H}^{i},x_{H}^{j},x_{H}^{W}\right).
\label{M2HH}
\end{eqnarray}

\noindent According to the Fierz Theorem \cite{FierzTheorem}, we can write

\begin{eqnarray}
\lefteqn{
\bar{v}(\bar{q}) \gamma^{\mu}\left(1 - \gamma^5\right) v(\bar{b})\cdot
 \bar{u}(q) \gamma_{\mu}\left(1 - \gamma^5\right) u(b)
}
\nonumber \\ & &
= - \bar{v}(\bar{q}) \gamma^{\mu}\left(1 - \gamma^5\right) u(b)\cdot
 \bar{u}(q) \ \gamma_{\mu}\left(1 - \gamma^5\right) v(\bar{b})
\label{Fierz_Theorem}
\end{eqnarray}

The total amplitude $M^{HH}$ is then
\begin{equation}
M^{HH} = 2 M_{1}^{HH}.
\label{MHH}
\end{equation}

Let's consider the amplitude:
\begin{equation}
\langle B^{0}|M^{HH}|\bar{B}^{0}\rangle = \frac{G_{F}^{2}m_{W}^{2}}
{4\pi^2}\cot^{4}\beta \sum_{i,j} \xi_{i}\xi_{j} A S^{HH}(x_{H}^{i},
x_{H}^{j},x_{H}^{W})
\label{BMHHBbar}
\end{equation}

\noindent where

\begin{equation}
A \equiv \langle B^{0}| \bar{v}_{L}(\bar{q})\gamma^{\mu}
u_{L}(b) \cdot \bar{u}_{L}(q)\gamma_{\mu} v_{L}(\bar{b}) |
\bar{B}^{0} \rangle,
\nonumber
\label{Amatrixelement}
\end{equation}

\begin{equation}
\bar{v}_{L}(\bar{q}) \gamma^{\mu} = \bar{v}(\bar{q}) \gamma^{\mu}
\frac{1}{2}\left(1 - \gamma^5\right),
\nonumber
\label{vlbar}
\end{equation}

\begin{equation}
v_{L}(\bar{b}) = \frac{1}{2}\left(1 - \gamma^5\right) v(\bar{b}),
\nonumber
\label{vL}
\end{equation}

\begin{equation}
\bar{u}_{L}(q) \gamma_{\mu} = \bar{u}(\bar{q}) \gamma_{\mu}
\frac{1}{2}\left(1 - \gamma^5\right),
\nonumber
\label{ulbar}
\end{equation}

\begin{equation}
u_{L}(b) = \frac{1}{2}\left(1 - \gamma^5\right) u(b),
\nonumber
\label{uL}
\end{equation}

\noindent For our model, \textquoteleft{free particles inside the meson}"
\cite{SMCMBH}, we have
\begin{equation}
A = \frac{1}{16}\beta_{B} f_{B}^{2} m_{B}
\label{Avalue}
\end{equation}

\noindent Thus, one obtains

\begin{equation}
\langle B^{0}|M^{HH}|\bar{B}^{0}\rangle = 
\frac{\beta_{B}G_{F}^{2}m_{W}^{2}f_{B}^{2}m_{B}}
{64\pi^2}\cot^{4}\beta \sum_{i,j} \xi_{i}\xi_{j} S^{HH}(x_{H}^{i},
x_{H}^{j},x_{H}^{W})
\label{BHHBbarfinal}
\end{equation}

\section{Invariant amplitude $M^{HW}$}
For the box diagram (HW1) of Figure  \ref{Feynman_dBox}, the corresponding 
invariant amplitude is given by:

\begin{eqnarray}
\lefteqn{
M_{1}^{HW} = -i\left(\frac{g}{2\sqrt{2}}\right)^4\frac{1}{m_{W}^2}
\sum_{i,j}\xi_{i}\xi_{j} \int\frac{d^{4}\mathrm{K}}{\left(2\pi\right)^4}
\bar{v}(\bar{q})\gamma^{\mu}\left(1 - 
\gamma^5\right)
}
\nonumber \\ & &
\left(\not{\mathrm{K}} + m_{j}\right)
\left[m_{b}\tan\beta\left(1 + \gamma^5\right) + m_{j} \cot\beta\left(1 - 
\gamma^5\right)\right] u(b)
\nonumber \\ & &
\cdot \bar{u}(q)\left[m_{q} \tan\beta\left(1 - \gamma^5\right) + 
m_{i}\cot\beta\left(1 + \gamma^5\right)\right]
\nonumber \\ & &
\left(\not{\mathrm{K}} + m_{i}\right)\gamma^{\nu}\left(1 - \gamma^5\right)
v(\bar{b})\left(\eta_{\mu\nu} - \frac{\mathrm{K}_{\mu}\mathrm{K}_{\nu}}
{m_{W}^{2}}\right)
\nonumber \\ & &
\times \frac{1}{\left(\mathrm{K}^{2} - m_{W}^{2}\right)\left(
\mathrm{K}^{2} - m_{H}^{2}\right)\left(\mathrm{K}^{2} - m_{i}^{2}\right)
\left(\mathrm{K}^{2} - m_{j}^{2}\right)}
\label{M1HW}
\end{eqnarray}

\noindent Using:

$\gamma^{\mu}\gamma^5 + \gamma^5\gamma^{\mu} = 0$ , 
$\left(1 + \gamma^5\right)^2 = 2\left(1 + \gamma^5\right)$ ,
$\left(1 - \gamma^5\right)^2 = 2\left(1 - \gamma^5\right)$ ,
\newline
$\not{\mathrm{K}}\not{\mathrm{K}} = \mathrm{K}^{2}$, and taking the limit
in which $m_{q} \rightarrow 0$, we can write the invariant amplitude as

\begin{eqnarray}
\lefteqn{
M_{1}^{HW} = - 4i\left(\frac{g}{2\sqrt{2}}\right)^4\frac{1}{m_{W}^{2}}
\sum_{i,j}\xi_{i}\xi_{j}\{m_{i}^{2}m_{j}^{2} \cot^{2}\beta 
\bar{v}(\bar{q}) \gamma^{\alpha}\left(1 - \gamma^5\right) u(b)
}
\nonumber \\ & &
\cdot \bar{u}(q)\left(\gamma_{\alpha} I^{HW}(i,j) - \frac{1}{m_{W}^{2}}
\gamma^{\beta} I_{\alpha\beta}^{HW}(i,j)\right)\left(1 - \gamma^5\right)
v(\bar{b})
\nonumber \\ & &
+ m_{i}^{2}m_{b}\bar{v}(\bar{q})\gamma^{\mu}\left(1 - \gamma^5\right)
\gamma^{\alpha}u(b)
\cdot \bar{u}(q) \gamma_{\mu}\left( 1 - \gamma^5\right) v(\bar{b})
\left(I_{\alpha}^{HW}\right)^{(1)}(i,j)
\nonumber \\ & &
- \frac{1}{m_{W}^{2}} m_{i}^{2} m_{b} \bar{v}(\bar{q})\left(1 +
\gamma^5\right) u(b)
\nonumber \\ & &
\cdot \bar{u}(q)\gamma^{\alpha}\left(1 - 
\gamma^5\right) v(\bar{b})\left(I_{\alpha}^{HW}\right)^{(2)}(i,j)\}
\label{M1HWs}
\end{eqnarray}
\noindent where
\begin{equation}
I^{HW}(i,j) \equiv \int\frac{d^{4}\mathrm{K}}{\left(2\pi\right)^4}
\frac{1}{\left(\mathrm{K}^{2} - m_{W}^{2}\right)\left(\mathrm{K}^{2}
- m_{H}^{2}\right)\left(\mathrm{K}^{2} - m_{i}^{2}\right)
\left(\mathrm{K}^{2} - m_{j}^{2}\right)}
\label{IHWij}
\end{equation}
\begin{equation}
I_{\alpha\beta}^{HW}(i,j) \equiv 
\int\frac{d^{4}\mathrm{K}}{\left(2\pi\right)^4}
\frac{\mathrm{K}_{\alpha}\mathrm{K}_{\beta}}{\left(\mathrm{K}^{2} - 
m_{W}^{2}\right)\left(\mathrm{K}^{2}
- m_{H}^{2}\right)\left(\mathrm{K}^{2} - m_{i}^{2}\right)
\left(\mathrm{K}^{2} - m_{j}^{2}\right)}
\label{IalphabetaHWij}
\end{equation}

\begin{equation}
\left(I_{\alpha}^{HW}\right)^{(1)}(i,j) \equiv
\int\frac{d^{4}\mathrm{K}}{\left(2\pi\right)^4}
\frac{\mathrm{K}_{\alpha}}{\left(\mathrm{K}^{2} -
m_{W}^{2}\right)\left(\mathrm{K}^{2}
- m_{H}^{2}\right)\left(\mathrm{K}^{2} - m_{i}^{2}\right)
\left(\mathrm{K}^{2} - m_{j}^{2}\right)}
\label{IalphaHWij}
\end{equation}

\begin{equation}
\left(I_{\alpha}^{HW}\right)^{(2)}(i,j) \equiv
\int\frac{d^{4}\mathrm{K}}{\left(2\pi\right)^4}
\frac{\mathrm{K}^{2}\mathrm{K}_{\alpha}}{\left(\mathrm{K}^{2} -
m_{W}^{2}\right)\left(\mathrm{K}^{2}
- m_{H}^{2}\right)\left(\mathrm{K}^{2} - m_{i}^{2}\right)
\left(\mathrm{K}^{2} - m_{j}^{2}\right)}
\label{IalphaHW2ij}
\end{equation}
\noindent After momentum integration (see Appendix C), we get

\noindent If $ i \neq j$:
\begin{eqnarray}
\lefteqn{
I^{HW}(i,j) = \frac{-i\pi^2}{\left(2\pi\right)^4}\frac{1}{m_{H}^4}
\frac{1}{\left(x_{H}^{W} - 1\right)}
}
\nonumber \\ & &
\times \left[\frac{x_{H}^{W}
\left(ln\left(x_{H}^{W}\right) - 1\right)}{\left(x_{H}^{i} -
x_{H}^{W}\right)\left(x_{H}^{j} - x_{H}^{W}\right)}
+ \frac{1}{\left(x_{H}^{i} - 1\right)\left(x_{H}^{j} - 1\right)}
\right.
\nonumber \\ & &
+ \frac{\left(x_{H}^{W} - 1\right)}{\left(x_{H}^{j} - x_{H}^{W}\right)
\left(x_{H}^{j} - x_{H}^{i}\right)\left(x_{H}^{j} - 1\right)}
x_{H}^{j}\left(ln\left(x_{H}^{j}\right) - 1\right)
\nonumber \\ & &
+\left.  \frac{\left(x_{H}^{W} - 1\right)}{\left(x_{H}^{i} - 
x_{H}^{W}\right)
\left(x_{H}^{i} - x_{H}^{j}\right)\left(x_{H}^{i} - 1\right)}
x_{H}^{i}\left(ln\left(x_{H}^{i}\right) - 1 \right) \right]
\label{IHWijvalue}
\end{eqnarray}

\noindent If $i = j$:
\begin{eqnarray}
\lefteqn{
I^{HW}(i,i) =  \frac{-i\pi^2}{\left(2\pi\right)^4}\frac{1}{m_{H}^4}
\frac{1}{\left(x_{H}^{W} - 1\right)}\left[\frac{x_{H}^{W}
\left(ln\left(x_{H}^{W}\right) - 1\right)}{\left(x_{H}^{i} - x_{H}^{W}
\right)^2}
+ \frac{1}{\left(x_{H}^{i} - 1\right)^2} \right.
}
\nonumber \\ & &
\left.
+ \frac{\left(x_{H}^{W} - 
1\right)\left[ln\left(x_{H}^{i}\right)\left(x_{H}^{W}
- \left(x_{H}^{i}\right)^2\right) + 
x_{H}^{i}\left(2x_{H}^{i} -
x_{H}^{W} -1 \right)\right]}
{\left(x_{H}^{i} - 
x_{H}^{W}\right)^2
\left(x_{H}^{i} - 1\right)^2}\right]
\label{IHWiivalue}
\end{eqnarray}

\noindent If $i \neq j$:

\begin{eqnarray}
\lefteqn{
I_{\alpha\beta}^{HW}(i,j) =  
\frac{-i\pi^2}{8\left(2\pi\right)^4}\eta_{\alpha\beta}
\frac{1}{m_{H}^2}\frac{1}{\left(x_{H}^{W} - 1\right)}
}
\nonumber \\ & &
\times \left[ - \frac{\left(x_{H}^{W} - 1\right)\left(x_{H}^{i}
\right)^2\left(2ln\left(x_{H}^{i}\right) - 1\right)}
{\left(x_{H}^{j} - x_{H}^{i}\right)\left(x_{H}^{i} - 1\right)
\left(x_{H}^{i} - x_{H}^{W}\right)}\right.
\nonumber \\ & &
-  \frac{\left(x_{H}^{W} - 1\right)\left(x_{H}^{j}
\right)^2\left(2ln\left(x_{H}^{j}\right) - 1\right)}
{\left(x_{H}^{i} - x_{H}^{j}\right)\left(x_{H}^{j} - 1\right) 
\left(x_{H}^{j} - x_{H}^{W}\right)}
\nonumber \\ & &
\left. + \frac{1}{\left(x_{H}^{i} - 1\right)\left(x_{H}^{j} - 1\right)}
+ \frac{\left(x_{H}^{W}\right)^2\left(2ln\left(x_{H}^{W}\right) - 
1\right)}{\left(x_{H}^{i} - x_{H}^{W}\right)\left(x_{H}^{j} -
x_{H}^{W}\right)} \right]
\label{IabHWijvalue}
\end{eqnarray}

\noindent If $i = j$:
\begin{eqnarray}
\lefteqn{
I_{\alpha\beta}^{HW}(i,i) =
\frac{-i\pi^2}{8\left(2\pi\right)^4}\eta_{\alpha\beta}
\frac{1}{m_{H}^2}\frac{1}{\left(x_{H}^{W} - 1\right)}
\left[\frac{\left(x_{H}^{W} - 1\right)\left(x_{H}^{i}\right)}
{\left(x_{H}^{i} - 1\right)^2\left(x_{H}^{i} - x_{H}^{W}\right)^2}
\right.
}
\nonumber \\ & &
\times\left[x_{H}^{i}\left(2x_{H}^{i} - x_{H}^{W} - 1\right) 
-2ln\left(x_{H}^{i}
\right)\left(x_{H}^{i}x_{H}^{W} + x_{H}^{i} - 2x_{H}^{W}\right)\right]
\nonumber \\ & &
\left. + \frac{1}{\left(x_{H}^{i} - 1\right)^2} 
+ \frac{\left(x_{H}^{W}\right)^2\left(2 ln\left(x_{H}^{W}\right) - 
1\right)}
{\left(x_{H}^{i} - x_{H}^{W}\right)^2}\right]
\label{IabHWiivalue}
\end{eqnarray}

\noindent For $i \neq j$ and $i = j$:
\begin{equation}
\left(I_{\alpha}^{HW}\right)^{(1)}(i,j) = 0
\label{IaHW1ikvalue}
\end{equation}

\noindent For $i \neq j$ and $i = j$:
\begin{equation}
\left(I_{\alpha}^{HW}\right)^{(2)}(i,j) = 0
\label{IaHW2ikvalue}
\end{equation}

\noindent Introducing these integrals in \ref{M1HWs}, we find

\begin{eqnarray}
\lefteqn{
M_{1}^{HW} = - \frac{G_{F}^{2}m_{W}^{2}\cot^{2}\beta}{8\pi^2}
\sum_{i,j}\xi_{i}\xi_{j} \bar{v}\left(\bar{q}\right)\gamma^{\mu}
\left(1 - \gamma^5\right) u(b)
}
\nonumber \\ & &
\cdot \bar{u}\left(q\right)\gamma_{\mu}\left(1 - \gamma^5\right)
 v\left(\bar{b}\right) S^{HW}\left(x_{W}^{i}, x_{W}^{j}, x_{H}^{i},
x_{H}^{j}, x_{H}^{W}\right)
\label{M1HWfinal}
\end{eqnarray}

\noindent where  $S^{HW}\left(x_{W}^{i}, x_{W}^{j}, x_{H}^{i},
x_{H}^{j}, x_{H}^{W}\right)$ is given in (\ref{SHW}) and (\ref{SHWeq}).

\noindent Note that 

\begin{equation}
\lim_{x_{H}^{i} \rightarrow 0} 
S^{HW}\left(x_{W}^{i}, x_{W}^{i}, x_{H}^{i},
x_{H}^{i}, x_{H}^{W}\right) = 0. 
\end{equation}

We have another three diagrams. In the limit $m_{q} \rightarrow 0$, the
Fierz transformation shows that all four diagrams contribute equally and 
then, the total invariant amplitude is:

\begin{equation}
M^{HW} = 4 M_{1}^{HW}
\end{equation}

\noindent Therefore, as in (\ref{BHHBbarfinal}), the matrix element
$\langle B^{0}|M^{HW}|\bar{B}^{0} \rangle$ can be expressed as

\begin{eqnarray}
\lefteqn{
\langle B^{0}|M^{HW}|\bar{B}^{0} \rangle = - \frac{\beta_{B}
G_{F}^{2}m_{W}^{2}f_{B}^{2}m_{B}}{8\pi^{2}}\cot^{2}\beta
}
\nonumber \\ & &
\sum_{i,j}\xi_{i}\xi_{j}S^{HW}\left(x_{W}^{i}, x_{W}^{j}, x_{H}^{i},
x_{H}^{j}, x_{H}^{W}\right).
\label{BMHWBbar}
\end{eqnarray}
\section{Invariant amplitude $M^{WW}$}
The calculation of the invariant amplitude for the box diagrams (WW) in 
Figure \ref{Feynman_dBox}, was performed in detail in reference 
\cite{SMCMBH}:

\begin{equation}
M^{WW} = \frac{G_{F}^{2}m_{W}^{2}}{\pi^2}\sum_{i,j}\xi_{i}\xi_{j}
\bar{v}_{L}(\bar{q})\gamma^{\mu} u_{L}(b) \cdot  \bar{u}_{L}(q) 
\gamma_{\mu} v_{L}(\bar{b}) S^{WW}(x_{W}^{i},x_{W}^{j})
\end{equation}
\label{MWW}
\noindent Therefore

\begin{equation}
\langle B^{0}|M^{WW}|\bar{B}^{0} \rangle =\frac{\beta_{B}G_{F}^{2}
m_{W}^{2}f_{B}^{2}m_{B}}{16\pi^{2}}\sum_{i,j}\xi_{i}\xi_{j}
S^{WW}\left(x_{W}^{i},x_{W}^{j}\right).
\label{BMWWMbar}
\end{equation}
\section{Mass difference $\Delta m_{B}$}
The mass difference between the two states that diagonalize the 
hamiltonian is:
\begin{eqnarray}
\lefteqn{
\Delta m_{B} = m_{B_{H}} - m_{B_{L}} = 2\vert M_{12}\vert
= 2 \vert \langle B^{0} \vert M^{HH} + M^{HW} + M^{WW} \vert 
\bar{B}^{0} \rangle \vert
}
\nonumber \\ & &
= 2 \vert \langle B^{0} \vert M^{HH} \vert \bar{B}^{0} \rangle
+ \langle B^{0} \vert M^{HW} \vert \bar{B}^{0} \rangle +
\langle B^{0} \vert M^{WW} \vert \bar{B}^{0} \rangle \vert
\label{DeltamB}
\end{eqnarray}

\noindent where H and L stand for Heavy and Light, repectively.

\noindent Introducing (\ref{BHHBbarfinal}), (\ref{BMHWBbar}),
and (\ref{BMWWMbar}) in Equation (\ref{DeltamB}) (after correcting
by a color factor 4/3), we arrive to Equation (\ref{mixing}).

\chapter{Integrals}
\typeout{Integrals}
\section{$I_{\alpha}^{HH}(i,j)$}

From the identity:
\begin{equation}
\frac{1}{abcd} = 3! \int_{0}^{1}\int_{0}^{1}\int_{0}^{1}
\frac{x^2 y dx dy dz}{\left[\left(1 - x\right)d + x\left(1 - y\right)c
+ x y\left(1 - z\right)b + x y z a\right]^4}
\label{feynidentity}
\end{equation}

\noindent where $a, b, c, d \neq 0$; and setting:

$a = \left(\mathrm{K}^2 - m_{H}^{2}\right)$, 
$b = \left(\mathrm{K}^2 - m_{H}^{2}\right)$,
$c = \left(\mathrm{K}^2 - m_{i}^{2}\right)$,
$d = \left(\mathrm{K}^2 - m_{j}^{2}\right)$,

\noindent it is found that
\begin{eqnarray}
\lefteqn{
\frac{1}{\left(\mathrm{K}^2 - m_{H}^{2}\right)^{2}
\left(\mathrm{K}^2 - m_{i}^{2}\right)
\left(\mathrm{K}^2 - m_{j}^{2}\right)}
}
\nonumber \\ & &
= 3! \int_{0}^{1}\int_{0}^{1}\int_{0}^{1}
\frac{x^2 y dx dy dz}{\left[\mathrm{K}^2 + x\left(m_{j}^{2} - m_{i}^{2}
\right) + x y\left(m_{i}^{2} - m_{H}^{2}\right) - m_{j}^{2}\right]^4}
\label{identK}
\end{eqnarray} 

\noindent Introducing the last Equation in (\ref{IalphaHH}), we obtain
\begin{equation}
I_{\alpha}^{HH}(i,j) = 3!\int_{0}^{1}\int_{0}^{1}\int_{0}^{1}dxdydz
x^2 y 
\int\frac{d^{4}\mathrm{K}}{\left(2\pi\right)^4}
\frac{\mathrm{K}_{\alpha}}{\left[\mathrm{K}^{2} + M^{2}\right]^4}
\label{integIalphaHH}
\end{equation}
\noindent where
\begin{equation}
M^{2} = x\left(m_{j}^{2} - m_{i}^{2}\right) + x y \left(m_{i}^{2}
- m_{H}^{2} \right) - m_{j}^{2}
\label{M^2}
\end{equation}

\noindent Then using \cite{Pokorski}
\begin{equation}
I_{\mu} = \int\frac{d^{d}\mathrm{K}}{\left(2\pi\right)^{d}}
\frac{\mathrm{K}_{\mu}}{\left(\mathrm{K}^{2} + 2P\cdot \mathrm{K}
+M^{2} + i\epsilon\right)^{\alpha}} = - P_{\mu} I_{0}
\label{Imu}
\end{equation}

\begin{eqnarray}
\lefteqn{
I_{0} = \int\frac{d^{d}\mathrm{K}}{\left(2\pi\right)^{d}}
\frac{1}{\left(\mathrm{K}^{2} + 2 P\cdot\mathrm{K} + M^{2} +
i\epsilon\right)^{\alpha}}
}
\nonumber \\ & &
= \frac{i\left(-\pi\right)^{d/2}}{\left(2\pi\right)^{d}}
\frac{\Gamma\left(\alpha - \frac{d}{2}\right)}{\Gamma\left(\alpha\right)}
\frac{1}{\left(M^{2} - P^{2} + i\epsilon\right)^{\alpha - \frac{d}{2}}}
\label{I0}
\end{eqnarray}

\noindent With $n = 4$, $\alpha = 4$ and $P_{\mu} = 0$, we get
\begin{equation}
I_{\alpha}^{HH}(i,j) = 0.
\label{intIalphaHH}
\end{equation}
\section{$\left(I_{\alpha}^{HW}\right)^{(1)}(i,j)$}

Again, using the identity (\ref{feynidentity}) for:

$a = \left(\mathrm{K}^2 - m_{H}^{2}\right)$,
$b = \left(\mathrm{K}^2 - m_{W}^{2}\right)$,
$c = \left(\mathrm{K}^2 - m_{i}^{2}\right)$,
$d = \left(\mathrm{K}^2 - m_{j}^{2}\right)$,

\noindent we obtain the same integral (\ref{intIalphaHH}), but with 
\begin{equation}
M^{2} = x\left(m_{j}^{2} - m_{i}^{2}\right) + xy\left(m_{i}^{2} -
m_{W}^{2}\right) + x y z\left(m_{W}^{2} - m_{H}^{2}\right) - m_{j}^{2}
\label{M^22}
\end{equation}

\noindent Once more, (\ref{Imu}) implies
\begin{equation}
\left(I_{\alpha}^{HW}\right)^{(1)}(i,j) = 0.
\label{intIalphaHW}
\end{equation}
\section{$\left(I_{\alpha}^{HW}\right)^{(2)}(i,j)$}
On account of (\ref{feynidentity}), we can write
\begin{equation}
\left(I_{\alpha}^{HW}\right)^{(2)}(i,j) = 3! \gamma^{\mu}
\gamma^{\nu} \int_{0}^{1}\int_{0}^{1}\int_{0}^{1}dx dy dz x^2 y
\int\frac{d^{4}\mathrm{K}}{\left(2\pi\right)^{4}}
\frac{\mathrm{K}_{\mu} \mathrm{K}_{\nu} \mathrm{K}_{\alpha}}
{\left[\mathrm{K}^{2} + M^{2}\right]^4}
\label{inteIalphaHW2}
\end{equation}
\noindent then, from  \cite{Pokorski}
\begin{eqnarray}
\lefteqn{
I_{\mu \nu \alpha} = \int\frac{d^{d}\mathrm{K}}{\left(2\pi\right)^{d}}
\frac{\mathrm{K}_{\mu} \mathrm{K}_{\nu} \mathrm{K}_{\alpha}}
{\left[\mathrm{K}^{2} + 2 P \cdot \mathrm{K} + M^{2} + 
i\epsilon\right]^{\alpha}}
}
\nonumber \\ & &
= - I_{0}\left[P_{\mu}P_{\nu}P_{\alpha} +
\frac{\left(M^{2} - P^{2}\right)}
{2\left(\alpha - \frac{d}{2} - 1\right)}
\left(\eta_{\mu\nu} P_{\alpha} + \eta_{\mu \alpha}P_{\nu}
+ \eta_{\nu \alpha}P_{\mu}\right)\right]
\label{Imunualpha}
\end{eqnarray}
\noindent With $n = 4$, $\alpha = 4$ and $P_{\mu} = 0$, we get
\begin{equation}
\left(I_{\alpha}^{HW}\right)^{(2)}(i,j) = 0.
\label{intIalphaHW2}
\end{equation}

\section{$I^{HW}(i,j)$}
Equation (\ref{IHWij}) can be written as:
\begin{equation}
I^{HW}(i,j) = 3!\int_{0}^{1}\int_{0}^{1}\int_{0}^{1} dx dy dz x^2 y
\int\frac{d^{4}\mathrm{K}}{\left(2\pi\right)^4}
\frac{1}{\left(\mathrm{K}^2 + M^{2}\right)^{4}}
\label{integrIHWij}
\end{equation}

\noindent From (\ref{I0}), we have
\begin{equation}
\int\frac{d^{4} \mathrm{K}}{\left(2\pi\right)^{4}}
\frac{1}{\left(\mathrm{K}^{2} + M^{2}\right)^{4}}
= \frac{i\pi^{2}}{\left(2\pi\right)^4}\frac{1}{3!\left(M^{2}
\right)^{2}}
\label{integralc14}
\end{equation}
\noindent and then

\begin{equation}
I^{HW}(i,j) = \frac{i\pi^2}{\left(2\pi\right)^{4}}
\int_{0}^{1}\int_{0}^{1}\int_{0}^{1} f(x, y, z) dx dy dz
\label{IHWijintegral}
\end{equation}
\noindent where
\begin{equation}
f(x, y, z) = \frac{x^{2} y }{\left[x\left(m_{j}^{2} - m_{i}^{2}\right)
+ x y\left(m_{i}^{2} - m_{W}^{2}\right) + x y z\left(m_{W}^{2} -
m_{H}^{2}\right) - m_{j}^{2}\right]^{2}}
\label{f(x,y,z)}
\end{equation}

\noindent Integrating (\ref{IHWijintegral}), we obtain (\ref{IHWijvalue})
and (\ref{IHWiivalue}).
\section{$I_{\alpha\beta}^{HW}(i,j)$}
Equation (\ref{IalphabetaHWij}) can be written as:
\begin{equation}
I_{\alpha\beta}^{HW}(i,j) = 3!\int_{0}^{1}\int_{0}^{1}
\int_{0}^{1}dx dy dz x^2 y \int\frac{d^{4}\mathrm{K}}
{\left(2\pi\right)^{4}}\frac{\mathrm{K}_{\alpha}
\mathrm{K}_{\beta}}{\left(\mathrm{K}^{2} + M^{2}\right)^{4}}
\label{IabHWijintegral}
\end{equation}
\noindent Using the integral \cite{Pokorski}

\begin{eqnarray}
\lefteqn{
I_{\alpha\beta} = \int\frac{d^{d}\mathrm{K}}{\left(2\pi\right)^{d}}
\frac{\mathrm{K}_{\alpha}\mathrm{K}_{\beta}}
{\left(\mathrm{K}^{2} + 2 P \cdot \mathrm{K} + M^{2} + i\epsilon
\right)^{\alpha}}
}
\nonumber \\ & &
= I_{0}\left[P_{\alpha} P_{\beta} + \frac{1}{2}\eta_{\alpha\beta}
\left(M^{2} - P^{2}\right)\frac{1}{\left(\alpha - \frac{d}{2} - 1\right)}
\right],
\label{Ialphabetaintegral}
\end{eqnarray}

\noindent we obtain

\begin{equation}
\int\frac{d^{4}\mathrm{K}}{\left(2\pi\right)^{4}}
\frac{\mathrm{K}_{\alpha}\mathrm{K}_{\beta}}
{\left(\mathrm{K}^{2} + M^{2}\right)^{4}}
= \frac{i\pi^{2}}{2\left(2\pi\right)^{4}}
\frac{1}{3! M^{2}}\eta_{\alpha\beta}
\label{intHalphaKbeta}
\end{equation}

\noindent and therefore, (\ref{IabHWijintegral}) becomes 
\begin{equation}
I_{\alpha\beta}^{HW}(i,j) = \frac{i\pi^2}{2\left(2\pi\right)^{4}}
\eta_{\alpha\beta}
\int_{0}^{1}\int_{0}^{1}\int_{0}^{1} g(x, y, z) dx dy dz
\label{IabHWijintegralf}
\end{equation}
\noindent where
\begin{equation}   
g(x, y, z) = \frac{x^{2} y }{\left[x\left(m_{j}^{2} - m_{i}^{2}\right)
+ x y\left(m_{i}^{2} - m_{W}^{2}\right) + x y z\left(m_{W}^{2} -
m_{H}^{2}\right) - m_{j}^{2}\right]}
\label{g(x,y,z)}
\end{equation}

\noindent After some long, but trivial calculations, we arrive to
Equations (\ref{IabHWijvalue}) and (\ref{IabHWiivalue}).

\chapter{$A^{0} \rightarrow Z^{0} \gamma$ decay}
\typeout{$A^{0} \rightarrow Z^{0} \gamma$ decay}

As an example, we derive the expression for the width decay corresponding 
to the channel $A^{0} \rightarrow Z^{0} \gamma$. The Feynman diagrams are 
given in Figure \ref{FeynmanAZgamma}.

\begin{figure}
\begin{center}  
\scalebox{0.9}
{\includegraphics{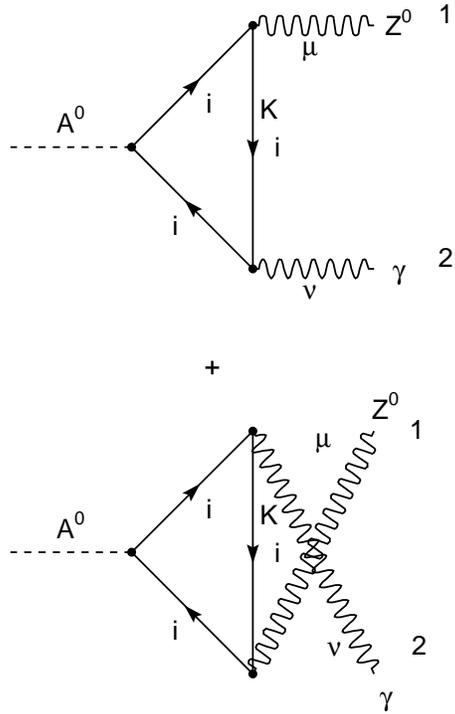}}
\caption{Feynman diagrams corresponding to the decay $A^{0} \rightarrow
Z^{0} \gamma$. $i = d, s, b, e^{-}, \mu^{-}, \tau^{-}$ or $u, c, t$.}
\label{FeynmanAZgamma}
\end{center}    
\end{figure}

\section{$\bf{(a)}$ $i = e^{-}, \mu^{-}, \tau^{-}, d, s, b$}

In the unitary gauge, the invariant amplitude corresponding to the first 
Feynman diagram
of Figure \ref{FeynmanAZgamma} is:

\begin{eqnarray}
\lefteqn{
M_{1a} = - \frac{g^{2}e\tan\beta}{4m_{W}\cos\theta_{W}}
\sum_{i = d, s, b, ..} N_{i} Q_{i} m_{i}
\int\frac{d^{d}\mathrm{K}}{\left(2\pi\right)^{d}} \left(T\right)
}
\nonumber \\ & &
\times \frac{1}{\left(\mathrm{K}^{2} - m_{i}^{2}\right)
\left[\left(\mathrm{K} - \mathrm{P}_{2}\right)^{2} - m_{i}^{2}\right]
\left[\left(\mathrm{K} + \mathrm{P}_{1}\right)^{2} - m_{i}^{2}\right]}
\epsilon_{1\mu}^{*}\epsilon_{2\nu}^{*} \mu^{*}
\label{M1aAZgamma}
\end{eqnarray}

\noindent where
\begin{equation}
T = Trace\left[\gamma^{\mu}\left(g_{V}^{i} - g_{A}^{i}\gamma^{5}\right)
\left(\not{\mathrm{K}} + m_{i}\right)\gamma^{\nu}
\left(\not{\mathrm{K}} - \not{\mathrm{P}}_{2} + 
m_{i}\right)\gamma^{5}
\left(\not{\mathrm{K}} + \not{\mathrm{P}}_{1} + m_{i}\right)\right]
\label{Tr}
\end{equation}

\noindent and  $\mu^{*} = \mu^{(4 - d)/2}$ is a mass parameter. 
$Q_{i}$ and $m_{i}$ are the charge and mass of the 
particle $i$, and $N_{i}$ is a color factor ($N_{i} = 1$ for leptons, 
$N_{i} = 3$ for quarks).
$\mathrm{P}_{1}$ and $\mathrm{P}_{2}$ are the momenta of the
$Z^{0}$ and the photon, respectively. Finally, $\epsilon_{1\mu}^{*}$ and
$\epsilon_{2\nu}^{*}$ are polarization vectors.  

\noindent Using

\begin{equation}
\gamma^{\mu}\gamma^{5} + \gamma^{5}\gamma^{\mu} = 0,
\label{id1}
\end{equation}

\begin{equation}
\gamma^{\mu}\not{\mathrm{K}} + \not{\mathrm{K}}\gamma^{\mu} = 
2\mathrm{K}^{\mu},
\label{id2}
\end{equation}

\begin{equation}
trace(odd \hspace{0.1in}number\hspace{0.1in} of\hspace{0.1in} \gamma^{'}s) 
= 0,
\label{id3}
\end{equation}

\begin{equation}
trace\left(\gamma^{\mu}\gamma^{\nu}\right) = d \eta^{\mu\nu},
\label{id4}
\end{equation}

\begin{equation}
trace\left(\gamma^{\mu}\gamma^{\nu}\gamma^{\rho}\gamma^{\sigma}
\right) = d\left(\eta^{\mu\nu}\eta^{\rho\sigma} -
\eta^{\mu\rho}\eta^{\nu\sigma} + \eta^{\mu\sigma}\eta^{\nu\rho}
\right),
\label{id5}
\end{equation}

\begin{equation}
trace\left(\gamma^{5}\gamma^{\mu}\gamma^{\nu}\gamma^{\alpha}
\gamma^{\beta}\right)
= - 4 i \epsilon^{\mu\nu\alpha\beta},
\label{id6}
\end{equation}

\begin{equation}
trace\left(\gamma^{5}\gamma^{\mu}\gamma^{\nu}\right) = 0,
\label{id7}
\end{equation}

\noindent and

\begin{equation}
\not{\mathrm{K}}\not{\mathrm{K}} = \mathrm{K}^{2},
\label{id8}
\end{equation}

\noindent we can show that
\begin{eqnarray}
\lefteqn{
T = d m_{i} \{ -2g_{A}^{i} P_{1}^{\mu} \mathrm{K}^{\nu}
- \frac{4i}{d} g_{V}^{i}\epsilon^{\mu\nu\alpha\beta}\mathrm{P}_{2\alpha}
\mathrm{P}_{1\beta}
}
\nonumber \\ & &
+ g_{A}^{i}\left[-2 \mathrm{K}^{\nu}\mathrm{P}_{2}^{\mu} 
+2 \left(\mathrm{P}_{2} \cdot \mathrm{K}\right)\eta^{\mu\nu}\right]
- g_{A}^{i}\mathrm{K}^{2}\eta^{\mu\nu}
\nonumber \\ & &
+g_{A}^{i}\left(\eta^{\mu\nu}\left(\mathrm{P}_{1} \cdot \mathrm{P}_{2}
\right) - \mathrm{P}_{2}^{\mu}\mathrm{P}_{1}^{\nu} + 
\mathrm{P}_{1}^{\mu}\mathrm{P}_{2}^{\nu}\right) + m_{i}^{2}
g_{A}^{i}\eta^{\mu\nu}\}
\label{tracevalue}
\end{eqnarray}

\noindent From the identity \cite{Ryder}

\begin{equation}
\frac{1}{abc} = 2 \int_{0}^{1} dx \int_{0}^{1 - x}
\frac{dy}{\left[a \left(1 - x - y\right) + b x + c y\right]^{3}}
\label{feynidentity2}
\end{equation}

\noindent where $a, b, c \neq 0$; and setting:

$a = \left(\mathrm{K}^2 - m_{i}^{2}\right)$,
$b = \left[\left(\mathrm{K} + \mathrm{P}_{1}\right)^{2} 
- m_{i}^{2}\right]$,
$c = \left[\left(\mathrm{K} - \mathrm{P}_{2}\right)^{2}
- m_{i}^{2}\right]$,

\noindent we have
\begin{eqnarray}
\lefteqn{
\frac{1}{ \left(\mathrm{K}^2 - m_{i}^{2}\right)
 \left[\left(\mathrm{K} + \mathrm{P}_{1}\right)^{2}
- m_{i}^{2}\right]\left[\left(\mathrm{K} - 
\mathrm{P}_{2}\right)^{2} - m_{i}^{2}\right]}
}
\nonumber \\ & &
= 2 \int_{0}^{1} dx \int_{0}^{1 - x} \frac{dy}
{\left[\mathrm{K}^{2} - m_{i}^{2} + m_{Z}^{2} x
+ 2 \mathrm{K} \cdot \left( \mathrm{P}_{1} x 
- \mathrm{P}_{2} y \right) \right]^{3}}
\label{feynident2}
\end{eqnarray}

\noindent putting $\mathrm{K}^{'} = \mathrm{K} +
\left(\mathrm{P}_{1} x - \mathrm{P}_{2} y\right)$, and working in the rest 
frame of $A^{0}$, we get

\begin{eqnarray}
\lefteqn{
M_{1a} = - \frac{2 g^{2} e \tan\beta d}{4 m_{W} \cos\theta_{W}}
\sum_{i = d, s, b, ...}N_{i}Q_{i} m_{i}^{2}\int_{0}^{1} dx \int_{0}^{1 - 
x} dy 
}
\nonumber \\ & &
 \int\frac{d^{d} \mathrm{K}^{'}}{\left(2\pi\right)^{d}}
\{- 2g_{A}^{i}\mathrm{P}_{1}^{\mu}\mathrm{K}^{'\nu}
- \frac{4 i}{d}g_{V}^{i}\epsilon^{\mu\nu\alpha\beta}\mathrm{P}_{2\alpha}
\mathrm{P}_{1\beta} - 2g_{A}^{i}\mathrm{P}_{2}^{\mu}\mathrm{K}^{'\nu}
\nonumber \\ & &
+ 2g_{A}^{i}\eta^{\mu\nu}\left(\mathrm{P}_{2} \cdot \left(\mathrm{K}^{'}
- \left(\mathrm{P}_{1}x - \mathrm{P}_{2}y\right)\right)\right)
-g_{A}^{i}\eta^{\mu\nu}\left(\mathrm{K}^{'} - \left(\mathrm{P}_{1}x
- \mathrm{P}_{2}y\right)\right)^2 
\nonumber \\ & &
+ g_{A}^{i}\eta^{\mu\nu}
\left(\mathrm{P}_{1} \cdot \mathrm{P}_{2}\right)
+ m_{i}^{2}g_{A}^{i}\eta^{\mu\nu}\}
\nonumber \\ & &
\times\left[\mathrm{K}^{'2} - m_{i}^{2} + m_{Z}^{2}x - m_{Z}^{2}
x^{2} + 2\left(\mathrm{P}_{1} \cdot \mathrm{P}_{2}\right)xy\right]^{-3}
\epsilon_{1\mu}^{*}\epsilon_{2\nu}^{*} \mu^{*}
\label{M1aK'}
\end{eqnarray}

\noindent where we have used:
$\mathrm{P}_{1}^{\mu}\epsilon_{2\mu}^{*} = 0$,
$\mathrm{P}_{1}^{\mu}\epsilon_{1\mu}^{*} = 0$
and $\mathrm{P}_{2}^{\nu}\epsilon_{2\nu}^{*} = 0$.

\noindent With the integrals \cite{Pokorski}:

\begin{equation}
\int d^{d}\mathrm{K}^{'} \frac{\mathrm{K}^{'\mu}}
{\left[\mathrm{K}^{'2} - m_{i}^{2} + m_{Z}^{2}x\left(1 - x\right)
+ 2\left(\mathrm{P}_{1} \cdot \mathrm{P}_{2} \right) xy\right]^{3}} 
= 0,
\label{intident1}
\end{equation}

\begin{equation}
\int d^{d}\mathrm{K}^{'} \frac{1}
{\left[\mathrm{K}^{'2} - m_{i}^{2} + m_{Z}^{2}x\left(1 - x\right)
+ 2\left(\mathrm{P}_{1} \cdot \mathrm{P}_{2} \right) xy\right]^{3}}
= I_{0}\left(x,y\right),
\label{intident2}
\end{equation}
\noindent and

\begin{eqnarray}
\lefteqn{
\int d^{d}\mathrm{K}^{'} \frac{\mathrm{K}^{'2}}
{\left[\mathrm{K}^{'2} - m_{i}^{2} + m_{Z}^{2}x\left(1 - x\right)
+ 2\left(\mathrm{P}_{1} \cdot \mathrm{P}_{2} \right) xy\right]^{3}}
}
\nonumber \\ & &
= I_{0}\left(x,y\right)\frac{d}{2}\frac{\left[
- m_{i}^{2} + m_{Z}^{2}x\left(1 - x\right) +
2\left(\mathrm{P}_{1} \cdot \mathrm{P}_{2}\right)xy\right]}
{\left(2 - \frac{d}{2}\right)}
\label{intident3}
\end{eqnarray}

\noindent where
\begin{equation}
I_{0}\left(x,y\right) = \frac{i\left(-\pi\right)^{d/2}
\Gamma\left(3 - \frac{d}{2}\right)}{2\left[- m_{i}^{2}
+ m_{Z}^{2}x\left(1 - x\right) + 2\left(\mathrm{P}_{1}
\cdot \mathrm{P}_{2}\right)xy\right]^{3 - d/2}}
\label{I0(x,y)};
\end{equation}
\noindent we get

\begin{eqnarray}
\lefteqn{
M_{1a} = - \frac{2 g^{2} e \tan\beta d}{4 m_{W} \cos\theta_{W}
\left(2\pi\right)^{d}}
\sum_{i = d, s, b, ...}N_{i} Q_{i} m_{i}^{2} \int_{0}^{1} dx 
\int_{0}^{1 - x} dy I_{0}\left(x, y\right)
}
\nonumber \\ & &
\times\left[- \frac{4i}{d}g_{V}^{i} \epsilon^{\mu\nu\alpha\beta}
\mathrm{P}_{2\alpha}\mathrm{P}_{1\beta} - 2 g_{A}^{i} \eta^{\mu\nu}
\left(\mathrm{P}_{1} \cdot \mathrm{P}_{2} \right) x 
+ g_{A}^{i}\eta^{\mu\nu}\left(\mathrm{P}_{1}\cdot \mathrm{P}_{2}\right)
\right.
\nonumber \\ & &
+ m_{i}^{2} g_{A}^{i} \eta^{\mu\nu}
- g_{A}^{i} \eta^{\mu\nu} m_{Z}^{2} x^{2}
+ 2 \left(\mathrm{P}_{1}\cdot \mathrm{P}_{2}\right)xy g_{A}^{i}
\eta^{\mu\nu}
\nonumber \\ & &
\left. - g_{A}^{i} \eta^{\mu\nu}\frac{d}{2}\frac{\left[- m_{i}^{2} + 
m_{Z}^{2}
x\left(1 - x\right) + 2\left(\mathrm{P}_{1}\cdot \mathrm{P}_{2}\right)
xy\right]}{\left(2 - \frac{d}{2}\right)}\right]
\epsilon_{1\mu}^{*}\epsilon_{2\nu}^{*}\mu^{*}
\label{M1afinal}
\end{eqnarray}

The invariant amplitude corresponding to the second Feynman diagram
of Figure  \ref{FeynmanAZgamma} (crossed diagram) is:

\begin{eqnarray}
\lefteqn{
M_{2a} = - \frac{g^{2} e \tan\beta}{4 m_{W} \cos\theta_{W}}
\sum_{i = d, s, b, ...}N_{i}Q_{i}m_{i} \int
\frac{d^{d}\mathrm{K}}{\left(2\pi\right)^{d}}\left(Tra\right)
}
\nonumber \\ & &
\times \frac{1}{\left(\mathrm{K}^{2} - m_{i}^{2}\right)
\left[\left(\mathrm{K} - \mathrm{P}_{1}\right)^{2} - m_{i}^{2}\right]
\left[\left(\mathrm{K} + \mathrm{P}_{2}\right)^{2} - m_{i}^{2}\right]}
\epsilon_{1\mu}^{*}\epsilon_{2\nu}^{*} \mu^{*} 
\label{M2a}
\end{eqnarray}

\noindent where

\begin{eqnarray}
\lefteqn{
Tra = Trace\left[\gamma^{\nu}\left(\not{\mathrm{K}} + m_{i}\right)
\gamma^{\mu}\left(g_{V}^{i} - g_{A}^{i}\gamma^{5}\right)\right.
}
\nonumber \\ & &
\left. \left[\left(\not{\mathrm{K}} - \not{\mathrm{P}}_{1}\right) + m_{i}
\right]\gamma^{5}\left[\left(\not{\mathrm{K}} + \not{\mathrm{P}}_{2}
\right) + m_{i}\right]\right]
\label{Tra}
\end{eqnarray}

\noindent In a similar way, and after performing
the calculation of the trace, we can show that

\begin{eqnarray}
\lefteqn{
M_{2a} = - \frac{2 g^{2} e \tan\beta d}{4 m_{W} \cos\theta_{W}
\left(2\pi\right)^{d}}
\sum_{i = d, s, b, ...}N_{i} Q_{i} m_{i}^{2} \int_{0}^{1} dx
\int_{0}^{1 - x} dy I_{0}\left(x, y\right)
}
\nonumber \\ & & 
\times\left[- \frac{4i}{d}g_{V}^{i} \epsilon^{\nu\mu\alpha\beta}
\mathrm{P}_{1\alpha}\mathrm{P}_{2\beta} + 2 g_{A}^{i} \eta^{\mu\nu} 
\left(\mathrm{P}_{1} \cdot \mathrm{P}_{2} \right) x
- g_{A}^{i}\eta^{\mu\nu}\left(\mathrm{P}_{1}\cdot \mathrm{P}_{2}\right)
\right.
\nonumber \\ & &
- m_{i}^{2} g_{A}^{i} \eta^{\mu\nu}
+ g_{A}^{i} \eta^{\mu\nu} m_{Z}^{2} x^{2}
- 2 \left(\mathrm{P}_{1}\cdot \mathrm{P}_{2}\right)xy g_{A}^{i}
\eta^{\mu\nu}
\nonumber \\ & &
\left. + g_{A}^{i} \eta^{\mu\nu}\frac{d}{2}\frac{\left[- m_{i}^{2} +
m_{Z}^{2}  
x\left(1 - x\right) + 2\left(\mathrm{P}_{1}\cdot \mathrm{P}_{2}\right)
xy\right]}{\left(2 - \frac{d}{2}\right)}\right]
\epsilon_{1\mu}^{*}\epsilon_{2\nu}^{*}\mu^{*}
\label{M2afinal}
\end{eqnarray}

\noindent Adding Equations (\ref{M1afinal}) and (\ref{M2afinal}), we 
obtain

\begin{eqnarray}
\lefteqn{
M_{1a} + M_{2a} = \frac{4ig^{2}e\tan\beta}{m_{W}\cos\theta_{W}
\left(2\pi\right)^{d}}\epsilon^{\mu\nu\alpha\beta}\mathrm{P}_{2\alpha}
\mathrm{P}_{1\beta}
}
\nonumber \\ & &
\sum_{i = d, s, b, ...}N_{i}Q_{i}m_{i}^{2}g_{V}^{i}
\int_{0}^{1} dx \int_{0}^{1 - x} dy I_{0}\left(x, y\right)
\epsilon_{1\mu}^{*}\epsilon_{2\nu}^{*} \mu^{*}
\label{M1a+M2a}
\end{eqnarray}

\noindent In the limit $d \rightarrow 4$, $I_{0}\left(x, y\right)$
is

\begin{equation}
I_{0}\left(x, y\right) = \frac{i\pi^{2}}{2}
\frac{1}{\left[-m_{i}^{2} + m_{Z}^{2} x\left(1 - x\right) +
2\left(\mathrm{P}_{1} \cdot \mathrm{P}_{2}\right) xy\right]}
\label{I0reduced}
\end{equation}

\noindent Introducing
\begin{equation}
I_{i} = \int_{0}^{1} dx \int_{0}^{1 - x}\frac{1}
{\left[-m_{i}^{2} + m_{Z}^{2} x\left(1 - x\right) +
2\left(\mathrm{P}_{1} \cdot \mathrm{P}_{2}\right) xy\right]} dy,
\label{Ii}
\end{equation}

\noindent we can write

\begin{eqnarray}
\lefteqn{
M_{1a} + M_{2a} = \frac{g^{2}e\tan\beta}{8\pi^{2} m_{W} \cos\theta_{W}}
\epsilon^{\mu\nu\alpha\beta}\mathrm{P}_{1\alpha}\mathrm{P}_{2\beta}
}
\nonumber \\ & &
\left(\sum_{i = d, s, b, ...}N_{i}Q_{i}m_{i}^{2}g_{V}^{i}I_{i}\right)
\epsilon_{1\mu}^{*}\epsilon_{2\nu}^{*}
\label{M1aM2a}
\end{eqnarray}

\section{$\bf{b)}$ $i = u, c, t$}
In this case, the invariant amplitude is obtained from Equation
(\ref{M1aM2a}) replacing $\tan\beta$ by $\cot\beta$, and then

\begin{eqnarray}
\lefteqn{
M_{1b} + M_{2b} = \frac{g^{2}e\cot\beta}{8\pi^{2} m_{W} \cos\theta_{W}}
\epsilon^{\mu\nu\alpha\beta}\mathrm{P}_{1\alpha}\mathrm{P}_{2\beta}
}
\nonumber \\ & &
\left(\sum_{i = u, c, t}N_{i}Q_{i}m_{i}^{2}g_{V}^{i}I_{i}\right)
\epsilon_{1\mu}^{*}\epsilon_{2\nu}^{*}
\label{M1bM2b}
\end{eqnarray}

\section{Width decay}

The total invariant amplitude corresponding to the process $A^{0}
\rightarrow Z^{0} \gamma$ is

\begin{equation}
M = M_{1a} + M_{2a} + M_{1b} + M_{2b} = \frac{g^{2} e}{8 \pi^{2} m_{W}
\cos\theta_{W}}\epsilon^{\mu\nu\alpha\beta}\mathrm{P}_{1\alpha}
\mathrm{P}_{2\beta} \epsilon_{1\mu}^{*} \epsilon_{2\nu}^{*}
L(\beta)
\label{Mtotal}
\end{equation}
\noindent where
\begin{equation}
L(\beta) =
\tan\beta\sum_{i = d, s, b, ...}N_{i}Q_{i}m_{i}^{2}g_{V}^{i}I_{i}
+ \cot\beta\sum_{i= u, c, t}N_{i}Q_{i}m_{i}^{2}g_{V}^{i}I_{i}
\label{Lbeta}
\end{equation}

\noindent The absolute value of the invariant amplitude squared
and summed over final polarizations is
\begin{eqnarray}
\lefteqn{
\overline{|M|^{2}} = 
\frac{g^{4}e^{2}}{64\pi^{4}m_{W}^{2}\cos^{2}\theta_{W}}
\epsilon^{\mu\nu\alpha\beta}\epsilon^{\rho\sigma\gamma\delta}
\mathrm{P}_{1\alpha}\mathrm{P}_{2\beta}\mathrm{P}_{1\gamma}
\mathrm{P}_{2\delta}
}
\nonumber \\ & &
\left(\sum_{\lambda}\epsilon_{1\mu}\epsilon_{1\rho}^{*}\right)
\left(\sum_{\lambda^{'}}\epsilon_{2\nu}\epsilon_{2\sigma}^{*}\right)
|L(\beta)|^{2}
\label{Msqav}
\end{eqnarray}

\noindent Since
\begin{equation}
\sum_{\lambda}\epsilon_{1\mu}\epsilon_{1\rho}^{*}
= -\eta_{\mu\rho} + \frac{\mathrm{P}_{1\mu}\mathrm{P}_{1\rho}}
{m_{Z}^{2}},
\label{sumpol1}
\end{equation}
\noindent 
\begin{equation}
\sum_{\lambda^{'}}\epsilon_{2\nu}\epsilon_{2\sigma}^{*}
= - \eta_{\nu\sigma},
\label{sumpol2}
\end{equation}
\noindent and because,
\begin{equation}
\epsilon^{\mu\nu\alpha\beta} \mathrm{P}_{1\alpha}
\mathrm{P}_{1\mu} = 0,
\label{idepsilon}
\end{equation}
\begin{equation}
\epsilon^{\mu\nu\alpha\beta}\epsilon_{\mu\nu}^{\hspace{0.15in}\gamma\delta}
= -2 \left(\eta^{\alpha\gamma}\eta^{\beta\delta} -
\eta^{\alpha\delta}\eta^{\beta\gamma}\right),
\label{idenepsilon}
\end{equation}
\noindent we obtain

\begin{equation}
\overline{|M|^{2}} = \frac{g^{2}e^{4}\left(\mathrm{P}_{1} \cdot
\mathrm{P}_{2}\right)^{2}}{32\pi^{4}m_{W}^{2}\sin^{2}\theta_{W}
\cos^{2}\theta_{W}}|L(\beta)|^{2}
\label{Msquaredaveraged}
\end{equation}

\noindent where we have used $g = e/\sin\theta_{W}$.

The differential decay rate corresponding to the channel
$A^{0} \rightarrow Z^{0} \gamma$ is

\begin{equation}
d\Gamma = \frac{\overline{|M|^{2}} |\vec{\mathrm{P}_{1}}|
d\Omega}{32\pi^{2}m_{A^{0}}^{2}}
\label{difdecayrate}
\end{equation}
\noindent From the kinematics we can show that
\begin{equation}
|\vec{\mathrm{P}_{1}}| = \frac{m_{A^{0}}^{2} - m_{Z}^{2}}{2m_{A^{0}}},
\label{kinem1}
\end{equation}

\begin{equation}
\mathrm{P}_{1} \cdot \mathrm{P}_{2} = \frac{1}{2}
\left(m_{A^{0}}^{2} - m_{Z}^{2}\right),
\label{kinem2}
\end{equation}
\noindent and because,
$\frac{G_{F}}{\sqrt{2}} = \frac{g^{2}}{8m_{W}^{2}}$ and
$\alpha_{em} = \frac{e^{2}}{4\pi}$, we can write

\begin{equation}
\Gamma\left(A^{0} \rightarrow Z^{0} \gamma \right) =
\frac{\sqrt{2}G_{F}\alpha_{em}^{2}m_{A^{0}}^{3}}
{32\pi^{3}\sin^{2}\theta_{W}\cos^{2}\theta_{W}}
\left(1 - \frac{m_{Z}^{2}}{m_{A^{0}}^{2}}\right)^{3}
|L(\beta)|^{2}.
\label{Gamma}
\end{equation}

\noindent With $\left(\mathrm{P}_{1} \cdot \mathrm{P}_{2}\right)$
given by \ref{kinem2}, the value of $I_{i}$ is

\begin{equation}
I_{i} = \frac{1}{4m_{i}^{2}}I\left(\tau_{i}, \lambda_{i}\right)
\label {Iivalue}
\end{equation}

\noindent where 
\begin{equation}
I\left(\tau_{i}, \lambda_{i}\right) \equiv
\frac{\tau_{i}\lambda_{i}}{\left(\lambda_{i} - \tau_{i}\right)}
\left(f\left(\tau_{i}\right) - f\left(\lambda_{i}\right)\right),
\label{Itaulambda}
\end{equation}

\begin{equation}
\tau_{i} = \frac{4 m_{i}^{2}}{m_{A^{0}}^{2}},
\label{taui}
\end{equation}

\begin{equation}
\lambda_{i} = \frac{4 m_{i}^{2}}{m_{Z}^{2}},
\label{lambdai}
\end{equation}

\noindent and

\begin{equation}
f(x) = \left\{ \begin{array}{ll}
-2 \left[ \arcsin \left(x^{-1/2} \right) \right]^2 &
\mbox{if $x > 1$} \\
\frac{1}{2} \left[ \ln \left( \frac{1 + \left( 1- x \right)^{1/2}}
{1 - \left( 1 - x \right)^{1/2}} \right) - i \pi \right]^2 &
\mbox{if $x \le 1$}
\end{array} \right.
\label{f(x)int}
\end{equation}

\noindent (note that for $x \ll 1$, we have
$f\left(x\right) \approx \frac{1}{2}\left[\ln\left(\frac{x}{4}\right)
+ i\pi\right]^{2})$

\noindent Thus, we obtain

\begin{eqnarray}
\lefteqn{
\Gamma\left(A^{0} \rightarrow Z^{0} \gamma\right) =
\frac{\sqrt{2}G_{F}\alpha_{em}^{2} m_{A^{0}}^{3}}{512\pi^{3}
\sin^{2}\theta_{W}\cos^{2}\theta_{W}}\left(1 - \frac{m_{Z}^{2}}
{m_{A^{0}}^{2}}\right)^{3}
}
\nonumber \\ & &
|\tan\beta\sum_{i = d, s, b, e^{-}, ..}N_{i}Q_{i}g_{V}^{i}
I\left(\tau_{i}, \lambda_{i}\right)
\nonumber \\ & &
+ \cot\beta\sum_{i= u, c, t}N_{i}Q_{i}g_{V}^{i}I\left(\tau_{i}, 
\lambda_{i}\right)|^{2}
\label{GammaAZgammafinal}
\end{eqnarray}

\noindent The coefficients $g_{V}^{i}$ are given in table \ref{c}
(Chapter three).

In Equation (\ref{GammaAZgammafinal}), the dominant contributions
correspond to the $\tau^{-}$ and the $b$ and $t$ quarks. Thus,
(\ref{GammaAZgammafinal}) reduces to Equation (\ref{A_Zgamma}).

\end{document}